\documentclass[12pt]{report}

\usepackage{suthesis-2e}

\usepackage{thesis-macros}
\usepackage{longtable}
\usepackage{lscape}
\usepackage{natbib}
\usepackage{graphicx}
\usepackage{epsfig}
\usepackage[breaklinks=true]{hyperref}

\submitdate{June 2007}

\title{Resolving the Formation of Protogalaxies}
\author{John H. Wise}
\dept{Physics} 
\principaladviser{Tom Abel}
\firstreader{Roger D. Blandford}
\secondreader{Robert V. Wagoner}

\begin{document}


\beforepreface
\prefacesection{Abstract}

Cosmic structure originated from minute density perturbations in an
almost homogeneous universe.  The first stars are believed to be very
massive and luminous, providing the first ionizing radiation and heavy
elements to the universe and forming 100 million years after the Big
Bang.  The impact from primordial stellar radiation is far reaching
and affects subsequent star and galaxy formation.  In this thesis, we
present results from adaptive mesh refinement calculations of the
formation of the first galaxies.  We gradually introduce important
physical processes, such as molecular hydrogen cooling and stellar
feedback, to base models that only consider atomic hydrogen and helium
cooling.  In these base models, we find that gas in dark matter halos
with masses $\sim$$10^8 \Ms$ centrally collapse before multiple
fragmentation occurs in a global disc.  We then investigate the
importance of molecular hydrogen cooling in early structure formation
in the presence of a soft ultraviolet radiation background.  We find
that molecular hydrogen plays an important role in star formation in
halos well below a virial temperature of 10,000 K even in the most
extreme assumptions of negative radiative feedback.  We also present
results from the first radiation hydrodynamics calculations of early
dwarf galaxy formation.  We develop a novel technique, adaptive ray
tracing, to accurately transport radiation from primordial stars.  We
find primordial stellar feedback alters the landscape of early galaxy
formation in that its angular momentum is increased and baryon
fractions are decreased.  We also describe the metal enrichment of the
intergalactic medium and early dwarf galaxies.  Finally we explore
cosmological reionization by these massive, metal-free stars and its
effects on star formation in early galaxies.

\prefacesection{Preface}

The purpose of this preface is to provide a synopsis of the original
methods and applications developed by the candidate for the work
presented in this thesis.  I also describe the most important
conclusions of the thesis that have not been demonstrated yet by
numerical simulations.

The first luminous objects are believed to have a profound impact on
its surroundings and future star and galaxy formation.  There have
been numerous numerical studies concerning (i) primordial star
formation, (ii) primordial stellar feedback, and (iii) early galaxy
formation.  However, nobody has been successful in accurately
simulating the evolution of ionized and heated regions around the
first stars until the epoch of galaxy formation.  In this thesis, I
connect these three topics together.  We approach this problem
methodically through an extensive suite of adaptive mesh refinement
simulations that focus on the formation of some of the earliest
galaxies in the universe while including important physical processes,
such as molecular hydrogen cooling and radiation transport.

Chapter 2, ``The Number of Supernovae from Primordial Stars in the
Universe'', describes results from a semi-analytic model of star
formation and feedback that was developed by the candidate.  Although
such models are widespread in the literature, our approach includes a
time-dependent and self-consistent approach to include negative
feedback.  It considers a soft ultraviolet radiation background that
is generated by ongoing star formation, which in turn suppresses
future star formation by requiring a larger dark matter halo mass to
form primordial stars.  The critical halo mass is generally set to be
constant in many semi-analytic models that focus on primordial star
formation and reionization.  This aspect sets our approach apart from
other studies.  This chapter was published in the \textit{The
  Astrophysical Journal}, 2005, Volume 629, 615.

Chapter 3, ``The Virialization of Baryons in Protogalaxies'', shows
that baryons experience an analog of violent relaxation that happens
in the virialization of collisionless dark matter.  Some groups assume
that pre-galactic gas clouds rotate like solid bodies that collapse
into galactic discs.  This assumption dates back to Crampin \& Hoyle
(1964), who showed that the rotation curves of spiral galaxies could
be explained by a uniform rotating spheroid.  We show that
virialization stirs turbulence to become supersonic.  In other words,
turbulent velocities dominate the rotational ones.  The properties of
a turbulent collapse differ greatly from a solid body rotator and
should affect calculations that rely on these idealized initial
conditions.  This chapter has been submitted to \textit{The
  Astrophysical Journal} for publication.

Chapter 4, ``Central Gas Collapse of a Atomic Hydrogen Cooling Halo'',
applies theories of rotational bar-like instabilities to the context
of a collapsing cosmological halo.  The instabilities are secular in
that turbulent viscosity amplifies perturbations created when
rotational energies are large enough.  We were the first to
demonstrate that these instabilities occur in the gaseous collapse of
turbulent clouds in numerical simulations.  I interpreted and
confirmed that our data exhibited such an instability.  These bars
transport angular momentum outward so that only the gas with the
lowest specific angular momentum falls to the center.  This partially
alleviates the difficulties imposed by angular momentum conservation
during a gaseous collapse to a dense central core if one considers the
initial gas cloud as a solid-body rotator.  This scenario is also
applicable to star formation in galactic molecular clouds.

Chapter 5, ``Suppression of H$_2$ Cooling in the UVB'', strengthens
the importance of molecular hydrogen cooling in dark matter halos with
masses well below the ones associated with galaxy formation.  We show
that molecular hydrogen cooling cannot be neglected when studying
structure formation in the early universe.

Chapter 6, ``The Nature of Early Dwarf Galaxies'', and Chapter 7,
``How Massive Metal-Free Stars Start Cosmological Reionization'',
focus on the radiative feedback from primordial stars and its impact
on early galaxy formation and the beginning of cosmological
reionization.  We have developed a novel technique, adaptive ray
tracing (Abel \& Wandelt 2002), to accurately compute radiation
transport from point sources.  Tom Abel originally integrated this
method into a single-processor version of the AMR code \textsl{enzo}.
The radiative transfer is coupled with the hydrodynamics, energy, and
chemistry solvers in \textsl{enzo}.  I then created and tested
routines to automatically create primordial star particles, whose
luminosities and lifetimes are determined from primordial stellar
models, in overdensities that will form such a such as shown by Abel
et al. (2002) and Yoshida et al. (2006b).  The star particles are
treated as point sources of radiation in our adaptive ray tracing
scheme.

More importantly, I developed a massively parallel version of the
adaptive ray tracing method that is functional on distributed and
shared memory machines.  The code is dynamically load-balanced so it
runs efficiently on many processors and minimizes inter-processor
communication of the ``photon packages''.  It is necessary to optimize
this method because approximately $10^6$ photons are traced through
$\sim$$10^7$ grid cells per timestep per point source.  Currently the
radiation transport consumes the same amount of computing time as the
hydrodynamics solver and increases the run time by approximately 50\%.
This additional time is a small cost to pay compared to the benefits
of the accurate depiction of radiative feedback in our simulations.
This code is one of two cosmology radiation hydrodynamics codes that
is parallelized and optimized for radiative feedback studies with
enough resolution to resolve Pop III hosting halos.  This is the first
time that a radiation hydrodynamics code has self-consistently
followed the formation of tens of primordial stars and their
associated \ion{H}{2} regions to the formation of an early dwarf
galaxy.

We first demonstrated our code in the article ``The \ion{H}{2} Region
of a Primordial Star'' that was authored by Tom Abel, John Wise, and
Greg Bryan in \textit{The Astrophysical Journal: Letters}, 2007,
Volume 659, L87.  Before the publication of our results, I thoroughly
tested the radiative transfer against analytical models of \ion{H}{2}
regions that propagate through, e.g., a uniform medium or a radially
decreasing density profile.  I further tested the star formation and
feedback algorithm with cosmological initial conditions.  For this
publication, I conducted the simulation and analysis in its entirety.

Additionally, I am involved in a collaboration with Ralf K{\"a}hler,
developing a photo-realistic volume renderer on graphics hardware for
astrophysical applications.  I created an accurate colormap generator
that is based on the blackbody spectrum of the temperature of the
rendered gas, which is then convolved with telescopic filters.  This
method is generalized so that any type of intrinsic spectrum,
e.g. H$\alpha$, [\ion{O}{3}], can be passed through an arbitrary set
of filters.  This effectively mimics an actual ``observation'' of our
simulation data.  It provides a good physical basis and standard for
imaging numerical datasets instead of using an arbitrary colormap.
Data can be easily compared between groups with this method.  This
work was published in the refereed article, ``GPU-Assisted Raycasting
of Cosmological Adaptive Mesh Refinement Simulations'' by Ralf
K{\"a}hler, John Wise, Tom Abel, and Hans-Christian Hege in the
\textit{Proceedings of Volume Graphics 2006}, in which I authored the
sections that describe the astrophysical basis and possible
applications of our method.


\prefacesection{Acknowledgements}

First and foremost, I would like to thank my adviser, Tom Abel, for
his guidance throughout my graduate career.  His enthusiasm for
astrophysics is contagious, and I have profited from the positive
environment that he creates.  The research that composes this thesis
would not have been possible without him.  I thank my thesis committee,
Roger Blandford, Bradley Efron, Robert Wagoner, and Risa Wechsler, for
their constructive criticism and involvement in this thesis.

This thesis has also benefited from many useful discussions with my
collaborators and scientific peers.  I thank Marcelo Alvarez, Greg
Bryan, Renyue Cen, Andr{\'e}s Escala, Ralf K{\"a}hler, Richard Klein,
Michael Kuhlen, Paul Kunz, Michael Norman, Brian O'Shea, Jeremiah
Ostriker, Ralph Pudritz, Darren Reed, Martin Rees, Tom Theuns, Peng
Wang, and Naoki Yoshida.  The simulations in this thesis were executed
mainly on the computers at SLAC for which I appreciate the continuous
support from the Scientific Computing and Computing Services (SCCS)
group and Stuart Marshall.  Furthermore, I could not have made it
through these six years without some entertainment outside the office.
In my first three years in graduate school at the Pennsylvania State
University, I want to thank in particular Simos Konstantinidis, Bret
Lehmer, Suvrath Mahadevan, Avi Mandell, Miroslav Mi{\'c}i{\'c}, Nikola
Milutinovi{\'c}, Kenneth Moody, Dave Morris, Manodeep Sinha, Michael
Sipior, Britton Smith, Junfeng Wang, and the whole international crew.
At Stanford, the people I thank are Teddy Cheung, Fabio Iocco, Ji-hoon
Kim, Kevin Schlaufman, Matthew Turk, and Fen Zhao.  I have also been
lucky to live close to fantastic mountain roads in both Central PA and
the Bay Area on which I have enjoyed ``spirited'' driving in my rotary
sports cars.

I would like to thank my parents for supporting me in whatever I
please to do and encouraging me to always to try my best and to never
give up.  Even though they can't read this, the attention and
compassion that my two cats, Jeremy and Sampson, give me after
everyday helped me make it through graduate school.  Most importantly,
I thank Emily Alicea-Mu{\~n}oz for being with me and loving me for the
last few years and for making me smile and laugh in even the toughest
moments and hopefully many more to come.

\afterpreface


\chapter{Introduction}

\begin{quote}
  \textit{``When judging a physical theory, I ask myself whether I
    would have made the Universe in that way had I been God''} \\ ---
  Albert Einstein
\end{quote}

\begin{quote}
  \textit{Pulchritudo splendor veritatis.} (Beauty is the splendor of
    truth.)
\end{quote}


It is human nature to be inquisitive of nature, which is the basis of
science in general.  Countless people have asked the long-standing
question, \textit{what is the origin of our world?}  Philosophers
dating back to ancient times have pondered this exact question in
detail.  As evidence grew that the universe is not geocentric, the
scope of this stimulating question then grew to the origin of the
Milky Way, which is ultimately influenced by the nature of the
universe.  This progression to the largest scales leads us to the
topic of cosmology, the study of the universe in its entirety.

\section{The Foundation of Galaxy Formation Theories}


Galaxy formation plays an important role in any cosmological theory
that must explain the observed dynamics and structures in the local
and distant universe.  We begin with a historical overview\footnote{The
  majority of the references in this overview originate from
  \citet{Jones76}.} of cosmological, in particular galaxy formation,
theories because it is beneficial to understand how the succession of
scientific advancements influences future scientists, which clarifies
its current status.

There were two main documented views of cosmology in ancient Greece.
Even two thousand and three hundred years later, the following
philosophies have modern counterparts.  The Epicurean view entailed
the universe beginning chaotic from which order arose by some means,
which parallels with the cosmic turbulence theory of galaxy formation.
Aristotle pictured the universe as initially uniform and orderly, and
departures from this state increase with time.  This philosophy is
strikingly similar to the gravitational instability theory of galaxy
formation.

Perhaps the birth of the study of galaxies is the discovery of spiral
structures in nebulae by \citet{Rosse1850a, Rosse1850b}.  At the time,
all nebulae were believed to exist within the galaxy, so this
instigated further work on the origins of the solar system.
\citet{Alexander1852} realized that this observation was evidence for
\citeauthor{Laplace1799}'s nebulae hypothesis
\citeyearpar{Laplace1799} that envisaged the Sun forming from a hot,
tenuous gas cloud that is nearly spherical and rotating like a solid
body \citep[see][for a review]{Aitken1906}.  Self-gravity causes a
gravitational contraction that continues until the system is
rotationally supported at its surface.  He postulated that an
equatorial ring was left behind at this stage, and the collapse
continues.  This repeats as neccessary to create several rings that
are not in dynamical equilibrium, and individual rings then aggregrate
into planets.  \citeauthor{Alexander1852} also suggested that the
initial angular momentum may differentiate ``elliptical nebulae'' and
``spiral nebulae''; unfortunately, this did not spark any further
investigations.

The next spurring moment in the history of galaxy formation models was
the publication of the first photographs of galaxies
\citep{Roberts1889, Keeler1900} that incited many theories about the
formation of the observed structures \citep[e.g.][]{Chamberlain1901,
  Jeans1902, Sutherland1911}.  The nebulae hypothesis was the main
underlying of these works.  This is where \citeauthor{Jeans1902}
formulated the conditions necessary for a cloud to gravitationally
overcome pressure forces, i.e. the Jeans length.  Here he invoked his
theory to suggest that the gas cloud fragments into planets.

The last foundation established for galaxy formation was the actual
discovery by Hubble of the extragalactic nature of these spiral
nebulae \citep{Hubble1925a, Hubble1925b, Hubble1925c}.  Now previous
theories had to be reformulated to length scales five orders of
magnitude larger and applied to billions of stars instead of one.
\citet{Jeans1918, Jeans1928} now applied his theory of gravitational
collapse to cosmological scales, theorizing that galaxies formed from
gravitational instabilities in an initially uniform universe.  This is
the main foundation for the gravitational instability picture of
galaxy formation that has dominated modern cosmology theories.
\citet{Hubble1929} then discovered that more distant galaxies were
receding faster than closer ones -- indicative of an expanding
universe.  Although Jeans did not consider this scenario in his later
works, \citet{Gamow1939}, who envisioned the universe as an enormous
primeval gas cloud, generalized the collapse problem to an expanding
universe, where material fragmented into galaxies that is the basis of
modern galaxy formation theories.

\section{Galaxy Formation prior to Dark Matter Cosmogonies}


The gravitational instability picture of galaxy formation was
established by \citet{Jeans1918, Jeans1928}, in which galaxies grow
from initial density perturbations in the early universe.  This line
of research has dominated the field for the last sixty years.

To illustrate Jeans' theory, consider an isolated, uniform gas cloud
within a gravitational potential.  The potential can be either be
generated by the gas itself or some other concentric system (e.g. a
dark matter halo).  Jeans postulated for the system to collapse, its
gravitational forces must overcome pressure forces.  For this
condition to be valid, we use the virial theorem of the system,
neglecting any bulk motions or magnetic fields,
\begin{eqnarray}
\langle E \rangle_{\rm{grav}} &>& -2 \langle E \rangle_{\rm{kin}} \nonumber\\
\frac{3}{5} \; \frac{GM^2}{R} &>& 3 NkT
\end{eqnarray}
where $G$ and $k$ are the gravitational and Boltzmann's constant,
respectively.  Here $M$, $R$, $N = M/\mu m_{\rm{H}}$, and $T$ are the
mass, radius, total particle number, and temperature of the system,
where $m_{\rm{H}}$ and $\mu$ is the mass of a hydrogen atom and mean
molecular weight in units of $m_{\rm{H}}$, respectively.  After some
simple algebra and putting the mass in terms of radius and density
$\rho$, we arrive at the length scale,
\begin{equation}
  \label{eqn:JeansLength}
  L_J = \sqrt{\frac{15kT}{4\pi G \mu m_{\rm{H}} \rho}},
\end{equation}
the Jeans' length, that a system becomes unstable to collapse.
Similarly the Jeans' mass
\begin{equation}
M_J = \frac{4\pi}{3} \rho L_J^3 \propto T^{3/2} \rho^{-1/2}
\end{equation}
is the critical mass of a system that gravitationally collapses.  For
collisionless dark matter, the temperature can be replaced by a term
proportional to the velocity dispersion squared, $v^2$.

Before further focusing on this preferred theory, we note that there
was the parallel effort of cosmic turbulence that attempted to explain
galaxy formation from an initial turbulent medium.  It was applied to
the formation of the solar system, galactic dynamics, and galaxy
formation \citep{vonWeizsacker43, vonWeizsacker48, Heisenberg48}.
However, there were some concerns expressed by \citet{Gamow52} and
\citet{Bonnor56} about the necessity of ab initio turbulence.  The
cosmic turbulence theory model dropped out of favor in the 1950's for
the gravitational instability theory but had a brief resurgence a
decade later by the Ozernoi group \citep[e.g.][]{Ozernoi68a,
  Ozernoi68b, Ozernoi71} after the discovery of the cosmic microwave
background (CMB).

Now we focus back on the gravitational instability theory.  There lies
a difficulty in this theory in that there is no exact solution to this
cosmological problem if there are no special symmetries.  Nevertheless
at early times in the universe, the growth of structure can be
computed using linear perturbation theory that evolves the initial
density fluctuations until they reach overdensities, $\delta \equiv
\rho/\bar{\rho}$, approach unity, which we explore next.

\subsection{Growth and Dampening of Perturbations}

The growth of adiabatic perturbations in an expanding universe was
first calculated by \citet{Lifschitz46}.  Two decades later, the CMB
was discovered and measured to be $\sim$3.5 K \citep{Penzias65}.  This
background was predicted in the seminal papers of \citet{Gamow46} and
\citet{Alpher48}, which outlined a cosmogony starting with a hot Big
Bang.  After this discovery, galaxy formation models and the
associated growth of perturbations were formulated in this new
cosmological model \citep[][being the first of many]{Peebles65}.

The growth of structure can be understood by three scales.  Firstly,
the horizon length scale ($R_{\rm{H}} = ct$) determines the amount of
matter 
\begin{equation}
  \label{eqn:horizonMass}
  M_{\rm{H}} = \frac{1}{6} \; \pi \rho (ct)^3
\end{equation}
that is causally connected.  Secondly, the Jeans length scale
(eq. \ref{eqn:JeansLength}) determines the scales that can
gravitationally collapse.  Lastly at $l \ll L_J$, perturbations are
subjected to dissipative forces, such as Thomson drag forces and
viscosity, and act as acoustic modes.

Ignoring any inhomogeneities on small scales, Einstein's general
relativistic equations can describe the dynamics of an isotropic and
homogeneous universe, using the Robertson-Walker metric,
\begin{equation}
  \label{eqn:friedmann}
  ds^2 = dt^2 - a^2(t) \left[ \frac{dr^2}{1 - kr^2} + r^2 (d\theta^2 +
    \sin^2 \theta d\phi^2) \right],
\end{equation}
where $a = (1+z)^{-1}$ is the scale factor that describes the
expansion of the universe, $k$ describes the curvature of the universe,
and $z$ is the redshift.  Appropriate choices for $k$ are 0, --1, and
+1 for a flat, open, and closed universes.  The solution to Einstein's
equations with this metric is the Friedmann equation
\begin{equation}
  \label{eqn:friedmannSoln}
  \frac{k}{a_0^2} = \frac{8\pi G}{3}\rho_0 - H_0^2,
\end{equation}
where the ``0'' subscripts denotes the present-day values, and $H =
\dot{a}/a$ is the Hubble constant.  We can now apply this to the
growth of structure.

Before recombination and at sufficiently early times, an adiabatic
perturbation will be larger than the horizon scale.  The growth rate
in this regime can be calculated by comparing a perturbed spherical
region to the background universe.  In this exercise, the background
is a spatially flat Friedmann universe, i.e. $k$ = 0, with a density
$\rho_a$.  The perturbation can be treated like a closed $k$ = 1
universe because the metric of a $k$ = 1 universe and a 3-sphere
embedded in an abstract 4-dimensional Euclidean space are exactly
equal.  Its density $\rho_b$ is perturbed with a small and positive
displacement $\delta\rho$ from $\rho_a$.  From equation
(\ref{eqn:friedmannSoln}), we have
\begin{equation}
  \label{eqn:perturb1}
  H_a^2 = \frac{8\pi G}{3} \rho_a; \quad
  H_b^2 + \frac{1}{a_b^2} = \frac{8\pi G}{3} \rho_b .
\end{equation}
To calculate the growth of the perturbation $\delta\rho$, we compare
the perturbation to its background when the expansion rates are equal,
$H_a = H_b$ at a given time.  This results in the relation
\begin{equation}
  \label{eqn:perturb2}
  \left(\frac{\rho_b - \rho_a}{\rho_a}\right) =
  \frac{\delta\rho}{\rho_a} =
  \frac{3}{8\pi G \rho_a a_b^2}
\end{equation}
We want the evolution of $\delta\rho$ in terms of $a$.  We therefore
state the density of the universe in terms of the scale factor.  In a
radiation dominated universe, $\rho \propto a^{-4}$, and in a matter
dominated universe, $\rho \propto a^{-3}$.  Combining these
proportionalities with equation (\ref{eqn:perturb2}), we obtain
\begin{equation}
  \label{eqn:growth1}
  \frac{\delta\rho}{\rho} \propto \left\{
    \begin{array}{l@{\quad}l}
      a^2 & \mathrm{(radiation \; dominated)} \\
      a   & \mathrm{(matter \; dominated)}
    \end{array}
  \right. .
\end{equation}

Once the density perturbation enters the horizon scale, it can be
suppressed by two processes.  The first suppresses it by pressure
support that was discussed previously.  The system is pressure
supported when $t_{\rm{pressure}} > t_{\rm{dyn}}$ or equivalently
\begin{equation}
  \label{eqn:pressureSupport}
  \frac{L_J}{c_s} < \sqrt{\frac{1}{G\rho_{\rm{dom}}}}.
\end{equation}
Here the ``dom'' subscript denotes the dominant mass-energy component
of the universe at a given scale factor, and $c_s$ is the sound speed.
If matter is non-relativistic, the velocity $v$ decays as $a^{-1}$
because of the redshifting of momentum; otherwise it is constant.  The
second process suppresses the growth when the expansion timescale,
$t_{\rm{exp}} \sim (G\rho_{\rm{dom}})^{-1/2}$, is less than the
collapse timescale, which is particularly true in the radiation
dominated phase.

\begin{figure}[t]
  \begin{center}
    \vspace{0.25cm}
    \includegraphics[width=0.6\textwidth]{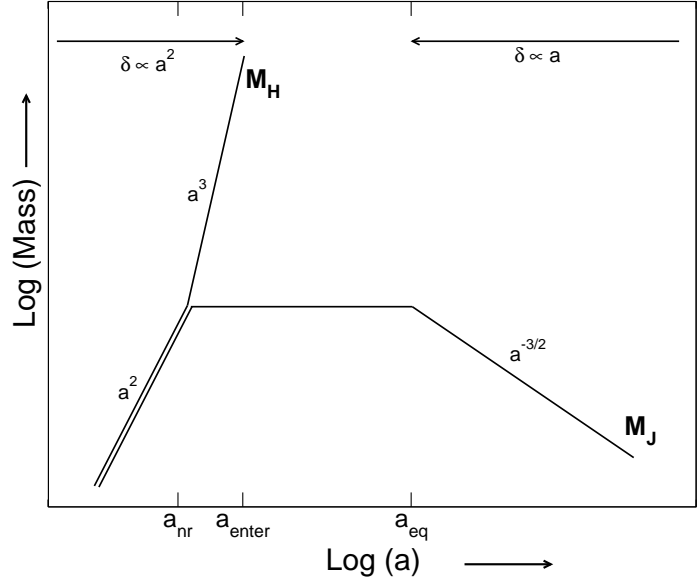}
    \caption[Density Perturbation Evolution]{\label{fig:growth} The
      evolution of a density perturbation in the early universe.  The
      scale factors $a_{\rm{nr}}$, $a_{\rm{enter}}$, and $a_{\rm{eq}}$
      denote when matter becomes non-relativistic, enters the horizon,
      and decouples from radiation.  This figure has been adapted from
      figure 4.1 in \citet{Padmanabhan93}.}
  \end{center}
\end{figure}

Now we consider the case when a perturbation enters the horizon scale
when the background is relativistic and radiation dominated.  The
Jeans length is first affected by the momentum redshifting when matter
is relativistic, and afterwards it follows the growth in a radiation
dominated universe, $\rho \propto a^{-4}$.  When it enters the matter
dominated phase, $\rho \propto a^{-3}$.  Putting this all together, we
have
\begin{equation}
  \label{eqn:JeansEvo}
  L_J \propto \frac{v}{\rho_{\rm{dom}}^{1/2}} \propto \left\{
    \begin{array}{l@{\quad}l}
      a^2     & \rm{(relativistic)} \\
      a       & \rm{(radiation \; dominated)} \\
      a^{1/2} & \rm{(matter \; dominated)}
    \end{array}
  \right. .
\end{equation}
Analogously the Jeans mass scales as
\begin{equation}
  \label{eqn:JeansMassEvo}
  M_J \propto \rho_{\rm{DM}} L_J^3 \propto \left\{
    \begin{array}{l@{\quad}l}
      a^2           & \rm{(relativistic)} \\
      \rm{constant} & \rm{(radiation \; dominated)} \\
      a^{-3/2}       & \rm{(matter \; dominated)}
    \end{array}
  \right. .
\end{equation}
Figure \ref{fig:growth} shows the evolution of $M_J$ and $M_H$.

Baryons behave similarly to dark matter with the major exception of a
significant drop in $L_J$ at recombination at $z \sim 1100$ when the
temperature drops to 3000 K in approximately 20\% of the Hubble time.
After recombination, baryons are regulated by gas pressure instead of
photon pressure.  This causes the pressure to drop by a factor
$n_\gamma/n_b \sim 10^9$, and the Jeans length decreases by a similar
factor.

Density perturbations come in two major types: adiabatic and
isothermal.  The former type retain a particular entropy and behave
like sound waves with equal photon and baryon fluctuations.  These
perturbations do not grow before recombination.  Additionally they are
damped below a mass scale $\sim$$10^{12} \Ms$ because of Thomson drag
forces, i.e. photon viscosity \citep{Silk74}.  The latter type are
density fluctuations immersed in a photon bath and do not conserve
entropy.  They cannot grow in amplitude before recombination because
they are coupled with radiation.  After recombination, the Jeans mass
is $\sim$$10^5 \Ms$ in a Einstein-deSitter (EdS) universe.  Both types
of fluctuations result in a smaller mass scales becoming non-linear
earlier, which is represented by the baryon fluctuation spectrum
\begin{equation}
  \label{eqn:growth2}
  \left(\frac{\delta\rho}{\rho}\right)_{\rm{baryon}} \propto 
  M^{-1/3 - n/6}.
\end{equation}
But as stated before, mass scales below $10^{12} \Ms$ are suppressed
in adiabatic fluctuations.

The physical picture painted by these two types of baryon fluctuations
differ greatly in nature.  Adiabatic ones create a scenario of
top-down structure formation, in which objects that are large as
galaxy clusters form first then fragment to form galaxies through
Jeans instabilities in ``pancake'' overdensities
\citep{Doroshkevich74}.  Isothermal fluctuations produce a bottom-up
scenario of structure formation, where smaller objects form first and
combine to form larger entities \citep{Peebles74}.  An analysis of the
covariance function of galaxies provides evidence for the latter
scenario because there is no preferred mass scale \citep{Totsuji69,
  Peebles74}.  Conversely, one would expect a strong feature at
$\sim$$10^{12} \Ms$ for a $\Omega = 1$ universe if structure had
formed top-down.  Furthermore, the three-point correlation function of
galaxies indicates that they are hierarchically clustered
\citep{Peebles75}.

\begin{figure}[t]
  \begin{center}
    \vspace{0.25cm}
    \includegraphics[width=0.6\textwidth]{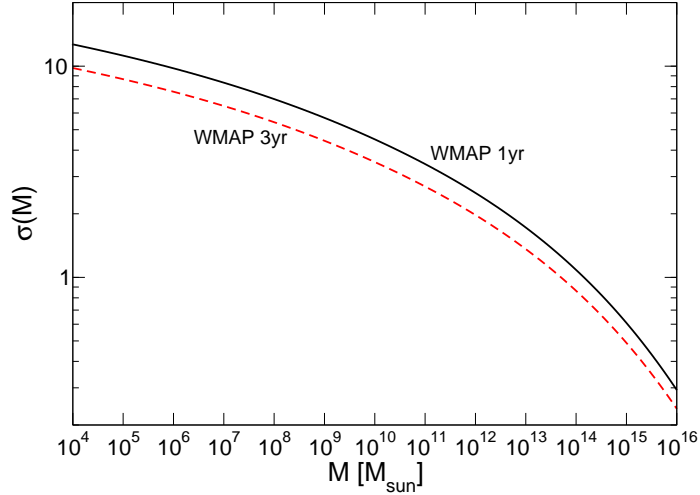}
    \caption[Variance of Density Fluctuations]{\label{fig:sigmaM} The
      variance of density fluctuations as a function of smoothing
      mass for first (\textit{black solid}) and third (\textit{red
        dashed}) year results from WMAP.}
  \end{center}
\end{figure}

Density perturbations may exist on all scales and follow a random
Gaussian distribution; therefore it is useful to consider the
statistics of a smoothed density field.  The power spectrum of
perturbations $P(k)$ that is extrapolated to $z = 0$ is smoothed with
a top-hat filter, whose Fourier transform is
\begin{equation}
  \label{eqn:tophat}
  \tilde{W}(x) = \frac{3(\sin x - x \cos x)}{x^3}.
\end{equation}
The resulting variance of the fluctuations at $z = 0$ is
\begin{equation}
  \label{eqn:sigma}
  \sigma^2(R) = \frac{1}{2\pi^2} \int^\infty_0 dk \, k^2 P(k)
  \tilde{W}^2(kR) .
\end{equation}
This is usually normalized to the variance of fluctuations $\sigma_8$
in a sphere with a radius 8 $h^{-1}$ Mpc that is derived from galaxy
clusters and the CMB.  Figure \ref{fig:sigmaM} shows $\sigma(M)$ for
cosmological parameters determined from the first and third year CMB
data of the Wilkinson Microwave Anisotropy Probe
\citep[WMAP;][]{Spergel03, Spergel06}.  In linear perturbation theory,
a top-hat collapses to a point when the overdensity $\delta_c$ = 1.686
\citep{Peebles93} that can be extrapolated to redshift $z$ by
$\delta_{\rm{crit}}(z) = \delta_c / D(z)$, where $D(z)$ is the growth
factor for linear perturbations.  It is customary to express the
rarity of a collapse with $\delta_{\rm{crit}}(z)$ in units of
$\sigma(M)$, i.e. a 3$\sigma$ fluctuation being more rare than a
typical 1$\sigma$ one.

\subsection{Tidal Torque Theory}

A previously outstanding problem in structure formation was the
acquistion of angular momentum that results in the universal rotation
in spiral galaxies.  How did it arise from an expanding, almost
uniform, universe in a way that angular momentum is conserved?
Kelvin's circulation theorem states that an irrotational system must
remain irrotational in the absence of dissipative forces, which is
true when structure is growing linearly.  During the collapse of
protogalactic gas clouds, there are certainly dissipative forces and
thus Kelvin's theorem is inapplicable.  This is also true in relaxing
stellar systems \citep[i.e. violent relaxation;][]{LB67}.

Angular momentum must exist before galaxies formed.  One might
conclude that there was an original vorticity field in the universe,
but \citet{Hoyle49} and \citet{Peebles69} independently suggested that
tidal torques from neighboring conglomerations induced galactic spins.
Within the gravitational instability theory, tidal torques have its
greatest influence on protogalactic clouds when they decouple from the
Hubble flow (``turn-around'').  When the overdensity turns around, the
radius is largest and $\delta\rho/\rho = 1$.  The tidal torques do not
violate angular momentum conservation as the two systems gain orbital
energy that is equal and opposite of the induced spins.  We note,
however, that both early analytic and numerical work by
\citet{Peebles69, Peebles71} and \citet{Efstathiou79} showed that
there is a deficiency of a factor of 5 in the predicted amount of
angular momentum in the Milky Way when compared with the observation
of \citet{Innanen66}.

\citet{Peebles69} predicted that the spin parameter,
\begin{equation}
  \label{eqn:spinParameter}
  \lambda \equiv \frac{\vert L \vert \sqrt{\vert E \vert}} {G M^{5/2}},
\end{equation}
which measures the net rotation of a halo, has a cosmological average
of $\sim$0.08.  He verified this through numerical simulations of a
cosmologically expanding system.  Here L, E, and M are the angular
momentum, energy, and mass of the object.  During the clustering of
systems, most of the angular momentum is acquired during its early
assembly.  Linear perturbation theory predicts that this angular
momentum scales as $a^{5/2} \propto t^{5/3}$.  \citet{Efstathiou79}
confirmed the results of \citeauthor{Peebles69} with higher resolution
$n$-body simulations and also concluded that galactic discs must have
dissipated a large amount of energy during the collapse from a large
initial radius.  More recent $n$-body simulations and analytical works
\citep[e.g.][]{Barnes87, Steinmetz95} have constrained $\langle
\lambda \rangle \approx 0.04$, and later it was found to have a
log-normal distribution
\begin{equation}
  \label{eqn:lambda_prob}
  p(\lambda)d\lambda = \frac{1}{\sigma_\lambda \sqrt{2\pi}} \exp
  \left[ -\frac{\ln^2 (\lambda/\lambda_0)}{2\sigma_\lambda} \right]
  \frac{d\lambda}{\lambda},
\end{equation}
where $\lambda_0 = 0.042 \; \pm \; 0.006$ and $\sigma_\lambda = 0.5 \;
\pm \; 0.04$ \citep[e.g.][]{Bullock01} in a CDM universe that has
become the currently favored model.  We discuss the nature of galaxy
formation in such a universe next.

\section{Galaxy Formation with Cold Dark Matter}

By the 1980's, there was growing evidence that the universe was
dominated by cold dark matter (CDM) that hierarchically assembles
cosmic structure.  CDM is collisionless and has a thermal velocity
smaller than peculiar velocities and the Hubble flow
\citep[see][]{Blumenthal84}.  The best candidate for CDM are
weakly-interacting massive particles (WIMPs)\footnote{For a recent
  review of detection techniques, see \citet{Wai07}}, whose radiation
from annihilation may be detected by the \textit{Gamma-ray Large-Area
  Space Telescope} \citep[GLAST;][]{Gehrels99}.

Previously popular theories of dark matter include hot DM that is
composed of light neutrinos ($m \sim 30$ eV) and warm DM that is
composed of more massive ($m \sim 1$ keV) neutrinos.

\subsection{Observational Motivation and Evidence}

The ultimate goal of galaxy formation theories is to explain the origin
and properties of all galaxies.  It is thus necessary to connect and
test theories with observations.  \citet{Jones76} gives a list of five
broad characteristics of galaxies that can be tested against, which is
still relevant 30 years later.
\begin{enumerate}
\item The masses of galaxies, galactic internal dynamics, interacting
  galaxies, dynamics of groups, aggregation of galaxy clusters, galaxy
  luminosity function.
\item The angular momenta and binding energies of galaxies
\item The origin of different morphologies of galaxies
\item The clustering of galaxies
\item An explanation of how young galaxies, if not all galaxies, form
  early in the universe.
\end{enumerate}

\subsubsection{Dark Matter}

\citet{Zwicky1933} observed that the galaxies in the Coma cluster were
moving faster than the velocity dispersion of a system, $\sigma^2 =
GM/R$, with a mass of the observed material.  This prompted the
``missing mass'' problem in galaxy clusters.  Further evidence
accumulated when the rotation velocity of M31 was observed to be
constant with radius out to 24 kpc \citep{Rubin70}.  Observations
established that there was a dark component to the universe, whereas
its mass and nature can be constrained by comparing observations with
galaxy formation models.

Mass scales below $10^{11}$ and $10^{15}$ solar masses are suppressed
in warm and hot DM cosmologies, respectively, because the neutrinos
freely stream from the perturbation when it enters the horizon.  To
explain the multitude of galaxies below these scales, structure must
have formed ``top-down'' as in the case of adiabatic perturbations.
Hot DM models predict that galaxy clusters form first at $z \sim 2$
with smaller objects fragmenting from them, conflicting with the ages
of globular clusters, stellar ages in galaxies, and dwarf galaxies.
Warm DM models are slightly more successful in this respect but still
cannot account for the properties and abundances of dwarf galaxies.
In contrast, the ``bottom-up'' structure formation model of CDM
explains the properties of galaxies and galaxy clusters over a mass
range from $10^7$ to $10^{15} \Ms$.

\subsubsection{Galaxies}

\citet{Hubble1926} first categorized the morphologies of galaxies into
elliptical, spiral, S0, and irregular types.  He also realized that
the surface brightness of elliptical galaxies is approximated well by
a decreasing function, $I(r) = I_0 / (1 + r/r_0)^2$, where $r_0$ is a
scale length.  \citet{deVaucouleurs59} provided another good fit with
an exponentially decreasing function that is proportional to
$\exp(r^{1/4})$.  However, these fits deviated in the centers of
galaxies, and the surface brightness was dynamically modeled by
\citet{King66}, who considered a Maxwellian velocity distribution of
the stars and a tidal cutoff radius.  Combining all of the previous
works, \citet{Faber76} then showed that the luminosity of elliptical
galaxies $L \propto \sigma^4$, now referred to as the Faber-Jackson
relation.  A similar relation for spiral and S0 galaxies was
established by \citet{Tully77}, $L \propto V_c^4$, where $V_c$ is the
rotational velocity of the disk.

However, there are peculiar outliers in galaxy catalogs that do not
fit with any of the above categorizations.  There are approximately 10
galaxies out of $\sim$4000 NGC galaxies that are undergoing mergers or
close encounters.  \citet{Toomre72} proposed that there had been a
total of $\sim$500 collisions in the NGC sample throughout the history
of the unvierse, based on their numerical simulations of spiral galaxy
mergers that resulted in a elliptical object.  From this, they
suggested that this could explain the $\sim$400 ellipticals in the
catalog.  In the next section, we will discuss hierarchical clustering
that induces these mergers.

We can also use the abundances and clustering of galaxies to refine
and develop galaxy formation models.  For the former, the abundances
of galaxies can be characterized with a luminosity function in the
form,
\begin{equation}
  \Phi(L)dL = \Phi_* \left(\frac{L}{L_*}\right)^\alpha 
  \exp\left(-\frac{L}{L_*}\right) \frac{dL}{L_*},
\end{equation}
where $L$ is the luminosity of the galaxy, $L_* \approx 4 \times
10^{10} L_\odot$, $\alpha \approx -1$, and $\Phi_*$ is the number of
galaxies per Mpc$^{-3}$ at the characteristic luminosity
\citep{Schechter76}.  This predicts that low-luminosity galaxies are
extremely abundant with an exponential cutoff at high luminosities.
This was first formulated for galaxies in rich clusters, but later it
has been confirmed to apply to all galaxies \citep[e.g.][]{Moore93,
  Benson03}.  The clustering of galaxies are explained by hierarchical
assembly that is the next topic in the introduction.

\subsection{Hierarchical Structure Formation}

\begin{figure}[t]
  \begin{center}
    \vspace{0.25cm}
    \includegraphics[width=0.6\textwidth]{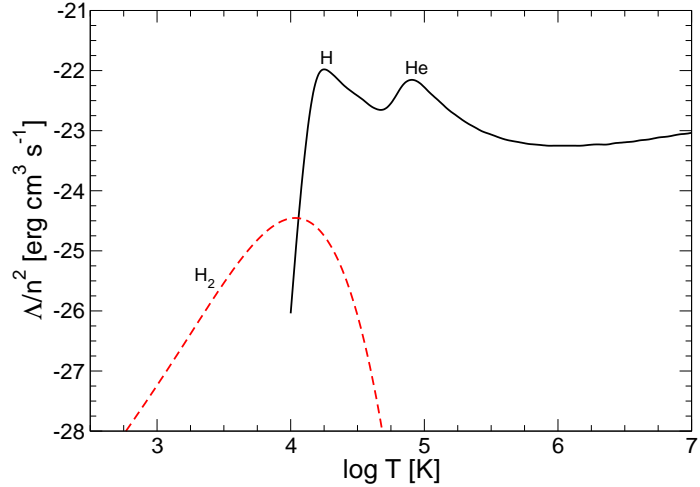}
    \caption[Cooling Function for Metal-free Gas]
    {\label{fig:coolingFn} The cooling function for atomic line
      cooling of metal-free gas (\textit{black solid}) and cooling
      from \hh~formation in the gas phase (\textit{red dashed}).  The
      humps in the atomic cooling function at $10^{4.3}$ K and $10^5$
      K are created by excitations of hydrogen and helium,
      respectively.  The cooling rates are taken from
      \citet{Sutherland93} and \citet{Galli98} for the atomic and
      molecular cases, respectively.}
  \end{center}
\end{figure}

Structure in a CDM cosmogony forms hierarchically, and this was first
proposed, albeit not in a CDM context, by \citet{Peebles65}.  Smaller
objects collapse first and merge with other objects to form larger
cosmological halos, and matter organizes in a self-similar nature that
grows with the scale factor.  \citet{White78} were the first to
consider galaxies forming within a hierarchical clustering of ``heavy
halos'', where the baryons are not the dominate source of gravity.
During the initial collapse of the protogalactic cloud, baryons are
shock-heated to the virial temperature,
\begin{equation}
  \label{eqn:intro_tvir}
  T_{\rm{vir}} = \frac{\mu m_h V_c^2}{2k},
\end{equation}
where
\begin{equation}
  \label{eqn:r200_vc}
  V_c = \sqrt{\frac{GM}{r_{200}}}
  \quad \textrm{and} \quad
  r_{200} = \left[ \frac{GM}{100\Omega_{\rm{CDM}}(z) H^2(z)} \right]^{1/3},
\end{equation}
are the circular velocity of the halo and $r_{200}$ is the radius
enclosing an average DM overdensity of 200.  Here $\Omega_{\rm{CDM}}$
is the mass-energy fraction of CDM in units of the critical density
$\rho_c = 3H_0^2/8\pi G$.  The first galaxies to form have \tvir~$>
10^4$ K because the cooling function of pristine atomic gas sharply
rises by several orders of magnitude.  At this temperature, the number
of free electrons able to excite hydrogen greatly increases, and these
excitations cool the gas \citep{Spitzer78}.  We plot the cooling
function for pristine, metal-free gas in Figure \ref{fig:coolingFn}.
Only when the gas can cool efficiently (\tdyn~$>$~\tcool), the gas is
able to radiate its energy away and rapidly collapse on a free-fall
time \citep{Rees77}.  Based on the same timescale argument, objects
above $10^{13} \Ms$ have virial temperatures $4 \times 10^6$ K, where
the gas cannot cool within a Hubble time.  This can explain the sharp
cutoff in the galaxy luminosity function above $10^{13} \Ms$.
Additionally this explains the hot, X-ray emitting gas in galaxy
clusters.

\citet{White78} also proposed that galaxies forming by hierarchical
assembly will likely have an imprint on present-day galaxies.  As
halos merge together, they suggested that the outer layers of low-mass
halos are stripped during the virialization of the parent halo on a
timescale greater than a dynamical time \tdyn.  Since low-mass
galaxies merge together to form Milky Way type galaxies, the luminous
cores of low-mass galaxies may survive to the present-day and exist in
identifable stellar systems.  Additionally, some fraction of dwarf
galaxies may have formed in the early universe \citep[see][]{Mateo98}.
Numerical calculations and cosmological simulations show that CDM
models with hierarchical clustering can account for the abundances and
characteristics of both galaxies and clusters
\citep[e.g.][]{Peebles82, Springel05}.  However, one must consider
feedback effects in low-mass galaxies to correctly describe galaxies
across the mass spectrum.

\subsection{Biased Galaxy Formation}

Dark matter constitutes more than 80\% of matter in the universe
\citep{Spergel06} and dominates the gravitational potential field of
large-scale structure and individual halos.  It is important to
quantify how galaxies and baryons trace these potential wells.  There
is a ``bias'' in galaxies in that galactic properties depend on the
formation environment \citep{Dekel87}.  Higher $\sigma$ objects
collapse earlier and form in a different environment than lower
$\sigma$ peaks of the same mass.  Since $\bar{\rho}(z) \propto
(1+z)^3$, the timescales of the cooling and dynamical processes differ
at higher redshift as they scale with $\rho^{-1}$ and $\rho^{-1/2}$,
respectively.  Additionally for $z > 10$, Compton cooling can be
dominant in ionized regions, in which the cooling time $t_C \propto
f_e^{-1} (1+z)^{-4}$, where $f_e$ is the electron fraction.

There is also bias in the suppression of galaxy formation from stellar
feedback and intergalactic medium (IGM) heating.  As low-mass galaxies
($M \sim 10^8 \Ms$) are more suspectible to stellar feedback and
outflow, the ultraviolet (UV) heating of the IGM from stellar sources
may inhibit gravitational collapse and star formation in such
galaxies.  Earlier forming galaxies are affected less than their
lower-$\sigma$ counterparts since the UV radiation background, which
affects cooling rates and can photo-evaporate small gas clouds,
steadily grows with time.  The clustering of halos also biases galaxy
formation, which is very relevant for the first stars and galaxies.
This creates biasing in most feedback processes in these first
structures because there could be more nearby stars.  This could lead
to the observed differences in the galaxy-environment correlation
\citep{Dekel87}.

\subsection{Semi-Analytic Models}

Semi-analytic galaxy formation models \citep[e.g.][]{Somerville99,
  Benson00, Cole00} are an alternative to studying galaxy formation
with explicit numerical cosmological simulations.  The philosophy of
their methods originated from the works of \citet{White78},
\citet{White91}, \citet{Cole91}, and \citet{Lacey91}.  They use
simplistic models of baryon physics and star formation that are set in
a dark matter merger tree that is usually produced by extended
Press-Schetcher formalism \citep{Press74, Bond91, Lacey93}.  Density
profiles and spins of dark matter halos in the merger tree are
accurately represented statistically by results of numerical
simulations \citep{Navarro97, Bullock01}.  Some of the simplifications
are however justified as some processes, in particular star formation
and feedback, are not well understood.  Furthermore, most of the free
parameters are tuned to match observations and analytics, which allows
the application of this method to other poorly understood galaxy
populations.

One main ingredient in semi-analytic models is disc formation.
Motivated by the observation that the Milky Way formed from a gaseous
collapse \citep{Eggen62} and tidal torque theory \citep{Hoyle49,
  Peebles69}, \citet{Fall80} investigated the properties of disc
formation in the collapse of a slowly rotating cloud within a ``heavy
halo'', i.e. dark matter dominated.  The system rotates like a
solid-body, i.e. $v \propto r$, in that all of the small-scale kinetic
motions have been dissipated into heat, which is represented by the
spin parameter $\lambda$ (eq. \ref{eqn:spinParameter}).  The resulting
disc model agrees remarkably well with observations, in particular the
exponentially decreasing surface density.  These models were improved
by \citet{Mo98}, who embed the collapsing gas cloud in a more accurate
dark matter density profile \citep{Navarro97}.  Furthermore, they only
allow discs to form if they are dynamical stable against bar formation
\citep[see][and references therein]{Christodoulou95} that results in
an excellent match with the Tully-Fisher relation.  Many modern
semi-analytic models use the formalism of \citet{Mo98} for disc
formation.

Semi-analytic models of galaxy formation have been broadly successful
in reproducing a large range of galaxy properties and their evolution.
They have successfully reproduced (i) the local field galaxy
luminosity function \citep{Kauffmann93}, (ii) the slope and scatter of
the Tully-Fisher relation and the formation of most stars at $z < 1$
\citep{Cole94}, (iii) the counts and redshift distribution of faint
galaxies \citep{Kauffmann94}, (iv) a sharp decline in elliptical
galaxies at high redshift \citep{Kauffmann96}, and (v) the strong
clustering of Lyman break galaxies at $z = 3$ \citep{Wechsler01}.

Despite their many accomplishments in galaxy formation, current
implemenations may lose accuracy in the early universe when star and
galaxy formation differed greatly from the properties observed at $z <
6$.  In the early universe, the first stars form from metal-free gas,
which affects the collapse dynamics of the parent gas cloud and
results in a large stellar mass.  Thus some of the smallest galaxies
may be greatly affected by radiative feedback that originate from
these stars.  We focus on metal-free Population III (Pop III) star
formation in the next section.

\section[Molecular Hydrogen Cooling]{Molecular Hydrogen Cooling in the
  Early Universe}

Big bang nucleosynthesis mainly produces hydrogen and helium with only
trace amounts of deuterium and lithium \citep{Alpher48, Wagoner67}.
The first stars in the universe must have been devoid of heavy
elements.  However, there is a metallicity floor of $10^{-4}$ of solar
metallicity in Galactic halo stars, residing in the outskirts of the
Galaxy \citep{Beers05}.  This prompts the following questions:
\textit{Where are the first stars?  What are their properties?}  To
answer these questions, we describe the collapse and cooling of their
parent gas clouds.

Galactic star forming regions condense through molecular hydrogen
(\hh) cooling that mainly occurs on dust grains composed of carbon,
silicon, oxygen, iron, and several other elements.  This cannot be the
case in metal-free, pristine gas.  At low metallicities and
temperatures above 1000 K, \hh~formation can occur in the gas-phase by
two processes \citep{McDowell61}:
\begin{eqnarray}
  \label{eqn:h2form1a}
  \rm{H} + e^- &\rightarrow& \rm{H}^- + \gamma \\
  \label{eqn:h2form1b}
  \rm{H}^- + \rm{H} &\rightarrow& \rm{H}_2 + e^-
\end{eqnarray}
or less efficiently
\begin{eqnarray}
  \label{eqn:h2form2}
  \rm{H} + \rm{H}^+ &\rightarrow& \rm{H}_2^+ + \gamma \\
  \rm{H}_2^+ + \rm{H} &\rightarrow& \rm{H}_2 + \rm{H}^+ .
\end{eqnarray}

\citet{Saslaw67} were the first to realize that \hh~formation in the
gas-phase was important in star formation in the early universe.  They
used the latter reactions to determine that \hh~cooling dominates the
collapse of a pre-galactic cloud above number densities $n$ of $10^4
\cubecm$.  These high density regions can cool to $\sim$300 K and
continue to collapse.  \citet{Peebles68} suggested that globular
clusters were the first bound objects in the universe with masses
$\sim$$5 \times 10^5 \Ms$.  In their calculations, they first compute
the properties of these objects from linear perturbation theory and
then follow the initial contraction of the cloud, including
\hh~cooling (eq. \ref{eqn:h2form1a}, \ref{eqn:h2form1b}).  They find
that molecular hydrogen cooling is indeed efficient enough to drive a
free-fall collapse, in which only a small fraction of the total gas
mass forms stars due to the inside-out nature of the collapse.  The
determination of the properties of Pop III stars ultimately relies on
the mass scale on which the structure fragments.  Gravitationally
induced fragmentation can occur when
\begin{equation}
  \label{eqn:fragment1}
  \Gamma_{\rm{frag}} = 1 + \frac{(d \ln T/dt)}{(d \ln \rho/dt)} < 4/3,
\end{equation}
and thermal instabilities produce fragmentation when
\begin{equation}
  \label{eqn:fragment2}
  \rho \left(\frac{\partial L}{\partial \rho}\right)_T - 
  T    \left(\frac{\partial L}{\partial T}\right)_\rho +
  L(\rho, T) > 0,
\end{equation}
where $L$ is the radiative cooling rate.  \citet{Yoneyama72} followed
the evolutionary track of a collapsing primordial gas cloud and found
that the central regions fragmented into 60 \Ms~clumps (for cloud mass
$M = 10^6 \Ms$), in which he suggested that the first stars were
massive.

Molecular hydrogen is photo-dissociated by photons in the Lyman-Werner
(LW) band, 11.26 -- 13.6 eV \citep{Field66, Stecher67}, through the
two-step Solomon process,
\begin{equation}
  \label{eqn:solomon}
  \rm{H}_2 + \gamma \rightarrow \rm{H}_2^* \rightarrow \rm{H} + \rm{H},
\end{equation}
where H$_2^*$ is an electronically or vibrationally excited state that
decays into the continuum state, dissociating the molecule.  In this
energy band, the universe is optically thin with the exception of
Lyman resonances, i.e. \lya, Ly$\beta$, etc, resulting in non-local
suppression of \hh~formation by a soft UV radiation background.  Most
Pop III stars will have to contend with such a background.  The
critical halo mass for a cool, dense core to form increases with the
LW radiation background intensity \citep{Machacek01} that is the main
basis for the results of Chapter \ref{chap:rates}.  We leave the
details of stellar and associated supernovae (SNe) properties to the
chapter introductions.

\section{Computational Techniques}


The astrophysical community have benefited from numerical cosmology
simulations that have matured over the last two decades.  They have
improved our understanding of the nature of many astrophysical
systems.  Modern simulation techniques have advanced through both
ingenious software algorithms and the continual increase in computing
power.

The first gravitational $n$-body simulation intriguingly was not
performed on a digital computer but by photocells and light bulbs,
where the inverse-square law of radiation emulated a gravitational
field, to study the interaction of two colliding galaxies
\citep{Holmberg41}.  After the advent of digital computers, the first
$n$-body simulations that computed the dynamics of up to 100 particles
were performed in the early 1960's \citep{vonHoerner60, vonHoerner63,
  Aarseth63}.  A decade later, the first true cosmological simulations
of structure formation were carried out to confirm an analytic
framework that predicted the abundances of bound objects in the
universe \citep{Press74}.  Several advances in cosmology allowed the
tremendous growth in activity in the following decade, such as (i)
plausible dark models (e.g. CDM, hot DM), (ii) cosmic inflation
\citep{Guth81} provided an explanation for the generation of
primordial density fluctuations and a scale-invariant
Harrison-Zel'dovich spectrum \citep{Harrison70, Zeldovich72}, (iii)
evolution of fluctuations during recombination \citep{Peebles70}, and
(iv) applications of random Gaussian fields and linear perturbation
theory to density fluctuations.

CDM became the favored cosmological model starting in the 1980's
\citep[e.g.][]{Peebles82, Blumenthal84}, and cosmological simulations
could readily test this cosmogony and compare it with observations and
expectations.  In the following years, baryonic physics were added to
the calculations, providing access to additional astrophysical
problems.  First we focus on the techniques to evolve gravity in these
simulations.  We then detail some hydrodynamics solvers in the
suceeding section.

\subsection{Gravity Solvers}

Particle trajectories are evolved in time by solving Newton's laws
written in comoving coordinates:
\begin{equation}
  \label{eqn:nbody}
  \frac{d\xv}{dt} = \frac{1}{a}\vv, \quad
  \frac{d\vv}{dt} + H\vv = \gv, \quad
  \nabla \cdot \gv = -4\pi G a[\rho(\xv,t) - \bar{\rho}(t)],
\end{equation}
where $\gv$ is the gravitational potential field \citet{Peebles93}.
The time integration is usually calculated with a second-order
accurate leapfrog method \citep{Efstathiou85}, which is more than
adequate for cosmological applications.  In practice, it is beneficial
to use the time variable $s = \int a^{-2} dt$ instead of proper time
because the equation of motion (second equation in
eq. [\ref{eqn:nbody}]) simplifies to $d^2\xv/ds^2 = a\gv$.  This
allows the use of a sympletic integrator, i.e. phase space volume
preserving \citep[see][]{Quinn97}.  For large numbers of particles,
the direct summation of gravitational forces for all particle pairs is
prohibitive for cosmological simulations.  Several clever methods have
thus been formulated to overcome this dilemma, which we describe below.

\begin{itemize}
\item \textbf{Barnes-Hut Tree Algorithm}--- This method recursively
  divides the volume into cells that contain one or more particles
  \citep{Appel85, Barnes86}.  The cells must be smaller than some
  solid angle, usually $\delta x/r < 1$, where $\delta x$ and $r$ are
  the cell size and distance.  It is fully spatially adaptive.  A
  low-order multipole expansion of these cells are used to calculate
  the gravitational field instead of direct summation.  Its biggest
  advantage is speed, which scales as O(N log N), where N is the
  number of particles.  However, its main shortcoming is the large
  memory footprint of $\sim$25N words \citep{Hernquist87}.  This
  method is publicly available and fully parallelized by many groups
  \citep[e.g.][]{Hillis87, Makino89, Dubinski96}.

\item \textbf{Particle-Mesh (PM) Algorithm\footnote{For a review of PM
      and P$^3$M, see \citet{Hockney88}.}}--- The density field is
  mapped to a Cartesian grid.  The main mapping techniques used are
  (i) Nearest Grid Point that only considers particles inside the
  cell, (ii) Cloud-in-Cell that uses multi-linear interpolation to the
  surrounding eight cells, and (iii) Triangular Shaped Cloud that
  considers the nearest 27 cells in a higher-order interpolation
  scheme.  Fast Fourier Transforms allow the efficient computation of
  the Poisson equation, which scales as O(N$_g$ log N$_g$), where
  N$_g$ is the number of grid points.  After the gravitational field
  has been calculated, the forces are interpolated back to the
  particles, using the same scheme to perform the first interpolation
  to reduce error propagation.  The advantage in this method is its
  speed and light memory consumption, but the force resolution is poor
  for particle separations under $\sim$5 grid cells.
  
\item \textbf{P$^3$M and Adaptive P$^3$M}--- This algorithm
  \citep{Efstathiou81} combines the advantages of PM and direct
  summation methods.  At long distances, gravitational forces are
  calculated with the PM method, but contributions from nearby
  particles are calculated by direct summation.  The particle pair
  summations limit the speed in strong overdensities, but this can be
  overcome by using the tree algorithm at short distances
  \citep{Xu95}.  However, \citet{Couchman91} devised a more elegant
  solution to adaptively place subgrids (adaptive P$^3$M) that have
  isolated boundary conditions in overdensities.  Then the PM method
  solves the Poisson equation in these subgrids.  In the finest
  subgrids, pair summation is only computed when the separation is
  less than a few cell widths, dramatically increasing the speed and
  accuracy of the calculation.

\item \textbf{Multi-resolution Mesh Methods}--- Some improved methods
  use the ideas of adaptive P$^3$M in which the resolution of the
  calculation increases with overdensities.  These include (i)
  adaptive mesh refinement of the density field
  \citep[e.g.][]{Jessop94, Kravtsov97}, (ii) splitting of particles in
  overdensities \citep{Splinter96}, and (iii) a moving mesh
  that deforms to sample overdensities at higher resolutions
  \citep{Gnedin95, Pen95}.

\end{itemize}

\subsection{Hydrodynamics Solvers}

Almost twenty percent of matter in the universe is baryonic, some of
which radiatively cools and condenses to form stars and galaxies.  The
hydrodynamics of cosmological structures have been studied since
\citet{Doroshkevich78} studied the one-dimensional collapse of
sheet-like structures (Zel'dovich pancakes).  The difficulty of
baryonic physics arises from the non-linearity and discontinuities in
strong shocks.  The solution of the fluid equations in an expanding
universe is aided by rewriting them in comoving coordinates: the
continuity equation,
\begin{equation}
  \label{eqn:hydroEqn1}
  \frac{\partial}{\partial t} \left(\frac{\rho_b}{\bar{\rho}_b}\right)
  + \frac{1}{a} \nabla \cdot \vv_b = 0,
\end{equation}
the momentum equation,
\begin{equation}
  \frac{\partial \vv_b}{\partial t} 
  + \frac{1}{a} \vv_b \cdot \nabla\vv_b
  + H \vv_b =
  -\frac{1}{a\rho_b} \nabla p
  + \gv ,
\end{equation}
and the energy equation,
\begin{equation}
  \label{eqn:energyEqn}
  \frac{\partial u}{\partial t}
  + \frac{1}{a} \vv_b \cdot \nabla u =
  -\frac{p}{a \rho_b} \nabla \cdot \vv_b
  + \frac{1}{\rho_b} (\Gamma - \Lambda).
\end{equation}
Here $\rho_b$, $\bar{\rho}_b$, $\vv_b$, $u$, $p$, $\Gamma$, and
$\Lambda$ are the gas density, mean gas density, peculiar velocity,
specific energy, pressure, and heating and cooling rates per unit
volume, respectively.  For a perfect gas, the specific energy $u =
p/[(\gamma-1)\rho_b]$, where $\gamma$ is the adiabatic index.  There
have been three major approaches that we discuss next in solving these
fluid equations.

\begin{itemize}
\item \textbf{Smoothed Particle Hydrodynamics (SPH)}--- This
  Lagrangian method \citep{Lucy77, Gingold77, Evrard88} describes
  fluid parcels with particles that have gas properties, such as
  density, temperature, and chemical abundances.  These quantities are
  calculated by smoothing over usually 20--30 of the nearest
  neighboring particles with a kernel.  For example, the density at a
  particular position $\xv$ is
  \begin{equation}
    \label{eqn:sph_rho}
    \rho_b(\xv) = \sum_{i=1}^N m_i W(\xv - \xv_i, h),
  \end{equation}
  where $W$ and $h$ are the smoothing kernel and smoothing length.
  The kernel $W$ is usually a spline with some cutoff radius
  proportional to $h$.  Then a modified set of fluid equations using
  the kernel are solved.  SPH benefits from the advances in
  collisionless particle algorithms because of its particle nature.
  It is inherently spatially adaptive but suffers from poor resolution
  in underdense regions.

\item \textbf{Eulerian Grid Algorithms}--- Grid algorithms with finite
  differencing have been utilized to solve the fluid equations since
  the 1960's \citep[see][for an early review]{Ritchmyer67}, who also
  established that additional terms (e.g. heating and cooling rates,
  gravity, cosmological expansion, etc.) could be solved by operator
  splitting.  In order to capture strong shocks, robust schemes must
  be used, which have been extensively tested in the computational
  fluid dynamics community \citep[e.g.][]{Sod85, Leveque92}.  There
  are two main approaches for shock captures: the total-variational
  diminishing method (TVD) and the piecewise-parabolic method (PPM).
  TVD uses an approximation to the Riemann solution, is second-order
  accurate outside of shocks, and can resolve shocks in two grids
  cells \citep{Ryu93}.  PPM is a third-order accurate Godunov solution
  for the Riemann problem, which is solved using a quadratic
  interpolation of the densities that minimizes post-shock
  oscillations \citep{Colella84, Woodward84}.

\item \textbf{Adaptive Mesh Refinement (AMR)}--- Eulerian methods can
  capture shocks well but cannot follow high density regions well
  since it is constrained to a fixed grid.  This drawback can be
  bypassed, however, by using AMR that dynamically places higher
  resolution grids in regions of interest (e.g. high densities,
  shocks, etc.).  AMR has been successfully implemented in several
  cosmology codes \citep[e.g.][]{Anninos94, Bryan97, Kravtsov97} to
  date and has demonstrated that it can evolve systems over 12 orders
  of magnitude in length \citep{Bryan01}.

\end{itemize}

\subsection{Additional Physical Models}
\label{sec:physicalModels}

The accurate treatment of gas dynamics involves not only gravity and
hydrodynamics but also additional physics such as heating and cooling,
chemical reactions, and radiation transport.  \citet{Cen90} was the
first to include radiative cooling in the form of a cooling function
that considered equilibrium between recombinations and collisional
ionization.  Considering radiative cooling permits the simulation to
follow the condensation of galaxies in virialized objects.  However,
the assumption of equilibrium chemsitry breaks down behind shocks and
in dense cooling regions.  Here one must solve a rate equation network
for non-equilibrium chemistry \citep{Cen92A, Haehnelt96}.  These
methods have also been successful in follwing the formation of the
first objects with non-equilbrium solvers that include molecular
hydrogen cooling \citep{Haiman96, Abel97, Anninos97, Gnedin97}.

Heating rates in cosmology emerge from stellar radiation, supernova
explosions, and active galactic nuclei (AGN).  These processes are
usually included by a phenomenological method that represents stars or
galaxies as collisionless particles that return thermal energy (and in
some cases, metals) to the hydrodynamical grid or SPH particles
\citep{Cen92a}.

Gas heating from stellar radiation can be represented better if one
considers radiative transfer, an essential ingredient to most
astrophysical scenarios.  However, its solution is most difficult
since it is a function of seven variables: position, frequency,
direction, and time.  This has not deterred the community from
appropriate solutions, in which we note three effective methods.
\begin{itemize}

\item A spatially homogeneous and isotropic radiation field in the
  computational volume can be used to study the effects of a radiation
  background.  Additionally, one can focus on the energy dependence of
  this background on structure formation \citep{Cen92A}.  This approach
  has also been extended to include an approximation to local
  absorption \citep{Gnedin97}.

\item Variable Eddington factors, the ratio of radiation stress to
  energy density, provides accurate solutions for radiation transport
  in both optically thin and thick regimes \citep{Ducloux92, Stone92,
    Gnedin01}.

\item Ray tracing from point sources of radiation accurately tracks
  ionization fronts and regions through the computational volume
  \citep[e.g.][]{Abel99, Abel02b, Mellema06}.  However, to obtain
  well-sampled radiation fields at large distances, one must trace an
  obscene amount of rays originating from the source.  Adaptive ray
  tracing avoids this problem by tracing only hundreds of rays from
  the source and splitting them as they propagate farther from the
  source \citep{Abel02b, Abel07}.

\end{itemize}

\subsection{Applications}

The first true application of cosmological simulations tested various
cosmology models and parameters, i.e. CDM, HDM, $\Omega$, $H_0$.  They
proved useful in refining the CDM model and ruling out the HDM model.
In addition to that application, there have been other significant
applications to astrophysics.  We highlight four of them here.
\begin{enumerate}

\item Galaxy clusters are the most recently virialized objects and
  provide a good problem for simulations \citep[see][]{Frenk99}, which
  have refined the relationships between mass, luminosity, and
  velocity dispersions.  Furthermore, comparing the substructure,
  morphologies, and radial profiles of clusters found in simulations
  to observations can constrain cosmological parameters.

\item They have constrained the nature of quasar absorption and
  \lya~forest lines.  Absorption in these lines occurs in a few
  different cases: well-defined clouds, filamentary and sheet-like
  structures, and velocity caustics.  Cosmological simulations can
  accurately account for the column density distribution of \lya~lines
  \citep{Zhang97, Bryan99}.

\item A universal density profile of DM halos has been discovered
  using cosmological simulations \citep{Navarro97}.  The density
  $\rho(r) \propto [(r/r_s)(1 + r/r_s)^2]^{-1}$, where $r_s$ is a
  scale radius, in all DM halos regardless of their formation redshift
  and mass, initial power spectrum, and cosmological parameters.
  
\item Cosmological simulations have verified that structure forms in a
  self-similar way that evolves as $a(t)^\alpha$, where $\alpha =
  2/(3+n)$ and $n$ is the spectral index of the density fluctuation
  power spectrum.

\item Additionally, they have fine-tuned the Press-Schechter mass
  function \citep{Jenkins01} and verified the nature of galaxy
  clustering \citep{Kravtsov04, Weinburg04} and the non-linear power
  spectrum of dark matter \citep{Wechsler06}.  Simulations have also
  investigated and confirmed the predicted halo merging rates and halo
  accretion histories from extended Press-Schechter formalism
  \citep{Wechsler02}.

\end{enumerate}

Clearly numerical simulations can be used as a powerful astrophysical
tool.  They will be the basis for this thesis, which we outline in the
next section.

\section{Thesis Overview}

Recent developments have shown that the first luminious objects in the
universe are metal-free massive ($M \sim 100 \Ms$) stars that form in
dark matter halos with masses $\sim$$10^6 \Ms$ \citep{Abel02a, Bromm02a,
  Yoshida06b}.  These stars form at redshifts between 10 and 50.
Closer to the present-day, galaxy formation models are well
established and have made several successful interpretations about the
nature of galaxy formation in the universe after reionization ($z <
6$).  This thesis aspires to bridge the gap between theories of the
first stars and galaxy formation.  Hence its focus is the formation of
the first galaxies in the universe, shortly after the birth of Pop III
stars.

Before delving into extensive numerical investigations, it is
beneficial to determine whether Pop III stars play a significant role
in the early universe through a semi-analytic model of reionization
that utilizes Press-Schechter formalism and results of numerical
simulations that study the negative feedback in the first bound
objects.  This is the subject of Chapter \ref{chap:rates}.

Galaxy formation involves many physical processes outlined in
\S\ref{sec:physicalModels}.  In order to isolate the importance of
each process, we gradually introduce new physics into our simulations.
All of the simulations in this thesis have the same initial conditions
but consider different physical models.  Chapter
\ref{chap:virialization} shows the simplest models to which we
compare all later simulations to assess the relevance of the added
process.  Here we study the baryonic virialization of cosmological
objects through an analysis of the virial theorem of such objects, in
particular the generation of turbulence.  Here we see supersonic
turbulence when radiative cooling is efficient.  This can have
applications to galaxy formation models and galactic molecular clouds
that are supersonically turbulent.

We next describe the collapse of halos with masses $\sim$$10^8 \Ms$,
undergoing atomic hydrogen cooling in Chapter \ref{chap:collapse}.
These objects are usually considered as the first galaxies and sources
of radiation in galaxy formation models.  We investigate this case to
study the nature of the first galaxy under this assumption.  We find
that it centrally collapses before fragmentation occurs in a global
disk.  This may, however, never occur in nature because we neglect
\hh~formation and cooling and stellar feedback from Pop III stars.
Nevertheless, it is a great testbed for theories of turbulent
collapses that are ubiquitous in the universe.

In Chapter \ref{chap:progenitors}, we consider \hh~formation and
cooling in the calculations.  We extend the work of
\citet{Machacek01}, who investigated the negative feedback of a
Lyman-Werner background in \hh~formation in the early universe.  We
add the models with no residual electrons from recombination, an
essential part to \hh~formation in the gas-phase, from recombination.
We find that collisional ionization in halos aids \hh~formation and
the collapse \citep{Shapiro87}.  These collapsing halos are three
times less massive than previously thought, which translates into
early galaxies being an order of magnitude more abundant.  We conclude
that \hh~plays an essential role in early galaxy formation and cannot
be neglected.

Finally we include radiative transfer, a vital component to
simulations of early structure formation.  This is the first time an
accurate radiation transport scheme has been applied to galaxy
formation at any epoch.  We use adaptive ray tracing to compute the
radiation transport from Pop III stars.  In one simulation, we include
the effects of pair-instability SNe from Pop III stars from which we
track the propagation of the first metals in the universe.  In Chapter
\ref{chap:nature}, we study the global properties of early dwarf
galaxies, including the metallicities of such galaxies and the IGM.
The baryon fractions of such galaxies are reduced up to a factor of
three, and the angular momentum is increased by a factor of two when
considering the effects of stellar feedback from Pop III on early
galaxy formation.  Chapter \ref{chap:reion} then focuses on star
formation rates at redshifts greater than 15, the properties of star
formation regions, and the start of cosmological reionization.

We state the main conclusions of this thesis in Chapter
\ref{chap:conclusion}.

\chapter[Number of Primordial Supernovae]{The Number of Supernovae
  from Primordial Stars in the Universe}
\label{chap:rates}

Cosmological structure assembles hierarchically.  If one follows this
logic, Milky Way type galaxies must have formed from the mergers of
thousands of smaller entities over the history of the universe.  This
leads to the following questions: \textit{What are the smallest and
  first objects with luminous counterparts that eventually compose
  most of the present-day structure?  What are their properties?  Are
  they observable?}

These questions have aroused much interest in the formation of the
first objects.  It has been established by both analytical work and
numerical simulations that the first stars form in dark matter halos
with masses $\sim$$10^6 \Ms$.  The gas in these objects condense by
molecular hydrogen cooling and form a single, central massive star.
Observations of primordial stars are currently out of reach even by
the most powerful telescopes.  In this chapter, we investigate the
feasibility of future observations of supernovae from primordial
stars, in particular their magnitudes and rates on the sky.  This work
involves semi-analytic models of structure formation, which are less
computationally intensive than explicit three-dimensional numerical
simulations.  This allows for a parameter-space study before
proceeding to three-dimensional computations of the formation of the
first stars and its feedback on galaxy formation.

This chapter was published in the \textit{The Astrophysical Journal},
2005, Volume 629, 615.  This paper is co-authored by Tom Abel, who
guided me in the development of the semi-analytic code and the
interpretation of its results.

\section{Introduction}
The properties of pre-reionization luminous objects are integral to
our comprehension of the process of reionization and their effect on
subsequent structure formation.  Observations of distant ($z = 6.28,
6.4$) quasars depict the relics of reionization with their
accompanying Gunn-Peterson troughs \citep{Becker02, Fan02}.
Furthermore, Lyman alpha forest carbon abundances of
$\sim$10$^{-2}$ Z$_\odot$ and 10$^{-3.7}$ Z$_\odot$ observed at
redshifts 3 and 5, respectively, indicate that numerous early
supernovae (SNe) enriched the intergalactic medium (IGM)
\citep{Songaila96, Songaila01}.  These early generations of stars are
at least partly responsible for ionizing the Universe.  A fraction of
these stars lie in protogalaxies, but the other fraction of early
stars are metal-free, form through molecular hydrogen cooling, and are
very massive (M $\sim$ 100$\Ms$) \citep{Abel00, Abel02a}.  In
$\Lambda$CDM cosmologies, these stars form at 10 $\lsim$ z $\lsim$ 50.
The Wilkinson Microwave Anisotropy Probe (WMAP) data further
constrains the epoch of reionization from the measurement of the
optical depth due to electron scattering, $\tau_{es}$ = 0.17 $\pm$
0.04 (68\% confidence), which corresponds to a reionization redshift
of 17 $\pm$ 5 when assuming instantaneous reionization
\citep{Kogut03}.  However in this paper, we shall show this epoch is
gradual and its effect on primordial star formation.

In hierarchical models of structure formation, small objects merge to
form more massive structures.  Eventually, a fraction of halos are
massive ($\sim 5 \times 10^5\Ms$ at $z \sim 20$) enough to host
cooling gas \citep{Couchman86, Tegmark97, Abel98, Fuller01}.  These
halos do not cool through atomic line cooling since $T_{vir} < 10^4$;
however, primordial gas contains a trace of \hh.  Free electrons allow
\hh~to form, which acts as an effective coolant at several hundred
degrees through rotational and vibrational transitions.  \hh~is easily
photo-dissociated by photons in the Lyman-Werner (LW) band,
11.26--13.6eV \citep{Field66, Stecher67}, thus \hh~can be destroyed by
distant sources in a neutral Universe.  Primordial stars produce
copious amounts of UV photons in the LW band, destroying the most
effective cooling process at $z \sim 20$.  This negative feedback from
the UV background significantly inhibits the primordial star formation
rate in the early Universe by requiring a larger potential well for
gas to condense.  Many groups have explored the effects of a UV
background on cooling and collapsing gas.  Firstly, \citet{Dekel87}
discovered that \hh~can be dissociated from large distances.  Then
more quantitatively, \citet{Haiman97b} questioned how collapsing,
homogeneous, spherical clouds are affected by a UV background.  In
more detail, \citet{Haiman00} determined whether a gas cloud collapsed
by comparing the cooling time with the current lifetime of the cloud,
which was calculated in the presence of solving the spherically
symmetric radiative transfer equation along with time-dependent
\hh~cooling functions.  However, these studies only considered
spherically symmetric cases while realistically these halos are
overdensities within filaments.  To combat this problem,
\citet{Machacek01} employed a three-dimensional Eulerian adaptive mesh
refinement (AMR) simulation to determine quantitative effects of a UV
background on gas condensation.

We consider lower mass stars to form in more massive halos that may
fragment via atomic line cooling as well as the metal lines from the
heavy elements expelled by earlier generations of stars.  We use the
prescription outlined in \citet{Haiman97a} (hereafter HL97) to model
the metal abundances and ionizing photon rates of these stars.

With infrared space observatories, such as the Spitzer Space
Telescope, Primordial Explorer \citep[PRIME;][]{Zheng03}, and James
Webb Space Telescope (JWST), sufficient sensitivity is available to
detect the SNe from primordial stars.  Although SNe remnants are
bright for short periods of time, they may be the best chance to
directly observe primordial stars due to their large intrinsic
luminosities.  If these events are recorded, many properties, such as
mass, luminosity, metallicity, and redshift, of the progenitors can be
calculated using metal-free or ultra metal-poor SNe models, which can
validate or falsify simulations of the first stars \citep{Abel02a}.  To
evolve our calculation through redshift, we need to retain information
about the UV background and number densities of primordial stars,
which can extend to determining such quantities as volume-averaged
metallicity and ionized fraction of the Universe.  Our model is
constrained with (a) the WMAP optical depth measurement, (b)
primordial star formation and suppression in high resolution
hydrodynamical simulations, and (c) local observations of dwarf
galaxies that constrain high redshift protogalaxies properties.

We organize the paper as follows.  In \S\ref{sec:method2}, we report
the semi-empirical method behind our calculations, which incorporates
effects from negative feedback and an ionized fraction and is
constrained from the WMAP result and local dwarf galaxy observations.
In \S\ref{sec:results2}, we present the number density of primordial
SNe in the sky.  We also explore the feasibility of observing these
SNe by calculating their magnitudes and comparing them with the
sensitivities of infrared space observatories.  Then in
\S\ref{sec:discussion2}, we discuss possible future observations and
numerical simulations that could further constrain our model.
Finally, \S\ref{sec:summary2} summarizes our results and the
implications of the first observations of SNe from primordial stars.

\section{The Method}
\label{sec:method2}
We use several theories and results from simulations of structure
formation and metal-free stars.  We use a $\Lambda$CDM cosmology with
$\Omega_\Lambda$ = 0.70, $\Omega_{\rm{CDM}}$ = 0.26, $\Omega_b$ =
0.04, h = 0.7, $\sigma_8$ = 0.9, and n = 1.  $\Omega_\Lambda$,
$\Omega_{\rm{CDM}}$, and $\Omega_b$ are the fractions of mass-energy
contained in vacuum energy, cold dark matter, and baryons,
respectively.  $h$ is the Hubble parameter in units of 100 km s$^{-1}$
Mpc$^{-1}$.  n = 1 indicates that we use a scale-free power spectrum,
and $\sigma_8$ is the variance of random mass fluctuations in a sphere
of radius 8h$^{-1}$ Mpc.  We use a CDM power spectrum defined in
\citet{Bardeen86}, which depends on $\sigma_8$ and h.

The remaining parameters are the primordial stellar mass and the
factors f$_{esc}$ and f$_\star$, which dictate the production and
escape of ionizing photons.  f$_{esc}$ is the photon escape fraction,
and f$_\star$ is the star formation efficiency.

To evolve the ionization and star formation behaviors of the Universe,
we must determine the density of dark matter halos that host early
stars.  In $T_{vir} < 10^4$ K halos (henceforth ``minihalos''),
\hh~cooling is the primary mechanism that provides means of
condensation into cold, dense objects.  However in $T_{vir} > 10^4$ K,
which corresponds to $M_{vir} > 10^8 \Ms [(1+z)/10]^{-3/2}$, halos,
hydrogen atomic line cooling allows the baryons to fragment and cool
into stars.

\subsection{Minihalo Star Formation}

The first quantity we need to begin our calculation is the minimum
halo mass that forms a cold, dense gas core due to \hh~cooling.
Radiation in the LW band photo-dissociates \hh, which inhibits star
formation in minihalos.  This negative feedback from a UV background
does not necessarily prohibit the formation of primordial stars in
minihalos, but only increases the critical halo mass in which
condensation occurs, and delays the star formation.  From their
simulations of pre-galactic structure formation, \citet{Machacek01}
determined the minimum mass of halo that hosts a massive primordial
star is
\begin{equation}
\label{minMass}
\frac{M_{min}}{\Ms} = \exp\left(\frac{f_{cd}}{0.06}\right)
  \left(1.25 \times 10^5 + 8.7 \times 10^5
  F_{LW,-21}^{0.47}\right),
\end{equation} 
where $M_{min}$ is the minimum halo mass that contains a cold, dense
gas core; $f_{cd}$ is the fraction of gas that is cold and dense;
$F_{LW}$ is the flux within the LW band in units of 10$^{-21}$ \flux .
For our calculation, we consider $f_{cd} = 0.02$, which is a
conservative estimate in which we have an adequate source of star
forming gas.  With our chosen $f_{cd}$ and no UV background, a 1.74
$\times$ 10$^5$ $\Ms$ halo will form a cold, dense core, which will
continue to form a primordial star.  In a typical UV background of J =
10$^{-21}$ \emis , the minimum mass is 4.16 $\times$ 10$^6
\Ms$.  Additionally, minihalos can only form a star within neutral
regions of the Universe since they are easily photo-ionized
\citep{Haiman01, Oh03}, and \ion{H}{1} is a necessary ingredient for
producing \hh.

Numerical simulations \citep[e.g.][]{Abel02a, Bromm02a} illustrated
that fragmentation within the inner molecular cloud does not occur and
a single massive (M $\sim$ 100$\Ms$) star forms in the central
regions.  These stars produce hard spectra and tremendous amounts of
ionizing photons.  For example, the ionizing photon to stellar baryon
ratio n$_\gamma$ $\sim$ 91300, 56700, and 5173 for H, He, He$^+$ in a
200$\Ms$ star \citep{Schaerer02}.  We calculate the ionizing photon
flux by considering
\begin{equation}
\label{ion_mini}
\left(\frac{dN_\gamma}{dt}\right)_{mini} = \bar{Q} \: T_{life} \:
\frac{d\rho_{mini}}{dz} \: \frac{dz}{dt},
\end{equation}
where $N_\gamma$ is the total number of ionizing photons in the
Universe; $\bar{Q}$ is the time-averaged photon flux; T$_{life}$ is
the stellar lifetime; and $\rho_{mini}$ is the comoving number density
of minihalos that we determine from an ellipsoidal variant of
Press-Schechter (PS) formalism \citep{PS74, Sheth02}.

The primordial initial mass function (IMF) is unknown, therefore, we
assume a fixed primordial stellar mass for each calculation.  We run
the model for primordial stellar masses, M$_{fs}$, of 100, 200, and
500 \Ms~and use the time-averaged emissivities from metal-free
stellar models with no mass loss \citep{Schaerer02}.  In our models,
the minihalo is quickly ionized and all photons escape into the IGM
\citep{Whalen04}.  Furthermore, we use a blackbody spectrum at 10$^5$
K to approximate the spectrum of the primordial star since surface
temperatures are virtually independent of mass at M $\gsim$ 80$\Ms$.

\subsection{Star Formation in Protogalaxies}
In $T_{vir} > 10^4$ K halos, baryons can cool efficiently through
atomic line cooling, thus fragmenting and forming stars.  As described
in HL97, we parametrize the properties of these stars by a couple of
factors.

\subsubsection{Star formation efficiency}
High redshift galaxies appear to have similar properties as local
dwarf galaxies.  We can constrain the star formation efficiency
f$_\star$ by letting local observations guide us.  In these galaxies,
f$_\star$ ranges from 0.02 to 0.08 \citep{Taylor99, Walter01}.  Using
the orthodox Schmidt star formation law in local dwarf galaxies,
\citet{Gnedin00} estimated f$_\star$ to be 0.04 if it were constant
over the first 3 Gyr but can also be as low as 0.022.  It should be
noted that f$_\star$ can be higher if star formation ceased after a
shorter initial burst.

Analyses of metallicities in local dwarf galaxies reveal their prior
star formation.  Both Type Ia and Type II SNe contribute iron to the
IGM, but Type II SNe provide most of the $\alpha$-process elements
(e.g. C, N, O, Mg) to the IGM.  Type II SNe occur on timescales $< 3
\times 10^7$ yr while Type Ia are delayed by $3 \times 10^7$ yr to a
Hubble time \citep{Matteucci01}.  Therefore, we expect an
overabundance of $\alpha$-process elements with respect to iron just
after the initial starburst.  [$\alpha$/Fe] versus [Fe/H]\footnote{We
use the conventional notation, [X/H] $\equiv$ log(X/H) -
log(X$_\odot$/H$_\odot$)} plots help us inspect the evolution of the
ISM/IGM metallicity.  If the prior star formation is inefficient
(i.e. spirals and irregulars), [$\alpha$/Fe] is only shortly
overabundant, which is characterized by a short plateau versus [Fe/H].
On the other hand, if the star formation is fast and occurs early in
the lifetime of the galaxy, [$\alpha$/Fe] remains in the plateau
longer due to the quick production of metals by Type II SNe \citep[see
Figure 1 in][]{Matteucci02}.  Also, \citet{Venn03} discovered no
apparent plateau in [$\alpha$/Fe] versus [Fe/H] comparisons in dwarf
spheroidal and irregular galaxies.  They conclude that star formation
must have been on timescales longer than Type Ia SNe enrichment, which
hints at a low and continuous star formation rates in these
protogalaxies when compared to recent star formation.

These low star formation efficiencies are further supported by the
galaxies contained in the Sloan Digital Sky Survey
\citep[SDSS;][]{York00}.  Star formation efficiencies within low mass
galaxies (M $<$ 3 $\times$ 10$^{10} \Ms$) decline as M$^{2/3}$
\citep{Kauffmann03}.  At high redshift and before reionization, most
of the protogalaxies tend to be few $\times$ 10$^7$ $\Ms$, which is
comparable to many local dwarf galaxies \citep{Mateo98}.  It should
also be noted that no star formation history is alike within
individual Local Group galaxies, but it is worthwhile to adopt a
global star formation efficiency and observe the consequences on
reionization and primordial star formation.  With the stated
constraints, we set f$_\star$ = 0.04 in our main model in concordance
with the Schmidt Law \citep{Gnedin00} and stellar abundances
\citep{Venn03}.

\begin{center}
\renewcommand{\thefootnote}{\alph{footnote}}
\begin{longtable}{ccccc}
\caption[Ionizing luminosities for metal-poor IMFs]{Ionizing photon
  production per stellar baryon and luminosities for metal-poor
  IMFs} \label{imf} \\

\hline\hline \\[-3ex]
Z & n$_{\gamma,\:H}$ & n$_{\gamma,\:He}$ & n$_{\gamma,\:He^+}$ & 
log $\cal{L}$ \\
 & & & & [erg s$^{-1}$ M$_\odot^{-1}$] \\
\hline
\endhead

10$^{-7}$	& 20102	& 8504	& 8	& 36.20\\
10$^{-5}$	& 15670	& 5768	& 0.2	& 36.16\\
0.0004	        & 13369	& 3900	& 0	& 36.28\tablenotemark{a}\\
\hline
\footnotetext[1]{See text for a discussion on values.}
\end{longtable}
\renewcommand{\thefootnote}{\arabic{footnote}}
\end{center}

\subsubsection{Stellar luminosities}
We consider multiple IMFs for star formation within protogalaxies.
Our main model uses a Salpeter IMF with a slope $\alpha$ = --2.35,
metallicity Z = 10$^{-7}$, 10$^{-5}$, and 0.0004, and (M$_{low}$,
M$_{up}$) = (1, 100)$\Ms$.  We take the continuous starburst
spectrum evolution from these particular IMFs that were calculated
with Starburst99 \citep{Schaerer03, Leitherer99}.  To estimate the
ionizing photon per stellar baryon ratio, n$_\gamma$, and luminosities
of the IMF at a particular metallicity, we interpolate in log$_{10}$
space between the two adjacent IMFs.  When Z $<$ 10$^{-7}$, we
consider the Z = 10$^{-7}$ IMF.  The properties of these IMFs are
listed in Table 1.  Note that more metal-rich starbursts result in a
softer spectrum and less ionizing photons in which n$_{\gamma,\:H}$
decreases almost by a factor of 2.

In the Z = 0.0004 IMF, we retain the luminosity from the Z = 10$^{-5}$
and set n$_{\gamma,\:He^+}$ = 0.  We choose to do so because of the
theoretical uncertainty of hot Wolf-Rayet (WR) stars and their
presence \citep[for a review, see][]{Schaerer00}.  In
\citet{Schaerer03}, the luminosity and spectral hardness increases due
to the presence of WR stars at higher metallicities.  However,
\citet{Smith02} calculate WR spectra that are significantly softer.

\subsubsection{Ionizing photon escape fraction}
Radiation emitted by these stars have a probability $f_{esc}$ to
escape from the protogalaxy and ionize the IGM.  The protogalaxy ISM
density and composition plays the biggest role in determining this
factor.  \citet{Heckman01} showed that local and distant starburst
galaxies, including the gravitationally lensed galaxy MS 1512-cB58 (z
= 2.7), have $f_{esc}$ $\lsim$ 0.06.  This result agrees with previous
analyses of the $f_{esc}$ in the Lyman continuum \citep{Leitherer95,
Hurwitz97}.  However, Lyman break galaxies may have $f_{esc} \gsim
0.2$ \citep{Steidel01}.  In conjunction with f$_\star$ = 0.04 and
M$_{fs}$ = [100, 200, 500]$\Ms$, the choice of f$_{esc}$ = [0.050,
0.033, 0.028], respectively, in our calculation results in the same
optical depth from WMAP, $\tau_{es}$ = 0.17, thus we use these values
in the main models.  \citet{Wood00} and \citet{Ricotti00} have argued
that the escape fraction may be very small due to the much higher
densities at high redshift. However, the shallow potential wells and
their small size make them susceptible to photo-evaporation and
effects of radiation pressure \citep{Haehnelt95}.  Other
uncertainties such as dust content, metallicity, and whether the ISM
density scales as (1+z)$^3$ blur our intuition about the escape of
ionizing radiation.  Therefore, we allow $f_{esc}$ to vary from
0.001--0.25 since it is unclear whether high-redshift, low-mass
protogalaxies allow photons to escape due to self-photoevaporation or
absorb the photons due to a higher proper gas density when compared to
starburst galaxies.

\subsubsection{Ionizing Photon Rates}
These factors are multiplicative in the amount of radiation that is
available from protogalaxies to ionize the IGM.  The rate of photons
emitted that can ionize species X is
\begin{equation}
\label{dn_dt}
\left(\frac{dN_\gamma}{dt}\right)_{proto} = \rho_0 \frac{f_{esc}
\: f_\star \: n_{\gamma,X}}{\mu \: m_p} \: \frac{d\psi_{proto}}{dz}
\frac{dz}{dt},
\end{equation}
where $\mu$ is the mean molecular weight; $\rho_0$ =
$\Omega_b$(3H$_0^2$/8$\pi$G); $m_p$ is the mass of a proton;
$\psi_{proto}$ is the mass fraction contained in protogalaxies that is
calculated by PS formalism.  We restrict the product of $f_{esc}$ and
$f_\star$ to be in a range from 10$^{-4}$--10$^{-2}$ since the
resulting reionization histories fall within the measured $\tau_{es}$
error bars.  The reionization epoch greatly depends on the factors
$f_{esc}$ and $f_\star$.  In order to explore the consequences of
different values of $\tau_{es}$ and their resulting SNe rates, we vary
the factors f$_{esc}$ and f$_\star$.
%
\begin{figure}[t]
  \begin{center}
    \includegraphics[width=0.6\textwidth]{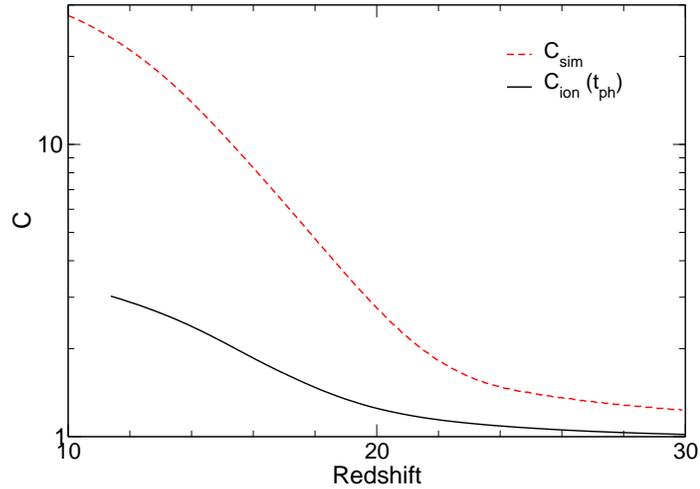}
    \caption[Clumping Factor]{\label{clumping}
      The dashed line is the gas clumping factor as calculated in
      our adiabatic hydrodynamical AMR simulation.  Using equation
      (\ref{clumpchange}), we estimate the clumping factor in the ionized
      region to be solid line.}
  \end{center}
\end{figure}
\subsection{Clumping Factor}
Overdense regions experience an increased recombination rate.
Overdensities are characterized by the gas clumping factor C =
$\langle n_H^2 \rangle / \langle n_H \rangle^2$, and the recombination
rate is increased by this factor.  To calculate this parameter in our
model, we utilize the same cosmological Eulerian AMR code {\em enzo}
\citep{Bryan99} as in \citet{Machacek01} in a 256$^3$ simulation with
a comoving box side of 500kpc, eight levels of refinement, and the
previously specified cosmological parameters.  The top grid has a
total (gas) mass resolution of 1013 (135) $\Ms$, which is smaller
than the cosmological Jeans mass,
\begin{equation}
\label{jeans}
M_J \approx 10^4 \left(\frac{\Omega_M h^2}{0.15}\right)^{-1/2} 
	\left(\frac{\Omega_b h^2}{0.02}\right)^{-3/5}
	\left(\frac{1+z}{11}\right)^{3/2} \Ms .
\end{equation}
Thus, we account for all collapsed halos in our clumping factor
calculation.  The simulation is purely adiabatic with no background
radiation or atomic/molecular cooling.  We show the clumping factor in
our simulation, C$_{sim}$, in Figure \ref{clumping}.  Cooling only
affects localized regions of star formation and does not contribute
greatly in the boosting of C.  However, ionizing radiation causes
photoevaporation of halos, which decreases C in the process
\citep{Haiman01}.  We correct for this process by considering all
minihalos with M$_J$ $<$ M $<$ M$_{min}$ are photo-evaporated in the
ionized regions of the Universe.  For simplicity, we assume the IGM
has a clumping factor of unity although underdense regions in the IGM
correspond to C$_{IGM}$ $\lsim$ 1.  We concentrate on the gas clumping
factor in the ionizing regions since the goal is to calculate the
increase in recombination rates.  Consider an ionized region whose
volume filling fraction F$_H$ is increasing.  The evolution of the gas
clumping factor is 
\begin{equation}
\label{clumpchange}
\dot{C} = C_{sim}\dot{F}_H - \frac{C-1}{t_{ph}}.
\end{equation}
The first term accounts for the expansion that will incorporate
unaffected, clumpy material into the region, and the second term
represents the photoevaporation of overdensities.  Any overdensities
in the ionized region will be photo-evaporated by a background UV flux
in approximately a sound crossing time of the halo,
\begin{equation}
t_{ph} \approx \beta\frac{R_{vir}}{10 \kms},
\end{equation}
where $\beta$ is a normalization factor that accounts for the
differences in halo densities at various redshifts and masses, and
R$_{vir}$ is the mean virial radius of a halo with M$_J$ $<$ M $<$
M$_{min}$ \citep{Haiman01}.  Since $\beta$ remains within $\sim$15\%
of unity for minihalos, we infer $\beta$ = 1.

The effect from photoevaporation is illustrated in Figure
\ref{clumping}.  Overdensities that are engulfed by ionized regions
cannot sufficiently increase the gas clumping factor to overcome the
photoevaporation that occurs, and C remains within the range 1--3 for
the entire calculation in our main model.  This equation is weakly
sensitive to t$_{ph}$ since varying t$_{ph}$ by a factor of 6 alters C
only by a factor of 2.  In a later paper, we shall computationally
address the effect of radiation on the clumping factor.

\subsection{The Evolution of UV Background and Halo Densities}
We initialize the following method at z = 75 with no UV background,
evolve the UV background at the questioned redshift, and repeat the
described procedure until cosmological hydrogen reionization occurs.

Given a minimum mass of a star-forming halo, we can exploit PS
formalism to calculate number densities of these halos.  For
minihalos, we calculate their number densities for halos with masses
above M$_{min}$ (eq. \ref{minMass}) and virial temperatures below
10$^4$ K.  Likewise, the protogalaxy mass fraction is calculated by
considering all halos more massive than a corresponding T$_{vir}$ =
10$^4$ K.

We choose a variant of PS formalism, which is an ellipsoidal collapse
model that is fit to numerical simulations \citep{Sheth99, Sheth01,
Sheth02}.  This model is concisely summarized in \citet{Mo02}.  In
minihalos, it is reasonable to assume one star forms per halo since
E$_{bind}$ $\lsim$ E$_{SNe}$, where E$_{bind}$ and E$_{SNe}$ are the
binding energy of the host halo and kinetic energy of the primordial
SNe, respectively.  The gas is totally disrupted in the halo and
requires $\sim$100 Myr, which is approximately the Hubble time at
cosmological reionization, to re-collapse (Abel, Bryan, \& Norman
2002, unpublished).  Therefore,
\begin{equation}
\label{rho_star}
\frac{d\rho_\star}{dz} = \frac{d\rho_{mini}}{dt} \frac{dt}{dz},
\end{equation}
where $\rho_\star$ is the comoving density of primordial stars.

We evolve the spectrum from early stars to investigate how the UV
background behaves, particularly in the LW band, with the cosmological
radiative transfer equation,
\begin{equation}
\label{cosmo_rte}
\left( \frac{\partial}{\partial t} - \nu H \frac{\partial}{\partial
  \nu} \right) J = -3HJ - c\kappa J +
  \frac{c}{4\pi}\epsilon,
\end{equation}
where $H = H(z) = 100h E(z)$ is the Hubble parameter, and $E(z) =
\sqrt{\Omega_\Lambda + \Omega_m (1+z)^3}$.  $J = J(\nu,z)$ is specific
intensity in units of \emis.  $\kappa$ is the continuum
absorption coefficient per unit length, and $\epsilon = \epsilon(\nu)$
is the proper volume-averaged emissivity in units of \emissivity
\citep{Peebles93}.  We define the luminosity of the objects as
\begin{displaymath}
L(\nu) = \left\{ \begin{array}{r@{\quad}l} 
4\pi R^2 B_\nu(T=10^5K) & {\rm (mini)}\\
f_{O\star} f_\star \mathcal{L} \frac{B_\nu(T=23000K)}{B_\nu(\nu = 2.7kT/h_p; \: T=23000K)} & {\rm (proto)}
\end{array} \right.,
\end{displaymath}
where B$_\nu$ is a blackbody spectrum, and $\cal{L}$ is listed in
Table 1.  R is the radius of the primordial star \citep{Schaerer02}.
f$_{O\star}$ is the fraction of O stars in the starburst
\citep{Schaerer03}.  k and h$_{\rm{p}}$ are Boltzmann's constant and
Planck's constant, respectively.  For the protogalaxies, T $\sim$
23000K because the spectrum in the LW band will be dominated by OB
stars, and we weight the luminosity by this blackbody spectrum.
Emissivity will be nearly zero above 13.6eV in the neutral Universe
due to hydrogen and helium absorption.

Photons from primordial stars and protogalaxies between 13.6eV and
several keV ionize the surrounding neutral medium.  Using the
intrinsic $\geq$13.6eV ionizing photon rates from primordial stars
(eq. \ref{ion_mini}) and protogalaxies (eq. \ref{dn_dt}), we calculate
a volume-averaged neutral fraction.  The change of ionized hydrogen
comoving density due to photo-dissociation and recombination is
\begin{equation}
\label{neutral_h}
\frac{dn_{HII}}{dt} = -\frac{dN_\gamma}{dt} + C k_{rec} (1+z)^3 n_H^2
\left(1-\frac{n_{HII}}{n_H}\right)^2,
\end{equation}
where n$_{HII}$ and n$_H$ = 0.76$\rho_0$/m$_p$ are the ionized and
total hydrogen comoving density; dN$_\gamma$/dt is the rate of
ionizing photons from primordial stars and protogalaxies; C is the
clumping factor; and k$_{rec} =$ 2.6 $\times$ 10$^{-13}$ s$^{-1}$
cm$^{-3}$ is the case B recombination rate of hydrogen at T $\approx$
10$^4$.

In the case for once-ionized helium, $k_{rec}$ is approximately equal
to the hydrogen case, but since the helium number density is less than
hydrogen, less recombinations occur.  Na\"{\i}vely, this will result
in a higher He$^+$ fraction than H$^+$ due to the hardness of the
primordial radiation.  Realistically when the ionizing photons reach
the ionization front, they will ionize either helium or hydrogen since
hydrogen still has a finite photo-ionization cross-section above
23.6eV.  We consider the He$^+$ regions to be equal to the H$^+$
regions.  Then we add N$_{\gamma,\:He}$ to N$_{\gamma,\:H}$ to
compensate for this effect.  We perform a similar analysis on doubly
ionized helium in which k$_{rec,\: He^+}$ = 1.5 $\times$ 10$^{-12}$
s$^{-1}$ cm$^{-3}$, but we evaluate equation (\ref{neutral_h})
directly with N$_{\gamma, \:He^+}$.
\begin{figure}[t]
  \begin{center}
    \includegraphics[width=0.6\textwidth]{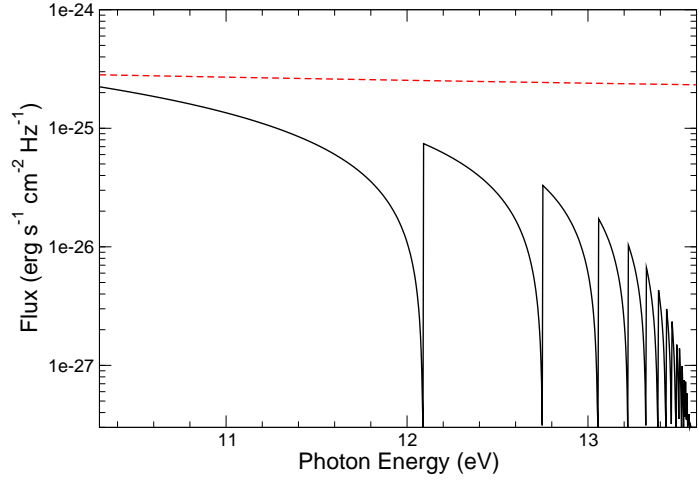}
    \caption[Sawtooth Spectrum in the Lyman-Werner band]{\label{saw}
      The red dashed line is an example unprocessed spectrum of a
      continuous source of radiation.  After being absorbed and re-emitted
      by Lyman-series transitions, it transforms into the ``sawtooth''
      spectrum ({\em black solid line}).  This particular spectrum is for
      z$_{on}$ = 30 and z$_{obs}$ = 20.}
  \end{center}
\end{figure}
The absorption coefficient in equation (\ref{cosmo_rte}) is ignored
since we take into account the Lyman series line absorption by the
following procedure.  Photons with 11.26eV $<$ E $<$ 13.6eV escape
into the IGM, which will photo-dissociate \hh.  To calculate the flux
within the LW band, we must consider the processing of photons by the
Lyman series transitions \citep{Haiman97b}.  Before reionization,
these transitions absorb all photons at their respective energies.
These absorbers can be visualized as optically thick ``screens'' in
redshift space, for which the photon must have been emitted after the
farther wall in redshift.  This process creates a sawtooth spectrum
with minimums at redshift screens, which is illustrated in Figure
\ref{saw}.  For instance, an observer at $z = 20$ observes photons at
12.5eV; it must have been emitted by the Ly$\gamma$ line at 12.75eV at
$z = 20.4$.  The fraction of photons that escape from these walls
during an integration step is
\begin{equation}
\label{lyman_abs}
f = \frac{1 - [(1+z)/(1+z_{screen})]^{1.5+\alpha}}
{1 - [(1+z)/(1+z+\Delta z)]^{1.5+\alpha}},
\end{equation}
where $\alpha$ = --1.8 is the slope of a power-law spectrum ($F_\nu
\propto \nu^{-\alpha}$) and
\begin{equation}
\label{abs_wall}
z_{screen} = \frac{\nu}{\nu_{Lyi}}(1+z_{obs}) - 1.
\end{equation}
$\nu_{Lyi}$ is the nearest, blueward Lyman transition to $\nu$.  Inherently,
$z_{screen}$ must be between $z$ and $z+\Delta z$.

Considering the processes described above, we finally evolve the UV
background with the volume averaged emissivity,
\begin{equation}
\label{ems_eqn1}
\epsilon_{mini}(\nu) = \rho_{mini} \times L(\nu) \times \left(1 -
   \frac{n_{HII}}{n_H}\right),
\end{equation}
\begin{equation}
\label{ems_eqn2}
\epsilon_{proto}(\nu) = \nu^{-1} \rho_0 \frac{d\psi_{proto}}{dt}
\times L(\nu, f_\star),
\end{equation}
\begin{equation}
\label{ems_eqn3}
\epsilon(\nu) = (1+z)^3 \left[\epsilon_{mini}(\nu) +
\epsilon_{proto}(\nu)\right] \times f(\nu).
\end{equation}

\section{Results}
\label{sec:results2}
In this Section, we present the results of our calculation of the
evolution of SNe rates and the UV background.  We also present the
variance of optical depth to electron scattering and SNe rates with
different primordial and protogalaxy star formation scenarios.  In
Figure \ref{evo_plot}, we plot minimum halo masses that host
primordial stars, densities of those halos, SNe rates, and the UV
background in the LW band for the main models.

\begin{figure}[t]
  \begin{center}
    \vspace{0.4cm}
    \includegraphics[width=\textwidth]{chapter2/fig/f3.eps}
    \caption[Primordial SNe rates]{\label{evo_plot} ({\em Clockwise from
        upper right}) Primordial SNe rates (yr$^{-1}$ deg$^{-2}$) per
      unit redshift; Cumulative primordial SNe rate (yr$^{-1}$
      deg$^{-2}$); Specific intensity (\emis) in the LW band; Comoving
      density (Mpc$^{-3}$) of halos above the critical star formation
      mass in neutral regions; Critical halo mass ($\Ms$) for primordial
      star formation.  The solid, dotted, and dashed lines correspond to
      calculations run with a fixed primordial stellar mass of 100, 200,
      and 500 $\Ms$, respectively.}
  \end{center}
\end{figure}
\subsection{Optical Depth to Electron Scattering}
Optical depth due to electron scattering, 
\begin{equation}
\label{thomson}
\tau_{es}(z) = \int^z_0 \bar{n}_e \sigma_{TH} c \left(\frac{dt}{dz}\right)
dz,
\end{equation}
where $n_e$ and $\sigma_{TH}$ are the proper electron density and
Thomson cross-section, respectively, is a good observational test to
determine the neutral fraction of the Universe before reionization.
The WMAP satellite measured $\tau_{es}$ = 0.17 $\pm$ 0.04 at 68\%
confidence.

\begin{figure}[t]
  \begin{center}
    \includegraphics[width=0.6\textwidth]{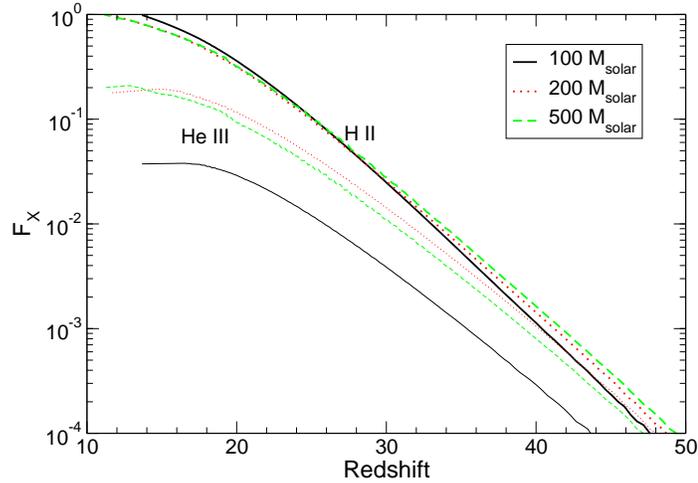}
    \caption[Evolution of ionized filling factors]{\label{filling}
      Evolution of filling factors of ionized hydrogen (top) and
      doubly ionized helium (bottom).  The legend is the same as Figure
      \ref{evo_plot}.}
  \end{center}
\end{figure}

To accurately calculate $n_e$, we must
consider all ionizations of primordial gas, which includes H$^+$,
He$^+$, and He$^{++}$.  Therefore, the proper electron density is
\begin{equation}
n_e = (1+z)^3 \left(n_{H}F_{H^+} + n_{He}F_{He^+} +
2n_{He}F_{He^{++}}\right),
\end{equation}
where F$_{H^+}$ F$_{He^+}$, and F$_{He^{++}}$ are the ionized volume
fraction of H$^+$, He$^+$, and He$^{++}$, respectively.  The effect of
more luminous primordial stars is evident in Figure \ref{filling} as
they ionize the Universe faster at high redshifts.  To match the WMAP
result, less ionizing photons are required from protogalaxies if the
primordial IMF is skewed toward higher masses.  Although these models
have the same total $\tau_{es}$, cosmological reionization occurs at z
= 13.7, 11.6, 11.2 for M$_{FS}$ = 100, 200, and 500$\Ms$,
respectively.  However, these reionization redshifts are not
consistent with observed Gunn-Peterson troughs in z $\sim$ 6 quasars
\citep{Becker02, Fan02} and the high IGM temperatures at z $\sim$ 4
inferred from Ly$\alpha$ clouds \citep{Hui03}.  Perhaps portions of
the Universe recombine after complete reionization, which will match
the most distant quasar observations \citep{Cen02}.  If this were
true, the first reionization epoch has to be faster and earlier to
compensate for this partial recombination and to match the WMAP
result.  This would lower our SNe rates slightly since primordial star
formation will be further suppressed by the quicker reionization.

The ionizing history of the Universe is directly related to the number
of ionizing photons that are produced and escape into the IGM.  If we
fix $\tau_{es}$ = 0.17, we constrain f$_\star$ and f$_{esc}$ to a
power law
\begin{equation}
f_{esc} = B \: f_\star^{-a},
\end{equation}
where B = [0.00307, 0.00239, 0.00217] and a = [0.906, 0.839, 0.806]
for primordial stellar masses 100, 200, and 500$\Ms$, respectively.
The flattening of the power law with increasing M$_{FS}$ indicates the
increasing ionizing contribution from primordial stars.

\begin{center}
\begin{longtable}{ccccc}
\caption{Gaussian fits to M$_{{\rm min}}$ for the main
  models} \label{gaussFit}\\

\hline\hline \\[-3ex]
M$_\odot$ & A & z$_0$ & $\sigma$ & C \\
\hline
\endhead

100 & 1.67 $\pm$ 0.01 & 10.0 $\pm$ 0.2 & 21.9 $\pm$ 0.1 & 5.19 $\pm$ 0.01 \\
250 & 1.71 $\pm$ 0.00 & 10.5 $\pm$ 0.0 & 23.3 $\pm$ 0.0 & 5.16 $\pm$ 0.00 \\
500 & 1.72 $\pm$ 0.01 & 11.3 $\pm$ 0.2 & 23.5 $\pm$ 0.1 & 5.15 $\pm$ 0.00 \\
\hline
\end{longtable}
\tablecomments{Coefficients are for the function log(M) = A $\times$
  exp[-(z-z$_0$)$^2$/(2$\sigma^2$)] + C.}
\end{center}

\subsection{Primordial Supernovae Rates}
The natural unit in our computation is comoving density, yet a more
useful unit is observed SNe yr$^{-1}$ deg$^{-2}$.  First we assume
these SNe are bright for 1 yr and then correct for time dilation.  We
consider the equation
\begin{equation}
  \label{rates}
  \frac{d^2N}{dtdz} = \frac{dV_c}{dz} \frac{d\rho_\star}{dt} (1 + z)^{-1},
\end{equation}
where the (1+z)$^{-1}$ converts the proper SNe rate into the observer
time frame, and
\begin{equation}
  \label{volume}
  \frac{dV_c}{dz} = D_H \frac{(1+z)^2 D_A^2}{E(z)} \: \Omega
\end{equation}
is the comoving volume element.  $\Omega$ (deg$^2$ = 3.046 $\times$
10$^{-4}$ sr) is the solid angle of sky that we want to sample.  $D_H
= c/H_0$ is the Hubble distance.  $D_A = D_M/(1+z)$ is the angular
diameter distance, and $D_M = D_H \int^z_0 E^{-1}(z') dz'$ is the
comoving distance \citep{Peebles93}.  The above equations are only
valid for a flat $\Lambda$CDM universe.

Since we only allow primordial stars to form in neutral regions, SNe
rates from these stars are highly dependent on the f$_{esc}$ and
f$_\star$, but less sensitive to the stellar primordial mass,
M$_{fs}$.  In our main models, the SNe rate varies little with
primordial stellar mass and is 0.34 yr$^{-1}$ deg$^{-2}$.  Even if we
vary f$_\star$ in a range 0.01--0.1 and fix $\tau_{es}$, primordial
SNe rates do not vary more than 10\% from the main models when
constrained by WMAP.  As f$_{esc}$ increases, the SNe rates decrease
due to a higher ionized volume filling factor.  The effect of a higher
f$_\star$ squelches SNe rates by two processes, a higher ionized
filling factor and higher UV background, which restricts primordial
stars to form in more massive halos.

The primordial SNe rate peaks at z $\sim$ 14--20, earlier epochs for
larger primordial stellar masses, and falls sharply afterwards due to
the ensuing cosmological reionization.  Primordial star formation
ceases after z $\sim$ 12--16.  The combination of an increasing UV
background, reionizing Universe, and disruption of minihalos from
primordial SNe suppresses all primordial star formation.  For each
main model, we fit a Gaussian curve to the minimum mass of a minihalo
that can host a primordial star, and the parameters are listed in
Table \ref{gaussFit} and are valid for z $>$ z$_0$.

\begin{figure}[t]
  \begin{center}
    \includegraphics[width=0.6\textwidth]{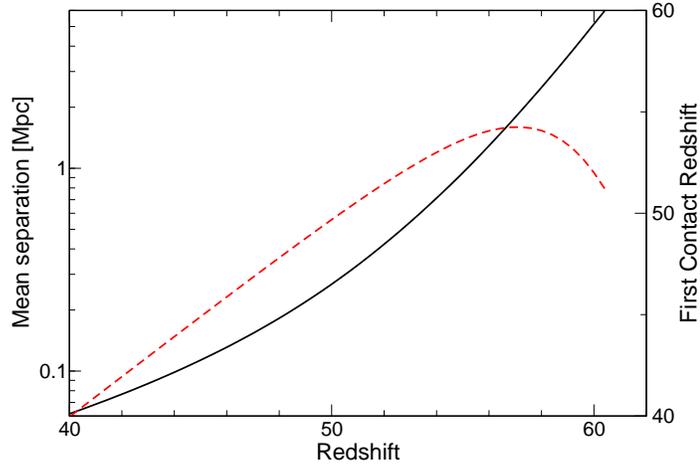}
    \caption[Earliest epoch of negative feedback]{\label{contact} The
      {\em solid} line is the mean proper separation between
      minihalos.  The {\em dashed} line is the redshift where
      minihalos first receive radiation from neighboring sources
      forming at the same epoch.}
  \end{center}
\end{figure}

It is fascinating that some rare primordial SNe occur at z $\gsim$ 40.
As an exercise, we estimate the earliest epoch of minihalo star
formation in the visible Universe with PS formalism and by considering
it takes $\sim$9.33 Myr for a halo to form a protostellar core
\citep{Abel02a}.  With PS formalism, the ``first'' epoch equals where
the minihalo density is the inverse of the comoving volume inside z =
1000 (10523 Gpc$^3$).  Then we include an additional 9.33 Myr for the
ensuing star formation.  The halo masses of 1.74 $\times$ 10$^5$ and
10$^6$ $\Ms$ correspond to formation times of z $\sim$ 71 and 64,
respectively.  Another interesting event to calculate is when the SNe
rate equals one per sky per year, which occurs at z $\sim$ 51 when
considering the collapse and star formation timescales.  This epoch is
in agreement with the \citet{Escude03} estimate of z $\simeq$ 48.

Furthermore using the proper minihalo density, we calculate the
average light travel time between sources, which indicates when
negative \hh~feedback first affects star formation.  Star formation
occurs $\sim$9.33 Myr after halo formation, so radiation escapes into
the IGM at a time t$_{\rm{rad}}$ = t$_{\rm{H}}$(z$_{\rm{form}}$) +
9.33 Myr, where z$_{\rm {form}}$ is the halo formation redshift.
These sources have a mean proper separation of d = $\rho_\star^{-1/3}$
/ (1+z); therefore the time where radiation influences other halos is
t$_{\rm{rad}}$ + d/c.  The minimum of this time, corresponding to z =
54, is where radiation reaches other halos for the first time.  We
illustrate the mean proper separation and epoch of first radiation
effects in Figure \ref{contact}.

\begin{figure}[t]
  \begin{center}
    \includegraphics[width=0.6\textwidth]{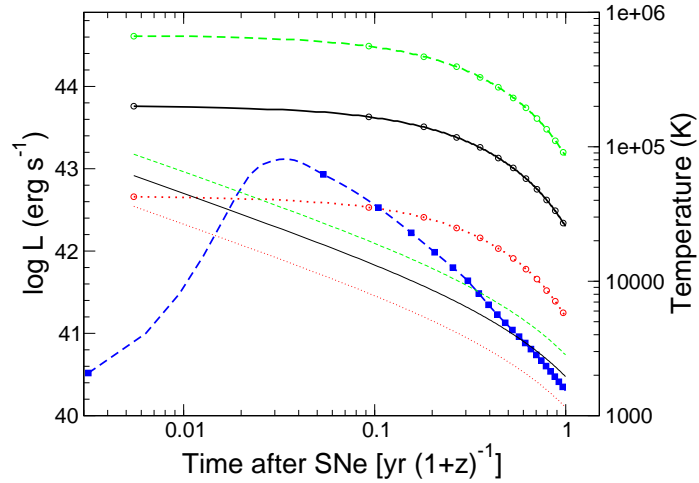}
    \caption[Primordial SNe Luminosities]{\label{sne_lum}Using the decay
      of $^{56}$Ni and Equation (\ref{sne_temp}), we calculate the
      luminosity of primordial SNe and their effective temperature.  Heavy
      ({\em top}) lines with circles are luminosities and light ({\em
        bottom}) lines are temperature.  The blue dashed line with
      filled squares is a Type Ia luminosity evolution for comparison
      \citep{Woosley86}.  The lines from top to bottom (\textit{ dotted,
        solid, dashed}) are for stellar masses 175, 200, and 250 \Ms.}
  \end{center}
\end{figure}

\subsection{Magnitudes and Observability}
The magnitudes of primordial SNe are as important as their occurrence
rates to catch these events unfolding in the distant universe.  We
exploit the $^{56}$Ni output from metal-free SNe models to calculate
luminosities L(t) from the two-step decay of $^{56}$Ni to $^{56}$Fe
\citep{Heger02a}.  We consider the emission spectrum to be a blackbody
spectrum with a time-dependent temperature,
\begin{equation}
\label{sne_temp}
T(t) = \left[\frac{L(t)M}{8\pi\sigma Et^2}\right]^{1/4} (1+z),
\end{equation}
where M and E are the stellar mass and kinetic energy of the SNe,
respectively, calculated using free expansion arguments.  The
temperature and luminosity of a 175, 200, and 250$\Ms$ primordial
star SNe are depicted in Figure \ref{sne_lum}.  For comparison, we
plot the typical light curve for a Type Ia SN.  Kinetic energy is
taken from the SNe models \citep{Heger02a}.  We consider the SNe
remnant to be in free expansion for the first year because radiation
from the primordial star should expel most of surrounding medium to
create a low density, highly ionized region of approximately 100 pc in
size for a 120$\Ms$ primordial star \citep{Whalen04}.

\begin{figure}[t]
  \begin{center}
    \includegraphics[width=0.6\textwidth]{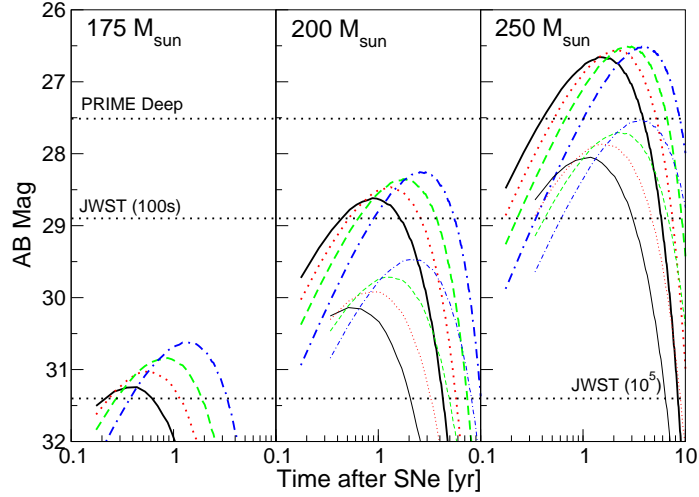}
    \caption[Primordial SNe Observability]{\label{mags} ({\em Left to
        right}) Magnitudes for M$_{FS}$ = 175, 200, 250 $\Ms$.  The
      dashed horizontal lines indicate limiting magnitudes of space
      infrared observatories.  The heavy ({\em top}) and light ({\em
        bottom}) lines are SNe at z = 15 and 30, respectively.  The
      magnitudes for the 175 $\Ms$ and z = 15 case is not shown since
      it is too dim.  The solid, dotted, dashed, and dash-dotted lines
      are for Spitzer wavebands centered at (3.56, 4.51, 5.69, 7.96)
      $\mu$m. {\em Note:} The detection limits of Spitzer and PRIME
      Medium are not shown since they are too high at 24.5 and 25.6,
      respectively.}
  \end{center}
\end{figure}

In the upper range (M $>$ 200$\Ms$) of pair-instability SNe, these
events produce 1--57$\Ms$ of $^{56}$Ni \citep{Heger02a}, which will
produce tremendous luminosities when compared to typical Type II SNe
outputs of only 0.1--0.4 $\Ms$ \citep{Woosley86}.  However,
uncertainty in the primordial IMF places doubt on the frequency of
primordial star SNe with high $^{56}$Ni yields.  Using conventional
Type II SNe parameters of L $\simeq$ 3 $\times$ 10$^{42}$ erg
s$^{-1}$, T = 25000K (2 days $\lsim$ t $\lsim$ 7 days), and T = 7000K
(7 days $\lsim$ t $\lsim$ 2 months) that are tuned by observed light
curves, \citet{Jordi97} determine apparent magnitudes that are 1--3
mag lower than our values, which are brighter due to the greater
nickel production of pair-instability SNe models.  Their model should
apply to the pair-instability SNe with little $^{56}$Ni ejecta.
Furthermore, our temperatures are higher due to the greater amount of
$^{56}$Ni decay in pair-instability SNe.

In typical Type II light curves, radioactive decay does not
significantly contribute to the luminosity in the plateau stage
\citep{Popov93}.  In primordial SNe, the $^{56}$Ni decay may overwhelm
the typical sources of energy within the expanding fireball and create
a totally different light curve.  We stress that our light curve is a
very rough estimate of the processes occurring within a primordial
SNe.

In Figure \ref{mags}, we compare maximum AB magnitudes of low- and
high-mass pair-instability SNe at various redshifts and sensitivities
of space infrared observatories.  We consider the sensitivities of
SIRTF at 3.5\micron, PRIME medium and deep surveys, and JWST with
exposure times of 100s and 10$^5$ s.  Our simple estimate leads to
very high predicted luminosities, and more detailed numerical models
of these explosions are clearly desirable.

\section{Discussion}
\label{sec:discussion2}
Other factors could alter the feasibility of observing primordial
SNe. For instance, $\sim$2\% of primordial stars might die in
collapsar gamma-ray bursts (GRBs) \citep{Heger02b}.  \citet{Lamb99}
also provides a ratio of GRBs to Type Ib/c SNe rates that is $\sim$3
$\times$ 10$^{-5}$ ($\theta_b$ / 10$^{-2}$)$^{-1}$, where $\theta_b$
is the beaming angle.  However, this estimate is for a normal Salpeter
IMF, and as mentioned before, the primordial star IMF could be skewed
toward high masses, which would increase the probability of a GRB.
Zero- and low-metallicity massive stars outside the pair-instability
mass range die as GRBs or jet-driven SNe, which are similar to GRBs
but not as energetic and spectrally hard.  If we consider a
proportionality constant, $f_{GRB}$, and beaming angle, we estimate
that the all-sky primordial GRB rate,
\begin{equation}
R_{GRB} = 2.8 \left(\frac{\theta_b}{0.01}\right)
\left(\frac{f_{GRB}}{0.02}\right) \; {\rm GRBs} \; {\rm yr}^{-1}.
\end{equation}
In M $\lsim$ 140$\Ms$ stars, a collapsar results when it forms a
proto-neutron stars, cannot launch a supernova shock, and directly
collapses to form a black hole in $\sim$1 s \citep{Woosley93,
MacFadyen99}.  In M $\gsim$ 260$\Ms$ stars, it does not create a
proto-neutron star and directly collapses into a massive black hole
\citep{Fryer01}.  Afterwards, these black holes can accrete gas and
produce X-rays that can further ionize the Universe
\citep{Ricotti03b}.  A few of these X-ray sources could be detected in
the Chandra deep field \citep{Ricotti05, Alexander03}.  In most GRB
models \citep[for an overview, see][]{Meszaros02}, the radiation is
beamed to small opening angles due to relativistic effects, which
would render a fraction of GRBs to be unobservable in our perspective.
Finally, gravitational lensing will significantly decrease the
apparent magnitudes for selected primordial SNe for very small survey
areas.  However, the overall magnitude distribution is slightly dimmed
by gravitational lensing \citep{Marri98, Marri00}.

Deviations in the amplitudes of fluctuations, $\sigma_8$, and a
running spectral model \citep{Liddle92, Liddle93} also affect
primordial SNe rates.  We ran a set of models with $\sigma_8$ = 0.8,
and predictably, the rates decreased by a factor of $\sim$2.5 due to
lesser powers at small mass scales.  When we match $\tau$ = 0.17,
rates range from 0.12 to 0.2 yr$^{-1}$ deg$^{-2}$ with the rates
peaking earlier at z $\sim$ 20, but now the redshift peaks are nearly
independent of redshift with only $\delta$z $\approx$ 1 separating the
100 and 500$\Ms$ models.  If we keep the protogalaxy parameters from
the main models, rates do not change; however, the rates peak later
than our main models by $\delta$z $\approx$ 4, and $\tau$ lowers to
0.14.  According to inflationary models, the spectral index of
fluctuations should be slowly varying with scale.  While analyzing
WMAP data, \citet{Peiris03} determined that the fluctuation amplitude
is significantly lower at small scales.  With lesser powers at small
scales, primordial SNe rates decrease such as the case of a lower
$\sigma_8$.  In principle, studying primordial SNe occurrences with
respect to redshift could furnish direct constraints on the power
spectrum at these small scales.

We have demonstrated our model's dependence on the free parameters,
$f_{esc}$, $f_\star$, and M$_{fs}$.  We now suggest several
observational and numerical methods in order to constrain our models.
To constrain early star formation history of dwarf irregulars with the
[$\alpha$/Fe] versus [Fe/H] comparison, metal analyses of stars
with [Fe/H] $<$ --1.5 are needed to determine early star formation
rates of particular systems \citep[see Figure 1.5 in][]{Venn03}.
Dwarf spheroidals exhibit greater variance in star formation histories
from system to system, but a similar study will increase our knowledge
of high redshift star formation in these small galaxies to further
constrain $f_\star$ at high redshifts.  To improve on these studies,
the advances in multi-object spectrographs (e.g.
VIMOS\footnote{http://www.eso.org/instruments/vimos/},
FLAMES\footnote{http://www.eso.org/instruments/flames/}, and
GMOS\footnote{http://www.gemini.edu/sciops/instruments/gmos/gmosIndex.html})
enable the gathering of many stellar spectra in one exposure.  This
will greatly increase the stellar population data of dwarf galaxies.

Since some massive stars die as a long-duration GRB, these events can
convey information from the death of Pop III stars.  The propagation
of the initial burst and afterglow provide information about the total
energy, gas density in the vicinity, and the Lorentz factor of the
beam.  The host galaxy ISM properties will help constrain the
$f_{esc}$ in high redshift galaxies.  As in the case of GRB 030329
\citep{Stanek03}, the power-law spectra of the afterglow can be
subtracted to obtain a residual that resembles a typical SN spectrum,
which may be used to roughly determine the mass of the progenitor.
Observing the afterglow is necessary to determine its redshift.  The
prospect of observing prompt afterglows will be accomplished easier
with {\em Swift}\footnote{http://swift.gsfc.nasa.gov/}, which can
possibly detect GRB afterglows to z = 16 and 33 in the K and M bands,
respectively \citep{Gou04}.  Furthermore, $\gsim$50\% of GRBs occur
earlier than z = 5, and 15\% of those high redshift GRBs are
detectable by {\em Swift} \citep{Bromm02b}.  The comparison to nearby
SNe can provide crucial information about the high redshift ISM.
Finally, it is a possibility to explore the intervening absorption with
the fast-pointing and multi-wavelength observations of {\em Swift}
\citep{Vreeswijk03, Loeb03, Barkana04}.  This IGM absorption would
constrain the reionization history of the Universe better, which may
change the primordial SNe rates, but more specifically the rate per
unit redshift, which would roughly conform to the filling factor
evolution.

The radiation from protogalaxies and primordial stars will not only
ionize the Universe but also contribute to the near-infrared
background (NIRB).  Calculations have shown that primordial stars can
provide a significant fraction of radiation to the NIRB
\citep{Santos02, Salvaterra03}; however, the paradigm of early
reionization set by WMAP was not considered at the time.  f$_\star$ of
protogalaxies and densities of primordial stars can be further
constrained if future studies of the NIRB consider the large
emissivities of zero- and low-metallicity sources at z $\gsim$ 15.

Other outlooks include detecting high redshift radio sources and
searching for 21cm absorption and emission \citep{Hogan79, Scott90,
Iliev03, Furlanetto04}.  By looking for 21cm signatures, observations
would be directly probing the neutral regions of the Universe since
the Gunn-Peterson trough saturates only at a neutral filling factor of
$\sim$10$^{-5}$.  Lastly, additional searches for high redshift
starbursts \citep[e.g.][]{Ellis01, Pello04} in lensed fields will
furnish an understanding of the characteristics of these objects and
help tighten our models of early ``normal'' and primordial star
formation.

Metal-free stars with masses between $\sim$140--260$\Ms$ result in a
SNe that is visible and eject heavy elements into the IGM.  When a
star is within this range, the stellar core has sufficient entropy
after helium burning to create positron/electron pairs.  These pairs
convert the gas energy into mass while not greatly contributing to
pressure.  This creates a major instability, where the star contracts
rapidly until oxygen and silicon implosive burning occurs.  Then the
star totally disrupts itself by these nuclear explosions and leaves no
remnant \citep{Barkat67, Bond84}.  Stars between $\sim$100--140$\Ms$
experience this instability.  It is not violent enough to disrupt the
star, but pulsations and mass loss occur until equilibrium is reached
and the hydrogen envelope is ejected \citep{Bond84, Heger02a}.

Metal production from primordial stars cannot account for the volume
filling factor of metals as seen in Ly$\alpha$ clouds in current
models \citep{Norman04}.  To further test this work, we determine a
volume-averaged metallicity of [Fe/H] $\lsim$ --4.1 using our SNe
densities and the metal production of pair-instability SNe
\citep{Heger02a}.  Therefore, ubiquitous star formation in
protogalaxies most likely polluted the sparse regions of the Universe.
Combining the primordial and protogalaxy metal output, simulations
with proper metal transport should agree with the abundances observed
in Ly$\alpha$ clouds.  Such simulations that takes into account both
of these metal sources and matches the results of \citet{Songaila96}
and \citet{Songaila01} is needed to constrain star formation before
reionization.

Numerical simulations are also needed to investigate radiative
transfer from a protogalaxy stellar population.  This scenario will
contain more complexities than a single primordial star in a spherical
halo as in \citet{Whalen04}.  With a protogalactic simulation with the
Jeans length resolved, star formation and feedback, and radiative
transfer, it will be possible to study the evolution of the ISM in a
protogalaxy, which will produce insight and better constraints on the
photon escape fraction, $f_{esc}$, in the Lyman continuum at high
redshifts.  \citet{Ricotti02a, Ricotti02b} thoroughly study radiative
transfer around protogalaxies; however, $f_{esc}$ is still a parameter
when it should be determined from the radiative transfer results in
the simulations.  Ideally, the analytical ideas about ionizing the
Universe in \citet{Madau95, Haardt96, Madau99} should be realized in
such simulations but in the context of the current paradigm of a high
redshift reionization as indicated by WMAP.  Also we may hope by using
numerical radiative transfer techniques for line and continuum
radiation in three dimensions to push the simulations of \citet{Abel02a}
to follow the entire accretion phase of the first stars.  This would
lead to stronger constraints on the masses of the very first stars.

\section{Summary}
\label{sec:summary2}
The WMAP measurement of optical depth to electron scattering places a
constraint of early cosmological reionization of the universe, which we
show to be mainly from star formation in protogalaxies.  This general
result is in agreement with other studies of reionization after WMAP
\citep[e.g.][]{Cen03, Somerville03, Ciardi03, Ricotti03a, Sokasian04}.
\begin{itemize}
\item The radiation from protogalaxies squelches primordial star
formation, and {\em $\sim$0.34 primordial SNe deg$^{-2}$ yr$^{-1}$}
are expected.  SNe rates can vary from 0.1 to $>$1.5 deg$^{-2}$
yr$^{-1}$ depending on the choice of primordial stellar mass and
protogalaxy parameters while still constrained by the WMAP result.
The peak of SNe rate occurs {\em earlier} with {\em increasing
primordial stellar masses}.  These results are upper limits since the
rate of visible primordial SNe depends on the IMF because only a
fraction will lie in the pair instability SNe mass range.  The other
massive primordial stars might result in jet-driven SNe or long
duration GRBs.
\item Stellar metal abundances and star formation in local dwarf
galaxies aid in estimating protogalaxy characteristics.  We choose
f$_\star$ = 0.04 and f$_{esc}$ = 0.050 in our 100 $\Ms$ model.  In
protogalaxies, the star formation efficiency is slightly lower than
local values, but the photon escape fraction is within local observed
fractions.
\item Primordial stars enrich the IGM to a maximum volume-averaged
[Fe/H] $\simeq$ --4.1 if the IMF is skewed toward the pair-instability
upper mass range.  A proper IMF will lessen this volume-averaged
metallicity since only a fraction of stars will exist in this mass
range.
\item The entire error bar of the WMAP measurement of optical depth to
electrons can be explained by a higher/lower f$_\star$ and f$_{esc}$.
Protogalaxies can ionize the IGM easily since low metallicity
starburst models produce 20--80\% more ionizing photons than previously
used (Z = 0.001) IMFs.  Massive primordial stars provide $\sim$10\% of
the necessary photons to achieve reionization.  No exotic processes or
objects are necessary.
\item Only the upper mass range of pair instability SNe will be
observable with JWST since the low mass counterparts do not produce
enough $^{56}$Ni to be very luminous.
\end{itemize}

Although the IMF of primordial stars is unknown, simulations have
hinted that a fraction of metal-free stars may exist in the
pair-instability mass range.  When observational rates, light curves,
and spectra are obtained from future surveys, these data would provide
very stringent constraints on the underlying CDM theory as well as our
understanding of primordial star formation.

\chapter[Baryonic Virialization]{The Virialization of Baryons in
  Protogalaxies}
\label{chap:virialization}

Galaxy formation entails a complex interplay of many processes,
e.g. star formation, hydrodynamics, central black holes, hierarchical
merging, and non-local radiative feedback.  The philosophy of the rest
of this thesis aims to isolate the effects of each relevant process in
galaxy formation.  In this light, we start with a set of control
simulations that use a simple model of galaxy formation.  We consider
the hydrodynamics and gas chemistry of early galaxy formation.  These
calculations provide a basis for comparison against more complicated
simulations that include primordial star formation and feedback.
Nevertheless even in the absence of star formation and feedback, the
gas dynamics of a hierarchically assembling and virializing object
proves to be a rich subject, which we detail in this chapter.

This chapter has been submitted to \textit{The Astrophysical Journal}
for publication.  This paper is co-authored by Tom Abel, who suggested
many of the analysis techniques and their physical interpretations.

\section{Introduction}

The process of virialization is clearly fundamental to all scales of
galaxy formation.  \citet{LB67} demonstrated that violent relaxation
occurs during the virialization of a dissipationless system, but does
the equivalent occur for the baryonic matter?  If it does, how it
achieves virial equilibrium should be inherently different because of
hydrodynamical effects and radiative cooling.  Additionally, this
would create a Maxwellian velocity distribution for the baryons as
well.  Turbulent velocities would exceed rotational ones.  This would
be at odds with the standard galaxy formation theories, which
generally assume smooth rotating gaseous distributions
\citep[e.g.][]{Crampin64, Fall80, Mo98}.  The first occurrence of
widespread star formation can be regarded as the commencement of
galaxy formation, and its feedback on its host will affect all
subsequent star formation.  It is crucial to model the initial stage
of galaxy formation accurately.  Differences in initial configurations
of a collapsing halo may manifest itself in different types of central
luminous objects, whether it be a stellar disk \citep{Fall80, Mo98}, a
starburst \citep[see \S4 in][for a review]{Kennicutt98}, or a massive
black hole \citep{Bromm03, Volonteri05, Spaans06, Begelman06}.  These
differences may result from varying merger histories and the ensuing
virial heating or turbulence generation of the new cosmological halo.

For galaxy clusters, cosmological virialization has been studied
extensively \citep{Norman99, Nagai03a, Nagai03b, Schuecker04,
  Dolag05}.  It is customary to connect the velocity dispersion to a
temperature through the virial theorem for a collisionless system,
where the potential energy equals twice the kinetic energy.  However
X-ray observations and such cosmological simulations of galaxy
clusters have indicated that turbulent energies are comparable to
thermal energies.  Central turbulent pressure decreases the density,
but the temperature is largely unchanged.  This leads to an increased
entropy and a flatter entropy radial profile \citep{Dolag05} that is
in better agreement with X-ray observations \citep[e.g.][]{Ponman99}.
Simulations of merger dynamics suggest that turbulence is mostly
generated in Kelvin-Helmholtz instabilities between bulk flows and
virialized gas during minor mergers \citep[e.g.][]{Ricker01,
  Takizawa05}.  Alternatively turbulence can be generated by
conduction \citep{Kim03, Dolag04} or acoustic transport of energy
\citep{Norman99, Cen05}.

In standard galaxy formation models \citep{Rees77, Silk77, White78,
  Blumenthal84, White91, Mo98}, gas shock-heats to the virial
temperature as it falls into DM halos.  These models succeed with
considerable accuracy in matching various observables, such as star
formation histories, galaxy luminosity functions, and the Tully-Fisher
relationship \citep{White91, Lacey91, Cole94, Cole00}.  Galaxy
formation models depend on the virial temperature, most notably
through the cooling function that controls star formation rates and
their associated feedback mechanisms.  Atomic hydrogen and helium
radiative cooling is efficient in halos with masses between $10^8$ and
$10^{12} \Ms$ as cooling times can be less than the dynamical time of
the system, a condition that galaxy clusters do not satisfy.  This
strong cooling suggests that in galaxies thermal energy may be less
important for virialization than turbulent kinetic energy.  Motivated
by the results of galaxy cluster turbulence, we investigate this
potentially important role of turbulence and radiative cooling in
galaxy formation, using a series of high resolution numerical
simulations of protogalactic halos in this work.

We consider kinetic energy and pressure forces in our virial analysis
of protogalactic halos \citep[cf.][]{Shapiro99, Iliev01}.  This allows
us to investigate the equilibrium throughout the entire halo and
determine the importance of each energy component in the virial
theorem.  Kinetic energy can be decomposed into radial and azimuthal
motions along with turbulence, which affects the collapse of gas
clouds primarily in three ways. First as seen in galactic molecular
clouds, turbulence plays an integral part in current theories of star
formation as the density enhancements provide a favorable environment
for star formation \citep{Larson79, Larson81, Myers99, Goldman00}.
Second if the turbulence is supersonic, gas dissipates kinetic
energy through radiative cooling, which aids the gaseous collapse
\citep{Rees77}.  Conversely turbulent pressure adds an additional
force for the collapsing object to overcome and can delay the collapse
into a luminous object.  Last, turbulence provides an excellent
channel for angular momentum transport as the halo settles into
rotational equilibrium to satisfy Rayleigh's inviscid rotational
stability argument \citep{Rayleigh20, Chandra61} in which the specific
angular momentum must increase with radius.  Cosmological hydrodynamic
simulations have just begun to investigate angular momentum transport
within turbulent collapsing objects, and turbulence seems to play a
large role in segregating low (high) angular momentum gas to small
(large) radii \citep{Norman99, Abel02a, Yoshida06b}.

We study idealized cases of structure formation where stellar feedback
is ignored because it provides a convenient problem to focus on the
interplay between cosmological merging, hydrodynamics, and cooling
physics during the assembly of early halos.  Some of the discussed
physical principles should, however, be applicable to galaxies of all
masses.  These simulations provide the simplest scenario to which we
can incrementally consider further additional physics, such as \hh~and
HD cooling physics \citep{Saslaw67, Palla83, Flower00}, primordial
stellar feedback \citep{Whalen04, Kitayama04, Alvarez06, Yoshida06a,
  Abel07}, metal enrichment from primordial stars \citep{Heger02,
  Tumlinson06}, AGN feedback \citep{Springel05, Kuhlen05}, and
``normal'' metal-enriched star formation \citep[see][for a
review]{Larson03}.

We present a suite of adaptive mesh refinement simulations that are
described in \S2.  Then we analyze the local virial equilibrium and
shocks in halos in \S3.  There we also differentiate between infall
through voids and filaments and its associated virialization.  We
discuss the situations in which virial heating and turbulence occur.
Next in \S4, we decompose the velocity distribution in principle axes
to explore virialization in both the DM and baryonic components.
Furthermore we decompose velocities into shear and compressible flows
to study turbulent flows in the virialized gas.  We discuss the
implications of these results on star and galaxy formation in \S5.
Finally we summarize in the last section.

\section{The Simulations}

To investigate protogalactic (\tvir~$>10^4$ K) halo virialization in
the early universe, we utilize an Eulerian structured, adaptive mesh
refinement (AMR), cosmological hydrodynamical code, \enzo\footnote{See
  http://http://lca.ucsd.edu/portal/software/enzo} \citep{Bryan97,
  Bryan99, OShea04}.  \enzo~solves the hydrodynamical equations using
the second order accurate piecewise parabolic method
\citep{Woodward84, Bryan95}, while a Riemann solver ensures accurate
shock capturing with minimal viscosity.  Additionally \enzo~uses an
adaptive particle-mesh $n$-body method to calculate the dynamics of
the collisionless dark matter particles \citep{Couchman91}.  Regions
of the simulation grid are refined by two when one or more of the
following conditions are met: (1) Baryon density is greater than 3
times $\Omega_b \rho_0 N^{l(1+\phi)}$, (2) DM density is greater than
3 times $\Omega_{\rm{CDM}} \rho_0 N^{l(1+\phi)}$, and (3) the local
Jeans length is less than 4 cell widths.  Here $N = 2$ is the
refinement factor; $l$ is the AMR refinement level; $\phi = -0.3$
causes more frequent refinement with increasing AMR levels,
i.e. super-Lagrangian behavior; $\rho_0 = 3H_0^2/8\pi G$ is the
critical density; and the Jeans length, $L_J = \sqrt{15kT/4\pi\rho G
  \mu m_H}$, where $H_0$, $k$, T, $\rho$, $\mu$, and $m_H$ are the
Hubble constant, Boltzmann constant, temperature, gas density, mean
molecular weight in units of the proton mass, and hydrogen mass,
respectively.  The Jeans length refinement ensures that we meet the
Truelove criterion, which requires the Jeans length to be resolved by
at least 4 cells on each axis \citep{Truelove97}.

We conduct the simulations within the concordance $\Lambda$CDM model
with WMAP first year parameters (WMAP1) of $h$ = 0.72, \Ol~= 0.73,
\Om~= 0.27, \Ob~= 0.024$h^{-2}$, and a primordial scale invariant ($n$
= 1) power spectrum with $\sigma_8$ = 0.9 \citep{Spergel03}.  $h$ is
the Hubble parameter in units of 100 km s$^{-1}$ Mpc$^{-1}$.  \Ol,
\Om, and \Ob~are the fractions of critical energy density of vacuum
energy, total matter, and baryons, respectively.  Last $\sigma_8$ is
the rms of the density fluctuations inside a sphere of radius
8$h^{-1}$ Mpc.

Using the WMAP1 parameters versus the significantly different WMAP
third year parameters \citep[WMAP3;][]{Spergel06} have no effect on
the evolution of individual halos as are considered here.  However
these changes play an important role in statistical properties.  For
example, halos with mass $10^6 \Ms$ at redshift 20 correspond to
$2.8\sigma$ peaks with the WMAP1 but are $3.5\sigma$ peaks for WMAP3.
The \Om/\Ob~ratio also only changed from 6.03 to 5.70 in WMAP3.  

We also have verified that there is nothing atypical about the mass
accretion rate histories of the objects we study.  The mass accretion
history of these objects exhibit smooth growth during minor mergers
and accretion and dramatic increases when a major merger occurs.  This
behavior is consistent with typical halo assemblies in extended
Press-Schechter calculations \citep{Bond91, Bower91, Lacey93, Lacey94,
  vdBosch02b} and cosmological numerical simulations
\citep[e.g.][]{DeLucia04, Gao05}.  The mass accretion histories in our
simulations are well described by the fitting function of
\citeauthor{vdBosch02b} with $M_0 = 3 \times 10^7 \Ms$, $z_f = 17$,
and $\nu = 12.5$.  We also compare our data against the mass accretion
histories of \citeauthor{Gao05}, who tested their data against an
extended Press-Schechter calculation of the growth history of the
halos.  We find no major discrepancies between the two histories.

The initial conditions of this simulation are well-established by the
primordial temperature fluctuations in the cosmic microwave background
(CMB) and big bang nucleosynthesis (BBN) \citep[][and references
therein]{Hu02, Burles01}.

We perform two realizations with different box sizes and random
phases.  In the first simulation (simulation A), we set up a
cosmological box with 1 comoving Mpc on a side, periodic boundary
conditions, and a 128$^3$ top grid with three nested child grids of
twice finer resolution each.  The other simulation is similar but with
a box side of 1.5 comoving Mpc (simulation B).  We provide a summary
of the simulation parameters in Table \ref{tab:sims}.  These volumes
are adequate to study halos of interest because the comoving number
density of $>$10$^4$ K halos at $z=10$ is $\sim$6 Mpc$^{-3}$ according
to an ellipsoidal variant of Press-Schechter formalism
\citep{Sheth02}.  We use the COSMICS package to calculate the initial
conditions at $z$ = 129 (119)%
\renewcommand{\thefootnote}{\fnsymbol{footnote}}%
\footnote{To simplify the discussion, simulation A will always be
  quoted first with the value from simulation B in parentheses.}%
\renewcommand{\thefootnote}{\arabic{footnote}} %
\citep{Bertschinger95, Bertschinger01}, which calculates the linear
evolution of matter fluctuations.  We first run a dark matter
simulation to $z=10$ and locate the DM halos using the HOP algorithm
\citep{Eisenstein98}.  We identify the first dark matter halo in the
simulation with \tvir~$>$ 10$^4$ K and generate three levels of
refined, nested initial conditions with a refinement factor of two,
centered around the Lagrangian volume of the halo of interest.  The
nested grids that contain finer grids have 8 cells between its
boundary and its child grid.  The finest grid has an equivalent
resolution of a 1024$^3$ unigrid.  This resolution results in a DM
particle mass of 30 (101) $\Ms$ and an initial gas resolution of 6.2
(21) $\Ms$.

%
%
\begin{center}
\begin{longtable}{ccccccc}
\caption{Simulation Parameters} \label{tab:sims} \\

\hline\hline\\[-3ex]
Name & l & z$_{end}$ & N$_{part}$ & N$_{grid}$ & N$_{cell}$ & Cooling model \\
   & [Mpc] &  &  &  &  & \\
\hline
\endhead

A0 & 1.0 & 15.87 & 2.22 $\times$ 10$^7$ & 30230 & 9.31 $\times$ 10$^7$
(453$^3$) & Adiabatic \\

A6 & 1.0 & 15.87 & 2.22 $\times$ 10$^7$ & 40486 & 1.20 $\times$ 10$^8$
(494$^3$) & H,He \\

A9 & 1.0 & 18.74 & 2.22 $\times$ 10$^7$ & 45919 & 1.21 $\times$ 10$^8$
(495$^3$) & H,He,\hh \\

B0 & 1.5 & 16.80 & 1.26 $\times$ 10$^7$ & 23227 & 6.47 $\times$ 10$^7$
(402$^3$) & Adiabatic \\

B6 & 1.5 & 16.80 & 1.26 $\times$ 10$^7$ & 21409 & 6.51 $\times$ 10$^7$
(402$^3$) & H,He \\

B9 & 1.5 & 23.07 & 1.26 $\times$ 10$^7$ & 20525 & 5.59 $\times$ 10$^7$
(382$^3$) & H,He,\hh \\

\hline
\end{longtable}
\tablecomments{Col. (1): Simulation name. Col. (2): Number of dark
  matter particles. Col. (3): Number of AMR grids. Col. (4): Maximum
  number of unique grid cells. Col. (5): Maximum level of refinement
  reached in the simulation. Col. (6): Resolution at the maximum
  refinement level. Col. (7): Cooling model.}
\end{center}

\enzo~employs a non-equilibrium chemistry model \citep{Abel97,
  Anninos97}.  We conduct three simulations for each realization with
(i) the adiabatic equation of state with an adiabatic index $\gamma =
5/3$, (ii) a six species chemistry model (H, H$^{\rm +}$, He, He$^{\rm
  +}$, He$^{\rm ++}$, e$^{\rm -}$), and (iii) a nine species chemistry
model that adds H$_2$, H$_2^{\rm +}$, and H$^{\rm -}$ to the six
species model.  In the nine species model, we use the molecular
hydrogen cooling rates from \citet{Galli98}.  These models are
differentiated in the text by denoting 0, 6, and 9, respectively,
after the simulation name (e.g. simulation B0).  Compton cooling and
heating of free electrons by the CMB and radiative losses from atomic
and molecular cooling are also computed in the optically thin limit.

To restrict the analysis to protogalactic halos in the \hh~models, we
suppress \hh~formation in halos that cannot undergo \lya~cooling by
reducing the residual electron fraction to $10^{-12}$ instead of a
typical value of $\sim10^{-4}$ only at the initial redshift
\citep{Shapiro94}.  This mimics an extreme case where all \hh~is
dissociated by an extremely large radiation background, and the halo
can only collapse and form stars when free electrons from ionized
hydrogen can catalyze \hh~formation.

We end the simulations with non-equilibrium cooling when the gas
begins to rapidly cool and collapse.  We choose a final resolution
limit of $\sim$3000 (4000) proper AU, corresponding to a refinement
level of 15.  We end the adiabatic simulations at the same redshift.
In a later paper, we will address the collapse of these halos to much
smaller scales.

\section{Virial Analysis}

The equation of motion for an inviscid gas in tensor notation reads:
\begin{equation}
\rho \frac{D v_i}{Dt} = -\frac{\partial}{\partial x_i} p  + \rho g_i \nonumber
\label{mest3.39}
\end{equation}
where $D/Dt = \partial /\partial t + v_j \partial / \partial x_j $ is
the total derivative. Here $v$ is velocity; $p$ is pressure; $\rho$ is
density; and $g = \nabla\Phi$ where $\Phi$ is the gravitational
potential.  From this \citet{Chandra53} derived the general virial
theorem for a region contained within a surface $\Sv$, in scalar form,
\begin{equation}
  \frac{1}{2} \frac{D^2 I }{Dt^2} =  2 \mathcal{T} + \mathcal{V}
  + 3 (\gamma-1)\mathcal{E} - \int p  \, \rv \cdot   d\Sv,
\label{vt}
\end{equation}
where 
\begin{equation}
  \mathcal{V} = -\frac{1}{2}G \int_V
  \frac{\rho(\xv)\rho(\xv')}{|\xv - \xv' |} d\xv d\xv' ,
\end{equation}
$\mathcal{T} = \frac{1}{2} \int \rho \vv^2 d\xv $, $I = \int \rho
\xv^2 d\xv$, and $\mathcal{E}=\int \varepsilon \; d\xv $ denote the
gravitational potential energy, the trace of the kinetic energy and
inertia tensor, and the total internal thermal energy, respectively.
The surface term $E_s$ (the last term in eq. [\ref{vt}]) is often
negligible in the outer regions of the halo.  The system is not
necessarily in virial equilibrium if $\ddot{I} = 0$, but the
time-averaged quantity is zero when the entire system is in virial
equilibrium.  A system is expanding or contracting when $\ddot{I}$ is
positive or negative, respectively, based on energy arguments.
\citet{Ballesteros06} gives counterexamples to this simple
interpretation.  However, in the cases presented here, spherically
averaged radial velocities are always negative.

We define the halo as the material contained in a sphere with a radius
$r_{200}$ enclosing an average DM overdensity of 200 and as such
relates to mass by
\begin{equation}
\label{eqn:r200}
r_{200} = \left[ \frac{GM}{100\Omega_{\rm{CDM}}(z) H^2(z)} \right]^{1/3},
\end{equation}
where $M$ is the mass of the halo, $\Omega_{\rm{CDM}}(z)$ is evaluated
at a redshift $z$, and $H$ is the Hubble parameter at $z$.  The region
where the cooling time is shorter than a Hubble time is denoted as the
cooling radius \rcool,
\begin{equation}
\label{eqn:rcool}
t_{\rm{cool}} (r_{\rm{cool}}) \equiv H(z)^{-1}
\end{equation}
\citep{White91}.  Mass and radius define a circular velocity and
virial temperature, which are
\begin{equation}
\label{eqn:tvir3}
V_c = \sqrt{\frac{GM}{r_{200}}}
\quad \textrm{and} \quad
T_{vir} = \frac{\mu m_p V_c^2}{2k},
\end{equation}
for a singular isothermal sphere \citep[see][with $\beta$ = 1 and
$\Delta_c$ = 200]{Bryan98}.  0.59 and 1.22 in units of the proton mass
are appropriate values for $\mu$ for the fully ionized and completely
neutral states of a primordial hydrogen and helium mixture of gas,
respectively.  We use $\mu$ = 1.22 throughout this paper.  We note
that \citet{Iliev01} considered non-singular, truncated isothermal
spheres, and the resulting virial temperature is $\sim$15\% lower than
the one calculated in equation (\ref{eqn:tvir3}).

For $\gamma = 5/3$, \tvir~is the temperature at which an ideal
adiabatic gas reaches virial equilibrium with the specified potential.
Please note that for an isothermal gas where $\gamma$ is close to
unity virial equilibrium is established between the turbulent energies
and the gravitational potential as the $3 (\gamma-1) \mathcal{E}$ term
in equation~(\ref{vt}) goes to zero.

\subsection{Local Analysis}
\label{sec:local}

\begin{figure}
\begin{center}
\vspace{0.4cm}
\plotone{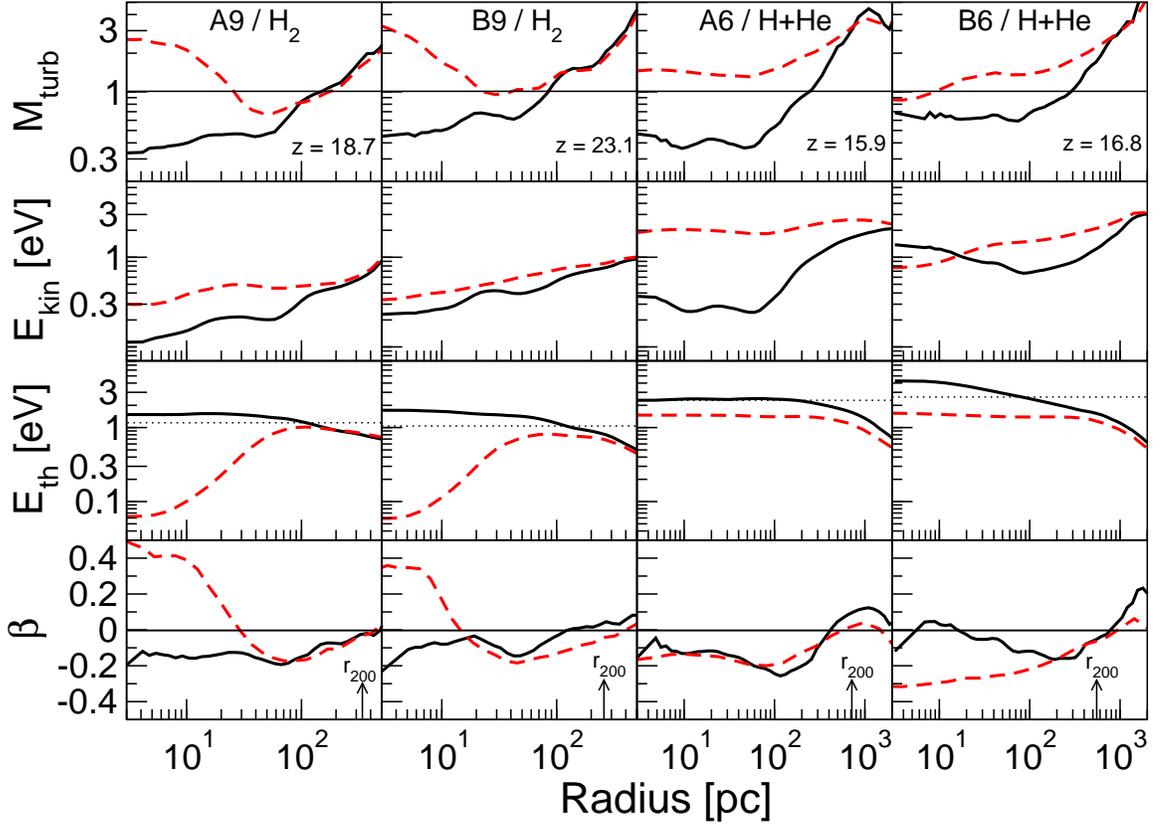}
\caption[Radial profiles of energies in the virial
theorem]{\label{fig:energies} \footnotesize{A comparison of
    (\textit{top to bottom}) turbulent Mach numbers, turbulent and
    thermal energies, and virial parameters between simulations with
    radiative cooling (\textit{dashed}) and adiabatic models
    (\textit{solid}).  The main coolant is listed at the top of each
    column.  The \textit{first} and \textit{second} columns display
    the state of these variables at z = \zhh~= 18.7 (23.1) for
    Simulation A and B, respectively.  The \textit{third} and
    \textit{fourth} columns are the data at z = \zlya~= 15.9 (16.8).
    The \textit{top} row depicts the importance of radiative cooling
    in generating trans- and super-sonic turbulence throughout the
    halo during virialization.  The \textit{middle} two rows show that
    when radiative cooling is efficient the halo cannot virialize
    through heating but must virialize by increasing its kinetic
    (turbulent) energies.  The dotted line in the \textit{third} row
    marks \tvir~(eq. [\ref{eqn:tvir3}]) with $\mu$ = 1.22, estimated
    from the total halo mass.  We plot the virialization parameter
    $\beta$ (eq. [\ref{eqn:beta}]) to investigate the local virial
    equilibrium ($\beta = 0$), particularly at \rr.  Furthermore,
    $\beta$ allows us to determine the mass-averaged dynamics of the
    system at a given radius, where $\beta > 0$ and $< 0$ correspond
    to decelerating and accelerating collapses, respectively.
    \rr~(eq. [\ref{eqn:r200}]) is marked on the bottom of each
    column.}}
\end{center}
\end{figure}

We evaluate the terms of equation (\ref{vt}) with respect to radius
(i.e. the volume contained in a radius $r$).  Figure
\ref{fig:energies} illustrates the radial structure of (a) the
turbulent Mach number,
\begin{equation}
  \label{eqn:mach}
  \mathcal{M}_{turb} = \frac{v_{rms}}{c_s}; \quad
  c_s = \sqrt{\frac{\gamma kT}{\mu m_h}}
\end{equation}
(b) turbulent and (c) thermal energies per $m_h$, and (d) a
``virialization'' parameter\footnote{This is a modified version of the
  $\beta$ used in \citet{Shaw06} to account for kinetic energies and
  so that it does not diverge when $\mathcal{V} \rightarrow 0$ in the
  center.  It still has the same behavior of $\beta \rightarrow 0$ as
  $\ddot{I} \rightarrow 0$.}
\begin{equation}
  \label{eqn:beta}
  \beta = \frac{3(\gamma - 1)\mathcal{E} + 2\mathcal{T}}{E_s -
    \mathcal{V}} - 1,
\end{equation}
of the adiabatic and radiative cooling simulations when the cooling
halo collapses.  Here $v_{rms}$ is the three-dimensional rms velocity
and is assessed using the gas velocities relative to the mean gas
velocity of each spherical shell.  In the top row of Figure
\ref{fig:energies}, we also plot the Mach number, using $v_{rms}$ with
respect to the mean velocity of the gas within $r_{200}$.  In the six
and nine species simulation, this occurs at \zhh~= 18.7 (23.4) and
\zlya~= 15.9 (16.8), respectively.  The radial profiles are centered
on the densest point in the simulation with the collapsing halo.
Several properties of the most massive halo in each simulation are
detailed in Table \ref{tab:halos}.  The sections of the Table compare
the halo in the adiabatic, \lya, and \hh~simulations.

%
%

%
%
\begin{figure}[t]
\begin{center}
\vspace{0.6cm}
\includegraphics[width=0.8\textwidth]{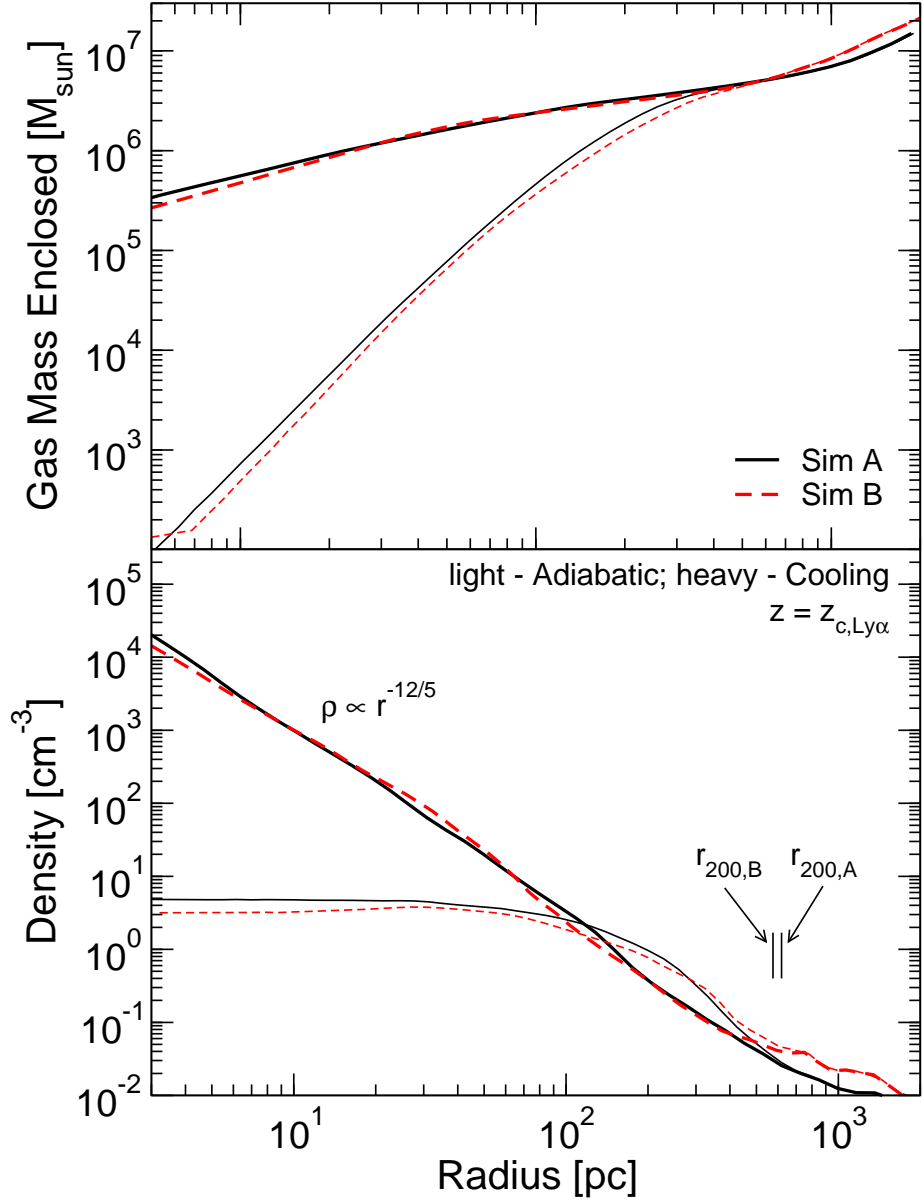}
\caption[Density and Enclosed Gas Mass Profiles]{ Mass-weighted radial
  profiles for gas mass enclosed (\textit{top}) and number density
  (\textit{bottom}) for Simulations A (\textit{solid black}) and B
  (\textit{dashed red}) at z = \zlya~= 15.9 (16.8).  The
  \textit{light} and \textit{heavy} lines represent data for adiabatic
  and cooling models, respectively.  The virial shock in the cooling
  halos occurs at $\frac{2}{3}$\rr, illustrated by the density
  increasing at smaller radii.}
\label{fig:dens}
\end{center}
\end{figure}


Figure \ref{fig:dens} shows the mass-weighted radial profiles of gas
mass enclosed and gas density at z = \zlya~in the adiabatic and
cooling cases.  Both realizations are remarkably similar.  Halos in
the adiabatic case have a central core with a radius $\sim$50 pc and
gas density of $\sim$5.0 (3.5)\cubecm.  Core densities in simulation A
are slightly higher than simulation B, which has larger thermal and
turbulent pressures (see Figure \ref{fig:energies}).
%
%
%
\clearpage
\begin{center}
\renewcommand{\thefootnote}{\alph{footnote}}
\begin{longtable}{ccccccc}
\caption{Halo Properties} \label{tab:halos} \\

\hline\hline \\[-3ex]
  Name & z$_{coll}$ & M$_{tot}$ & $\rho_c$ & \tvir\tablenotemark{a} &
  T$_c$ & $\langle T \rangle$ \\
   &  & [$\Ms$] & [cm$^{-3}$] & [K] & [K] & [K] \\
\hline
\endhead

\multicolumn{7}{c}{\hh~Induced Collapse}\\
\hline
A0 & 18.74    & 9.8 $\times$ 10$^6$ & 8.1 & 9200 & 10000 & 5700 \\
A6 & 18.74    & 9.8 $\times$ 10$^6$ & 17 & 9200 & 7700 & 5500 \\
A9 & 18.74    & 9.8 $\times$ 10$^6$ & 1.6 $\times$ 10$^6$ & 9100 & 590 & 5000 \\

B0 & 23.07    & 6.2 $\times$ 10$^6$ & 13 & 8300 & 12000 & 6000 \\
B6 & 23.07    & 6.2 $\times$ 10$^6$ & 15 & 8300 & 8900 & 5600 \\
B9 & 23.07    & 6.7 $\times$ 10$^6$ & 3.0 $\times$ 10$^6$ & 8700 & 580 & 4200 \\

\hline
\multicolumn{7}{c}{\lya~Induced Collapse} \\
\hline
A0 & 15.87 & 3.6 $\times$ 10$^7$ & 4.9 & 19000 & 17000 & 12000 \\
A6 & 15.87 & 3.6 $\times$ 10$^7$ & 1.8 $\times$ 10$^6$ & 19000 & 8700 & 7300 \\

B0 & 16.80 & 3.5 $\times$ 10$^7$ & 3.8 & 19000 & 31000 & 11000 \\
B6 & 16.80 & 3.6 $\times$ 10$^7$ & 4.0 $\times$ 10$^6$ & 20000 & 9000 & 7500 \\

\hline
\footnotetext[2]{Virial temperatures are calculated with $\mu$ = 1.22
  in all cases.}
\end{longtable}

\tablecomments{Col. (1): Simulation name. Col. (2): Redshift of
  collapse through \hh~or \lya~cooling. Col. (3): Total
  mass. Col. (4): Central density. Col. (5): Virial temperature
  (i.e. eq. \ref{eqn:tvir3}). Col. (6): Central temperature. Col. (7):
  Mass-averaged temperature of the entire halo.}

\renewcommand{\thefootnote}{\arabic{footnote}}
\end{center}
\clearpage
%
\noindent With radiative cooling, gas infalls rapidly as it cools and
undergoes a self-similar collapse with $\rho \propto r^{-12/5}$.

\subsubsection{Virial Radius}

We define the virial radius \rvir~when $\beta = 0$ and $d\beta/dr <
0$.  When we radially average the \lya~halo, \rvir~= 419 (787) pc in
the adiabatic simulations where the corresponding $r_{200}$ value is
615 (576) pc.

A well defined shock exists on the interface between voids and
the halo.  This material shock-heats to \tvir~and virializes at a
radius comparable to \rr~in the adiabatic cases.  When we include
radiative cooling, this radius decreases everywhere around the
halo-void virial shock and is low as \rr/2. In contrast to
the voids, the filamentary gas shock-heats at an even smaller radius.
\citet{Dekel06} also studied the stability of cold inflows within a
hot virialized medium and found similar results.  Figures
\ref{fig:dens} and \ref{fig:entropy} illustrate these changes.  At
\rr, densities in the adiabatic case begin to increase more rapidly
than the cooling case as material accretes at the virial shock.  No
significant increase in $d\rho/dr$ is seen in the cooling case,
indicative of a self-similar collapse.

\subsubsection{Adiabatic Model}
\label{sec:adiabatic}

We start with the discussion of the adiabatic model as it is the
simplest case and later compare the calculations with radiative
cooling to this model.  Virialization should transfer potential energy
to kinetic energy that dissipates in shocks to thermal energy, which
is the implication of the dissipationless virial theorem.  The solid
lines in Figure \ref{fig:energies} represent the energies in adiabatic
models.  The physics illustrated in this Figure are as follows:

\medskip

1. \textit{Thermal energy}--- The gas shock-heats to the virial
temperature at the virial shock.  Virial heating continues with
decreasing radius as the surface term becomes significant in the
interior.  The resulting central temperature of the halo is 10000
(12000) K, which is 1.2 (1.5) \tvir, at a redshift of \zhh~when the
\hh~model collapses.  At the time (z = \zlya) of collapse caused by
\lya~cooling, the central temperature is 17000 (31000) K,
corresponding to 0.9 (1.6) \tvir.

\medskip

2. \textit{Kinetic energy}--- It increases along with the thermal
energy during virialization.  The gas is generally turbulent,
appearing as a velocity dispersion with a bulk radial inflow.  At \rr,
the kinetic energy is equivalent to the thermal energy,
$\mathcal{T}/\mathcal{E} \sim 1$.  This ratio steadily drops toward
the center, where $\mathcal{T} / \mathcal{E} \sim 1/3$.  This decrease
in kinetic energy is apparent in all the calculations except
simulation B at \zlya, increasing by a factor of two in the center.

\medskip

3. \textit{Turbulent Mach number}--- At \rr, the turbulent Mach number
\mturb~is maximal and varies from 1--3 in all adiabatic simulations.
\mturb~decreases to subsonic values $\sim$0.15 but never below in the
interior.  Note that \mturb~does not increase as the turbulent energy
towards the center in simulation B because of the also growing sound
speed there.

\medskip

4. \textit{Virialization parameter}--- Virial equilibrium is
quantified by the virialization parameter $\beta$, where the collapse
is retarding or accelerating when it is negative or positive,
respectively.  At \zhh~and \zlya~and in both simulations, $\beta$ is
within 20\% of being virialized ($\beta = 0$).  For comparison
purposes, this corresponds to a halo having 80\% of the required
velocity dispersion for virialization in the dissipationless case.  At
\rvir, $\beta$ is nearly zero which defines the virialized object.
Within \rvir, the values decrease to values around --0.1 but stays
$\lsim$ 0.

\medskip

Characteristics of turbulence in our adiabatic models are similar to
ones found in galaxy cluster simulations \citep{Norman99, Dolag05}.
Both groups find that turbulence provides $\sim$5--30\% of the total
pressure, i.e. $\mathcal{T} / (\mathcal{T} + \mathcal{E}$), in the
cluster cores.  Our protogalactic halos have $\sim$25\% of the
pressure in the turbulent form.  Also the galaxy clusters in
\citet{Norman99} have comparable Mach numbers of $\sim$1.6 at \rvir,
$\sim$0.5 at \rvir/3, and $\sim$0.3 in the core.  These similarities
suggest that virial turbulence is generated over a large range of mass
scales.

\subsubsection{\lya~Cooling Model}
\label{sec:lyacollapse}

At \zlya, halos in calculations with the H+He cooling model start to
rapidly collapse.  The dashed lines in the third and fourth columns of
Figure \ref{fig:energies} illustrate the energies of this model.

\medskip

1. \textit{Thermal energy}--- Compared to the adiabatic models, the
gas can radiatively cool through \lya~emission to T $\sim$ 8000 K
within \rcool~$\sim$ \rvir.  The entire halo is isothermal at this
equilibrium temperature.  Below this temperature, the cooling function
of pristine gas drops by several orders of magnitude, and the gas can
no longer cool efficiently.  The thermal energy is $\sim$65\% lower
than the adiabatic case.

\medskip

2. \textit{Kinetic energy}--- In response to the lesser thermal
energy, the system tends toward virial equilibrium by increasing
kinetic (turbulent) energy.  The gravitational potential and surface
terms do not appreciably change with the inclusion of radiative
cooling.  Turbulent energy within \rvir~increases as much as a factor
of 5 when compared to the adiabatic case.

\medskip

3. \textit{Turbulent Mach number}--- The changes in thermal and
kinetic energies equate to a increase of \mturb~by a factor of 2--3 to
values up to 1.5.  The turbulence is supersonic in all cases at the
virial shock, but when we include radiative cooling, this trait
emanates inward as the halo begins to rapidly cool.  When the central
core becomes gravitationally unstable, the entire halo is
supersonically turbulent.

\medskip

4. \textit{Virialization parameter}--- The increased kinetic
energies compensate for the loss in thermal energy and the halo
remains in a similar virial state.  This is apparent in the remarkably
similar radial characteristics of $\beta$ in the adiabatic and H+He
models of simulation A.

\subsubsection{\hh~Cooling Model}
\label{sec:h2collapse}

The collapses caused by \hh~cooling at z = \zhh~have very similar
dynamics as the halos described in the previous section.  The dashed
lines in the first and second columns of Figure \ref{fig:energies}
illustrate the energies of this model.

\medskip

1. \textit{Thermal energy}--- \hh~cooling is efficient down to 300 K,
so gas can depose a much larger fraction of its thermal energy.
Inside \rcool~$\sim$ 0.32 (0.19) \rvir, thermal energies are only 5\%
of the values in the adiabatic models.

\medskip

2. \textit{Kinetic energy}--- The turbulent energies must increase as
in the \lya~case, and they increase by 93\% (44\%) on average inside
\rcool.

\medskip

3. \textit{Turbulent Mach number}--- Similarly, \mturb~increases up to
a factor of 10 to become supersonic at values up to 3 throughout the
halo.  They are somewhat larger than the \lya~cases since \hh~can cool
to significantly lower temperatures than the virial temperature.

\medskip

4. \textit{Virialization parameter}--- The virial equilibrium of the
halos are also similar to the other models.  $\beta$ smoothly
transitions from nearly equilibrium at \rvir~to an increased radial
infall with $\beta$ = --0.2 at 70 pc.  Then it increases to 0.4 inside
10 pc, which corresponds to the gas decelerating from the rapid infall
as it encounters the central molecular cloud.

\subsubsection{Model Summary}

Baryons are close to virial equilibrium over three orders of magnitude
in length scale by gaining both thermal and kinetic energies
independent of cooling physics.  Central temperatures of the adiabatic
simulations are up to twice the nominal virial temperature.  Similar
to galaxy cluster studies, turbulence in the adiabatic model
contributes $\sim$25\% to the energy budget with Mach numbers $\sim$
0.3 in the center.  In cooling cases, atomic and molecular cooling
inhibit virialization through heating, therefore the object must
virialize by gaining kinetic energy up to five times the energy seen
in the adiabatic models.  This translates into the flow becoming
supersonically turbulent with Mach numbers ranging from one to three.

%
%
\begin{figure}
\begin{center}
\plotone{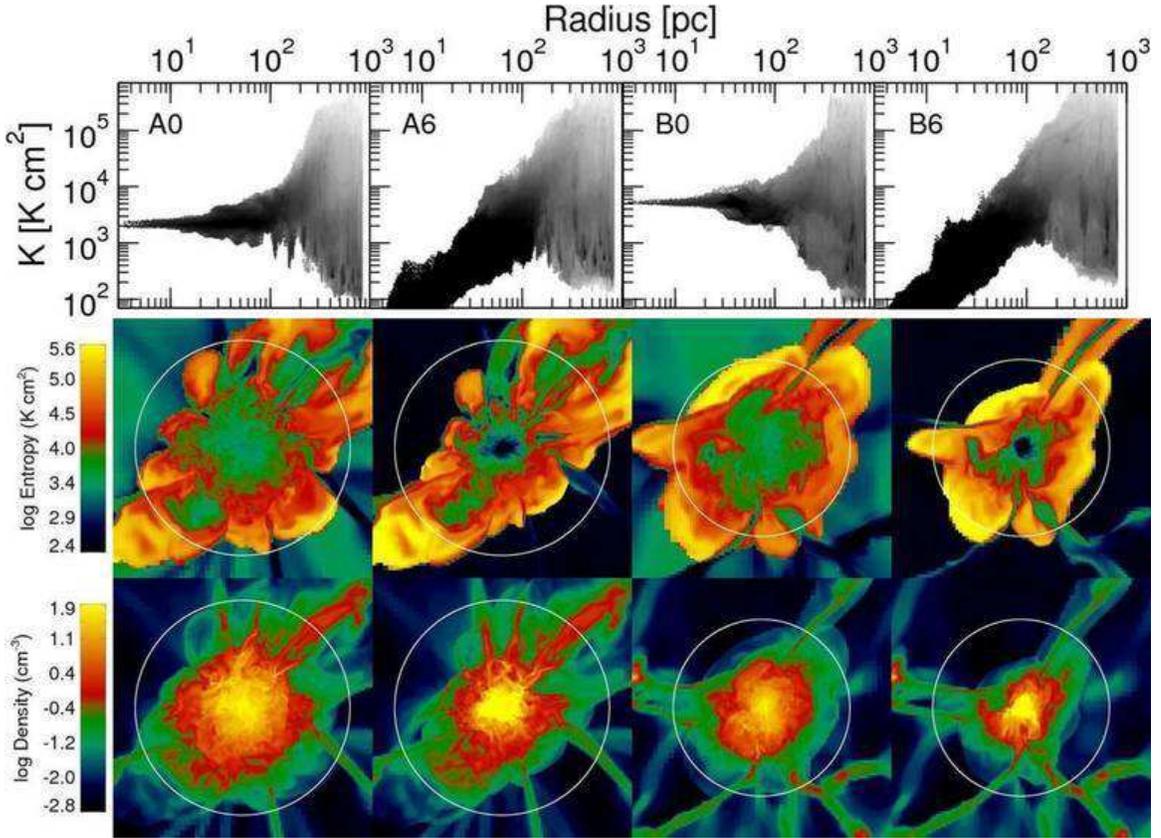}
\caption[Entropy and Density Projections]{\label{fig:entropy}
  Differences in entropy and density in a protogalactic halo at z =
  \zlya.  \textit{Left to right}: Simulations A0, A6, B0, B6.
  \textit{Top row}: Mass-averaged radial scatter plots of the
  adiabatic invariant $K = T / n^{2/3}$.  Radiative cooling allows the
  gas to cool and collapse in the center that accounts for the
  decrease in Simulations A6 and B6.  The material at $r \gsim 100$ pc
  and $K \lsim 10^{3.3}$ K cm$^2$ corresponds to the cold flows inside
  filaments that illustrates that virialization occurs at different
  radii depending on its origin.  \textit{Middle row}: Two-dimensional
  slices of entropy.  The circle denotes \rr = 615 pc and 576 pc for
  Simulation A and B, respectively.  The virial shock exists at
  approximately \rr~in the adiabatic models; however it shrinks to
  $\sim$2/3 of \rr~when we consider radiative cooling.  \textit{Bottom
    row}: Two-dimensional slices of number density of baryons.}
\end{center}
\end{figure}

\subsection{Variations in the Virial Shock}
\label{sec:shock}

Using the adiabatic invariant, $K = T/n^{2/3}$, which we label
``entropy'', allows us to differentiate between gas accreting from
voids and filaments.  As a precaution, we note that $K$ is not an
invariant when $\gamma$ varies; however, this is not the case in our
simulations in which we permit molecular hydrogen cooling.  Here
molecular fractions remain low, $< 10^{-3}$, and $\gamma \approx 5/3$
even in the densest regions.  The top row of Figure~\ref{fig:entropy}
depicts the variance of $K$ with respect to radius in two-dimensional
histograms, where the intensity of each pixel represents the mass
having the corresponding $K$ and $r$.  The middle and bottom rows
display two-dimensional slices of $K$ and density, respectively,
through the densest point in the halo.  The virialized gas from the
voids has low density and does not significantly contribute to the
mass averaged radial profiles.  Figure \ref{fig:entropy} illustrates
this gas at $r \sim r_{200}$ and $K \gsim 10^{4.5}$ K cm$^{2}$.  The
gas in filaments has lower entropy than the rest of the halo at $r >$
150 pc and $K \lsim 10^{3.3}$ K cm$^{2}$.  In equation (\ref{vt}), the
pressure in the surface term is the constant at a given radius.  The
accreting, denser, unshocked gas in filaments has lower temperatures
than the more diffuse accreting gas.  The gas remains cool until it
shocks and mixes well inside \rvir~and as small as $\sim$\rvir/4 in
the most massive filaments.  Similar characteristics of cold accretion
flows have been noted and discussed by \citet{Nagai03a},
\citet{Keres05}, and \citet{Dekel06}.

Entropy in the exterior of the halo differ little between adiabatic
and cooling runs outside of \rcool.  But as the gas falls within
\rcool, it cools and condenses, which gives a lower entropy, and the
$r$-$K$ histograms and entropy slices display this clearly.  Another
significant difference in the cooling simulations is the contraction
of the virial shock by a factor of 1/3 when compared to adiabatic
runs.  This is caused by the contraction of the cooling gas.  Here the
cold filaments penetrate to even smaller radii.  This is also evident
in the radial density profiles of Figure \ref{fig:dens}.

\subsection{Virial Heating and Turbulence}
\label{sec:virial_heat_turb}

In order for a system to remain in virial equilibrium as it grows in
mass, additional gravitational energy is balanced through two possible
mechanisms: heating of the gas ($\dot{\mathcal{E}} > 0$) and
increasing the kinetic energy of the gas ($\dot{\mathcal{T}} > 0$).
We differentiate between two main cases of virialization by comparing
the cooling time, \tcool~= kT/n$\Lambda$, of the system to the heating
time, \theat~= \tvir/$\dot{T}_{\rm{vir}} \sim \frac{3}{2}
M_{\rm{vir}} / \dot{M}_{\rm{vir}}$ in the case of rapid mass
accretion.  \citet{Birnboim03}, \citet{Dekel06}, and \citet{Wang07}
find that radiative cooling rates are greater than heating rates from
virialization for halos with masses below $10^{12} \Ms$.

\medskip

1. \textit{Thermalization} (\tcool~$>$ \theat)--- When no efficient
radiative cooling mechanisms (e.g. \hh, \lya, \ion{He}{1}) exist, the
system virializes by injecting energy into heat $\mathcal{E}$ and
partly into kinetic energy $\mathcal{T}$.  In the process, the halo
becomes pressure supported and virialized.  Traditional galaxy
formation scenarios only consider this thermalization while neglecting
the kinetic energy term of equation (\ref{vt}).  However it is
important to regard kinetic energy, even in adiabatic models, as the
gas violently relaxes.  Turbulence velocities are similar to the
velocity dispersion of the system and contributes notably to the
overall energy budget as seen in adiabatic cases in Figure
\ref{fig:energies}.

\medskip

2. \textit{Turbulence generation} (\tcool~$<$ \theat)--- When a
cooling mechanism becomes efficient, the system now dispenses its
thermal energy and loses pressure support within \rcool.  The gas
will cool to a minimum equilibrium temperature.  As the
cooling halo collapses and radial velocities increase, the gas still
lacks enough kinetic and thermal energy to match the gravitational
energy and surface term in equation (\ref{vt}).  The gas becomes more
turbulent in order to virialize.  We see this in the second row of
Figure \ref{fig:energies}, where turbulent energies are significantly
increased well inside the halo in the cooling models as compared to
the adiabatic calculations.

\medskip

Through virial analyses, we have shown that turbulent energies are
comparable, if not dominant, to thermal energies in galaxy formation.
In the next Section, we further investigate the significance and
nature of the turbulence through velocity distributions and
decompositions in order to study any small-scale anisotropies in the
internal flows.

\section{Velocity Distributions}

In CDM cosmogony, collisionless dark matter dominates the
gravitational potential and oscillates as it stabilizes.  \citet{LB67}
showed how a collisionless system undergoes violent relaxation if
embedded within a rapidly time-varying potential.  Individual mass
elements do not conserve energy during violent relaxation, only the
entire system conserves energy.  This behavior randomizes the energies
of the mass elements, and statistical mechanics makes the resulting
energy (velocity) distribution to tend to Maxwellian.  Furthermore,
the system ``forgets'' its original configuration during virialization
or the incorporation of a lesser halo.  Later studies have inferred
two baryonic scenarios of virialization. First, violent relaxation and
the accompanying phase mixing also applies to the gaseous component
\citep{vdBosch02, Sharma05}.  In the other case, the gaseous component
dissipates all turbulent motions and finally rigidly rotates as a
solid body with a velocity appropriate to the overall spin parameter
\citep[e.g.][]{Loeb94, Mo98, Bromm03}.

%
%
\begin{figure}
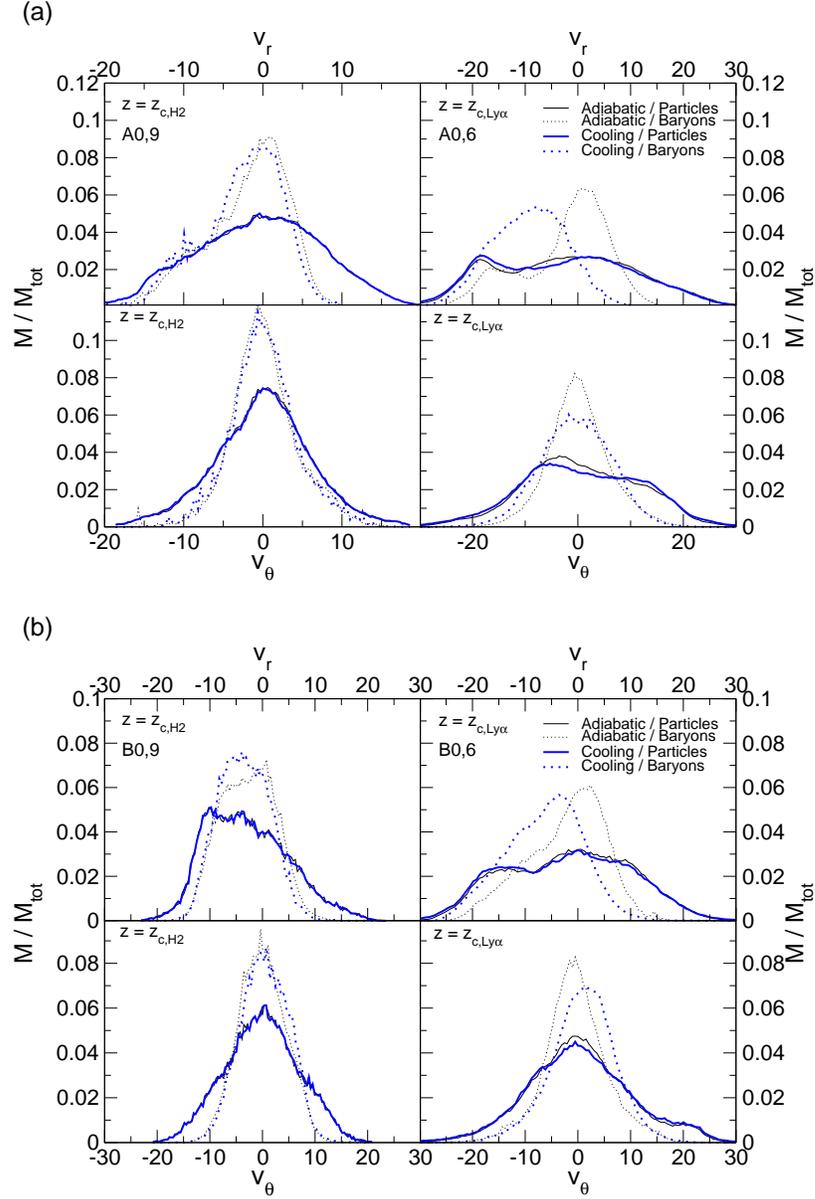

\centering
\includegraphics[width=0.7\textwidth]{chapter3/f4a.eps} \\[1em]
\includegraphics[width=0.7\textwidth]{chapter3/f4b.eps}
\caption[Velocity Distributions]{\label{fig:veldistA} (a) Simulation
  A. (b) Simulation B.  Radial (\textit{top}) and tangential
  (\textit{bottom}) velocity distributions of the most massive halo at
  z = \zhh~(\textit{left}) and z = \zlya~(\textit{right}).  The
  \textit{heavy, blue} lines are the distributions of the adiabatic
  models, and the \textit{light, black} lines are from the radiative
  cooling models.  \textit{Solid} and \textit{dotted} lines correspond
  to the velocity distributions of DM and baryons, respectively.
  These distributions can be decomposed into single or multiple
  Gaussians, depending on substructure.  This demonstrates that
  violent relaxation occurs for the baryons as well as the DM. The
  narrower distributions of the baryons is due to the dissipation in
  shocks.}
\end{figure}

Figure \ref{fig:veldistA} shows the velocity distributions of the dark
matter and baryonic components of the halo at \zhh~in the left column
and \zlya~in the right column.  It also overplots the simulations with
adiabatic and radiative cooling.  We plot the radial and tangential
velocity distribution on the top and bottom rows, respectively.  The
velocities are taken with respect to the bulk velocity of the halo.
We also transform the velocity components to align the z-axis and
total angular momentum vector of the DM halo.

The radial velocity distributions at z = \zhh~are approximately
Maxwellian in both dark matter and gas with a skew toward infall.  The
infall distributions are shifted by $\sim$1\kms~in the cooling case
when compared to adiabatic.  However at z = \zlya, the effects of
\lya~cooling become more prevalent in the halo when compared to
\hh~cooling, shifting the radial velocity distribution by
$\sim$5\kms~that is caused by faster infall.  These distributions have
two components that represent virialized gas and infalling gas in
filaments.  We further discuss this in the next Section.

The tangential velocity distributions are nearly Maxwellian in all
cases except for the dark matter in simulation A at z = \zlya~(right
panels in Figure \ref{fig:veldistA}a).  This deviation from Maxwellian
arises from two major mergers that occur between 25 and 85 Myr (z =
17--21) before the final collapse.  The residual substructure from the
major merger causes three distinct populations with Gaussian
distributions centered at --0.2, +13.6, and --6.7\kms~with $\sigma$
= 11.6, 4.2 and 3.6\kms, respectively.  These distributions clearly
do not resemble a solid body rotator, whose velocity distribution
would contain all positive velocities.  In other words the turbulent
velocities exceed the typical rotational speeds.

Distributions in dark matter are broader than the gas in both
simulations and collapse redshifts as expected because for the gas we
only give the bulk velocities and do not add the microscopic
dispersion \citep[cf.][]{vdBosch02, Sharma05}.

\subsection{Halo and Filament Contrasts}

The dark matter velocity distributions are typical of a virialized
system with the majority of the matter having a Maxwellian
distribution with a dispersion corresponding to the main halo
\citep{Boylan04, Dieman04, Kazantzidis04}.  Substructure appears as
smaller, superposed Gaussians, which are stripped of its outer
material as it orbits the parent halo.  Dynamical friction acts on the
substructure and decreases its pericenter over successive orbits, and
the subhalo is gradually assimilated in the halo.

The filaments penetrate deep into the halo and provide mostly radial
infall inside \rr.  They do not experience a virial shock at \rr, and
this contrast is apparent in the radial velocity distributions.  When
we restrict our analysis scope to the filaments (i.e. $r > 150$ pc and
$K < 10^{3.3}$ K cm$^{-2}$), the radial velocity distribution (Figure
\ref{fig:veldistA}) is skewed toward infall, centered at --15 km/s,
which is approximately the circular velocity of the halo.  The rest of
the gas outside of this region in $r-K$ space has already been
virialized, shock heated, and roughly exhibits a Maxwellian
distribution, centered at zero, with its associated substructures.
Hence the mass in filaments dominate the radial velocity distributions
at negative values in Figure \ref{fig:veldistA}.

\subsection{Turbulence}

Radial inflows can create turbulence in the halo.  Filaments provide
an influx material with distinct angular momentum.  This gas
virializes in the presence of an already turbulent medium that has a
relatively high specific angular momentum at $r > r_{200}/4$.  The
Rayleigh inviscid instability criterion requires
\begin{equation}
\label{eqn:rayleigh}
  \frac{dj^2}{dr} > 0 \quad \textrm{for rotational stability,}
\end{equation}
where $j$ is the specific angular momentum.  If this is not satisfied,
the system will become unstable to turbulence.  The onset of
turbulence can be delayed if viscosity were large enough so that
Reynolds numbers are below the order of 10$^2$ or 10$^3$.  However
there are many modes of instability if equation (\ref{eqn:rayleigh})
is not met, and even a gas with low Reynolds number will eventually
become fully turbulent \citep{Shu92}.

Velocity dispersions can characterize the general magnitude of a
turbulent medium, but its local nature is better detailed by applying
the Cauchy-Stokes decomposition,
\begin{equation}
  \uv^\prime = \uv + \frac{1}{2} \: {\bf \omega} \times \hv +
  \frac{1}{3} \: (\nabla \cdot \uv)\hv + 
  \frac{1}{2} \: \mathcal{D} \cdot \hv 
\end{equation}
that decomposes the velocity field into bulk motion $\uv$, vorticity
${\bf \omega} = \nabla \times \uv$, expansion and contraction $\nabla
\cdot \uv$, and a distortion $\mathcal{D}$ without change in volume.
Here the vector $\hv$ describes the separation between gas parcels at
position $\xv$ and $\xv^\prime$, and $\uv^\prime$ is the velocity at
$\xv^\prime$.

\begin{figure}
\begin{center}
\plotone{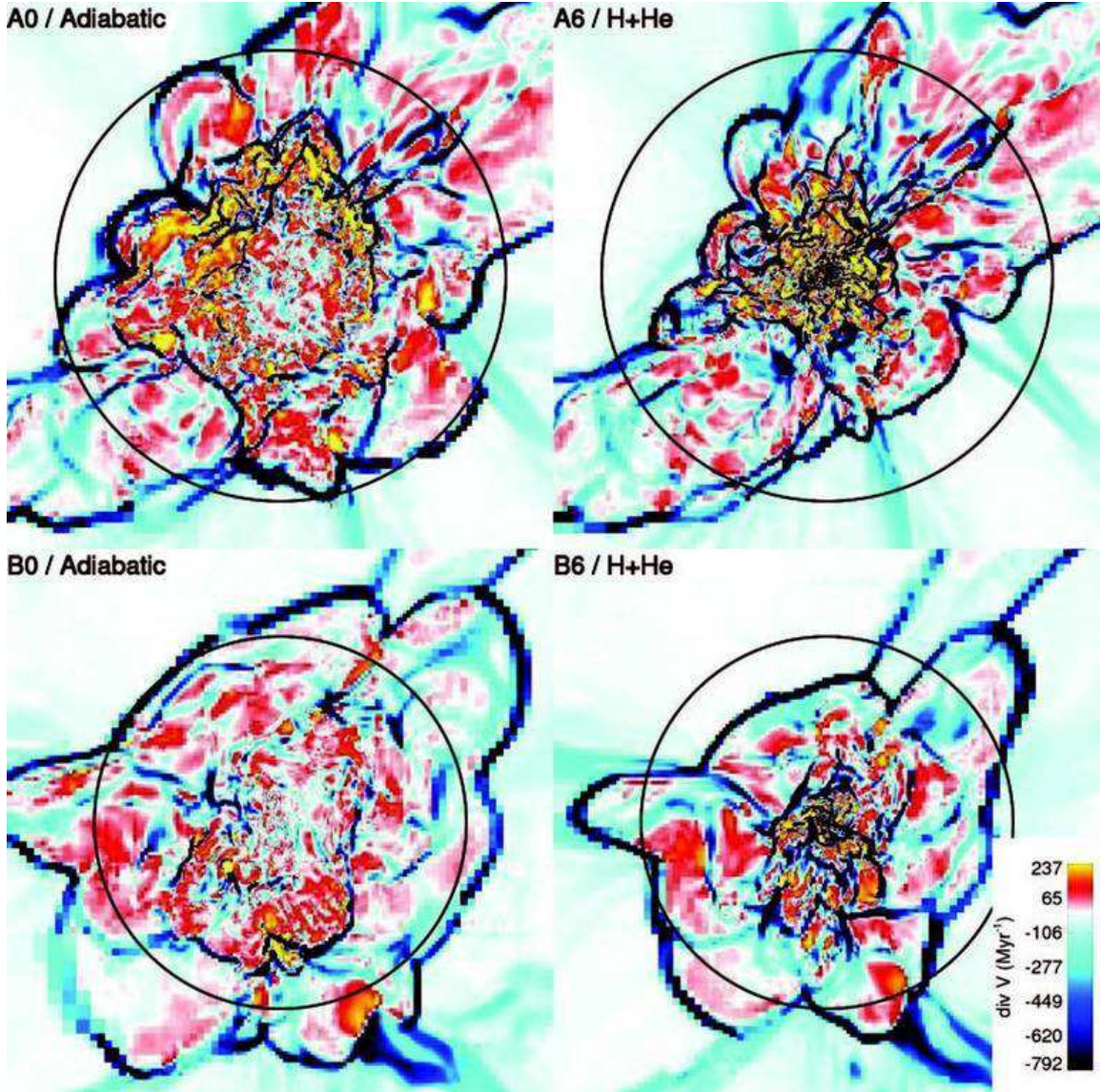}
\caption[Velocity Convergence]{\label{fig:divV} Two-dimensional slices
  of velocity divergence ($\nabla \cdot \vv$) at z = \zlya~for
  Simulations A0, A6, B0, and B6.  The fields of view are 1.49 and
  1.69 kpc for Simulations A and B, respectively.  Shocks are clearly
  denoted by large, negative convergence values.  In the adiabatic
  cases, these shocks mainly exist at large radii where the gas from
  the voids and filaments virializes.  When we consider radiative
  cooling, supersonic turbulence increases the frequency of shock
  fronts in the interior of the halo.}
\end{center}
\end{figure}

We relate $\nabla \cdot \uv$, which is plotted in Figure
\ref{fig:divV}, to a convergence timescale through the continuity
equation,
\begin{equation}
\label{eqn:continuity}
\dot{\rho} + \nabla(\rho\uv) = 0,
\end{equation}
that can be rewritten in terms of the total derivative $D/Dt$ as
\begin{equation}
\label{eqn:lct}
\frac{1}{\rho} \frac{D\rho}{Dt} = - \nabla \cdot \uv .
\end{equation}
$D\rho/Dt$ describes density changes along the fluid flow lines, and
the $1/\rho$ factor converts this change into an inverse timescale on
which local densities e-fold.  We denote $-(\nabla \cdot \uv)^{-1}$ as
the ``Lagrangian convergence timescale'' (LCT).  Converging flows
($\nabla \cdot \uv < 0$) are ubiquitous within the halo.  On large
scales, the smallest LCTs on the order of 20 kyr exist at the virial
shock, adjacent to both the filaments and voids.  In the cooling
models, these timescales are also small in turbulent shocks well
within \rr.  Analogous to the dynamical time, typical shocked LCTs
decrease toward the center as density increases.

%
%

%
%
\begin{figure}
\begin{center}
\plotone{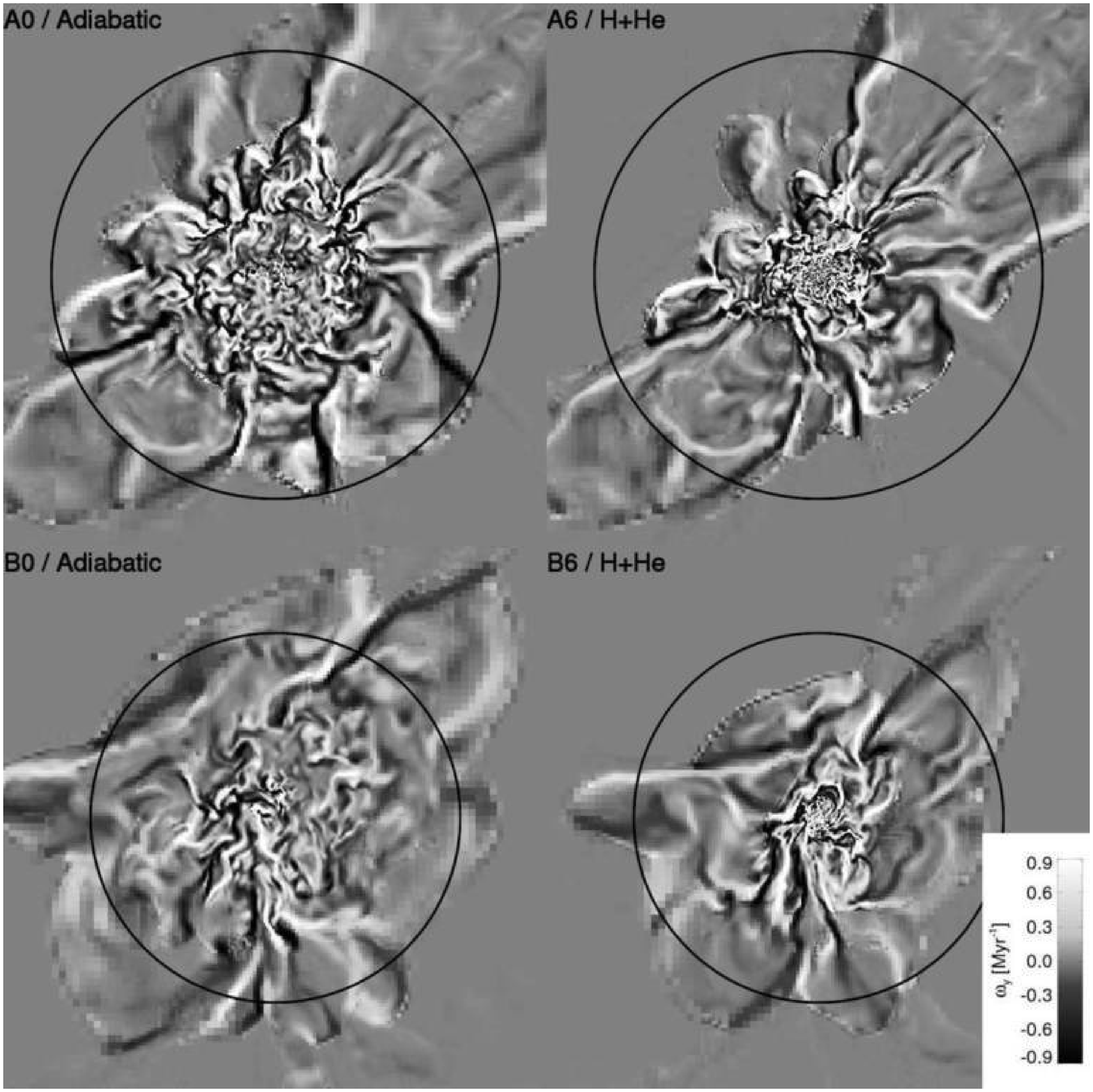}
\caption[Vorticity]{\label{fig:curlV} Two-dimensional slices of the
  component perpendicular to the slice of vorticity ($\nabla \times
  \vv$) at z = \zlya~for Simulations A0, A6, B0, and B6.  The fields
  of view are the same as in Figure \ref{fig:divV}.  This quantity
  emphasizes the large- and small-scale turbulent eddies in the halo.}
\end{center}
\end{figure}

The local magnitude and nature of turbulence is further illustrated by
the vorticity ${\bf \omega}$, whose component perpendicular to the
slice, $\omega_y$, is shown in Figure \ref{fig:curlV}.  The local
rotation period, $4\pi / \vert \omega \vert$ is also helpful to
quantify and visualize the nature of the flow.  The vorticity equation
reads
\begin{equation}
\frac{\partial\omega}{\partial t} + \nabla \times (\omega \times \uv) = 
\frac{1}{\rho^2} \nabla\rho \times \nabla p ,
\end{equation}
where the source term is non-zero when the density and pressure
gradients are not aligned, i.e. baroclinic vorticity generation.  This
occurs at and near shocks throughout the halo, regardless of radiative
cooling.  In the adiabatic models, vorticity exists even at modest but
sufficient resolution in the large pressure-supported cores (see
\mturb~for $r < 100$ pc in Figure \ref{fig:energies}) and generates a
turbulent medium with \mturb~= 0.3.  In the cooling models, this
large-scale vorticity is still present but increases in the collapsing
core.  As shocks become abundant in the center, we do not see any
dampening of kinetic energy.  Perhaps this mechanism maintains
turbulent motions during virialization, even in the presence of
dissipative shocks.  In Figure \ref{fig:curlV}, adjacent, antiparallel
fluid flows, i.e. a sign change in $\omega_y$, are ubiquitous, which
visually demonstrates that turbulence exists throughout the halo.  The
length scale of these eddies decrease with increasing density as with
the LCTs.

Hence we believe significant turbulence generated during virialization
should be present in all cosmological halos.  The cooling efficiency
of the gas, the total halo mass, and partly the merger history
determines the magnitude of turbulence.  We discuss some implications
of virial turbulence in the following section.

\section{Discussion}

We have investigated the virialization of early pre-galactic
cosmological halos in this paper with a suite of AMR simulations with
varying chemistry and cooling models and collapse epochs.  When
analyzing the local virial equilibrium of the halo, we do not assume
that it is in equilibrium but explicitly calculate all of the relevant
terms in the virial theorem.  In both adiabatic and radiative cooling
cases, we find that the kinetic (turbulent) energy is comparable, if
not dominant, with the thermal energy.  Turbulence appreciates as
radiative cooling becomes efficient because thermal energy alone
cannot bring the system into virial equilibrium.  In this case, the
gas attempts to virialize by increasing and maintaining its kinetic
energy.

Besides violent relaxation, at least two other hydrodynamic processes
will augment virial turbulence.  The first occurs when radial inflow
interacts with the virialized gas.  Due to the Rayleigh criterion, the
high angular momentum gas creates an instability when it is deposited
by filaments at small radii.  The second happens when minor and major
mergers create Kelvin-Helmholtz instabilities and drives additional
turbulence \citep[e.g.][]{Ricker01, Takizawa05}.  Our results show
that this turbulence is acting to achieve close to virial equilibrium
at all stages during assembly and collapse.

Virial turbulence may be most important in halos which can cool
rapidly when compared to virial heating from mass accretion.
Interestingly all halo masses below $\sim$10$^{12} \Ms$ that can cool by
\lya~emission satisfy this condition \citep{Birnboim03, Dekel06,
  Wang07}.

Turbulence appears to mix angular momentum efficiently so that it
redistributes to a radially increasing function, and thus only the
lowest specific angular momentum material sinks to the center.  This
segregation allows a collapse to proceed if it were self-similar and
basically non-rotating.  Similar results have been reported in
cosmological simulations of collapses of the first stars
\citep{Abel00, Abel02a, Yoshida03, Yoshida06b, OShea07}.  Here and in
protogalactic collapses, the turbulent velocities become supersonic.
One would expect even higher Mach numbers in larger potential wells
that still have \tcool~$<$~\tdyn.

The inclusion of the surface term allows us to study the virial
equilibrium in the halo's interior where the gravitational potential
is not influential.  Here our simulations show thermal and kinetic
energies balancing the surface term and potential energy to achieve
virial equilibrium.  Before cooling is efficient, gas virially heats
and its temperature can exceed the traditional virial temperature
within \rr/2~as seen in our adiabatic simulations.  The consequences
of this additional heating is substantial because halos can collapse
and form stars before the virial temperature reaches the critical
temperature, such as $\sim$10,000K for \lya~cooling.

The timescale at which the center collapse occurs is crucial to the
type of object that forms there.  If the collapse occurs faster than a
Kelvin-Helmholtz time for a massive star ($\sim$ 300 kyr), a black
hole might form from the lowest angular momentum gas.  Conversely if
the collapse is delayed by turbulent pressure, star formation could
occur in the density enhancements created by turbulent shocks.  The
ensuing radiative feedback may create outflows and thus slow further
infall and possibly prevent the formation of a central black hole.
The nature of the first galaxies poses an important question in the
high-redshift structure formation, and to address this problem we must
consider their progenitors -- the first stars.

\subsection{Pop III Feedback}

Numerical simulations have shown that the first, metal-free (Pop III)
stars form in isolation in its host halo.  They are believed to have
stellar masses $\sim$100 \Ms~\citep{Abel02a, Omukai03, Tan04,
  Yoshida06b} and produce $\sim$$10^{50}$ photons s$^{-1}$ that can
ionize hydrogen and dissociate \hh~\citep{Schaerer02}.
One-dimensional radiative hydrodynamical calculations \citep{Whalen04,
  Kitayama04} and recently three-dimensional radiative hydrodynamical
AMR \citep{Abel07} and SPH \citep{Yoshida06a} simulations found that
pressure forces from the radiatively heated gas drive a
$\sim$30\kms~shock outwards and expels the majority of the gas in the
host halo.  Additionally the star ionizes the surrounding few kpc of
the intergalactic medium (IGM).

Pop III stellar feedback invalidates some of our assumptions in the
calculations presented here, but the general aspect of kinetic energy
being dominant should hold in the presence of these feedback
processes.  In a later paper, we will expand our simulations to
include radiative feedback from primordial stars \citep[cf.][]{
  Yoshida06a, Abel07} and the metal enrichment from pair instability
supernovae \citep{Barkat67, Bond84, Heger02} of the IGM and subsequent
star formation.

\section{Summary}

We have investigated the process of virialization in pre-galactic gas
clouds in two cosmology AMR realizations.  Our virial analyses
included the kinetic (turbulent) energies and surface pressures of the
baryons in the system.  The significance of each energy component of
the gas varies with the effectiveness of the radiative cooling, which
we quantify by performing each realization with adiabatic, hydrogen
and helium, and \hh~cooling models.  We highlight the following main
results of this study as:

\medskip

1. Inside \rcool, gas cannot virialize alone through heating but must
gain kinetic energy.  It is up to a factor of five greater than thermal
energy throughout the protogalactic halos.  This manifests itself in a
faster bulk inflow and supersonic turbulent motions.

2. In the radiative cooling models, supersonic turbulence
($\mathcal{M}$ = 1--3) leads to additional cooling within turbulent
shocks.  We expect turbulence in larger galaxies, up to 10$^{12} \Ms$,
to be even more supersonic.

3. Baryonic velocity distributions are Maxwellian that shows violent
relaxation occurs for gas as well as dark matter.  Turbulent
velocities exceed typical rotational speeds, and these halos are
only poorly modeled as solid body rotators.

4. Virial shocks between the void-halo interface occur between
\rr/2~and \rr.  Dense, cold flows in filaments do not
shock-heat until well within \rr~and as small as \rr/4.

5. Turbulence generated during virialization mixes angular momentum so
that it redistributes to a radially increasing function (the Rayleigh
criterion).

\medskip

After the halo virializes, its central part will undergo turbulent
collapse, such as in primordial star formation and galactic molecular
clouds.  These collapses should be ubiquitous in early structure
formation as turbulence can be generated through virialization,
merging, and angular momentum segregation.  We conclude that
\textit{turbulence plays a key role in virialization and galaxy
  formation.}

\chapter[Central Gas Collapse]
{Central Gas Collapse of a Atomic Hydrogen Cooling Halo}
\label{chap:collapse}

Classical and modern galaxy formation models involve the gravitational
collapse of an initially homogeneous cosmological gaseous cloud.  In
the 1970's, evidence from theoretical and observational studies arose
that galaxies were embedded in a dark halo that outweighed baryons by
a factor of 10.  The gas condensed until it was rotationally supported
and formed a gaseous disk, where stars formed.  We study this
long-standing scenario with adaptive mesh refinement simulations that
follows the evolution of the collapse of cosmological halo.  We
achieve a dynamical range of $10^{15}$ in length scale and $10^{25}$
in density that permits the investigation of the collapse to stellar
scales in a cosmological volume.

These simulations are continuations of simulations with hydrogen and
helium radiative cooling discussed in the previous chapter.  We
neglect primordial star formation and feedback in these simulations,
which we will later show to be an important factor in early galaxy
formation.  However, these simulations provide an excellent testbed
for studying turbulent collapses, which are also applicable to current
theories of star formation.

This chapter is in preparation for publication in \textit{The
  Astrophysical Journal}.  It is co-authored by Matthew Turk, who
modified the cosmological hydrodynamics code \enzo~to accurately
follow the collapse of such systems to dynamic ranges greater than
$10^{10}$ in length scale, and Tom Abel, who helped with the
organization and clarity of this chapter and suggested some of the
analysis techniques.

\section{Motivation \& Previous Work}

Since the first investigations of galaxy interactions
\citep{Holmberg41} using light bulbs, the use of numerical simulations
in galaxy formation has developed dramatically.  Not only gravity but
also hydrodynamics and cooling are standard ingredients in the
sophisticated computer models studying galaxy formation and
interactions.  In hierarchical structure formation, dark matter (DM)
halos merge to form larger halos while the gas infalls into these
potential wells \citep{Peebles68, White78}.  \citeauthor{White78}
provided the basis for modern galaxy formation.  In this picture,
small galaxies form early and continuously merge into larger systems.

As more high redshift galaxies were observed in the following 10
years, \citet{White91} expanded and refined the ideas in White \& Rees
to explain and model the observed characteristics in these galaxies.
In their model, the halo accumulates mass until the gas cools faster
than a Hubble time, \tH, which usually occurs when atomic hydrogen
line, specifically \lya, cooling is efficient.  This happens when the
halo has \tvir~$>$~10$^4$ K, where the cooling function sharply rises
by several orders of magnitude because the number of free electrons
able to excite hydrogen greatly increases at this temperature
\citep{Spitzer78}.  One can define a cooling radius, \rcool, in which
the interior material is able to cool.  Once the halo reaches this
first milestone, \rcool~ increases through additional accretion and
cooling.  A rapid baryonic collapse ensues when \tcool~$<$~\tdyn~
\citep{Rees77}.  The material accelerates towards the center, and its
density quickly increases.  In the model discussed in White \& Frenk,
this collapse will halt once a gaseous disk forms stars.  In the first
scenario, angular momentum prevents the gas from collapsing further
and becomes rotationally supported.  Afterwards, this disk fragments
and star formation follows.  In the latter scenario, star formation
does not necessarily develop in a disk component, but the energy
released by stars during their lifetimes and associated supernovae
(SNe) terminates the collapse.

These concepts have been applied also to the earliest galaxies in the
universe \citep{Mo98, Oh02, Begelman06, Lodato06}.  Many studies
\citep[e.g.][]{Ostriker96, Haiman97b, Cen03, Somerville03, Wise05}
demonstrated that OB-stars within protogalaxies at $z > 6$ produce the
majority of photons required for reionization.  These protogalaxies
contain an ample gas reservoir for widespread star formation, and the
accompanying radiation propagates into and ionizes the surrounding
neutral intergalactic medium.  Several high redshift starburst
galaxies have been observed that support ubiquitous star formation at
$z > 6$ \citep{Stanway03, Mobasher05, Bouwens06}.  Additionally,
supermassive black holes (SMBH) more massive than 10$^8 \Ms$ are
present at these redshifts \citep[e.g.][]{Becker01, Fan02}.  Finally,
a reionization signature at z $\sim$ 10 \citep{Page06} further
supports and constrains stellar and SMBH activity at high redshifts.

Stars formed in these earliest galaxies are part of present day
galaxies.  Thus constraints on protogalactic environments may be
derived from local dwarf galaxies \citep[e.g.][]{Tolstoy03, Tolstoy04,
  Gnedin06, Helmi06}.  The distinction between SMBH formation and a
starburst galaxy should depend on the initial ingredients (e.g. seed
BHs, metallicity, merger histories) of the host halo, but the
evolution of various initial states is debatable.  It is essential to
study the hydrodynamics of high redshift halo collapses because the
initial luminous object that emerges will chemically and thermally
alter its surroundings.  For example, as the object emits ultraviolet
radiation, the nearby gas heats and thus the characteristic Jeans mass
increases, which may inhibit the accretion of new gas for future star
formation \citep{Efstathiou92, Thoul96}.

The following work will attempt to clarify early galaxy formation by
focusing on \tvir~$>10^4$ K halos and following their evolution in the
early universe.  \citet[][hereafter Paper I]{Wise07a} studied the
virialization of protogalactic halos and the virial generation of
supersonic turbulence.  In this paper, we only address the gas
dynamics of the continued, turbulent collapse of a halo and study the
evolution and characteristics of the central object.  In later
studies, we will introduce the effects from star and SMBH formation
and feedback and \hh~cooling.  The progressive introduction of new
processes is essential to understand the relevance of each mechanism.
We argue that our results are highly relevant for scenarios that
envisage SMBH formation from gaseous collapses.

\citet{Loeb94} and \citet{Bromm03} conducted smoothed particle
hydrodynamics (SPH) simulations that focused on the collapse of
idealized, isolated protogalactic halos.  The former group concluded
that a central $10^6 \Ms$ SMBH must exist to stabilize the thin gaseous
disk that forms in their calculations.  \citeauthor{Bromm03}
considered cases with and without \hh~chemistry and a background UV
radiation field.  They observed the formation of a dense object with a
mass $M \sim 10^6$, or $\gsim 10\%$ of the baryonic matter, in
simulations with no or suppressed \hh~formation.  \citet{Spaans06}
analytically studied the collapse of 10$^4$ K halos with an atomic
equation of state.  They find that $\sim$0.1\% of the baryonic mass
results in a pre-galactic BH with a mass $\sim$$10^5 \Ms$.
Additionally, \citet{Lodato06} also found that $\sim$5\% of the gas
mass in $M = 10^7 \Ms$ halos at $z \sim 10$ become unstable in a
gaseous disc and form a SMBH.  These calculations without metal
cooling and stellar feedback are useful to explore the hydrodynamics
of the collapse under simplified conditions.

A runaway gaseous collapse requires angular momentum transport so
material can inflow to small scales and form a central object.  The
stability of rotating gaseous objects have been subject of much
interest over the last four centuries and was thoroughly detailed by
the work of \citet[][hereafter EFE]{Chandra69}.  In the 1960's and
1970's, studies utilizing virial tensor techniques
\citep[EFE;][]{Lebovitz67, Ostriker69, Ostriker73a}, variational
techniques \citep{LyndenBell67, Bardeen77}, and N-body simulations
\citep{Ostriker73b} all focused on criteria in which a stellar or
gaseous system becomes secularly or dynamically unstable.  The first
instability encountered is an $m = 2$ bar-like instability, which is
conducive for angular momentum transport in order to form a dense,
central object.  \citet{Begelman06} investigated the conditions where
a gaseous disc would become rotationally unstable to bar formation
\citep[see][]{Christodoulou95a, Christodoulou95b}.  They adapt the
``bars within bars'' scenario \citep{Shlosman89, Shlosman90}, which
was originally formulated to drive SMBH accretion from a gaseous bar
that forms in a stellar galactic bar, to the scenario of pre-galactic
BH formation.  Here a cascade of bars form and transport angular
momentum outwards, and the system can collapse to small scales to form
a quasistar with runaway neutrino cooling, resulting in a central
SMBH.  In one of our realizations, a central bar-like instability
forms, for which we reveal the causing mechanism.

As briefly mentioned before, angular momentum plays a dominant role in
halo contractions.  The standard picture of galaxy formation inside
this paradigm includes the following elements: tidal torques from
cosmological neighbors create angular momentum \citep{Hoyle49,
  Peebles69}; baryonic angular momentum is conserved while cooling
\citep{Mestel63}; both DM and baryonic components of a virialized halo
have equivalent initial angular momentum distributions \citep{Fall80}.
The halo is subject to turbulence from gravitational instabilities
\citep{Elmegreen93} and angular momentum transport \citep[see Paper
I;][]{Toomre64, Lin87} as it collapses, and cosmological virialization
(see Paper I).

In \S2 we describe our simulations and their cosmological context.  In
the following section, we detail the analytical models of rotational
instabilities.  Next in \S4, we present our analysis of the halo
collapse simulations.  Here we investigate the structural and
hydrodynamical evolution, the initial halo collapse, rotational
instabilities, and the importance of turbulence.  In \S5, we discuss
its implications on the field of early galaxy formation.  There we
also examine the applicability and limitations of our results and
desired improvements in our simulations.  Finally we conclude in the
last section.

\section{Simulation Techniques}

To investigate protogalactic (T$_{{\rm vir}} >10^4$ K) halo collapses
in the early universe, we utilize an Eulerian structure, adaptive mesh
refinement (AMR), cosmological hydrodynamical code, \enzo\footnote{See
  http://cosmos.ucsd.edu/enzo/} \citep{Bryan97, Bryan99, OShea04}.
\enzo~solves the hydrodynamical equations using a second order
accurate parabolic method \citep{Woodward84, Bryan94}, while a Riemann
solver ensures accurate shock capturing with minimal viscosity.
Additionally \enzo~ uses a particle-mesh N-body method to calculate
the dynamics of the collisionless dark matter particles
\citep{Couchman91}.  Regions of the simulation grid are refined by two
when one of the following conditions are met: (1) Baryon density is
greater than 3 times $\Omega_b \rho_0 N^{l(1+\phi)}$, (2) DM density
is greater than 3 times $\Omega_{\rm{CDM}} \rho_0 N^{l(1+\phi)}$, and
(3) the local Jeans length is less than 16 cell widths.  Here $N = 2$
is the refinement factor; $l$ is the AMR refinement level; $\phi =
-0.3$ causes more frequent refinement with increasing AMR levels,
i.e. super-Lagrangian behavior; $\rho_0 = 3H_0^2/8\pi G$ is the
critical density; and the Jeans length, $L_J = \sqrt{15kT/4\pi\rho G
  \mu m_H}$, where $H_0$, $k$, T, $\rho$, $\mu$, and $m_H$ are the
Hubble constant, Boltzmann constant, temperature, gas density, mean
molecular weight in units of the proton mass, and hydrogen mass,
respectively.  The Jeans length refinement insures that we meet the
Truelove criterion, which requires the Jeans length to be resolved by
at least 4 cells on each axis \citep{Truelove97}. Runs with a
refinement criterion of 4, 8, and 16 Jeans lengths have
indistinguishable mass weighted radial profiles.

We conduct the simulations within the concordance $\Lambda$CDM model
with WMAP 1 year parameters of $h$ = 0.72, \Ol~= 0.73, \Om~= 0.27,
\Ob~= 0.024$h^{-2}$, and a primordial scale invariant ($n$ = 1) power
spectrum with $\sigma_8$ = 0.9 \citep{Spergel03}.  $h$ is the Hubble
parameter in units of 100 km s$^{-1}$ Mpc$^{-1}$.  \Ol, \Om, and
\Ob~are the fractions of critical energy density of vacuum energy,
total matter, and baryons, respectively.  Lastly $\sigma_8$ is the rms
of the density fluctuations inside a sphere of radius 8$h^{-1}$ Mpc.

Using the WMAP1 parameters versus the significantly different WMAP
third year parameters \citep[WMAP3;][]{Spergel06} have no effect on
the evolution of individual halos that are considered here.  However
these changes play an important role in statistical properties.  For
example, halos with mass $10^6 \Ms$ at redshift 20 correspond to
$2.8\sigma$ peaks with the WMAP1 but are $3.5\sigma$ peaks for WMAP3.
The \Om/\Ob~ratio also only changed from 6.03 to 5.70 in WMAP3.  Also
we have verified that there is nothing atypical about the mass
accretion rate histories of the objects we study.

The initial conditions of this simulation are well-established by the
primordial temperature fluctuations in the cosmic microwave background
(CMB) and big bang nucleosynthesis (BBN) \citep[][and references
therein]{Hu02, Burles01}.

%
%
\begin{center}
\begin{longtable}{ccccccc}
\caption{Simulation Parameters} \label{tab:params} \\

\hline\hline \\[-3ex]
Name & l & N$_{part}$ & N$_{grid}$ & N$_{cell}$ & L$_{max}$ & $\Delta$x \\
   & [Mpc] &  &  & &  & [R$_\odot$] \\
\hline
\endhead

A & 1.0 & 2.22 $\times$ 10$^7$ & 44712 & 1.23 $\times$ 10$^8$
(498$^3$) & 41 & $9.3 \times 10^{-3}$ \\
B & 1.5 & 1.26 $\times$ 10$^7$ & 22179 & 7.40 $\times$ 10$^7$
(420$^3$) & 41 & $1.4 \times 10^{-2}$\\

\hline
\end{longtable}

\tablecomments{Col. (1): Simulation name. Col. (2): Number of dark
  matter particles. Col. (3): Number of AMR grids. Col. (4): Maximum
  number of unique grid cells. Col. (5): Maximum level of refinement
  reached in the simulation. Col. (6): Resolution at the maximum
  refinement level.}
\end{center}

We perform two simulations in which we vary the box size to study
different scenarios and epochs of halo collapse.  In the first
simulation, we setup a cosmological box with 1 comoving Mpc on a side
(Simulation A), periodic boundary conditions, and a 128$^3$ top grid.
The other simulation is similar but with a box side of 1.5 comoving
Mpc (Simulation B).  We provide a summary of the simulation parameters
in Table \ref{tab:params}.  These volumes are adequate to study halos
of interest because the comoving number density of $>$10$^4$ K halos
at $z=10$ is $\sim$6 Mpc$^{-3}$ according to Press-Schechter formalism
\citep{Press74, Sheth02}.  We use the COSMICS package to calculate the
initial conditions at $z$ = 129 (119)%
\renewcommand{\thefootnote}{\fnsymbol{footnote}}%
\footnote{To simplify the discussion, simulation A will always
  be quoted first with the value from simulation B in parentheses.}
\renewcommand{\thefootnote}{\arabic{footnote}}%
\citep{Bertschinger95, Bertschinger01}.  It calculates the linearized
evolution of matter fluctuations.  We first run a dark matter
simulation to $z=10$ and locate the DM halos using the HOP algorithm
\citep{Eisenstein98}.  We identify the first dark matter halo in the
simulation that has \tvir~$>$ 10$^4$ K and generate three levels of
refined, nested initial conditions with a refinement factor of two
that are centered around the Lagrangian volume of the halo of
interest.  The nested grids that contain finer grids have 8 cells
between its boundary and its child grid.  The finest grid has an
equivalent resolution of a 1024$^3$ unigrid and a side length of 250
(300) comoving kpc.  This resolution results in a DM particle mass of
30 (101) $\Ms$ and an initial gas resolution of 6.2 (21) $\Ms$.  These
simulations continue from the endpoints of Simulations A6 and B6 of
Paper I.  Table \ref{tab:runs} lists the parameters of the most
massive halo in each realization.  We evolve the system until the
central object has collapsed and reached our resolution limit.  There
are 1.23 $\times$ 10$^8$ (498$^3$) and 7.40 $\times$ 10$^7$ (420$^3$)
unique cells in the final simulation output of the 1 and 1.5 Mpc
simulation, respectively.  The finest grid then has a refinement level
of 41 and a spatial resolution of roughly 0.01 of a solar radius in
both simulations.

%
%
\begin{center}
\renewcommand{\thefootnote}{\alph{footnote}}
\begin{longtable}{ccccccc}
\caption{Halos of interest} \label{tab:runs} \\

\hline\hline \\[-3ex]
l & z & M$_{tot}$ & $\sigma$ & $\rho_c$ & T$_c$ & M$_{BE}$ \\
\[Mpc] &  & [$\Ms$]  &  & [cm$^{-3}$] &[K] & [$\Ms$] \\
\hline
\endhead

1.0 & 15.87 & 3.47 $\times$ 10$^7$ & 2.45 & 5.84 $\times$ 10$^{21}$ & 
8190 & 4.74 $\times$ 10$^5$ \\

1.5 & 16.80 & 3.50 $\times$ 10$^7$ & 2.59 & 7.58 $\times$ 10$^{21}$ & 
8270 & 1.01 $\times$ 10$^5$ \\
\hline
\end{longtable}

\tablecomments{The subscript ``c'' denotes central quantities.}
\tablecomments{Col. (1): Box size of the simulation. Col. (2): Final
  redshift of simulation. Col. (3): Total halo mass. Col. (4):
  $\sigma$ of the total mass compared to matter
  fluctuations. Col. (5): Central halo density. Col. (6): Central gas
  temperature. Col. (7): Gravitationally unstable central mass.}

\renewcommand{\thefootnote}{\arabic{footnote}}
\end{center}

\enzo~employs a non-equilibrium chemistry model \citep{Abel97,
  Anninos97}, and we consider six species in a primordial gas (H,
H$^{\rm +}$, He, He$^{\rm +}$, He$^{\rm ++}$, e$^{\rm -}$).  Compton
cooling and heating of free electron from the CMB and radiative losses
from atomic cooling are computed in the optically thin limit.  At high
densities in the halo cores, the baryonic component dominates the
material.  However, the discrete sampling of the DM potential by
particles can become inadequate, and artificial heating (cooling) of
the baryons (DM) can occur.  To combat this effect, we smooth the DM
particles in cells with a width $<$0.24 ($<$0.36) comoving pc, which
corresponds to a refinement level of 15.

%
%
%
\begin{figure}
\begin{center}
\includegraphics[width=\textwidth]{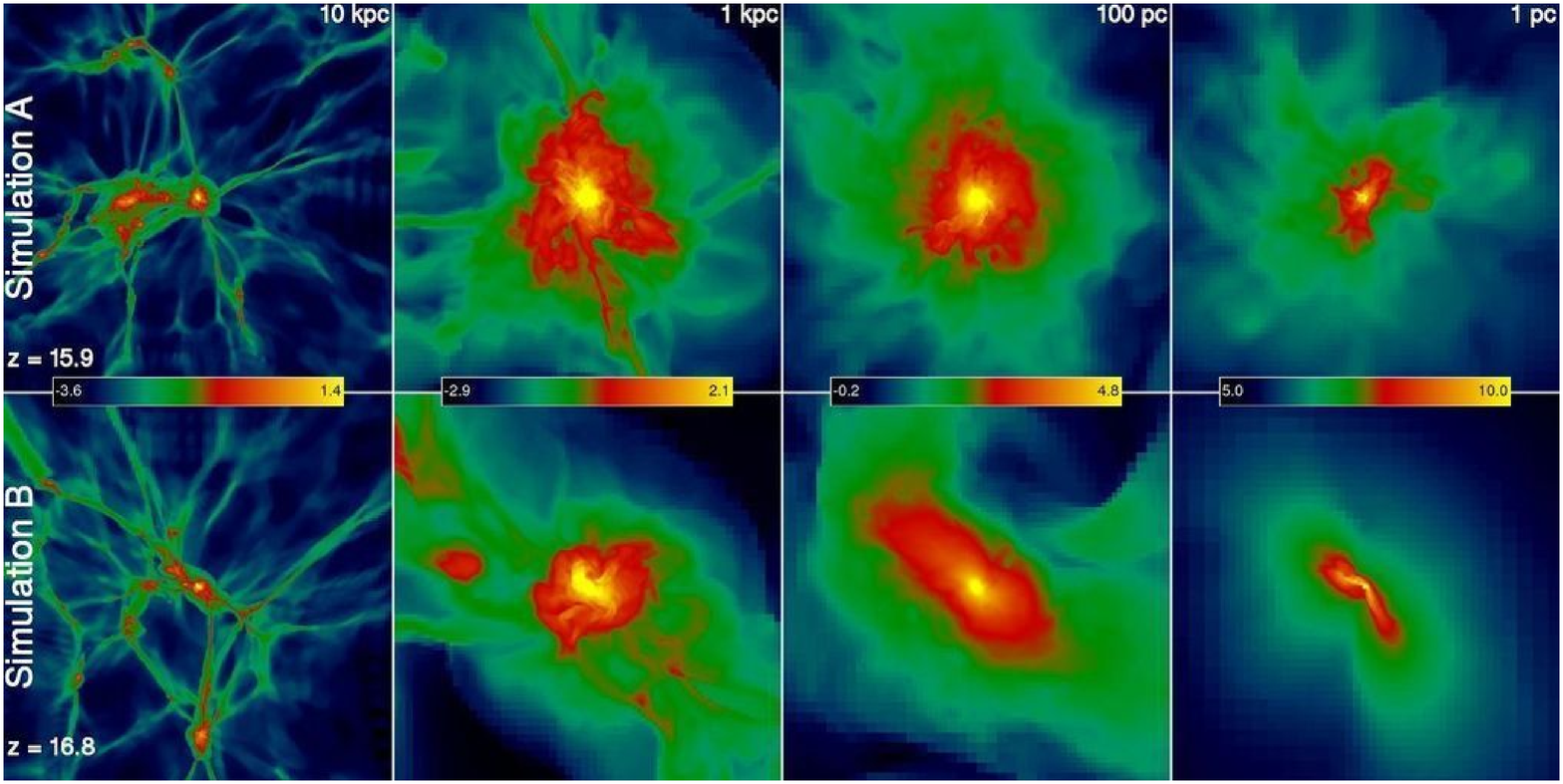}
\includegraphics[width=\textwidth]{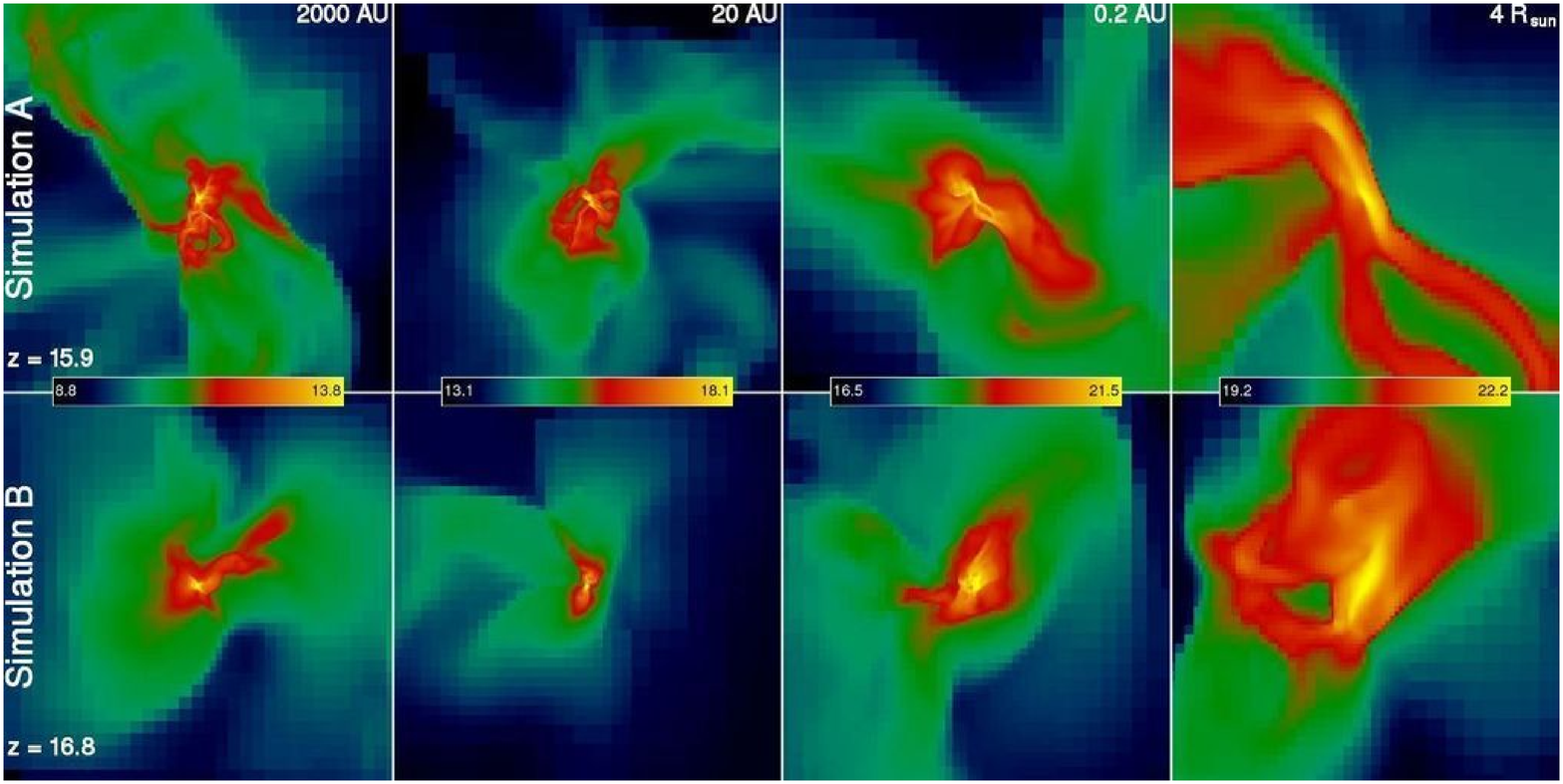}
\caption[Overview of density structures in a collapsing halo] {\scriptsize{An
  overview of the final state of the collapsing protogalactic gas
  cloud.  Slices of gas density in cm$^{-3}$ are shown through the
  densest point in the halo.  The \textit{first} and \textit{three}
  rows show Simulation A, and the \textit{second} and \textit{fourth}
  rows show Simulation B.  The columns in the top two rows from left
  to right are slices with a field of view of 10 kpc, 1 kpc, 100 pc,
  and 1 pc.  For the bottom two rows, the fields of view are 0.01pc,
  20AU, 0.2AU, and 4 R$_\odot$.  Note that each color scale is
  logarithmic (values increases with {\em black, blue, green, red, and
    yellow}), spans 5 orders of magnitude, and is unique for every
  panel.  At the 10 kpc scale, the filamentary large-scale structure
  is shown, and the protogalactic halo exists at the intersection of
  these filaments.  In the next scale, we show the protogalactic gas
  cloud.  At the 100 pc scale, a thick disk is seen in Simulation B.
  It is nearly edge-on in this view.  In the inner 10 pc, the panels
  focus on the gravitationally unstable central mass that has a radius
  of 7.9 pc and 1.5pc for Simulation A and B, respectively. In
  Simulation B at 1 pc, a bar forms from a rotational secular
  instability that transports angular momentum outwards.  Similar
  instabilities exist at radii of 0.2 pc, 90 AU, 0.7 AU, and 0.02 AU
  in Simulation B.  Simulation A also undergoes a secular bar
  instability at smaller scales at radii of 140 AU and 0.02 AU but
  shows a more disorganized medium at larger scales.}}
\label{fig:slices} 
\end{center}
\end{figure}

%
%
\begin{figure}[t]
\begin{center}
  \includegraphics[width=0.6\textwidth]{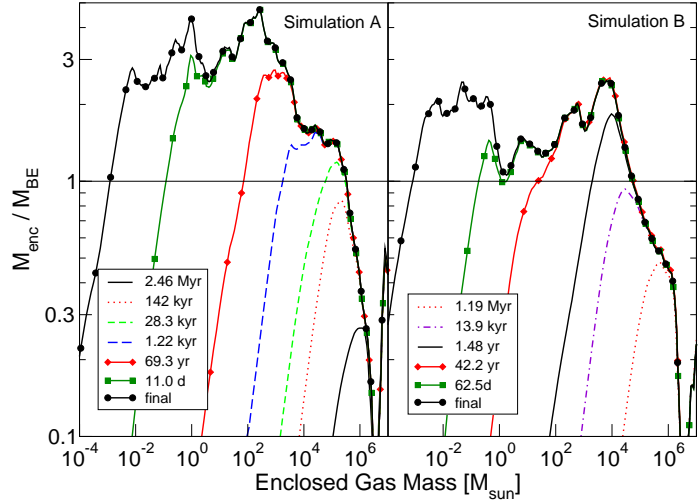}
  \caption[Bonner-Ebert mass]{The ratio of the enclosed gas mass and
    Bonnor-Ebert mass (eq. \ref{eqn:mbe}) for the final output {\em
      (black with circles)} and selected previous times that are
    listed in the legend.  Simulation A (\textit{left}) and B
    (\textit{right}).  For values above the horizontal line at
    $M_{\rm{enc}} / M_{\rm{BE}} = 1$, the system is gravitationally
    unstable.}
  \label{fig:mbe}
\end{center}
\end{figure}

\section{Results} 

In this section, we first describe how the halo collapses when it
starts to cool through \lya~line emission.  Then we discuss the role
of turbulence in the collapse.  Lastly we describe the rotational
properties and stability of the halo and central object.

\subsection{Halo Collapse}
\label{sec:collapse}


At z = 21.1 in simulation A, the progenitor of the final halo (\mvir~=
4.96 $\times$ 10$^6 \Ms$) starts to experience two major mergers,
which continues until z = 17.2 when \mvir~= 2.36 $\times$ 10$^7 \Ms$.
We define \mvir~as the mass M$_{200}$ in a sphere that encloses an
average DM overdensity of 200.  In simulation B, no major merger
occurs before the cooling gas starts to collapse, but it accumulates
mass by accretion and minor mergers.  Mergers disrupt the relaxed
state of the progenitor and create turbulence as these systems collide
and combine.  Additional turbulence arises during virialization, as
discussed in Paper I.  In the density slices of Figure
\ref{fig:slices}, more small scale density fluctuations are present in
simulation A.  These fluctuations penetrate farther into the potential
well in simulation A to scales%
\footnote{Note that all masses concerning the collapse are gas mass,
  not total mass.  The central regions of r $<$ 10 pc are baryon
  dominated so that $M_{{\rm enc,\; gas}} \approx M_{{\rm enc,\;
      tot}}$.  All length scales are in proper units unless otherwise
  noted.} of 5 pc, compared to simulation B that contains nearly no
fluctuations between 5 and 50 pc.  The virial temperatures are now
$\geq 10^4$ K, and therefore they can efficiently cool by atomic
hydrogen transitions.  The gas fulfills the critical condition for
contraction, \tdyn~$>$ \tcool, and proceeds to continuously collapse
in approximately a dynamical time.

%
%
%
\begin{figure}[t]
\plottwo{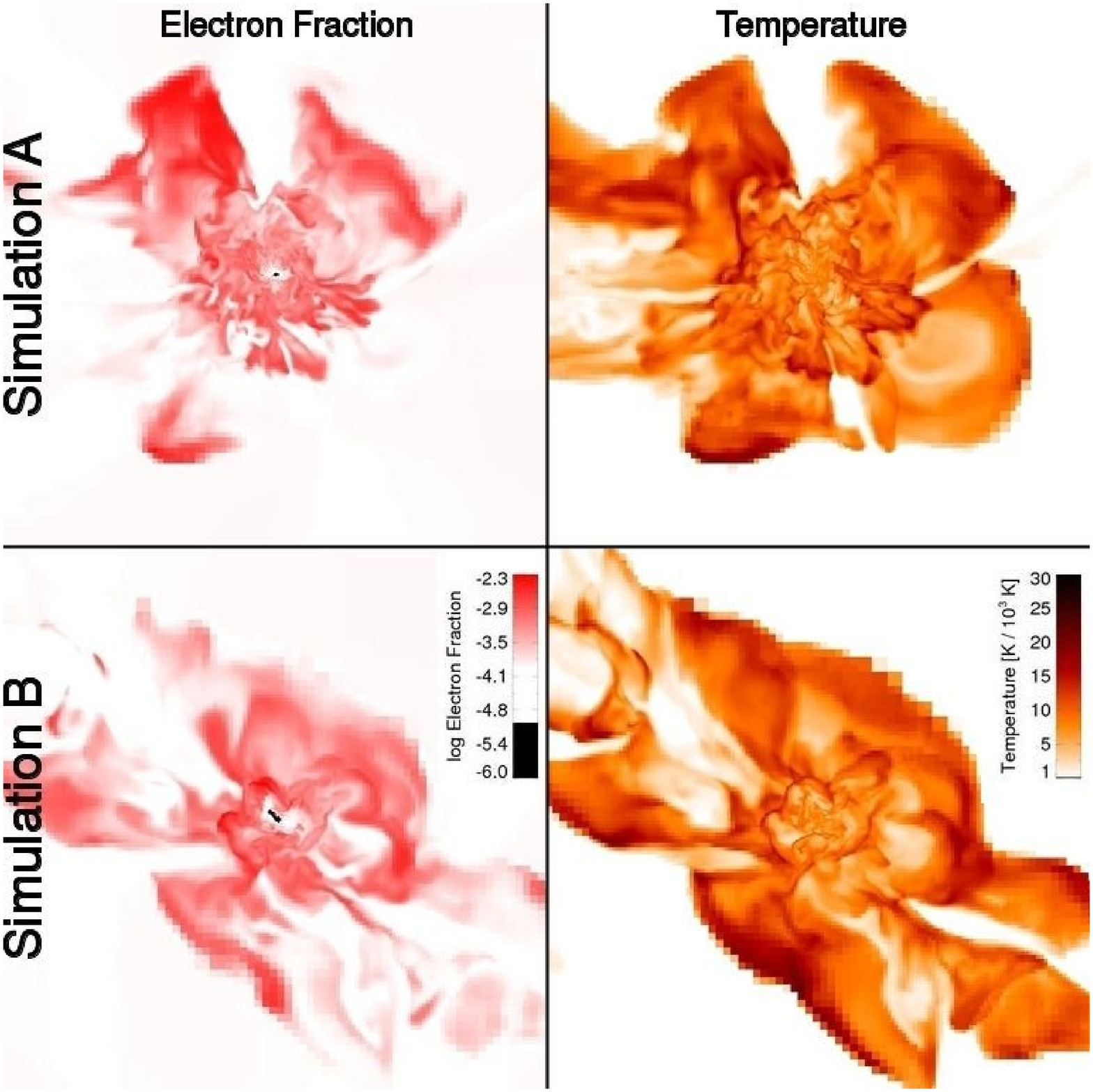}{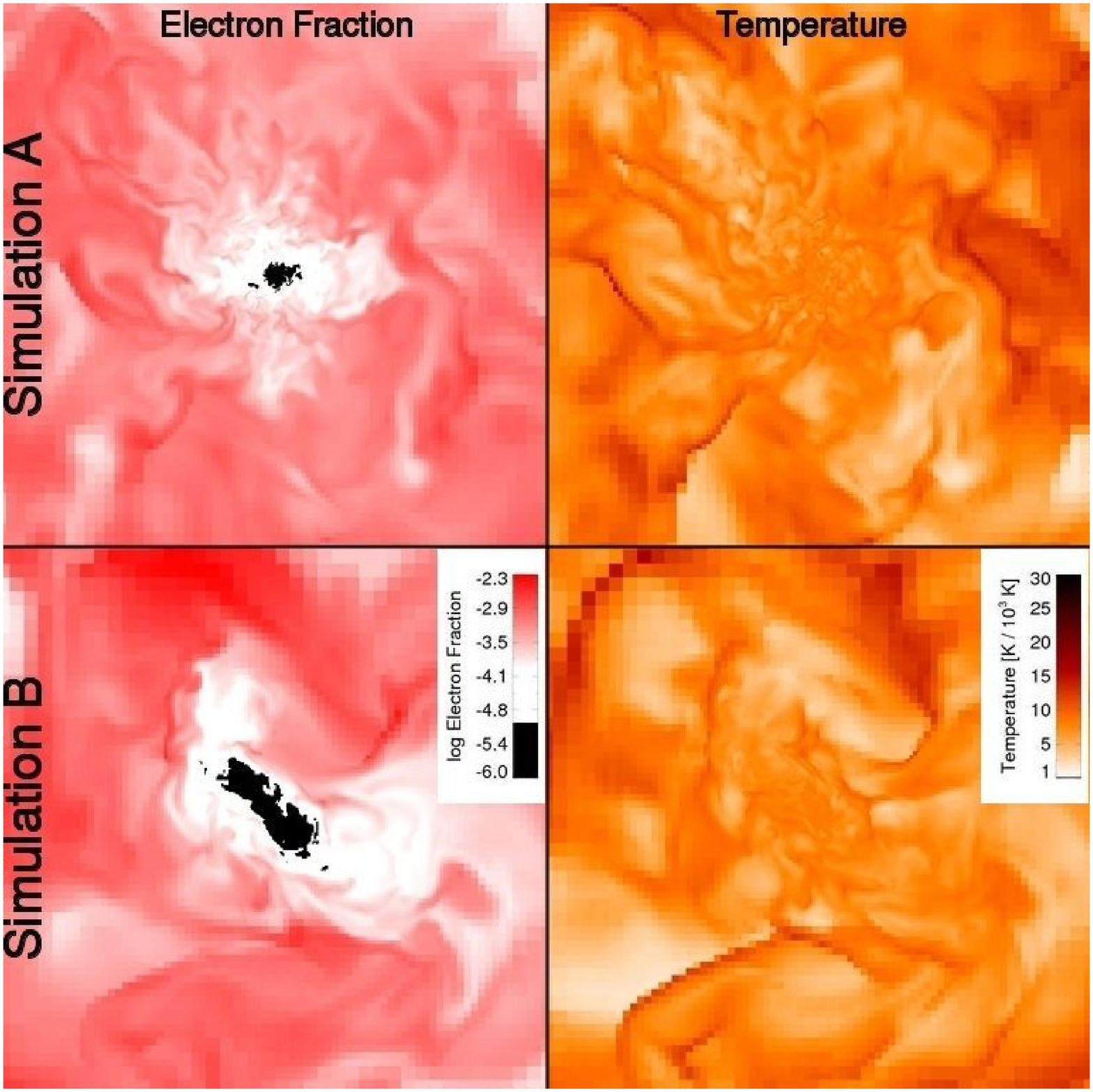}
\caption[Electron fraction and temperature of a collapsing
halo]{Slices of electron fraction (\textit{left}) and temperature
  (\textit{right}) of Simulation A (\textit{top}) and B
  (\textit{bottom}).  The field of view is 1.5 kpc (\textit{left
    panels}) and 200 pc (\textit{right panels}).  The color scale is
  logarithmic for electron fraction and linear for temperature.
  Electron fractions below 10$^{-5}$ are shown in black.  Supersonic
  turbulent shocks are ubiquitous throughout the halos.}
\label{fig:tempElec1} 
\end{figure}

%
%
\begin{figure}
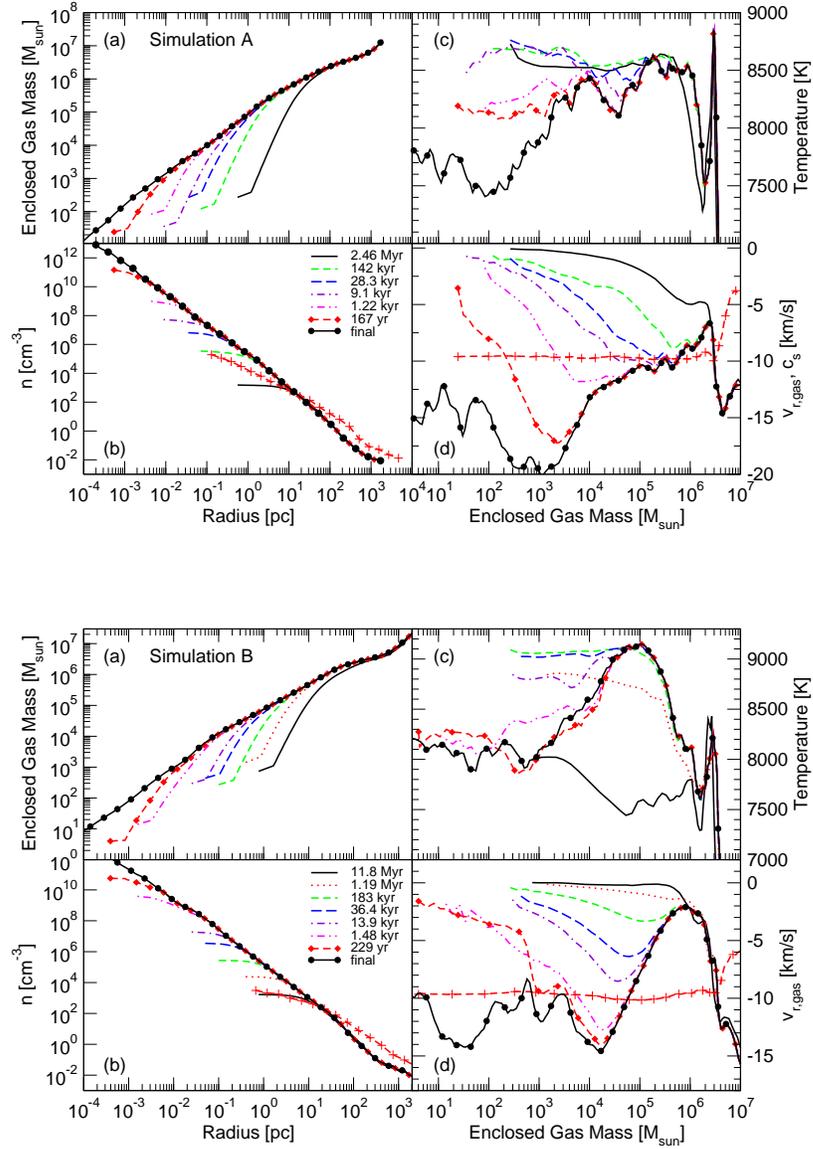

\begin{center}
\includegraphics[width=0.7\textwidth]{chapter4/fig/f3a} \\[3em]
\includegraphics[width=0.7\textwidth]{chapter4/fig/f3b}
\caption[Radial profiles at large scales]{Mass-weighted radial
  profiles at various times of (a) gas mass enclosed, (b) number
  density, (c) mass-weighted temperature, and (d) mass-weighted radial
  velocity for simulation A (\textit{upper panels}) and simulation B
  (\textit{lower panels}).  The quantities in the left and right
  panels are plotted with respect to radius and gas mass enclosed,
  respectively.  In (b), the dashed line with crosses is the dark
  matter density in units of $m_H$ cm$^{-3}$.  In (d), the dashed line
  with crosses is the negative of the sound speed in the final output.
  The times in the legends correspond to time before the end of the
  simulation.}
\label{fig:profilesA}
\end{center}
\end{figure}

%
%
%
\begin{figure}
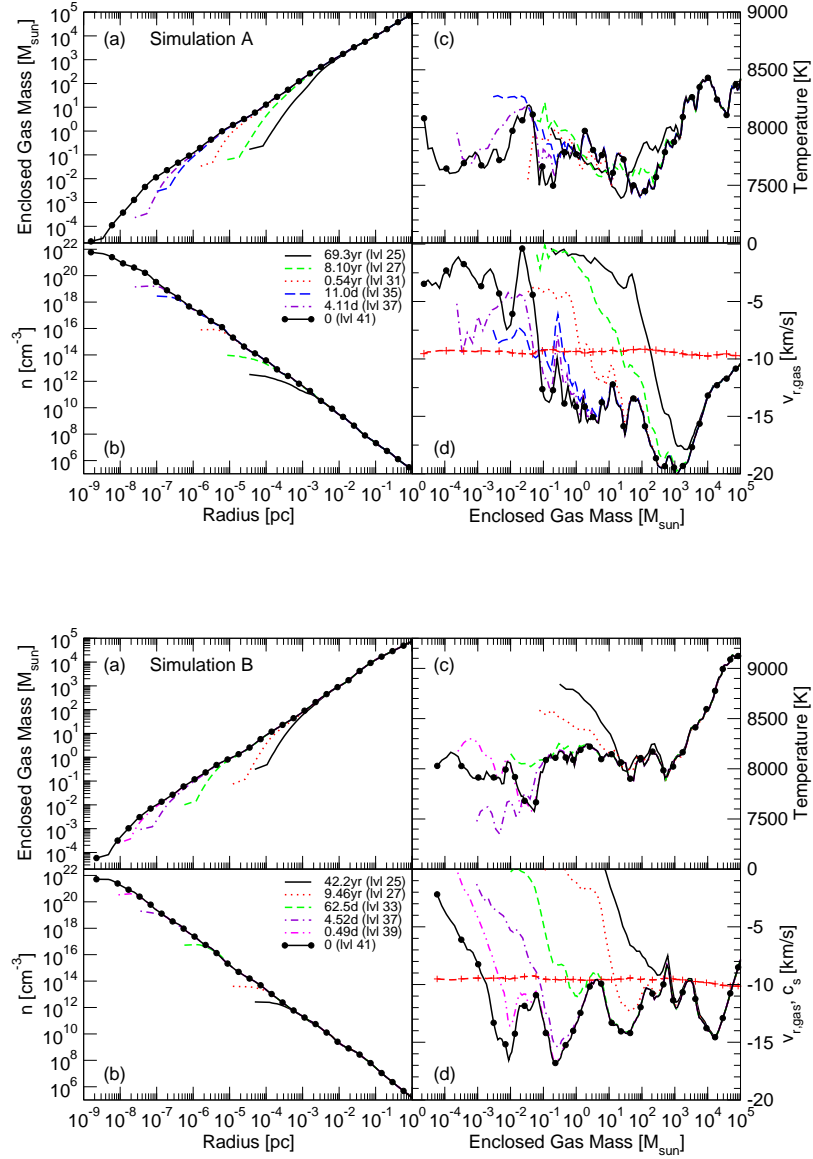

\begin{center}
\includegraphics[width=0.7\textwidth]{chapter4/fig/f3c} \\[3em]
\includegraphics[width=0.7\textwidth]{chapter4/fig/f3d}
\caption[Radial profiles at small scales]{Same as Figure
  \ref{fig:profilesA} for the inner parsec of Simulation A
  (\textit{upper panels}) and Simulation B (\textit{lower panels}).
  The maximum AMR level is listed next to the times in the legend.  In
  Simulation B, the local minima in radial velocities at $2 \times
  10^4$, 40, 0.3, and 0.01 $\Ms$ occur as angular momentum is
  transported outwards in secular bar-like instabilities.}
\label{fig:profilesA3} 
\end{center}
\end{figure}

The halo collapses in two stages.  We denote the beginning of the
first stage when \tdyn~$>$ \tcool~for the first time.  The second
stage begins when the central object becomes gravitationally unstable.

1. \textit{Cooling stage}--- As mass infalls toward the center, the
increased cooling rate, which is $\propto nn_e$, catalyzes the
collapse as atomic line transitions convert kinetic energy to
radiation.  Here $n$ and $n_e$ are the number density of baryons and
electrons, respectively.  The first stage starts 520 (36) kyr before
the last output.  The inner 100 pc have a steady decrease in electron
fraction that indicates atomic hydrogen cooling is now efficient in
this region.  However, only the gas within 1.5 (1.0) pc has
\tdyn~$\gsim$ \tcool~= 383 (100) kyr at this epoch.

2. \textit{Gravitationally unstable stage}--- We observe the start of
the second collapse stage when the central region becomes unstable to
gravitational collapse.  \citet{Ebert55} and \citet{Bonnor55}
investigated the stability of an isothermal sphere with an external
pressure $P_{ext}$ and discovered that the critical mass (BE mass
hereafter) for gravitational collapse is
\begin{equation}
\label{eqn:mbe}
M_{{\rm BE}} = 1.18 \frac{c_s^4}{G^{3/2}} P_{ext}^{-1/2} \,\Ms .
\end{equation}
If we set $P_{ext}$ to the local pressure, then
\begin{equation}
M_{{\rm BE}} \approx 20 T^{3/2} n^{-1/2} \mu^{-2} \gamma^2 \Ms .
\end{equation}
For both simulations, this stage occurs between 10 and 100 kyr before
we end the simulation.  We plot the ratio of the enclosed gas mass and
BE mass in Figure \ref{fig:mbe} for several epochs in the collapse.
When the clump becomes gravitationally unstable, the central 3.3
$\times$ 10$^5$ (5.5 $\times$ 10$^4$) $\Ms$ in the central
$r_{\rm{BE}}$ = 5.8 (0.9) pc exceeds the BE mass, and its \tdyn~= 520
(80) kyr.  Thus our numerical results agree with these analytic
expectations.

We follow the evolution of the accretion and contraction until the
simulation\footnote{We stop the simulation due to ensuing round-off
  errors from a lack of precision.  We use 80-bit precision arithmetic
  for positions and time throughout the calculation.} reaches a
refinement level of 41 (41) that corresponds to a resolution of 0.01
(0.014) R$_\odot$.  At this point, the central 4.7 $\times$ 10$^5$
(1.0 $\times$ 10$^5$) $\Ms$ are gravitationally unstable and
\textit{not} rotationally supported.  The central mass is nearly
devoid of free electrons where the electron fraction, $n_e / n <
10^{-6}$, and the temperature is $\sim8000$ K.  It has a radius of 7.9
(1.5) pc.  The central number density is 5.8 (7.6) $\times$ 10$^{21}$
cm$^{-3}$.  Two-dimensional slices of gas density in both simulations
are displayed at various length scales in Figure \ref{fig:slices}.
Figure \ref{fig:tempElec1} gives slices of electron fraction and
temperature with fields of view of 1.5 kpc and 200 pc.

Next we show the radial profiles of the final and preceding outputs in
Figures \ref{fig:profilesA} and \ref{fig:profilesA3}, where we plot
(a) enclosed gas mass, (b) number density, (c) mass-weighted
temperature, and (d) mass-weighted radial velocity.  Figure
\ref{fig:profilesA} focuses on length scales greater than 20 AU to $r
> \rvir$.  The halo collapses in a self-similar manner with $\rho(r)
\propto r^{-12/5}$.  We also overplot the DM density in units of
m$_{\rm{H}} \cubecm$ in the $b$ panels.  The DM density in simulation
A does not flatten as in simulation B with $\rho_{\rm{DM}} \propto
r^{-4/3}$ and $r^{-2/3}$, respectively.  In the $c$ panels, ones sees
that the entire system is isothermal within 10\% of 8000 K. In the $d$
panels, the sound speed in the final epoch is plotted, and one sees a
clear shock at a mass scale when $M_{{\rm enc}}$ first exceeded
$M_{{\rm BE}}$ where $v_r > c_s$.  Here $v_r$ is the radial velocity,
and $c_s$ is the local sound speed.

Figure \ref{fig:profilesA3} shows the data within 1 pc and later
times.  Here one sees the self-similar, isothermal collapse continues
to stellar scales.  However, the structure in the radial velocity in
simulation B exhibits a striking behavior with four successive
minima at mass scales $2 \times 10^4$, $10^3$, 6, and $10^{-3} \Ms$.
We attribute this to rotational bar-like instabilities that we discuss
later in the paper.  

\begin{figure}[t]
\begin{center}
\includegraphics[width=0.6\textwidth]{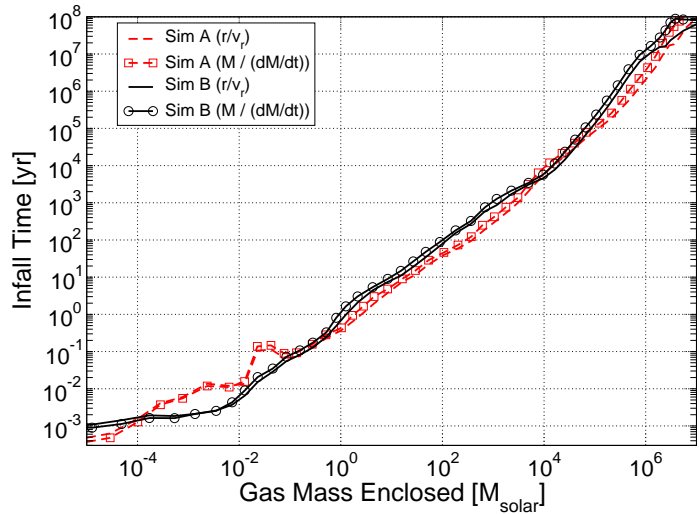}
\caption[Infall times]{Radial profiles of gas infall times.  To
  approximate a collapse timescale, the quantities $r/v_r$ {\em
    (solid)} and $M/(dM/dt)$ {\em (dashed)} are calculated and plotted
  here.}
\label{fig:infall} 
\end{center}
\end{figure}

If we consider $v_r$ constant from the last output, we can determine
the infall times, which are shown in Figure \ref{fig:infall}.  The
infall time, $t_{in} = r/v_r$, of the shocked BE mass is 350 (50) kyr.

%
%

\subsection{Turbulence}
\label{sec:turbulence}

\citet{Kolmogorov41} described a theory of the basic behavior of
incompressible turbulence where turbulence is driven on a large scale
that forms eddies at that scale.  Then these eddies interact to form
smaller eddies and transfer some of their energy to the smaller scale.
This cascade continues until the dissipation scale is reached, and
energy is dissipated through viscosity.  In supersonic turbulence,
most of the turbulent energy is dissipated through shock waves, which
removes the local nature of cascades found in incompressible
turbulence.  

In Paper I, we found that turbulence is stirred during virialization.
When radiative cooling is efficient, the gas cannot virialize by
gaining thermal energy and must increase its kinetic energy in order
to reach equilibrium, which it achieves by radial infall and turbulent
motions.

In addition to virial turbulence generation, mergers create
turbulence.  Here the largest driving scale will be approximately the
scale of the merging objects, and the turbulent cascade starts from
that length scale.  Additional driving may come from Kelvin-Helmholtz
instabilities as the mergers occur \citep{Takizawa05}.  Takizawa
considered mergers of galaxy clusters, however his work is still
applicable to the formation of protogalactic halos since similar
temperature contrasts also exist in this regime of mergers.  As the
lesser halo infalls into the massive halo, a bow shock and small-scale
eddies from the Kelvin-Helmholtz instability form between the two
interacting objects.  At later times, a dense, cool core remains in
the substructure of the lesser halo.  As they grow, these
instabilities destroy the baryonic substructure, and this gas mixes
with the existing gas in the massive halo and becomes highly
turbulent.

\begin{figure}[t]
\begin{center}
\includegraphics[width=0.6\textwidth]{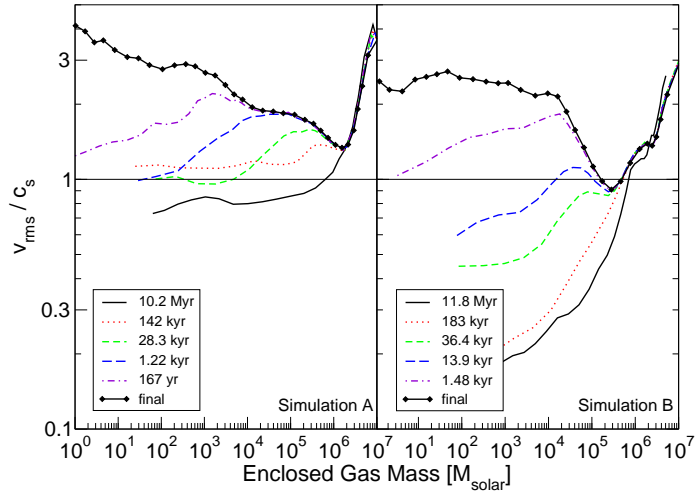}
\caption[Turbulent Mach Numbers]{The turbulent Mach number, $v_{rms} /
  c_s$, for the final output {\em (black with diamonds)} and selected
  previous times that are listed in the legend.  Simulation A
  (\textit{left}) and B (\textit{right}).}
\label{fig:mach}
\end{center}
\end{figure}

To quantify aspects of this turbulence, we inspect the turbulent Mach
number,
\begin{equation}
\mathcal{M} = \frac{v_{rms}}{c_s}; \quad 
c_s^2 = \frac{dP}{d\rho} = \frac{\gamma k T}{\mu m_H}.
\end{equation}
Here $P$ is pressure, $v_{rms}$ is the 3D velocity dispersion, and
$\gamma$ is the adiabatic index that we set to 5/3.  Radial profiles
of $\mathcal{M}$ are shown in Figure \ref{fig:mach}.  Before the core
becomes gravitationally unstable, the turbulence is subsonic within
the virial shock.  After the core becomes gravitationally unstable,
the turbulent Mach number rises to 2--3.  The collapse produces
turbulence on a timescale that is faster than it can be dissipated.



The initial turbulence may impact the nature of the central collapse.
In simulation A, the core initially has $\mathcal{M} \approx 1$, and
this results in a central object with 4.7 $\times$ 10$^5 \Ms$ and a
radius of 7.9 pc.  Alternatively, the core in simulation B has
$\mathcal{M} \approx 0.2$, and the central object is about five times
less massive and smaller, which corresponds to a free-fall time
approximately five times shorter as well.

%
%
%

\subsection{Spin Parameter Evolution}

\begin{figure}[t]
\begin{center}
\includegraphics[width=0.6\textwidth]{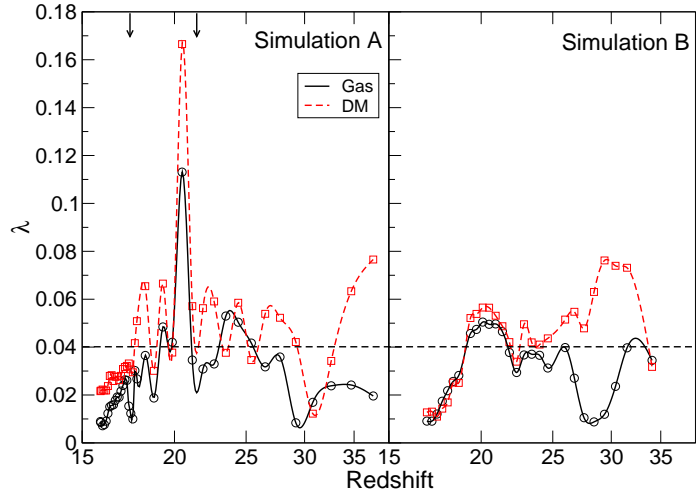}
\caption[Evolution of the spin parameter]{Spin parameter, $\lambda
  \equiv \vert L \vert \sqrt{\vert E \vert} / GM^{5/2}$, evolution of
  the main halo in the simulation.  {\em (left)} Simulation A.  {\em
    (right)} Simulation B.  The dashed and solid lines are the
  interpolated values for the DM and baryonic spin parameter.  The
  squares and circles correspond to the actual measurements from the
  DM and gas data, respectively.  The horizontal dashed line at
  $\lambda$ = 0.04 marks the mean cosmological spin parameter.  In
  Simulation A, two major mergers, whose beginning and end are marked
  by arrows, causes the large increase at z $\approx$ 21.  The
  oscillations occur as the merging halos orbit each other until they
  virialize.}
\label{fig:spin}
\end{center}
\end{figure}

During the hierarchical buildup of structure, tidal forces from
neighboring structures impart angular momentum to a halo, particularly
when its radius is maximal at the turn-around time \citep{Hoyle49,
  Peebles69}.  However in recent years, several groups have recognized
that the mergers impart a considerable fraction of angular momentum to
the system \citep{Steinmetz95, Gardner01, Vitvitska02, Maller02}.
Over many realizations of mergers, the net angular momentum change
would be zero.  In reality, an angular momentum residual remains after
the last major merger occurs because there are too few events to
cancel the randomization of halo spin.  Although each halo has unique
rotational properties, it is useful to define a dimensionless spin
parameter
\begin{equation}
\lambda \equiv \frac{\vert L \vert \sqrt{\vert E \vert}} {G M^{5/2}},
\end{equation}
where G is the gravitational constant and L, E, and M are the angular
momentum, energy, and mass of the object, that measures the rigid body
rotation of the halo \citep{Peebles71}.  In Figure \ref{fig:spin}, we
display the time evolution of $\lambda$ of the DM and baryons in our
simulations and mark the occurrence of the major merger in simulation
A. \citet{Eisenstein95b} \citep[preceded by][]{Barnes87} calculated
that the mean spin parameter, $\langle\lambda\rangle \approx 0.04$, is
weakly dependent on object mass and cosmological model, and this value
is also marked in Figure \ref{fig:spin}.  Also $\lambda$ weakly
depends on its merger history, where $\langle\lambda\rangle$ increases
in halos with the number of mergers.  Most of the angular momentum is
acquired from steady minor mergers and accretion because major mergers
only happen rarely (usually only once per logarithmic mass interval).
In 96\% of mergers, the majority of the internal spin originates from
the orbital energy of the infalling halo \citep{Hetznecker06}.

%
%

At $z \approx 22$ in simulation A, the spin parameter is large,
$\lambda = 0.06$, before the last major merger.  Then the spin
parameter increases by a factor of 3 during its major merger due to
the system being far from dynamical equilibrium.  The system becomes
virialized after approximately a sound crossing time, and the spin
parameter stabilizes at $\lambda \approx 0.03$ and proceeds to
decrease with time until $\lambda = 0.022$ at the time of collapse.
The above evolution of $\lambda$ agrees with the findings of
Hetznecker \& Burkert.  Simulation B describes a halo that does not
undergo a recent major merger and its final $\lambda$ = 0.013.

Both halos have less angular momentum than $\langle \lambda \rangle$
when the cooling gas collapses.  The probability distribution of
$\lambda$ can be described with the log-normal function
\begin{equation}
  \label{eqn:lambdaProb}
  p(\lambda)d\lambda = \frac{1}{\sigma_\lambda \sqrt{2\pi}} \exp
  \left[ -\frac{\ln^2 (\lambda/\lambda_0)}{2\sigma_\lambda} \right]
  \frac{d\lambda}{\lambda},
\end{equation}
where $\lambda_0 = 0.042 \; \pm \; 0.006$ and $\sigma_\lambda = 0.5 \;
\pm \; 0.04$ \citep[e.g.][]{Bullock01}.  From the cumulative
probability function resulting from equation (\ref{eqn:lambdaProb}),
11\% (1.2\%) of the cosmological sample of halos have larger spin
parameters than the halos described here.  \citet{Eisenstein95a}
demonstrated that halos with low spin parameters are candidates for BH
formation and quasar seeds.  However they argue that the angular
momentum needs to be at least an order of magnitude lower than the
mean.  Next we present further evidence that reveal that a gaseous
collapse is possible with not too atypical spin parameters.

\subsection{Global Disk}

In Simulation B, a thick disk with a radius of 50 pc and disk scale
height of $\sim$10 pc forms that is pressure supported and only
partially rotationally supported.  This equates to a shape parameter
$f = 0.89$ in Equation \ref{eqn:rotF}.  The circular velocities within
this disk achieve only a third of Keplerian velocities.  The lack of
full rotational support and large scale height suggests that a central
collapse occurs before any fragmentation in this large-scale disk.  In
contrast, we see a disorganized, turbulent medium and no large scale
disk formation in Simulation A that corresponds to $f = 2/3$, i.e. a
sphere.

\subsection{Instability of Maclaurin Spheroids}
\label{sec:analytics}

The dynamics of rotating systems is a classic topic in astrophysics
(see EFE \S\S1--6).  These self-gravitating systems are susceptible to
two types of instability.  Secular instability occurs when small
dissipative forces, e.g. viscosity, amplify perturbations to become
unstable in an otherwise stable inviscid configuration.  Dynamical
(also referred to as ordinary) instability results when some
oscillatory mode exponentially grows with time, regardless of any
dissipative forces.  Here we concentrate on Maclaurin spheroids
relevant for a uniform body rotating with a fixed angular velocity.
Maclaurin spheroids are a special case of Jacobi ellipsoids that are
axisymmetric.  The onset of the $m = 2$ bar-like instability in
gaseous Maclaurin spheroids happens for a given eccentricity,
\begin{equation}
  \label{eqn:eccentricity}
  e = \left( 1 - \frac{a_3^2}{a_1^2} \right)^{1/2} \geq \left\{ 
      \begin{array}{r@{\quad}l}
        0.8127 & \mathrm{(secular)} \\
        0.9527 & \mathrm{(dynamical)}
      \end{array}
  \right.,
\end{equation}
where $a_3$ and $a_1$ are the principle axes with $a_3 \leq a_1$ (EFE
\S33).  Eccentricity is related to the ratio, $t = T / \vert W \vert$,
of rotational kinetic energy to gravitational potential by
\begin{equation}
  \label{eqn:t_vs_e}
  t = \frac{1}{2} [ (3e^{-2} - 2) - 3(e^{-2} - 1)^{1/2} (\sin^{-1}
  e)^{-1} ],
\end{equation}
and the secular and dynamical instabilities correspond to $t =
(0.1375, 0.27)$, respectively \citep[e.g.][]{Ostriker73b}.

\begin{figure}[t]
\begin{center}
\includegraphics[width=0.6\textwidth]{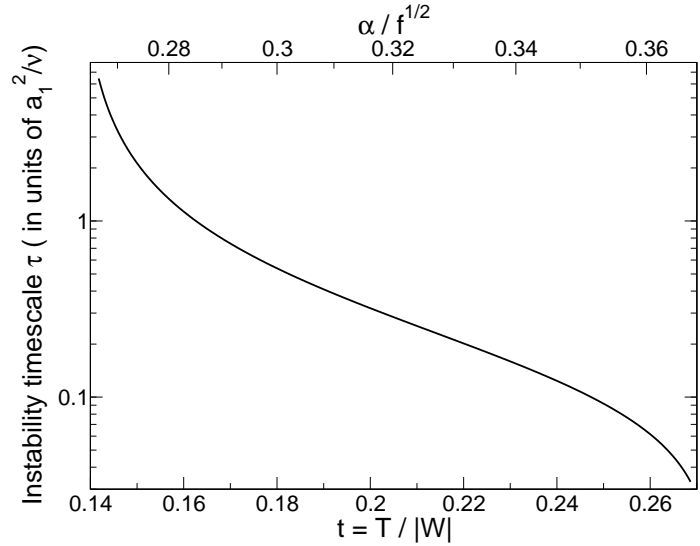}
\caption[Secular instability timescale]{Secular instability
  $e$-folding timescale in units of $a_1^2 / \nu$ as a function of t =
  $T / \vert W \vert$ and $\alpha = (tf/2)^{1/2}$
  (eq. \ref{eqn:alpha}).  At t $<$ 0.1375, the system is stable to all
  perturbations.  Above 0.27, the system is dynamically unstable, and
  this timescale is not applicable.}
\label{fig:secular_tau} 
\end{center}
\end{figure}

When $t$ is larger than 0.1375 but smaller than 0.27, both the
Maclaurin spheroid and Jacobi ellipsoid are perfectly stable against
small perturbations in the non-viscous case.  For a given $e$, the
Jacobi configuration has a lower total energy than its Maclaurin
counterpart and is therefore a preferred state.  Here any dissipative
force induces a secular bar-like instability.  The system slowly and
monotonically deforms through a series of Riemann S-type ellipsoids
until its final state of a Jacobi ellipsoid with an equal angular
momentum \citep{Press73} and lower angular velocity (EFE \S32) as
specific angular momentum is transported outward.  The instability
grows on an $e$-folding timescale
\begin{equation}
\label{eqn:secular_tau}
\tau = \phi a_1^2 / \nu,
\end{equation}
where $\phi$ is a constant of proportionality that asymptotes at $t$ =
0.1375, decays to zero at $t$ = 0.27, and is plotted in Figure
\ref{fig:secular_tau} (EFE \S37).  Here $\nu$ is the kinematic
viscosity.

%
%

\citet{Christodoulou95a, Christodoulou95b} generalized the
formulations for bar-like instabilities to account for self-gravity.
In addition, they consider different geometries, differential
rotation, and non-uniform density distributions.  They devised a new
stability criterion
\begin{equation}
  \label{eqn:alpha}
  \alpha \equiv \frac{T/\vert W \vert}{\Omega / \Omega_J} =
  \sqrt{\frac{f}{2} \frac{T}{\vert W \vert}}
\end{equation}
where $\Omega$ is the rotation frequency, 
\begin{equation}
\Omega^2_J = 2 \pi G \rho \left[ \frac{(1-e^2)^{1/2}}{e^3} \sin^{-1} e
  - \frac{1-e^2}{e^2} \right]
\end{equation}
is the Jeans frequency in the radial direction for a Maclaurin
spheroid, and
\begin{equation}
  \label{eqn:rotF}
  f = \frac{1}{e^2} \left[ 1 - \frac{e}{\sin^{-1} e} \sqrt{1 - e^2}
  \right]
\end{equation}
accounts for differing geometries\footnote{See
  \citet{Christodoulou95b} for more generalized geometries.} with $f =
2/3$ for a sphere and $f = 1$ for a disk.  Secular and dynamical
instabilities for Maclaurin spheroids occur above $\alpha$ = (0.228,
0.341), respectively.

From N-body simulations of disk galaxies, \citet{Ostriker73b} found
that a massive dark halo with comparable mass to the disk could
suppress secular instabilities.  In the case of a gaseous collapse
into a SMBH, the baryonic component dominates over the dark matter
component in the central 10 pc.  Secular instabilities cannot be
prevented through this process, which we demonstrate next.

\subsection{Rotational Instabilities}
\label{sec:rotInstab}

\begin{figure}[t]
\begin{center}
\includegraphics[width=0.6\textwidth]{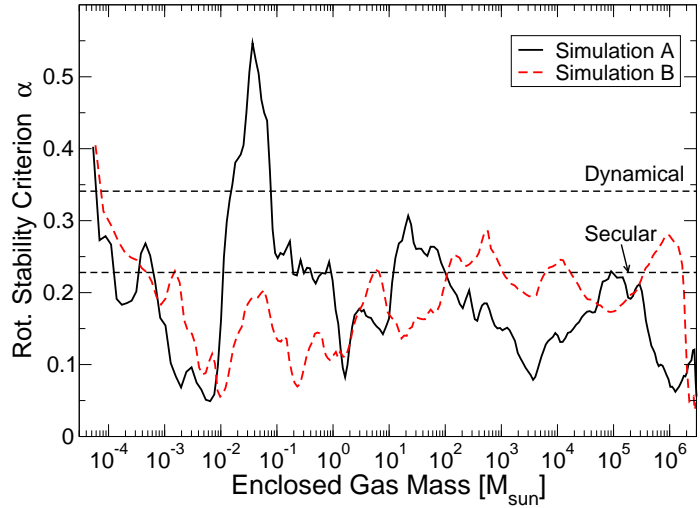}
\caption[Rotational instability parameter]{Rotational instability
  parameter $\alpha = \sqrt{fT/2\vert W \vert}$ for the thick disk
  with $r \simeq 50$ pc in simulations A (\textit{black solid line})
  and B (\textit{red dashed line}).  For $\alpha > 0.22$ denoted by
  the horizontal line, a secular instability occurs in the disk and
  leads to bar formation.  In simulation A, instabilities occur at
  mass scales of 100, 0.1, and $10^{-4} \Ms$.  In simulation B, the
  same happens at $2 \times 10^6$, $2 \times 10^4$, $10^3$, 6, and
  $10^{-3} \Ms$.  We also mark $\alpha = 0.341$ where a rotating
  system becomes dynamically unstable.  Only simulation A at 0.1 $\Ms$
  experiences a dynamical instability.}
\label{fig:diskB}
\end{center}
\end{figure}

In the $l$ = 1 pc panel of Simulation B in Figure \ref{fig:slices}, it
is apparent a bar-like instability exists in the gravitationally
unstable central object.  Figure \ref{fig:diskB} shows the instability
criterion $\alpha$ (eq. \ref{eqn:alpha}) against enclosed gas mass.
Here we transform the velocities to align the $z$-axis with the
baryonic angular momentum vector of the entire halo.  We use the
tangential velocities to calculate the rotational kinetic energy $T$.

%
%

As discussed before, Maclaurin spheroids are subject to secular $m =
2$ bar-like instabilities when $\alpha > 0.228$.  In simulation A, the
central object becomes unstable three approximate mass scales, 100,
0.1, and 10$^4 \Ms$.  The instability at 0.1 $\Ms$ is dynamically
unstable with $\alpha$ peaking at 0.55.  In simulation B,
instabilities occur at $2 \times 10^6$, $2 \times 10^4$, $10^3$, 6,
and $5 \times 10^{-4} \Ms$.

It is interesting to note that the innermost instability in both
simulations becomes dynamical ($\alpha > 0.341$), and $\alpha$
continues to increase rapidly toward the center.  However these
features should be taken with caution since it occurs near our
resolution limit, where the particular location used as the center
will influence the rotational energy one would calculate.

The $e$-folding time of secular instabilities $\tau$ is proportional to
$a_1^2$ (see eq. \ref{eqn:secular_tau}).  Hence small-scale
instabilities collapses on a faster timescale than its parent,
large-scale bar instability.  Viscosity, turbulent viscosity in this
case, is the main dissipative force that drives the instability.
$\tau$ is inversely proportional to the viscosity.  Therefore in the
data presented, this further shortens the $\tau$ because supersonic
turbulence is maintained to the smallest scales.  We also note that
the radial velocity (see Figure \ref{fig:profilesA3}) increases after
it becomes comparable to the sound speed.


%
%
\begin{figure}
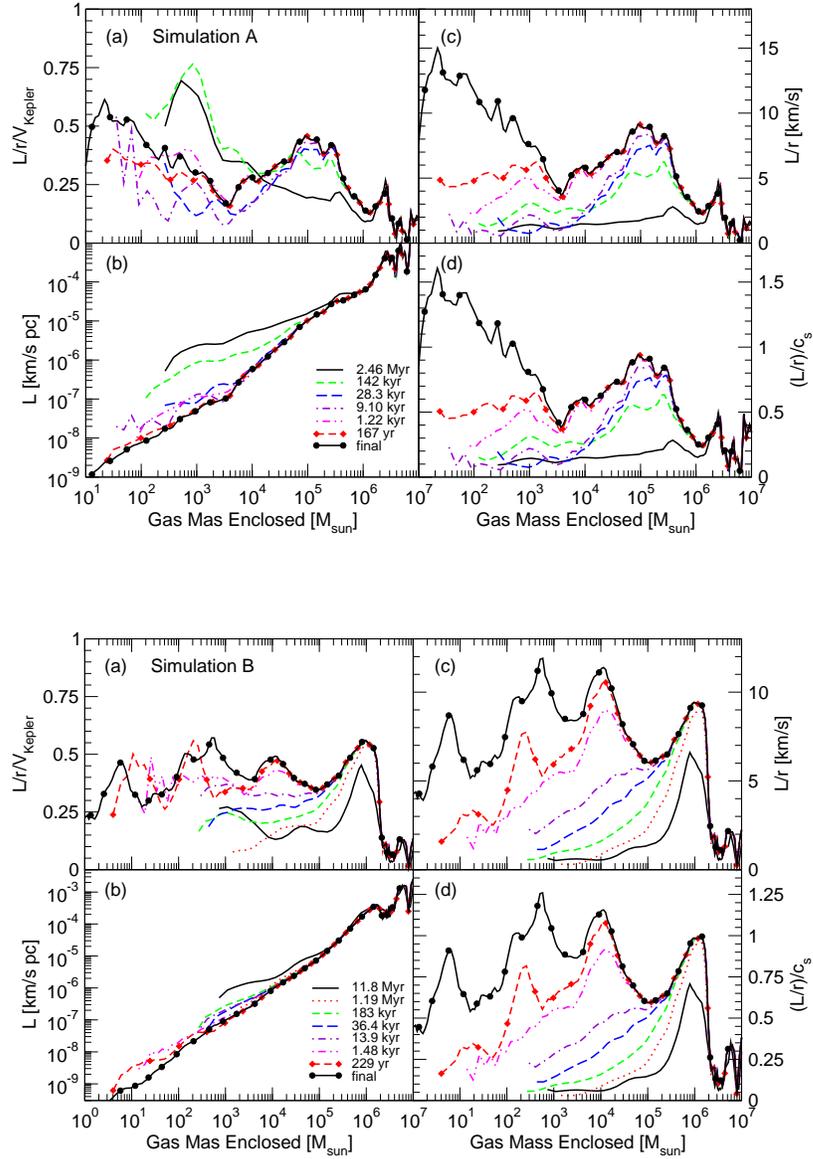

\begin{center}
  \includegraphics[width=0.7\textwidth]{chapter4/fig/f7a} \\[3em]
  \includegraphics[width=0.7\textwidth]{chapter4/fig/f7b}
  \caption[Radial profiles of rotational quantities at large
  scale]{Mass-weighted radial profiles of various rotational
    quantities in simulation A (\textit{left panels}) and simulation B
    (\textit{right panels}).  In panel A, we show the rotational
    velocity compared to the Kepler velocity = $\sqrt{GM/r}$.  In
    panel B, we display the specific angular momentum (in units of
    [km/s] pc).  In panels C and D, the typical rotational velocity
    and the ratio of the rotational velocity and sound speed are shown,
    respectively.  The line styles correspond to the same times in
    Figure \ref{fig:profilesA}.}
\label{fig:profilesB} 
\end{center}
\end{figure}

%
%
\begin{figure}
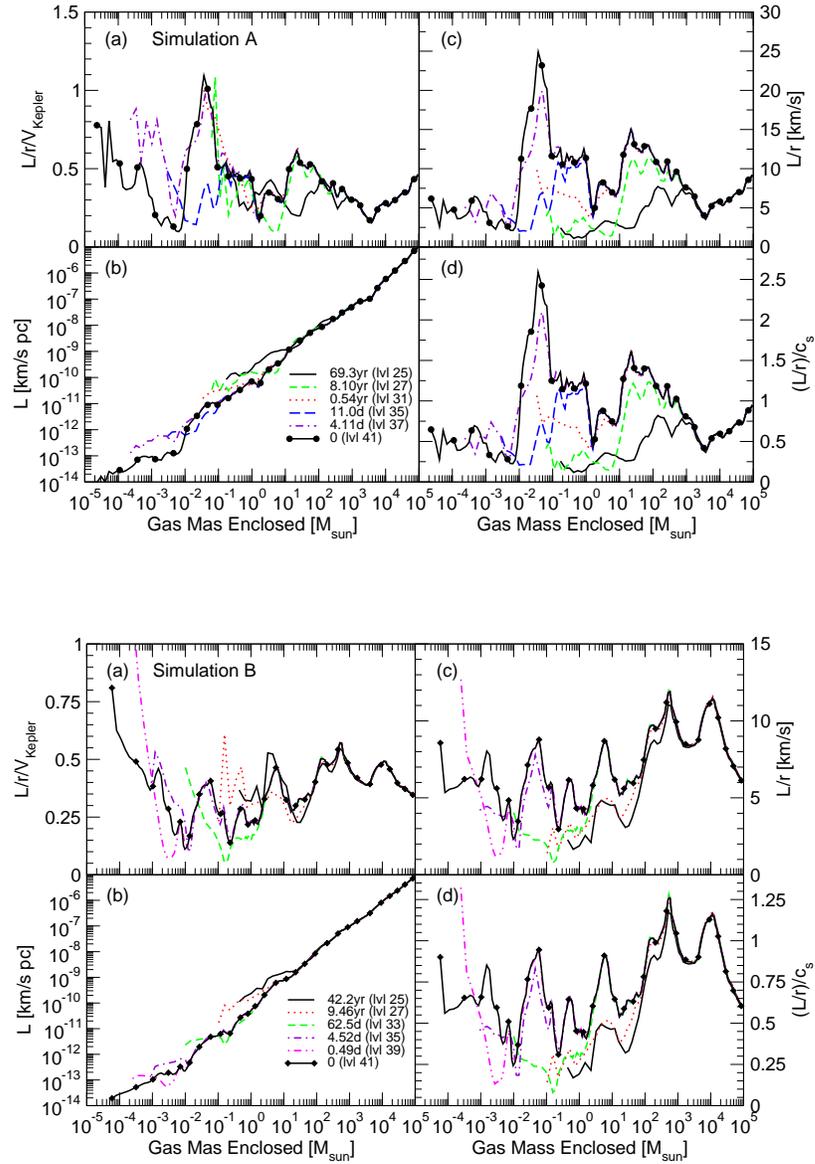

\begin{center}
  \includegraphics[width=0.7\textwidth]{chapter4/fig/f7c} \\[3em]
  \includegraphics[width=0.7\textwidth]{chapter4/fig/f7d}
  \caption[Radial profiles of rotational quantities at small
  scale]{The same as Figure \ref{fig:profilesB} but with the inner
    parsec of simulation A and B and the output times as listed in
    Figure \ref{fig:profilesA3}.}
  \label{fig:profilesB3} 
\end{center}
\end{figure}

\subsection{Rotational Properties}

During the collapse of the gas in our simulations, rotational support
never impedes the collapse.  In Figures \ref{fig:profilesB} and
\ref{fig:profilesB3}, we show (a) coherent rotational velocity divided
by Keplerian velocity $v_{kep}$, (b) rotational velocity, (c) specific
angular momentum, and (d) rotational velocity divided by the sound
speed.  We note that the rotational velocity $L/r$ plotted here is
different than organized rotation, i.e. a disk.  The radial profiles
only sample gas in radial shells, where the angular momentum vectors
are not necessarily parallel.

1. \textit{Simulation A}--- At $r > 1$ AU (M$_{\rm{enc}} = 1 \Ms$),
the typical rotational speed is two or three times lower than the
Kepler velocity, which is required for rotational support.  At this
radius, the infall becomes marginally rotationally supported,
i.e. $L/r \sim v_{kep}$, around 0.1 AU (M$_{\rm{enc}} = 0.07 \Ms$) in
Figure \ref{fig:profilesB3}.  The radial velocities react by slowing
from 15\kms~to around zero.  At late times, the rotational support is
steady at 0.1 AU, however the rotational velocities vary significantly
over time to the interior.

2. \textit{Simulation B}--- As discussed in Section
\ref{sec:rotInstab}, the collapse object experiences rotational bar
instabilities.  After an instability occurs, the radial velocities
increase due to angular momentum transport.  Then this material begins
to gain rotational velocities caused by conservation of angular
momentum.  However this causes another instance of a bar instability
to occur.  The increased infall velocity and associated decrease in
rotational velocities (i.e. the dips in Figures \ref{fig:profilesA3}d
and \ref{fig:profilesB3}d) depict this behavior.  At the final output,
the infalling material exhibits no rotational support similar to
Simulation A at $r > 1$ AU.  At the outputs that are 27 and 4.5 days
before the final output, there is a hint of rotational support at $r <
1.7 \Rs$ (M$_{\rm{enc}}$ = 0.33 Jupiter masses), but this decays in
the two later outputs.

We interpret the inner points where L/r/V$_{{\rm Kepler}}$
fluctuations greatly or increases above unity with caution due to the
nature of choosing a center in a turbulent medium.  If the central
sphere is smaller than a radius where the turbulent velocities average
to zero, we introduce errors into the angular momentum profiles by
sampling the turbulent gas incompletely.  Thus we do not conclude that
the inner regions are rotationally supported.


In the $b$-panels of Figure \ref{fig:profilesB}, one sees that
specific angular momentum inside M$_{\rm{enc}} < 10^6 \Ms$ decreases
over time and is transported outwards in the collapse.  As discussed
in Section \ref{sec:turbulence} and shown in Figure \ref{fig:mach},
supersonic turbulence typically exists at all radii.  Additionally,
the rotational speed never exceeds the local sound speed, and we
interpret this as further evidence for the efficient angular momentum
transport.  In a turbulent medium, transport is different than in a
disk.  At any radius, both high and low angular momentum gas exists,
whereas in a Keplerian disk all gas have equal amounts of angular
momentum.  Rayleigh's criterion \citep{Rayleigh20, Chandra61} states
that specific angular momentum in a system must be a monotonically
increasing function with radius.  If this does not hold, the lower
angular momentum gas at large radii will preferentially sink to
smaller radii and vice versa that is similar to Couette flow.  As this
occurs, the system becomes unstable to turbulence.  The onset of
turbulence can be delayed if viscosity is large enough so that Reynolds
numbers are below the order of 10$^2$ or 10$^3$.  However there are
many modes of instability if Rayleigh's criterion is not met, and a
gas with low Reynolds number will eventually become fully turbulent
\citep{Shu92}.

With a typical spin parameter, the thick disk with $r\sim50$ pc is
not rotationally supported.  In simulation A, a global disk does not
exist at all.  We attribute this behavior to the nature of angular
momentum transport in a turbulent medium.  Even with a higher spin
parameter, we do not expect a disk to fragment before the central
collapse due to the intrinsic amount of low angular momentum gas and
shorter dynamical timescales in the center.  This gas can collapse to
such small radii without fragmentation so that a central mass of
$\sim10^5 \Ms$ or 2\% of the halo gas mass forms.  After the initial
collapse, the thick disk may become rotationally supported as more
higher angular momentum gas infalls.

\section{Discussion}
\label{sec:discuss}


In our cosmological simulations, we find that a $\sim$10$^5 \Ms$ dense
object forms in the center of a metal-free protogalactic halo that
supports atomic hydrogen cooling.  Although we have neglected some
important processes, such as \hh~chemistry, star and BH formation and
feedback, our results show that rotational support cannot prevent
matter from collapsing to a dense, massive object with r $<$ 5 pc.
Before the central collapse, there is no global, rotationally
supported disk formation on the order of 50 pc \citep[cf.][]{Mo98}.
However this does not preclude this type of disk formation, but the
initial collapse into a central object must precede it.  Disk
formation and its associated star formation may also be dependent on
the feedback and outcome of the central object.  This additional stage
of galaxy formation should be included in the most frequently employed
models, where the earliest star formation occurs in high redshift disk
galaxies.

\subsection{Secular Instabilities}

Our calculations show that the ``bars within bars'' cascade
\citep{Shlosman89, Shlosman90} occurs for gaseous systems due to a
secular instead of dynamical instability \citep[cf.][]{Begelman06}.
For a given morphology, this allows bar-like structures to form in
systems having $T / \vert W \vert$ values 45\% smaller than the ones
calculated in \citeauthor{Begelman06}.  We also find that rotational
instabilities are possible without a global disk as in simulation A.

Our results should be applicable to any turbulent gaseous collapse, in
particular galactic molecular clouds.  Imagine the following scenario
of an initially turbulent, slowly rotating, and uniform sphere
beginning to collapse.  In a turbulent medium, the gas with low
specific angular momentum preferentially falls toward the center,
leaving the high specific angular momentum at the outskirts of the
cloud.  This is seen in our specific angular momentum profiles of
turbulent collapses \citep[also see][]{Abel02a, Yoshida06b}.  Material
with similar angular momentum now obtains some organized rotational
velocity.  Some shells become rotationally unstable to bar formation
as it gains rotational velocity in its attempt to conserve angular
momentum.  In this example, the instabilities are secular and occur at
$T/\vert W \vert \geq 0.14$ because turbulent viscosity provides the
dissipative force.  Then these rotational bar-like instabilities
transport high angular momentum outwards so that the gas with low
specific angular momentum can flow to a smaller radius.  Afterwards,
the ``bars within bars'' scenario drives gas to even smaller radius,
eventually forming a central object, whether it be a star or BH.
Angular momentum segregation in a turbulent medium and during the bar
instability cascade could alleviate ``the angular momentum problem''
of a gravitational collapse, in which a cloud must somehow shed
several orders of magnitude of angular momentum in order to form a
central star or, in this case, a pre-galactic SMBH.

Nevertheless, we do not advocate our simulations as evidence of
pre-galactic SMBH formation because we have neglected many important
processes related to \hh~cooling and primordial star formation that we
detail shortly afterwards.  We expect the stellar feedback from Pop
III stars to alter the ``initial conditions'' of early galaxy
formation, as shown by recent radiation hydrodynamical simulations of
the first stars \citep{Yoshida06a, Abel07}.

\subsection{The Fate of the Central Object}

Although we study the idealized collapse of a protogalactic collapse,
it is still possible to infer the fate of the central object.  In our
simulations, the merger history and the subsequent turbulence
influence the nature of the collapse.  In simulation A, a major merger
precedes the collapse, and the turbulence is significant throughout
the halo before and during the collapse.  In the final output,
turbulence extends inward to $r_{\rm{BE}}$ and may have caused the
central object to be more massive and less dense.  In simulation B, we
find a less massive but more compact central object, which leads to a
faster collapse.  Since this halo did not experience a major merger
before its collapse, the turbulence is smaller, and the medium is more
homogeneous and has coherent rotational motions in the thick disk.

These two different environments may result in different types of
objects.  If the crossing and dynamical timescales are greater than
the Kelvin-Helmholtz timescale, \tkh, for a massive star, star
formation may commence and provide sufficient feedback to prevent
further collapse.  Alternatively, the collapse could proceed faster
than \tkh~and initially forms a BH.  To further study the central
dynamics, it will be necessary to employ sink particles \citep{Bate95,
  Krumholz04} in order to avoid the decreasing timescales of the
collapsing object(s).

\subsection{The Role of Low Angular Momentum Gas}

To form a central dense object, a sufficient amount of gas needs to
have low angular momentum to infall to small radii.  The origin and
transport of this gas are integral to our comprehension of
protogalactic gas cloud collapses.  According to Rayleigh's inviscid
stability criterion, angular momentum segregates such that the lowest
angular momentum material piles up at the center of dark matter
potential wells.  We observe this behavior in the angular momentum
distributions.  The low angular momentum material may originate in
lower mass progenitors.  Here the gas resides in a shallower potential
well that results in lower turbulent and random thermal velocities
or equivalently an angular momentum distribution skewed to lower
values.  We argue that this effect is intimately linked to the gas
acting to achieve virial equilibrium at all stages during the collapse
(cf. Paper I).  As this gas infalls, it gains rotational energy as it
conserves angular momentum.  It then experiences a secular bar
instability as discussed in Section \ref{sec:rotInstab} that
transports angular momentum outward so only the lowest angular
momentum gas collapses toward the center.  This picture is true also
in simulations of the very first stars \citep{Abel02a}.

The angular momentum distribution also influences the manner in which
the halo collapses.  It should be noted that the angular momentum
distribution, not the total angular momentum of a halo, dictates the
nature of its collapse because gas parcels within the halo will
segregate into different regions of the halo.  In principle, the low
angular momentum material sinks inward and forms the central
structures.  If an insufficient amount of low angular momentum matter
exists in the halo, it cannot monolithically collapse and will form a
disk \citep{Fall80}.



\subsection{Applicability}

\subsubsection{Limitations of Current Approach}
\label{sec:limitation}

Our results depict the importance of turbulence, accretion, and the
hydrogen cooling in the initial collapse of these halos.  However we
are missing some essential processes, such as \hh~chemistry,
primordial and Population II stellar formation and feedback, SMBH
formation and feedback, and metal transport and cooling.  It was our
intention to study only the hydrogen and helium cooling case first and
gradually introduce other processes at a later time to investigate the
magnitude and characteristics of their effects, which we will present
in later papers.

Gas becomes optically thick to \lya~radiation above number densities
of $10^7 \mathrm{cm}^{-3}$, but we continue to use optically thin
cooling rates above this density.  We overestimate the cooling within
0.1 pc.  As a consequence, we do not suggest that these simulated
objects ever form in nature.  However this scenario poses an excellent
problem of a turbulent collapse.  This should be common in galaxy
formation, where turbulence is generated during virialization, and
star formation within turbulent molecular clouds.

\subsubsection{Desired Improvements}
\label{sec:improve}

Clearly local dwarf spheroidals contain stars with ages consistent
with formation at very high redshifts \citep{Ferrara00, Tolstoy02,
  Tolstoy03, Helmi06}.  To develop a model that desires to fit galaxy
luminosity functions down to the faintest observed galaxies one may
need a star formation and feedback model that follows molecular clouds
as small as one thousand solar masses in order to allow for the
dominant mode of star formation observed locally.  It should be
already technologically feasible with current cosmological
hydrodynamical models to simulate these galaxies one star at a time.

Correct initial conditions for early galaxy formation require prior
star and BH formation and feedback.  The typically adopted conditions
for phenomenological star formation are velocity convergence, a
density that exceeds some critical value, \tdyn~$>$ \tcool, and a low
temperature \citep{Cen92}.  Phenomenological primordial star formation
is possible if we include two additional conditions as utilized in
\citet{Abel07}.  First, the \hh~fraction must exceed 10$^{-3}$
\citep{Abel02a}, and second, the metallicity of the gas must not
exceed the ``critical metallicity'' of 10$^{-3}$ -- 10$^{-4}$ of the
solar value \citep{Bromm01}.  From prior studies
\citep[e.g.][]{Abel02a, Bromm03, OShea05}, we expect these stars to
form in halos that can support \hh~cooling and in relic \ion{H}{2}
regions.  The Lyman-Werner radiation from massive stars can dissociate
\hh~from large distances, suppress star formation in lower mass halos
\citep{Machacek01, Wise05}, and must be considered to accurately model
future star formation.

BH formation in the death of some primordial stars can also have a
profound effect on surrounding structure formation as it accretes
infalling matter during later mergers.  Therefore it is necessary to
include its effects from seed BHs from primordial stars with masses
outside of the range between 140 and 260 solar masses.  Also it is
possible to phenomenologically model SMBH formation in a similar
manner to the stellar case.  If the protogalactic collapse occurs
faster than the stellar formation timescale of a massive star, a SMBH
may form inside this region.  Using the stellar formation conditions
plus this condition and allowing the particle to accrete
\citep[i.e. sink particles;][]{Bate95, Krumholz04}, SMBH formation can
be followed in cosmological hydrodynamic simulations.  These sink
particles should regulate the accretion with an appropriate subgrid
model.  Important processes include an appropriate accretion rate
(e.g. Eddington or Bondi-Hoyle), turbulence \citep{Krumholz06},
rotational support of the infalling gas, and a viscosity timescale for
accretion disks.

For small galaxies radiative transfer effects can have a great impact
\citep[e.g.][]{Haehnelt95, Whalen04, Kitayama04, Alvarez06} and should
not be neglected.  The approach of \citet{Gnedin01}, although
promising, has not been implemented and coupled with an adaptive
hydrodynamics code that is capable of high dynamic range calculations.
However, the novel technique of adaptive ray tracing \citep{Abel02b}
has recently been implemented into \enzo~and used to study the
outflows and ionizing radiation from a primordial star \citep{Abel07}.
There ionization front instabilities create cometary small-scale
structure and shadowing effects as a result from the explicit
treatment of three-dimensional radiative hydrodynamics.  Lastly as
used in many stellar formation routines \citep{Cen92, Tassis03}, we
hope to include thermal and radiative feedback from Population II
stars in future studies.

\section{Conclusions}

We have simulated the hydrodynamics and collapse of a protogalactic
gas cloud in two cosmology AMR realizations.  Our focus on the
hydrodynamics presents a basis for future studies that consider
stellar and BH feedback.  In the idealized case presented, we find a
central dense object forms on the order of 10$^5 \Ms$ and $r \lsim 5$
pc.  This central object is not rotationally supported and does not
form a disk in our simulations.  However our results do not dismiss
disk formation in protogalaxies, and rotationally supported disk
formation may begin after the initial central collapse.  Disk
formation could also depend on the feedback and outcome of the central
object.  This initial central collapse adds another phase to the most
frequently employed galaxy formation models, where star formation
commences in a gaseous disk.

These simulations highlight the relevance of secular bar-like
instabilities in galaxy formation.  Similar bar structures are
witnessed in primordial star formation simulations.  As low angular
momentum infalls, it gains rotational energy as it conserves angular
momentum.  This induces an $m$ = 2, bar-like instability that
transports angular momentum outwards, and the self-similar collapse
can proceed without becoming rotationally supported and exhibits $\rho
\propto r^{-12/5}$.  This process repeats itself as material infalls
to small scales that is indicative of the ``bars within bars''
scenario.  We see three and five occurrences of embedded secular
instabilities in the two realizations studied here.

We also find that gas turbulence influences the collapse in two
different mechanisms.  First, supersonic turbulence provides a channel
for the gas to preferentially segregate according to specific angular
momentum.  The low angular momentum material sinks to the center and
provides the material necessary for a central collapse.  Second, the
degree of turbulence created from the previous mergers can delay the
central collapse.  The final characteristics of the central object
depend on the balance of collapse and star formation timescales.
Possible outcomes are BH formation, a starburst, or a combination of
both.

All of these cases are viable in the early universe, and the occurrence
of these cases depends on the merger history, local abundances in the
halo, and the existence of a seed BH.  Moreover, star formation should
occur whether a central BH exists or not.  Perhaps the frequency of
these different protogalactic outcomes may be traced with either 3D
numerical simulations that consider star and SMBH formation and
feedback along with metal transport or Monte Carlo merger trees that
trace Pop III star formation, metallicities, and BHs.  We will attempt
the former option in later studies to investigate protogalactic
formation in more detail.

\chapter{Suppression of H$_2$ Cooling in the UVB}
\label{chap:progenitors}

We have studied the standard model of galaxy formation, where dark
matter halos of masses greater than $10^8 \Ms$ are the first luminous
objects in the universe.  However, it has been known since the 1960's
that molecular hydrogen plays an important role in structure formation
in the early universe.  Molecular hydrogen is a delicate species that
can be dissociated from large distances since the universe is
optically thin to the \hh~dissociating radiation between 11.2 and 13.6
eV.  This effect has been argued that a soft ultraviolet background
renders the relevance of \hh~unimportant in this epoch.  To
investigate this claim, we include \hh~cooling in our simulations
while including such radiation backgrounds, adding yet more physics to
our study.  The implications of luminous objects existing before the
first galaxies are significant because their feedback processes will
change the initial density and temperature structures where the first
galaxies form.

This chapter is in preparation for publication in \textit{The
  Astrophysical Journal}.  It is co-authored by Tom Abel, who
suggested the initial motivation of this work and provided excellent
critique in the arguments used to stress the importance of \hh~cooling
in early structure formation.

\section{Motivation}

Cosmic structure forms hierarchically.  Any object in the universe
today started with copious numbers of small progenitors at redshifts
currently inaccessible to direct observations.  Traditionally in
galaxy formation \citep{Rees77, White78, Dekel87, White91, Baugh03}
\tvir~= $10^4$ K halos are assumed to be the first cooling
halos. Nevertheless since the late 1960's it has been known that
molecular hydrogen, formed in the gas phase, can dominate cooling in
objects of smaller virial temperature and mass \citep{Saslaw67,
  Peebles68b, Yoneyama72, Haiman96, Tegmark97, Abel98,
  Abel00}. Neglecting this early phase of \hh~cooling halos has been
justified by arguing that \hh~is destroyed via radiative feedback
effects \citep[cf.][]{Dekel87, Haiman97b, Haiman00, Glover01,
  Bromm03}. The photo-dissociation of \hh~via the Solomon process by
an early soft ultraviolet background (UVB) is generally assumed as the
main reason \citep{Oh02, Ciardi05, Haiman06}.

\begin{figure}[t]
\begin{center}
\includegraphics[width=0.6\textwidth]{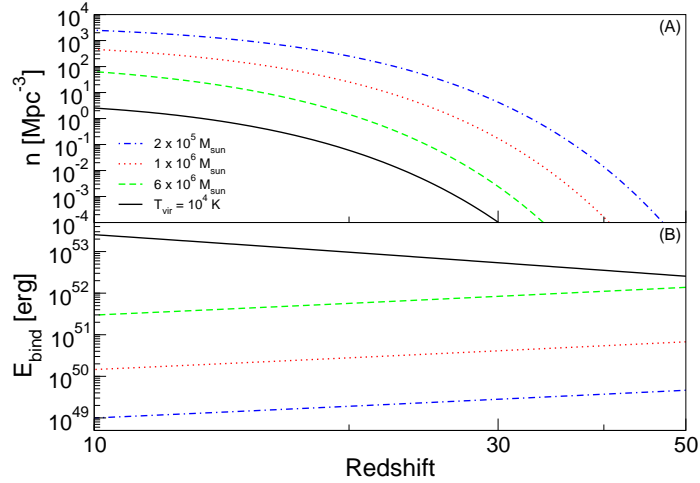}
\caption[Dwarf Galaxy Progenitor Abundances]{\label{fig:ps}
  \textit{Panel A}: Sheth Tormen number density of dark matter halos
  as a function of redshift for \tvir~= $10^4$ K, $M = 2 \times 10^5,
  10^6,~\rm{and}~6 \times 10^6~\Ms$ using WMAP 3 year data for
  parameters. \textit{Panel B}: Binding energies as a function of
  redshift for the corresponding halos (same line styles as in panel
  A)}
\end{center}
\end{figure}

The mass scale of halos considered enters exponentially in the
collapsed mass fraction and the abundance of halos.  Figure
\ref{fig:ps} shows the predicted abundances of the earliest building
blocks of galaxy formation as a function of redshift for the latest
concordance cosmology using the Sheth-Tormen formalism \citep{Press74,
  Sheth02}. The different lines correspond to different virial
masses. The solid line corresponds to halos with virial temperatures
of $10^4$ K, the temperature at and above which atomic hydrogen line
cooling is dominant. At redshift 30, e.g., the difference of
abundances of $2 \times 10^5$ \Ms~and \tvir~= $10^4$ K halos is five
orders of magnitude. Even at redshift 10 this disparity is still a
factor of a thousand. When studying reionization and chemical
evolution of galaxies and the intergalactic medium, one needs to
consider stellar feedback. The simple fact that the binding energy of
the gas of smaller mass halos is even less than the kinetic energy
deposited by even one supernova (SN) is illustrated in Figure
\ref{fig:ps}B. Surely whether the atomic hydrogen line (\lya) cooling
halos are formed from pristine primordial gas or are mergers of many
tens of progenitors that massive stars have enriched and expelled the
gas from should make a significant change in their further evolution.
The minimum mass of star forming halos is undoubtedly an important
issue independent of the techniques employed to study structure
formation.

%
%

Advances in cosmological hydrodynamics and its numerical methods
\citep{Cen92a, Zhang95, Katz96, Abel97, Anninos97, Bryan98, Gnedin01,
  Ricotti02a, Ricotti02b} allow now detailed investigations of all the
relevant physical processes. Modeling the expected negative feedback
from an early soft UVB is straightforward as a background flux only
causes a spatially constant photo-dissociation rate in the chemical
reaction network being solved when \hh~does not exist at high enough
abundances to self-shield. \citet[][MBA01 hereafter] {Machacek01} used
Eulerian adaptive mesh refinement (AMR) simulations to investigate the
role of such a \hh~dissociating (Lyman-Werner; LW) background on the
minimum mass of halos within which primordial gas can first cool for a
variety of radiation amplitudes.  In addition to a LW background, the
collapse of halos within relic \ion{H}{2} regions can be either
delayed or catalyzed.  \citet{Mesigner06} used AMR simulations with a
short-lived 3 Myr hydrogen ionizing UVB that simulates a nearby
massive, metal-free (Pop III) star.  They found that halo collapses
are prolonged if \juv~$\gsim$ 0.1 and catalyzed if below this critical
value, where \juv~is in units of $10^{-21}$ \emis~at a wavelength of
912 \AA.  In the case of a large UVB, the collapse is delayed due to
lower gas densities and higher cooling times.  In the small UVB
regime, excess free electrons in the relic \ion{H}{2} region
accelerate \hh~formation.  In both cases, feedback in relic \ion{H}{2}
subsides after $\sim$30\% of a Hubble time.  Strong suppression of
\hh~formation also occurs in $10^6 \Ms$ halos with a LW background
\jlw~$>$ 0.01.  \citet[][YAHS03 hereafter] {Yoshida03} similarly
addressed this issue using smoothed particle hydrodynamics (SPH). They
found an additional effect on the minimum collapse mass of dynamical
heating from the mass accretion history of the halo. As the heat input
increases, the virial temperature must rise before \hh~cooling can
start to dominate, and a cool phase develops in the center of the
potential well.

Self-consistent calculations in which the sources produce the
radiation backgrounds which in turn affect the number of new sources
are feasible so far only with semi-analytic approaches \citep[][WA05
hereafter] {Haiman00, Wise05} and small volume cosmological
simulations at low spatial resolutions \citep{Ricotti02a,
  Ricotti02b}. From these studies, one can derive realistic upper
limits on the amplitude of the expected soft UVB. In all studies that
include radiation sources in halos less than $10^4$ K halos, the
largest the soft UVB flux can get before the $T > 10^4 K$ halos
dominate the emission is $\jlw \sim 1$ \citep [cf.][WA05]{Haiman00,
  Ricotti02a, Ricotti02b}. Interestingly, for a LW intensity of $\jlw
\sim 0.1$, MBA01 found that $2 \times 10^6 \Ms$ halos were still able
to cool and collapse.  On the other hand at that \jlw, YAHS03 suggest
negative feedback should become so strong that the critical
\hh~fraction for cooling cannot be reached and cooling will not occur.
However, they did not explore this further with detailed higher
resolution simulations to check whether their analytical expectation
would hold.

We present a series of fourteen very high resolution Eulerian AMR
simulations designed to see how the largest possible feedback may
raise the minimum mass in which primordial gas will cool by molecular
hydrogen. The simulations techniques and details of the suite of
calculations is the topic of the next section. In the following
sections, we describe the results that show \hh~cooling cannot be
neglected in early structure formation. In the discussion, we describe
the nature of the UVB and why \hh~cooling can occur in such large
radiation backgrounds.  We also comment on the large range of
questions in cosmological structure formation that this conclusion
affects.

\section{Simulations and Assumptions}

We use the Eulerian AMR hydrodynamic code \enzo~\citep{Bryan97,
  Bryan99} to study the importance of \hh~cooling in early galaxy
formation.  \enzo~uses an $n$-body adaptive particle-mesh solver
\citep{Couchman91} to follow the dark matter (DM) dynamics.  We
perform two cosmological realizations with different box sizes and
random phases and WMAP 1 year parameters of ($h$, \Ol, \Om, \Ob,
$\sigma_8$, $n$) = (0.72, 0.73, 0.27, 0.024$h^{-2}$, 0.9, 1)
\citep{Spergel03}.  The significantly different third year WMAP
\citep[WMAP3;][] {Spergel06} results favor lesser small-scale power
that delays high-redshift structure formation by $\sim$40\% and alters
the statistical properties of DM halos \citep{Alvarez06}.  The ratio
\Om/\Ob~also only lowered by 5\% to 5.70.  However these differences
have no effect on the evolution and assembly of individual halos
studied here that have typical mass accretion histories.

The initial conditions are the same as in \citet{Wise07}.  Both
realizations have a top grid with a resolution of 128$^3$ with three
nested subgrids with twice finer resolution and are initialized at z =
129 (119)\footnote{To simplify the discussion, simulation A will
  always be quoted first with the value from simulation B in
  parentheses.} with the COSMICS package \citep{Bertschinger95,
  Bertschinger01}.  The box size is 1.0 (1.5) comoving Mpc.  The
innermost grid has an effective resolution of 1024$^3$ with DM
particle masses of 30 (101) \Ms and a side length of 250 (300)
comoving kpc.  We refine the AMR grids when either the DM (gas)
exceeds three times the mean DM (gas) density on the same level.  We
also refine so that the local Jeans length is resolved by at least 4
cells.

We focus on the region containing the most massive halo in the
simulation box and follow its evolution until it collapses to an
overdensity of $10^7$ that corresponds to a refinement level of 15 and
a spatial resolution of $\sim$3000 (4000) proper AU.

We perform each realization with seven sets of assumptions.  Table
\ref{tab:sims} summarizes them.  We use a nine-species (H, H$^{\rm
  +}$, He, He$^{\rm +}$, He$^{\rm ++}$, e$^{\rm -}$, H$_2$, H$_2^{\rm
  +}$, H$^{\rm -}$) non-equilibrium chemistry model \citep{Abel97,
  Anninos97} for all runs except the H+He runs that do not include
\hh~cooling.  The nine-species runs are specified by ``H2''.  Runs
with \hh~dissociating (Lyman-Werner; LW) radiation are denoted by
``LW'' followed by its negative log-flux.  We set \flw~to $10^{-22}$,
$10^{-21}$, and $10^{-20}$ \flux~because the first two are typical
values one finds in semi-analytic models of reionization and the
latter investigates the case of a very large UVB \citep[e.g.][]
{Haiman00, Wise05}.  We use the \hh~photo-dissociation rate
coefficient for the Solomon process from \citet{Abel97} of
$k_{\rm{diss}} = 1.1 \times 10^8 \flw$ s$^{-1}$.

%
%
\begin{center}
  \begin{longtable}{lccccc}
    \caption{Simulation Properties} \label{tab:sims5} \\

    \hline\hline \\[-3ex]
    Name & \hh & Residual e$^-$ & \flw & z$_a$ & z$_b$ \\
    \hline
    \endhead

    H2         \dotfill& Yes & Yes & 0         & 29.7  & 31.1 \\
    H2LW22     \dotfill& Yes & Yes & 10$^{-22}$ & 28.3  & 27.5 \\
    H2LW21     \dotfill& Yes & Yes & 10$^{-21}$ & 24.4  & 24.7 \\
    H2LW20     \dotfill& Yes & Yes & 10$^{-20}$ & 20.5 & 22.4 \\
    noe-H2     \dotfill& Yes & No  & 0         & 18.7  & 23.4 \\
    noe-H2LW20 \dotfill& Yes & No  & 10$^{-20}$ & 16.8  & 21.4 \\
    H+He       \dotfill& No  & Yes & 0         & 15.9  & 16.8 \\
    \hline
  \end{longtable}
  \tablecomments{These simulations are performed for both realizations.}
\end{center}

Free electrons are necessary to form \hh~in the gas phase.  In order
to restrict \hh~formation to \lya~line cooling halos in our ``noe-''
calculations, we reduce the residual free electron fraction from
$\sim10^{-4}$ \citep{Peebles68a, Shapiro94} to a physically low
$10^{-12}$ at the initial redshift.  This setup is designed to find
the first halos that can collapse and form stars once free electrons
from collisionally ionized hydrogen becomes available to catalyze
\hh~formation \citep{Shapiro87}.

This work is an extension of the original work of MBA01, adding the
calculations with \flw~= $10^{-20}$ and ones in which \hh~cannot cool
until \lya~cooling becomes efficient.  We consider these extreme cases
to strengthen the point made in MBA01 in which a UVB only increases
the critical halo collapse mass, never completely suppressing the
crucial importance of \hh~formation and cooling.  Our maximum spatial
resolution in the finest AMR level is a factor of four smaller than
MBA01; however, this does not cause any differences between our work
and MBA01 because these finest grid patches only exist in the dense,
central core during the final 150 kyr of the collapse.

\subsection{Virial Temperature}
\label{sec:tvir}

In galaxy formation models, the virial temperature is a key quantity
as it controls the cooling and star formation rates in a given halo.
We define a halo as the material contained in a sphere of radius
\rr~enclosing an average DM overdensity $\Delta_c$ of 200.  For an
isothermal singular sphere, the virial temperature
\begin{equation}
\label{eqn:tvir5}
T_{vir} = \frac{\mu m_p V_c^2}{2k},
\end{equation}
where $V_c^2 = GM/r_{200}$ is the circular velocity \citep[see][with
$\beta$ = 1]{Bryan98}.  Here $\mu$ is the mean molecular weight in
units of the proton mass $m_p$, and $k$ is Boltzmann's constant.  We
use this definition of \tvir~in this paper with $\mu$ = 0.59.  We
choose this value of $\mu$ to be consistent with the literature on
galaxy formation even though the halos presented in this paper are
neutral and have $\mu$ = 1.22.

%
%
\begin{figure}
  \begin{center}
    \includegraphics[width=0.75\textwidth]{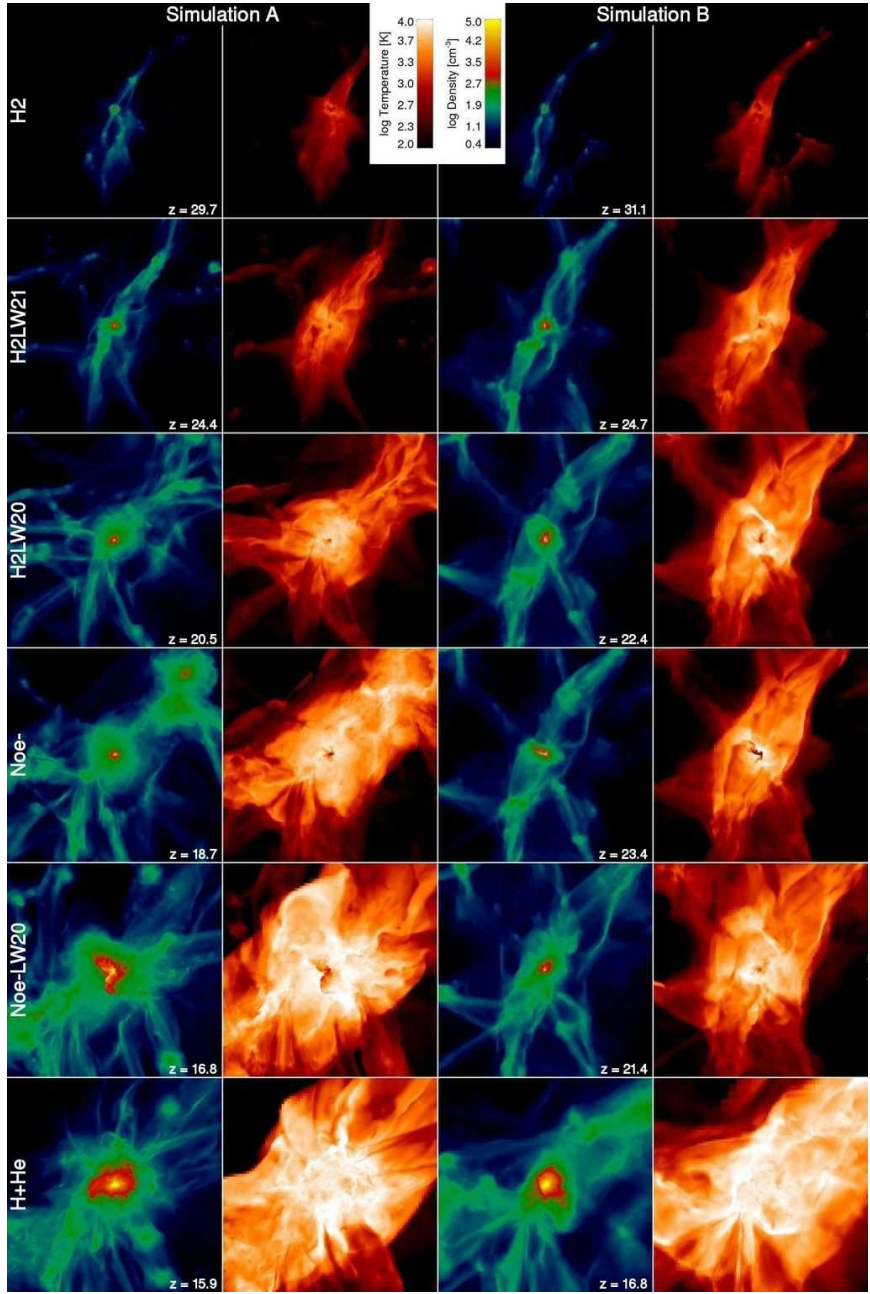}
    \caption[Density and Temperature Projections]{\label{fig:rho}
      \footnotesize{Projections in simulation A (left two columns) and
        B (right two columns) of the gas density (first and third
        columns) and temperature (second and fourth columns) at the
        times when the most massive halo starts to cool and collapse
        above an overdensity of $10^7$ in the models. The rows show
        the H2, H2LW21, H2LW20, and noe-H2, noe-H2LW20, and H+He runs
        from top to bottom, respectively.  Note the complex structure
        for the SimA-Noe-LW20 and SimB-Noe-H2 run in which central
        shocks lead to the formation of free electrons that promote
        the formation of H2 and trigger the collapse.  The field of
        view in all panels is 1.2 proper kpc.  The color maps are
        equal for all runs.}}
  \end{center}
\end{figure}

\section{Results}

We first describe the halo properties at collapse.  Then we compare
them to previous studies of collapsing halos in the presence of a soft
UVB.

\subsection{Halo Properties}

Figure \ref{fig:rho} shows density-squared weighted projections of gas
density and temperature when each calculation can cool and collapse to
an overdensity of 10$^7$.  It illustrates the large difference in the
sizes and morphologies of the collapsing halos in the various cases of
negative feedback.  All panels have the same field of view of 1.2
proper kpc and same color scales.  It is clear from the relative sizes
of the collapsing halos that the critical halo mass to cool increases
with the amount of negative feedback.  The virial shock and numerous
central shocks heat the gas to the virial temperature.  The central
shocks create fine structure seen in the temperature projections.  In
all of the \hh~cases, we see neither fragmentation nor large-scale
disk formation.  The internal structures of the halos with \hh~cooling
and residual free electrons are similar to previous studies of Pop III
star forming halos \citep[MBA01;][]{Abel00, Abel02a, Bromm02a,
  Yoshida03}, exhibiting a turbulent medium with a radially
monotonically decreasing density profile and a cool central core.

%
%
\begin{figure}[t]
\begin{center}
\includegraphics[width=0.6\textwidth]{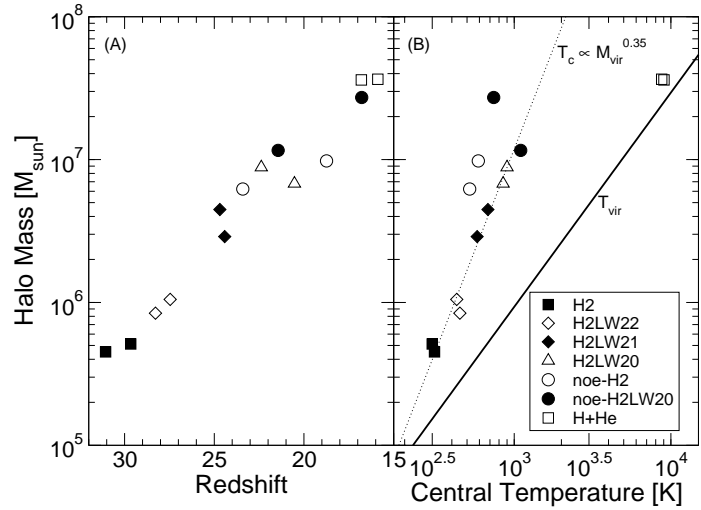}
\caption[Collapsing Halo Properties]{\label{fig:mass} \textit{Panel
    A}: Halo masses of the most massive halos as function of redshift
  when they reach a central overdensity of $10^7$. This allows to
  translate the mass values in Panel B to be converted to cooling
  redshifts.  It marks the runs with H+He (\textit{filled squares}),
  \hh~(\textit{open squares}), $\flw=10^{-22}$ (\textit{open
    diamonds}), $\flw=10^{-21}$ (\textit{filled diamonds}), no
  residual electrons (\textit{open circles}), and extreme feedback
  noe-LW20 (\textit{filled circles}) runs.  The two data points for
  each symbol represent simulations A and B.  Even for the most
  extreme cases of feedback cooling occurs much earlier than in the
  atomic line cooling only case. \textit{Panel B}: Central
  temperatures of the most massive halo in the simulation as a
  function of its mass at different redshifts.  The virial temperature
  computed form the dark matter halo mass at redshift 20 is the solid
  line.  The dotted line is the fitted relationship between the
  central gas temperatures and the halo mass in models with residual
  electrons and \hh~cooling.}
\end{center}
\end{figure}

Figure \ref{fig:mass}A depicts the halo mass and redshift when the
halo collapses for all of the runs, and Figure \ref{fig:mass}B shows
their central temperature at the same epoch.  The collapse redshifts,
$z_a$ and $z_b$, are also listed in Table \ref{tab:sims} for
simulations A and B, respectively.  As seen in other studies (MBA01,
YAHS03), the minimum DM halo mass to collapse increases with the
background intensity.  The H+He case predictably collapses at
\tvir~$\sim 10^4$ K, and all of the halos with \hh~cooling collapse at
much smaller masses.  The temperature of the central core increases
with halo mass from 300 K to 1000 K for halo masses $4 \times 10^5
\Ms$ and $10^7 \Ms$.  Restricting the data to models with residual
electrons, the central temperature increases as a power-law, $T_c = A
\mvir^B$, where $A = 3.1^{+1.3}_{-0.9}$, $B = 0.355 \pm 0.024$, and
\mvir~is in units of solar masses.  This relationship is plotted in
Figure \ref{fig:mass}.

\begin{figure}[t]
  \centering
  \vspace{0.4cm}
  \includegraphics[width=\textwidth]{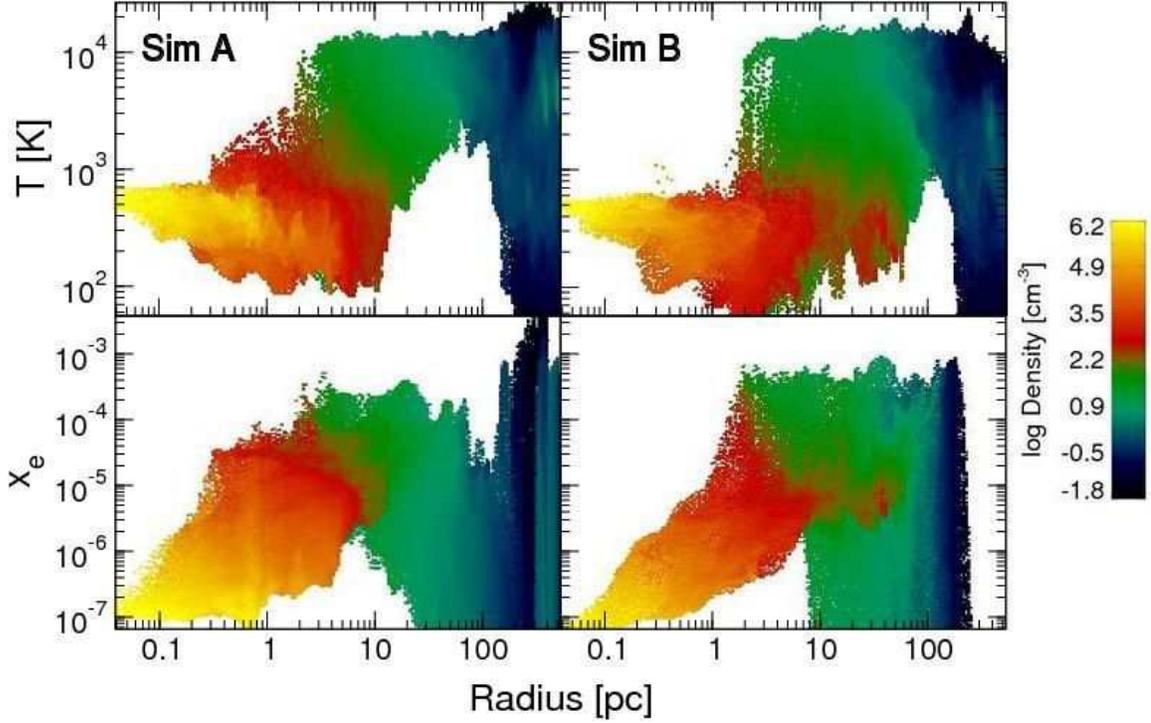}
  \caption[Temperature and Electron Fraction Radial Profiles]
  {\label{fig:efrac} Radial profiles of temperature (\textit{top}) and
    electron fraction (\textit{bottom}) colored by density for the
    ``noe-'' simulations with no residual free electrons or UVB in
    simulation A (\textit{left}) and B (\textit{right}).  The virial
    temperatures of these halos are 4600K and 4200K for simulation A
    and B, using equation (\ref{eqn:tvir5}) with $\mu$ = 0.59.}
\end{figure}

With neither residual electrons nor an UVB (noe-H2), the most massive
halo collapses at $9.8 \; (6.2) \times 10^6 \Ms$ at z = 18.7 (23.4).
Here \hh~formation in the gas phase can only become important when
sufficient free electrons are created by collisional ionization.
Virial heating in the center of halos can increase temperatures up to
twice the virial temperature \citep{Wise07} that collisionally ionizes
hydrogen in the central shocks and initiates \hh~cooling
\citep{Shapiro87} in halos well below virial temperatures of 10$^4$ K.
These shocks are abundant throughout the central regions.  Figure
\ref{fig:efrac} shows radial profiles of temperature and electron
fraction for both simulations and depicts gas shock-heating up to $2
\times 10^4$ K and electron fractions up to 10$^{-3}$.  The electron
fractions remain at unrealistically low values less than 10$^{-6}$ in
low density regions where gas has not been collisionally ionized.  The
higher density regions have condensed to densities above $3 \times
10^2 \cubecm$ after free electrons in protogalactic shocks induced
\hh~cooling.

A similar but extreme model demonstrates that even in the presence of
a very large UVB of \flw~= 10$^{-20}$ gas is able to form a cool and
dense central molecular core at a mass of $2.7 \; (1.1) \times 10^7
\Ms$ at redshift 16.8 (21.4).  Two major mergers in simulation A occur
between z = 17--21, and the associated heating allows the halo to
begin cooling by \hh.  A central core only forms once the system is
adequately relaxed after the mergers, which causes the collapse mass
difference between the realizations.

The combination of a recent major merger and collisional ionization
produces complex structures as seen in the density and temperature
projections of the SimA-Noe-LW20 and SimB-Noe- calculations in Figure
\ref{fig:rho}, unlike the other H2 models with a single cool central
core.

%
%

\subsection{Comparison to Previous Studies}

Through a series of AMR calculations with varying UVB intensities,
MBA01 found the minimum DM halo mass to be $2.5 \times 10^5 + 1.7
\times 10^6 (\flw/10^{-21})^{0.47} \; \Ms$ in order to cool and
condense 4\% of the baryons.  This fraction of cool and dense gas
agrees with simulations of the formation of Pop III stars
\citep{Abel02a, Yoshida06b}.  There is some scatter of $\sim$0.5 dex in
this threshold mass (see also YAHS03).  For the UVB intensities used
in our models (\flw~= 0, 10$^{-22}$, 10$^{-21}$, 10$^{-20}$), the
critical collapse masses are $2.5 \times 10^5$, $8.4 \times 10^5$,
$2.0 \times 10^6$, and $5.4 \times 10^6$\Ms.  Our calculations with
\hh~cooling and residual free electrons agree with the results of
MBA01.

YAHS03 studied the minimum collapse mass but also included the effects
of self-shielding.  Through their SPH simulations and arguments using
equilibrium \hh~abundances, they conclude that an UVB intensity of
\jj~= 0.1 nearly prevents halo collapses below \tvir~$\simeq$ 7000 K
where \lya~cooling becomes efficient.  They also deduce that \jj~= 1.0
completely prevents any \hh~cooling in these low-mass halos, based on
\hh~dissociation timescales.  We find the contrary in our H2LW21 and
H2LW20 calculations where the most massive halo collapses with a mass
of $4.5\;(2.9) \times 10^6 \Ms$ and $8.8\;(6.8) \times 10^6 \Ms$,
respectively.  Even in our noe- runs, the halo collapses when
\tvir~$\sim$ 4000 K, i.e. before \lya~cooling becomes important, which
is around the same mass scale that the H2LW20 runs condense.  We
ignore self-shielding in our calculations, but this would only
decrease the critical collapse mass.  This difference is most likely a
consequence of our high resolution shock capturing numerical method
and high spatial resolution as compared to the SPH approach of YAHS03
that was designed to study the statistical aspects of these objects.

The halo characteristics and the collapse redshift will likely depend
on halo merger histories as seen in these two realizations.  The
better statistics of MBA01 sampled this effect well.  Here the scatter
of threshold mass is $\sim$0.5 dex and is smaller than the mass
difference between halos with virial temperatures of 4000K and 10000K.
Thus our limited sample of halos should not change our result of the
importance of \hh~cooling in halos well below \tvir~= 10$^4$ K, even
with very large LW radiation backgrounds.

\section{Discussion}

Structure formation in the high-redshift universe is contained within
shallow potential wells that are sensitive to negative feedback from a
UVB.  Additionally local positive and negative feedback will influence
star formation and further complicate estimates of halo mass scales.
Some examples include

\begin{itemize}
\item \textit{Positive feedback}--- Enhanced \hh~formation in relic
  \ion{H}{2} regions \citep[e.g.][]{Ferrara98, OShea05, Johnson07} and
  ahead of the \ion{H}{2} ionization front \citep{Ricotti01, Ahn07},
  dust and metal line cooling \citep{Glover03, Schneider06,
    Jappsen07},
\item \textit{Negative feedback}--- Baryonic expulsion from host halos
  \citep{Whalen04, Kitayama04, Yoshida06a, Abel07}, photo-evaporation
  \citep{Susa06}, entropy floors \citep{Oh03}.
\end{itemize}

These processes are not within the scope of this paper and will be
considered in later publications that utilize three-dimensional
radiation hydrodynamic simulations with Pop III star formation.  Here
we only focused on the effects of a UVB on low-mass halos.

\subsection{The Nature of the UVB}

The intensity of the UVB is a monotonically increasing function of
redshift as more halos form stars.  The UVB increases on the order of
a Hubble time, which is much shorter than a dynamical time of a
collapsing halo and justifies the use of a constant intensity in our
calculations.

Self-consistent studies that evolve the UVB according to star
formation rates only find \jlw~to be in the range of 0.01 and 0.1 at
redshifts 15--20 (YAHS03, WA05).  WA05 calibrated their model against
the WMAP1 measurement of $\tau$ = 0.17.  With the WMAP3 result of the
electron scattering optical depth $\tau$ = 0.09 and less small-scale
power, UVB intensities will be even lower at these redshifts.

We can relate reionization to LW radiation by equating \jj~in the LW
band to a common quantity in reionization models, the ratio of emitted
hydrogen ionizing photons to baryons, $n_{\gamma, \rm{HI}} /
\bar{n}_b$, where $\bar{n}_b \simeq 2 \times 10^{-7} (1+z)^3 \cubecm$
is the cosmic mean of the baryon number density.  Assuming that
$J_{\rm{LW}}$ is constant in the LW band, the number density of LW
photons is
\begin{eqnarray}
  n_{\gamma, \rm{LW}} &=& \frac{4\pi}{c} \int^{\nu_2}_{\nu_1} 
  \frac{J_{\rm{LW}}}{h_p \nu} \; d\nu\nonumber \\
  &=& 1.19 \times 10^{-5} \jlw \; \textrm{cm}^{-3},
\end{eqnarray}
where $h_p$ is Planck's constant and $\nu_1, \nu_2$ = $2.70 \times
10^{15}$ Hz, $3.26 \times 10^{15}$ Hz bound the LW band.  To relate
\jj~to $n_{\gamma, \rm{HI}} / \bar{n}_b$, we must consider the
intrinsic ionizing spectrum and absorption from the IGM and host halo.
At redshift 20, the majority of star forming halos host Pop III stars
that emit a factor $\phi_{\rm{HI}} \simeq 10$ more hydrogen ionizing
photons than LW photons because of its $\sim10^5$~K surface
temperature.  Since the number density of sources exponentially
increases with redshift, the majority of the early UVB at a given
redshift originates from cosmologically nearby ($\Delta z / z \sim
0.1$) sources.  Lyman line resonances absorb a fraction $f_{\rm{abs}}
\sim 0.1$ of the LW radiation in the intergalactic medium in this
redshift range, producing a sawtooth spectrum \citep{Haiman97a}.
Additionally, absorption in the host halo reduces the number of
ionizing photons that escape into the IGM by a fraction
$f_{\rm{esc}}$.  For Pop III halos, this factor is close to unity
\citep{Yoshida06a, Abel07}.  By considering these multiplicative
processes, we now estimate
\begin{eqnarray}
  \label{eqn:ph_ratio}
  \frac{n_{\gamma, \rm{HI}}}{\bar{n}_b} &=& 
  \frac{n_{\gamma, \rm{LW}}}{\bar{n}_b} \;
  \left( \frac{1+z}{20} \right)^{-3} \;
  \phi_{\rm{HI}} \; f_{\rm{esc}} \; f_{\rm{abs}}^{-1}
  \nonumber \\
  &=& 0.64 \; J_{21} \left(\frac{1+z}{20}\right)^{-3}
  \left(\frac{\phi_{\rm{HI}}}{10}\right)
  \left(\frac{f_{\rm{esc}}}{1}\right) \\
  & & \times \left(\frac{f_{\rm{abs}}}{0.1}\right)^{-1} \nonumber
\end{eqnarray}
This estimate is in agreement with the reionization models of
\citet{Haiman00} and WA05 \citep[see also][]{Gnedin97}.  These models
find that sources produce a large UVB of \jj~$\sim$ 1 prior to
reionization.  When Pop III stars dominate the UVB, the LW radiation
will be small in comparison to the volume averaged hydrogen ionizing
emissivity because of the intrinsically hard Pop III spectra that
peaks at $\sim$300\AA.  Hence high-redshift halos should not be
exposed to a large UVB, i.e. \jj~$\gsim$ 0.1, and \hh~formation will
remain important before reionization.

Nearby star formation can boost the LW radiation over its background
value, but these bursts are short-lived as Pop III lifetimes are only
$\sim$3 Myr \citep{Schaerer02}.  For example, a 100 \Ms~star produces
10$^{50}$ LW photons s$^{-1}$ and will produce $\jlw > 0.1$ in the
surrounding 3 proper kpc, neglecting any
\hh~self-shielding.\footnote{LW self-shielding may be unimportant up
  to column densities of $10^{20} - 10^{21}$ cm$^{-2}$ if the medium
  contains very large velocity gradients and anisotropies
  \citep{Glover01}.}


The LW background is uniform outside these spheres of influence.  The
bursting nature of Pop III star formation does not affect the time
evolution of the background.  The intensity only depends on the number
of sources in a redshift range $\Delta z / z$ = 13.6 eV / 11.18 eV --
1, where the two energies bound the LW band, because any radiation
redward of the Lyman break contributes to the LW background.  Using a
conservative minimum halo mass for Pop III star forming halos of 3
$\times$ 10$^6 \Ms$ at redshift 20, there are $\sim$42000 halos that
have hosted a Pop III star in the volume contained within $\Delta z$,
using WMAP3 parameters with Sheth-Tormen formalism.  Clearly the
background is uniform considering the sheer number of sources within
this optically thin volume.  Local perturbations from Pop III star
formation should only affect the timing of nearby star formation but
not the global star formation rate.

\subsection{\hh~Cooling within a UVB}

YAHS03 used cosmological SPH simulations and \hh~formation and
dissociation timescales to argue that a LW background intensity of
$\jlw > 0.1$ suppresses \hh~formation so halos cannot cool before
virial temperatures of 7000 K are reached.  Employing the same
argument, we see that the \hh~formation timescale
\begin{equation}
  \label{eqn:h2form}
  t_{\rm{H_2}} = \frac{n_{\rm{H_2}}}{k_{\rm{H^-}} n_{\rm{H}} n_{\rm{e}}}
  = \frac{f_{\rm{H_2}}}{0.92 k_{\rm{H^-}} f_{\rm{e}} n}
  \approx 30 \; \textrm{kyr} ,
\end{equation}
with typical central values found in high-redshift halos before any
radiative cooling becomes efficient \citep[see][]{Wise07}.  Here
$f_{\rm{H_2}} = 10^{-6}$ and $f_{\rm{e}} = 10^{-4}$ are the \hh~and
electron number fraction, respectively, and $n = 10\cubecm$ is the
baryon number density.  $k_{\rm{H^-}} \approx 10^{-15}$ cm$^3$
s$^{-1}$ is the H$^-$ formation rate coefficient by electron
photo-attachment at $T = 1000$ K \citep{Abel97}.  This timescale is a
factor of 1000 smaller than the value calculated in YAHS03 because we
use the quantities from the halo center as compared to the mean
values.  The \hh~dissociation timescale is $k_{diss}^{-1}$ =
23/\jj~kyr, which is comparable with $t_{\rm{H_2}}$ using the values
above.

%
%
\begin{figure}
\begin{center}
\includegraphics[width=\textwidth]{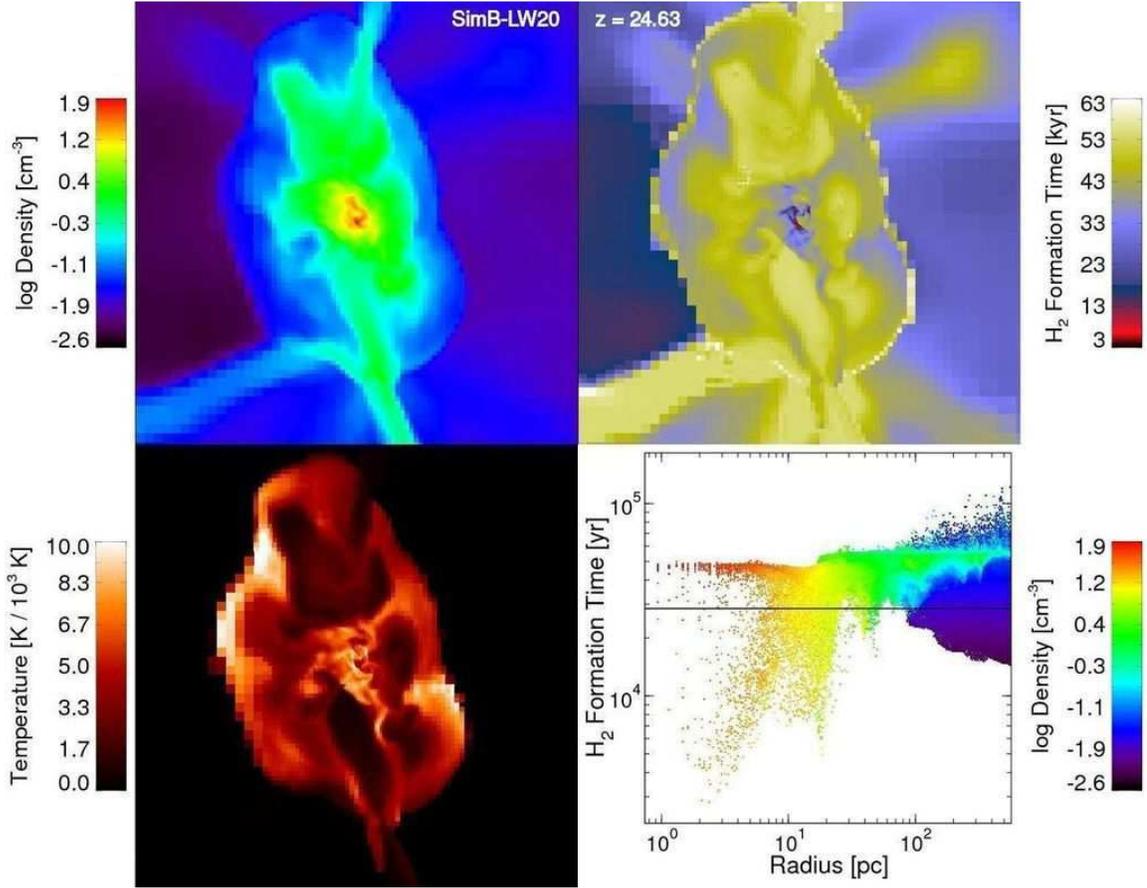}
\caption[\hh~Formation Conditions with an Ultraviolet
  Background]{\label{fig:h2form} The most massive halo in SimB-LW20
  (\flw~= $10^{-20}$) twenty million years before the cooling core
  collapses.  Slices of gas density (\textit{left}) and the
  \hh~formation timescale (Eq. [\ref{eqn:h2form}]; \textit{right})
  through the densest point in the halo are in the \textit{top} row.
  The \textit{bottom} row contains a slice of temperature
  (\textit{left}) and a radial profile of the \hh~formation timescale,
  colored by gas density.  The phase diagram and density slice have
  the same color scale.  The slices have a field of view of 500 proper
  pc.  In the central shocks, \hh~formation timescales are lower than
  the dissociation timescale of 28 kyr with \flw~= 10$^{-20}$ that is
  denoted by the horizontal line in the radial profile.  The central
  core is now efficiently cooling and will collapse 20 Myr after these
  data.}
\end{center}
\end{figure}

Figure \ref{fig:h2form} shows SimB-LW20 twenty million years before
the central core condenses.  At this time, the core is just beginning
to cool by \hh, catalyzed by the free electrons created in the central
shocks.  In these shocks, temperatures reach $1.4 \times 10^4$ K and
electron fractions up to $10^{-3}$ exist there.  These conditions
result in \hh~formation timescales less than 25 kyr, which is
necessary to cool in a UVB of \jj~$\sim$ 1.  Within the central 10 pc,
hot and cold gas phases exist.  The hot phase exists behind the shocks
that have lower densities around 10\cubecm~and $t_{\rm{H_2}} <$ 25
kyr.  This is where \hh~cooling is catalyzed by collisional ionization
in these shocks.  The cold phase has already cooled through \hh~and
has high densities and larger $t_{\rm{H_2}}$ values.  Both phases are
apparent in the panels of Figure \ref{fig:h2form}.  Similar conditions
create \hh~in the collapses in the ``noe-'' calculations, which have
sufficient gravitational potential energy, resulting in temperatures
above $10^4$ in central shocks.

Hence \hh~formation is possible in the centers of high-redshift halos
with virial temperatures below 10$^4$ K, even with a UVB of intensity
$\jlw \sim$ 1, larger than expected from semi-analytic models of
reionization.

\subsection{Impact on Semi-analytic Models}

Two consequences of a lower critical \lya~cooling halo mass are more
frequent and earlier galaxy formation and higher mass fractions in
cooling halos.  At redshift 20, e.g., abundances of \tvir~= 4000 K
halos are an order of magnitude larger than \tvir~= 10$^4$ K halos,
resulting from the exponential nature of Press-Schechter formalism.
The mass fraction contained in these halos is three times higher than
$10^4$ K halos.  In semi-analytic models of reionization and chemical
enrichment, the star formation rate (SFR) is linearly dependent on the
collapsed mass fraction since the SFR is usually a product of mass
fraction and star formation efficiency, which is the fraction of gas
collapsing into stars \citep[e.g.][]{Haiman97a}.  The star formation
efficiency for primordial stars is $\sim10^{-3}$ with a single massive
star forming in dark matter halos with mass $\sim10^6 \Ms$
\citep{Abel02a, Bromm02a, Yoshida06b}.  This fraction may rise to a few
percent in dwarf galaxies as widespread star formation occurs
\citep{Taylor99, Gnedin00, Walter01}.  Various studies predict that a
majority of the reionizing flux originates from dwarf galaxies
\citep[e.g.][]{Cen03, Sokasian04, Haiman06}.  If the mass contained in
star forming halos is three times greater than previously thought,
some of the predicted attributes, e.g. photon escape fractions and
star formation efficiencies, of high-redshift dwarf galaxy will
require appropriate adjustments to match observations, such as the
WMAP3 measurement of optical depth to electron scattering
\citep{Page06} and Gunn-Peterson troughs at $z \sim 6$
\citep{Becker01, Fan02}.

\section{Summary}

We conducted a suite of fourteen cosmology AMR simulations that focus
on the importance of \hh~cooling with various degrees of negative
feedback.  We summarize the findings of each model below.

\medskip

1. The calculations with a UVB of \flw~= (0, $10^{-22}$, $10^{-21}$)
agree with the results of MBA01, where the critical collapse halo mass
increases as a function of UVB intensity.  

2. Above \flw~= $10^{-21}$, it had been argued that an \hh~dissociating
background would inhibit any \hh~formation until the halo could cool
through \lya~cooling.  We showed that central shocks provide
sufficient free electrons from collisional ionization to drive
\hh~formation faster than dissociation rates even in a \flw~=
$10^{-20}$ background.

3. In our ``noe-'' models, we explored when collisional ionization
becomes important and conducive for \hh~formation.  This occurs at
\tvir~$\sim$ 4000 K.  Recent major mergers above this mass scale
create complex cooling structures, unlike the non-fragmented central
cores in smaller halos.

4. Even our most extreme assumptions of \jj~= 1 (\flw~$\simeq
10^{-20}$) and no residual free electrons cannot defeat the importance
of \hh~cooling in the early universe.  

\medskip

In any case, \hh~cooling triggers collapses in halos with virial
temperatures well below 10$^4$ K.  The lower critical halo mass,
corresponding to \tvir~$\sim$ 4000 K, increases mass fraction
contained in these halos by three times at redshift 20 and the number
density of high-redshift star forming halos by an order of magnitude!
By considering additional cases of extremely large negative feedback,
we have strengthened the results of MBA01 that \hh~cooling plays a key
role in high-redshift structure formation.  We conclude that a UVB
only delays and never completely suppresses \hh~formation and cooling
and subsequent star formation in these low-mass halos.

\chapter{The Nature of Early Dwarf Galaxies}
\label{chap:nature}

All observed galactic and extragalactic stars and intergalactic Lyman
forest clouds have some fraction of gas enriched by elements heavier
than helium.  This applies to even the oldest stars, whose ages are
comparable to the age of the universe, in our galaxy and surrounding
dwarf galaxies.  Big Bang nucleosynthesis produces hydrogen and helium
with trace amounts of deuterium and lithium \citep{Alpher48,
  Wagoner67}.  The first stars formed from this pristine gas, but this
prompts the question: \textit{How did the first generation of stars
  enrich the universe with the first metals?}


After quantifying the relevance of the hydrodynamics, hierarchical
assembly, and cooling physics in the previous chapters, we can finally
include the main focus of this thesis: the impact of primordial
stellar feedback on early galaxy formation.  These models probe the
same questions that were addressed in Chapter \ref{chap:rates} but
with a self-consistent treatment of primordial star formation in
adaptive mesh refinement simulations.  We have developed an accurate
radiation transport model that utilizes adaptive ray tracing
\citep{Abel02b} and has been parallelized and optimized for shared and
distributed computing systems.  This allows us to follow the radiative
feedback from the first stars and to investigate its impact on the
formation of the first galaxies.  We also follow the propagation of
metal ejecta from the supernovae of massive, metal-free stars.  We
investigate the global nature of the early dwarf galaxies that
ultimately form in our calculations.  These results may result in a
better understanding of the origin of the oldest stars in our galaxy
and some of the neighboring dwarf galaxies.

This is the final iteration of the gradual inclusion of physical
models in this thesis; however we note that many other processes may
be relevant in early galaxy formation, e.g. metal-line and dust
cooling, self-shielding of \hh~dissociating radiation, and magnetic
fields.

This chapter is in preparation for publication in \textit{The
  Astrophysical Journal}.  It is co-authored by Tom Abel, who
developed the original version of the adaptive ray tracing code used
in these simulations.

\section{Motivation}

The majority of galaxies in the universe are low-luminosity, have
masses of $\sim$$10^8$ solar masses, and are known as dwarf galaxies
\citep{Schechter76, Ellis97, Mateo98}.  Galaxies form hierarchically
through numerous mergers of smaller counterparts \citep{Peebles68,
  White78}, whose properties will inevitably influence the parent
galaxy.  Dwarf galaxies are the smallest galactic building blocks, and
this leads to the question on even smaller scales: how were dwarf
galaxies influenced by their progenitors?  To answer this intriguing
question, we let observations of local dwarf galaxies and numerical
simulations guide us.

A subset of dwarf galaxies, dwarf spheroidals (dSph), have the highest
mass-to-light ratios \citep{deBlok97, Mateo98} and contain a
population of metal-poor stars that are similar to Galactic halo stars
\citep{Tolstoy04, Helmi06}.  Stellar metallicities increase with time
as previous stars continually enrich the interstellar medium
(ISM). There is a metallicity floor of $10^{-3}$ and $10^{-4}$ of
solar metallicity in dSph and halo stars, respectively \citep{Beers05,
  Helmi06}.  Hence the lowest metallicity stars are some of the oldest
stars in the system and can shed light on the initial formation of
dwarf galaxies.  This metallicity floor also suggests that metal
enrichment was widespread in dark matter halos before low-mass stars
could have formed \citep[e.g.][]{Ricotti02b}.  Supernovae (SNe) from
metal-free (Pop III) stars generate the first metals in the universe
and may supply the necessary metallicity to form the most metal-poor
stars observed \citep{Ferrara98, Madau01a, Norman04}.

Dwarf galaxy formation can be further constrained with observations
that probe reionization and semi-analytic models.  Observations of
luminous quasars powered by supermassive black holes (SMBH) of mass
$\sim$$10^9 \Ms$ \citep{Becker01, Fan02, Fan06} and low-luminosity
galaxies \citep{Hu02, Iye06, Kashikawa06, Bouwens06, Stark07} at and
above redshift 6 indicate that active star and BH formation began long
before this epoch.  Semi-analytic models have shown that cosmological
reionization was largely caused by low-luminosity dwarf galaxies
\citep{Haiman97, Cen03, Somerville03, Wise05, Haiman06}.  Some of the
most relevant parameters in these models control star formation rates,
ionizing photon escape fractions, metal enrichment, and the minimum
mass of a star forming halo.  They are usually constrained using (i)
the cosmic microwave background (CMB) polarization observation from
WMAP that measures the optical depth of electron scattering to the CMB
\citep{Page06}, (ii) Gunn-Peterson troughs in $z \sim 6$ quasars, and
(iii) numerical simulations that examine negative and positive
feedback of radiation backgrounds \citep{Machacek01, Yoshida03,
  Mesigner06}.  Radiation hydrodynamical \textit{ab initio}
simulations of the first stars \citep{Yoshida06a, Abel07} and galaxies
can further constrain the parameters used in semi-analytic models by
analyzing the impact of stellar feedback on star formation rates and
the propagation of \ion{H}{2} regions in the early universe.
Moreover, these simulations contain a wealth of information pertaining
to the properties of Pop III star forming halos and early dwarf
galaxies that can increase our understanding of the first stages of
galaxy formation.

First we need to consider Pop III stars, which form in the progenitor
halos of the first galaxies, to fully understand the initial
properties of dwarf galaxies.  Cosmological numerical studies have
shown that massive Pop III stars form in dark matter halos with masses
$\sim$$10^6 \Ms$.  Recently, \citet{Yoshida06b} followed the gaseous
collapse of a molecular cloud that will host a Pop III star to
cosmologically high number densities of $10^{16} \cubecm$.  They
thoroughly analyzed the gas dynamics, cooling, and stability of this
free-fall collapse.  They found no fragmentation in the fully
molecular core that collapses into a single, massive $\sim$100\Ms~star
\citep[see also][]{Abel02a}.  Furthermore, \citet{Omukai03} determined
that accretion will halt at the same mass scale, using protostellar
models while considering different mass accretion histories.

Pop III stars with stellar masses roughly between 140 and 260 \Ms~end
their life in a pair-instability SN that releases $10^{51} - 10^{53}$
ergs of energy and tens of solar masses of heavy elements into the
ambient medium \citep{Barkat67, Bond84, Heger02}.  These explosions
are an order of magnitude larger than typical Type II SNe in both
quantities \citep{Woosley86}.  Pair-instability SN energies are larger
than the binding energies of their low-mass hosts, e.g., $2.8 \times
10^{50}$ ergs for a $10^6 \Ms$ halo at redshift 20.  Gas structures in
the host halo are totally disrupted and expelled, effectively
enriching the surrounding intergalactic medium (IGM) with the SN
ejecta \citep{Bromm03, Kitayama05}.  The combination of the shallow
potential well and large explosion energy suggests that these events
are good candidates for enriching the first galaxies and IGM.  Outside
of the pair-instability mass range, Pop III stars die by directly
collapsing into a BH \citep{Heger03}, possibly providing the seeds of
high-redshift quasars in galaxies that are associated with the rarest
density fluctuations \citep[e.g.][]{Madau01b, Volonteri05}.

One-dimensional calculations \citep{Whalen04, Kitayama04, Kitayama05}
and recent three-dimensional radiation hydrodynamical simulations
\citep{Yoshida06a, Abel07} have investigated how the Pop III stellar
feedback affects its host halo and nearby cosmic structure.  In
addition to SNe, stellar radiation from Pop III stars, which have
luminosities $\sim$$10^6 L_\odot$ \citep{Schaerer02}, alone can
dynamically affect gas at distances up to a few proper kpc, for which
ionization fronts are mainly responsible.  Ionization fronts and
\ion{H}{2} regions \citep[see][for a review]{Yorke86} have been
extensively studied in literature on star formation since
\citet{Stroemgren39}.  Stellar radiation generates an ionization front
that begins as a R-type front and transforms into a D-type front when
its speed slows to twice the sound speed of the ionized gas.  Then a
strong shock wave forms at the front and recedes from the star at
$\sim$ 30\kms.  The ionization front decouples from the shock wave and
creates a final \ion{H}{2} region that is 1 -- 3 proper kpc in radius
for massive Pop III stars residing in low-mass halos.  The ionized gas
is warm ($\sim$$3 \times 10^4$ K) and diffuse ($\sim$$1 \cubecm$).
The shock wave continues to accumulate gas and advance after the star
dies.  Eventually it stalls in the IGM, but in the process, it reduces
the baryon fraction of the halo below one percent \citep{Yoshida06a,
  Abel07}.

Clearly the number of progenitors of a given galaxy as well as the
star formation and feedback history of the progenitors will play a
role in shaping all of its properties.  But how much?  If most stars
of a galaxy are formed later, will the earliest episodes not be
entirely negligible?  To start addressing these questions, we have
carried out a suite of simulations that include accurate three
dimensional radiative transfer and the SN explosions of Pop III stars
and have followed the buildup of several dwarf galaxies from those Pop
III star hosting progenitors. The Pop III radiative and SN feedback
dramatically alters the properties of high redshift dwarf galaxies,
and we discuss some of the most striking differences here.  We leave a
more detailed exposition of star formation rates, star forming
environments, and the beginning of cosmic reionization for a later
paper.


In the following section, we detail our cosmological, radiation
hydrodynamics simulations and the star formation algorithm.  Then we
describe the global characteristics of dwarf galaxies that form in
our simulations in \S\ref{sec:reionResults}.  There we also focus on
metal enrichment of star forming halos and the IGM, arising from
pair-instability SNe.  In \S\ref{sec:reionDiscuss}, we discuss the
implications of our findings on the paradigm of high-redshift galaxy
formation by including \hh~chemistry and Pop III star formation and
feedback.  We summarize in the last section.

\section{Radiation Hydrodynamical Simulations}

We use the Eulerian AMR hydrodynamic code \enzo~\citep{Bryan97,
  Bryan99} to study the importance of primordial stellar feedback in
early galaxy formation.  \enzo~uses an $n$-body adaptive particle-mesh
solver \citep{Couchman91} to follow the dark matter (DM) dynamics.  We
first describe the setup of our simulations.  We then detail our star
formation recipe for primordial star formation.  Also we have
implemented adaptive ray tracing into \enzo~that concludes this
section.

\subsection{Simulation Setup}

We perform two cosmological realizations with different box sizes and
random phases and WMAP 1 year parameters of ($h$, \Ol, \Om, \Ob,
$\sigma_8$, $n$) = (0.72, 0.73, 0.27, 0.024$h^{-2}$, 0.9, 1)
\citep{Spergel03}.  Table \ref{tab:sims7} summarizes the details of
these simulations.  The characteristics of the individual halos
studied here are not affected by the significantly different WMAP
third year parameters \citep[WMAP3;][]{Spergel06}, which do affect the
statistical properties of such halos.  We have verified that nothing
atypical occurs during the assembly of the halos studied here.

%
%
\begin{center}
  \begin{longtable}{lcccccc}
    \caption{Simulation Parameters} \label{tab:sims7} \\

    \hline\hline \\[-3ex]
    Name & $l$ & SF & SNe & N$_{\rm{part}}$ & N$_{\rm{grid}}$ & N$_{\rm{cell}}$ \\
    & [Mpc] &  &  & &  &  \\
    \hline
    \endhead

    SimA-HHe & 1.0 & No & No & 2.22 $\times$ 10$^7$ & 40601 & 1.20
    $\times$ 10$^8$ (494$^3$) \\
    SimA-RT & 1.0 & Yes & No & 2.22 $\times$ 10$^7$ & 44664 & 1.19
    $\times$ 10$^8$ (493$^3$) \\
    SimB-HHe & 1.5 & No & No & 1.26 $\times$ 10$^7$ & 21409 & 6.51
    $\times$ 10$^7$ (402$^3$) \\
    SimB-RT & 1.5 & Yes & No & 1.26 $\times$ 10$^7$ & 24013 & 6.54
    $\times$ 10$^7$ (403$^3$) \\
    SimB-SNe & 1.5 & Yes & Yes & 1.26 $\times$ 10$^7$ & 24996 & 6.39
    $\times$ 10$^7$ (400$^3$) \\
    \hline
  \end{longtable}
  \tablecomments{Col. (1): Simulation name. Col. (2): Box
    size. Col. (3): Star formation. Col. (4): Supernova
    feedback. Col. (5): Number of dark matter particles. Col. (6):
    Number of AMR grids. Col. (7): Number of unique grid cells.}
\end{center}

The initial conditions are the same as in \citet{Wise07a}.  They both
have a top grid with a resolution of 128$^3$ with three nested
subgrids with twice finer resolution and are initialized at z = 129
(119) with the COSMICS package\footnote{To simplify the discussion,
  simulation A will always be quoted first with the value from
  simulation B in parentheses.} \citep{Bertschinger95,
  Bertschinger01}.  The box size is 1.0 (1.5) comoving Mpc.  The
innermost grid has an effective resolution of 1024$^3$ with DM
particle masses of 30 (101) \Ms and a side length of 250 (300)
comoving kpc.  We refine the AMR grids when either the DM (gas)
exceeds three times the mean DM (gas) density times a factor of $2^l$,
where $l$ is the AMR refinement level.  We also refine so that the
local Jeans length is resolved by at least 16 cells.  Refinement only
occurs in the initial innermost grid that has a comoving side length
of 250 (300) kpc.  We enforce a maximum AMR level of 12 in these
simulations that corresponds to a resolution limit of 2.9 (1.9)
comoving parsecs.

We use the nine species (H, H$^{\rm +}$, He, He$^{\rm +}$, He$^{\rm
  ++}$, e$^{\rm -}$, H$_2$, H$_2^{\rm +}$, H$^{\rm -}$)
non-equilibrium chemistry model in \enzo~\citep{Abel97, Anninos97}.
Compton cooling and heating of free electrons by the CMB and radiative
losses from atomic and molecular cooling are computed in the optically
thin limit.

We focus on the region containing the most massive halo in the
simulation box.  We perform three calculations -- simulation A with
star formation (SimA-RT), simulation B with star formation (SimB-RT),
and simulation B with star formation and SNe (SimB-SN).  We end
the calculations at the same redshift the halo with a virial
temperature of 10$^4$ K collapses at z = 15.9 (16.8) in the hydrogen
and helium cooling runs (HHe) of \citet{Wise07a}.

\subsection{Star Formation Recipe}

Star formation is modelled through an extension \citep{Abel07} of the
\citet{Cen92b} algorithm that automatically forms a star particle when
a grid cell has 
\begin{enumerate}
\item an overdensity exceeding $5 \times 10^5$
\item a converging velocity field ($\nabla \cdot \mathbf{v} < 0$)
\item rapidly cooling gas (\tcool~$<$ \tdyn)
\item an \hh~fraction greater than $5 \times 10^{-4}$.
\end{enumerate}
Then we remove half of the gas from the grid cells in a sphere that
contains twice the stellar mass, which is a free parameter.  Once
these criteria are met, \citet{Abel02a} showed that a Pop III star
forms within 10 Myr.  For this reason, we do not impose the Jeans
instability requirement used in Cen \& Ostriker and do not follow the
collapses to stellar scales.  We allow star formation to occur in the
Lagrangian volume of the surrounding region out to three virial radii
from the most massive halo at $z = 10$ in the dark matter only runs as
discussed in \citet{Wise07a}.  This volume has a side length of 195
(225) comoving kpc at $z = 30$ and 145 (160) comoving kpc at the end
of the calculation.

Runs with star formation only model all Pop III stars with M$_\star$ =
100 \Ms~that live for 2.7 Myr and emit $1.23 \times 10^{50}$ hydrogen
ionizing photons per second.  After its death, the star particle is
converted into an inert 100 \Ms~tracer particle.  The SNe runs use
M$_\star$ = 170 \Ms~that results in a lifetime of 2.3 Myr and $2.57
\times 10^{50}$ ionizing photons per second, in accordance with the no
mass loss stellar models of \citet{Schaerer02}.

When a 170 \Ms~star dies, it injects $E_{\rm{SN}} = 2.7 \times
10^{52}$ erg of thermal energy and 81 \Ms~of metals, appropriate for a
pair-instability SN of a 170 \Ms~star \citep{Heger02}, into a sphere
with radius $r_{\rm{SN}}$ = 1 pc centered on the star's position.  The
mass contained in the star particle and associated metal ejecta are
evenly distributed in this sphere.  The mass of the star particle is
changed to zero, and we track its position in order to determine the
number of stars associated with each halo.  We also evenly deposit the
SN energy in the sphere, which changes the specific energy by
\begin{equation}
  \label{eqn:e_inject}
  \Delta\epsilon = \frac{ \rho_0 \epsilon_0 + \rho_{\rm{SN}}
    \epsilon_{\rm{SN}} } { \rho_0 + \rho_{\rm{SN}} } - \epsilon_0,
\end{equation}
where $\rho_0$ and $\epsilon_0$ are the original gas density and
specific energy, respectively.  Here $\rho_{\rm{SN}} = M_\star /
V_{\rm{SN}}$ is the ejecta density; $\epsilon_{\rm{SN}} = E_{\rm{SN}}
/ M_\star / V_{\rm{SN}}$ is the ejecta specific energy; $V_{\rm{SN}}$
is the volume of a sphere with radius $r_{\rm{SN}}$.  In order not to
create unrealistically strong shocks at the blast wave and for
numerical stability reasons, we smoothly transition from this energy
bubble to the ambient medium, using the function
\begin{equation}
f(r) = A \left\{ 0.5 - 0.5 \tanh \left[ B \left( \frac{r}{r_{\rm{SN}}}
        - 1 \right) \right] \right\}.
\end{equation}
Here $A$ is a normalization factor that ensures $\int f(r) dr = 1$,
and $B$ controls the rate of transition to the ambient medium, where
the transition is steeper with increasing $B$.  We use $A$ = 1.28 and
$B$ = 10 in our calculations.

We continue to use the nine-species chemistry model as we do not
consider the additional cooling from metal lines.  We follow the
hydrodynamic transport of metals from the SNe to the enrichment of the
surrounding IGM and halos.

\subsection{Radiative Transfer}

For point sources of radiation, ray tracing is an accurate method to
calculate and evolve radiation fields.  However millions of rays must
be cast in order to obtain adequate ray sampling at large radii.  We
use adaptive ray tracing \citep{Abel02b} to overcome this dilemma
associated with ray tracing \citep[cf.][]{Abel07}.  We initially cast
768 rays, i.e. level three in HEALPix \citep{Gorski05}, from the
radiation source.  The photons contained in the initial rays are
equal, and their sum is the stellar luminosity.  Rays are split into 4
child rays, whose angles are calculated with the next HEALPix level,
if their associated solid angle is greater than 20\% of the cell area
$\Delta x^2$.  Photons are distributed evenly among the children.
This occurs if the ray travels to a large distance from its source, or
the ray encounters a highly refined AMR grid, in which adaptive ray
tracing accurately samples and retains the fine structure contained in
high resolution regions.

The rays cast in these simulations have an energy $E_{\rm{ph}}$ that
is the mean energy of hydrogen ionizing photons from the stellar
source.  For 100 \Ms~and 170 \Ms, this energy is roughly equal at 28.4
and 29.2 eV, respectively, due to the weak dependence of the surface
temperature of primordial stars on stellar mass.  The rays are
transported at the speed of light on constant timesteps $\Delta
t_{{\rm ph}}$ = 800 years, which is always less than the
hydrodynamical timesteps at the finest resolution.

To model the \hh~dissociating (Lyman-Werner; LW) radiation between
11.2 and 13.6 eV, we use an optically thin 1/$r^2$ radiation field
with luminosities calculated from \citet{Schaerer02}.  We use the
\hh~photo-dissociation rate coefficient for the Solomon process of
$k_{diss} = 1.1 \times 10^8 \flw$ s$^{-1}$, where \flw~is the LW flux
in units of \flux~\citep{Abel97}.

The radiation transport is coupled with the hydrodynamical, chemistry,
and energy solvers of \enzo.  Here we only consider hydrogen
photo-ionization.  We first calculate the photo-ionization and heating
rates caused by each ray and then sub-cycle the chemistry and heating
solvers with these additional rates.  Next we advance the
hydrodynamics of the system with the usual adaptive timesteps.

The hydrogen photo-ionization rate is computed by
\begin{equation}
\label{eqn:kph}
k_{ph} = \frac{ P_0 (1 - e^{-\tau}) } { n_{\rm{HI}} \; V_{\rm{cell}} \; \Delta
  t_{{\rm ph}} },
\end{equation}
where $P_0$ is the incoming number of photons, $n_{\rm{HI}}$ is the
number density of neutral hydrogen, $V_{\rm{cell}}$ is the volume of
the computational grid cell, and $\tau = n_{\rm{HI}} \;
\sigma_{\rm{HI}} \; dl$ is the optical depth.  Here $\sigma_{\rm{HI}}$
is the cross section of hydrogen, and $dl$ is the distance travelled by
the ray through the cell.  The heating rate is computed from the
excess photon energies of the photo-ionizations by
\begin{eqnarray}
\label{eqn:heating}
\Gamma_i &\;=\;& k_{ph} (E_{\rm{ph}} - E_i) \nonumber \\
&=& \frac{ P_0 (1 - e^{-\tau}) } {
  n_{\rm{HI}} \; V_{\rm{cell}} \; \Delta t_{{\rm ph}} } \;
(E_{\rm{ph}} - E_i).
\end{eqnarray}
In the case of hydrogen ionizing photons, $E_i$ = 13.6 eV.  In each
radiation timestep, the number of photons absorbed, i.e. $P_0 (1 -
e^{-\tau})$, is subtracted from the ray.  The ray is eliminated once
most of the associated photons (e.g. 99\%) are absorbed or the ray
encounters a highly optically thick region (e.g. $\tau > 20$).

\section{Results}
\label{sec:reionResults}

In this section we first discuss star formation in dwarf galaxy
progenitors.  Then we focus on the global characteristics of the most
massive halo.  We detail the different ISM phases.  Metal transport
from pair-instability SNe and the associated metal-enriched star
formation history are discussed lastly.

\subsection{Number of Star Forming Halos}

Gas in halos with masses $\lsim 10^6$ \Ms~is evacuated by a
$\sim$30\kms~D-type front, leaving a diffuse (1\cubecm) and warm ($3
\times 10^4$ K) medium \citep{Whalen04, Kitayama04, Yoshida06a,
  Abel07}.  The aftermath of SN explosions in relic \ion{H}{2} regions
is explored in spherical symmetry in  one-dimensional calculations by
\citet{Kitayama05}.  Even without SNe, star formation is suppressed
for $\sim$100 Myr before gas is reincorporated into the potential
well.  Pair-instability SNe provide an extra $\sim 10^{52}$ erg of
thermal energy and can evacuate halos up to 10$^7$ \Ms.  Hence star
formation within these low-mass halos are highly dependent on their
star formation and merger histories, as illustrated by
\citet{Yoshida06a}.

%
%
\begin{center}
  \begin{longtable}{lcccc}
    \caption{Global halo properties with star formation} \label{tab:sf} \\

    \hline\hline \\[-3ex]
    Name & N$_\star$($<$\rvir) & N$_\star$($<$3\rvir) &
    M$_{\rm{gas}}$/M$_{\rm{tot}}$ & $\lambda_{\rm{g}}$ \\
    \hline
    \endhead

    SimA-HHe   & \dots & \dots & 0.14  & 0.010 \\
    SimA-RT     & 14    & 16    & 0.081 & 0.053 \\
    SimB-HHe   & \dots & \dots & 0.14  & 0.010 \\
    SimB-RT     & 13    & 19    & 0.11  & 0.022 \\
    SimB-SN & 7     & 13    & 0.049 & 0.097 \\
    \hline
  \end{longtable}

  \tablecomments{Col. (1): Simulation name. Col. (2): Number of stars
    hosted in the halo and its progenitors. Col. (3): Number of stars
    formed in the Lagrangian volume within 3\rvir~of the most massive
    halo. Col. (4): Baryon fraction within \rvir. Col. (5): Baryonic
    spin parameter [Eq. \ref{eqn:spin}].}

\end{center}

Table \ref{tab:sf} summarizes the global properties of the most
massive halo at the time of collapse in the HHe calculations.
Approximately ten Pop III stars form in the progenitors, whose
original gas structures are nearly destroyed by radiative feedback, of
the 10$^4$ K halo.  More specifically at z = 15.9 (16.8), there are
22, 27, and 24 stars that form after redshift 30 in the SimA-RT,
SimB-RT, and SimB-SN runs, respectively.  The most massive halo
and its progenitors have hosted 14, 13, and 7 stars in the same
simulations.  When we look at the Lagrangian volume contained within
three times the virial radius of the most massive halo, there have
been 16, 19, and 13 instances of star formation.  

%
%
\begin{figure}
  \begin{center}
    \includegraphics[width=0.75\textwidth]{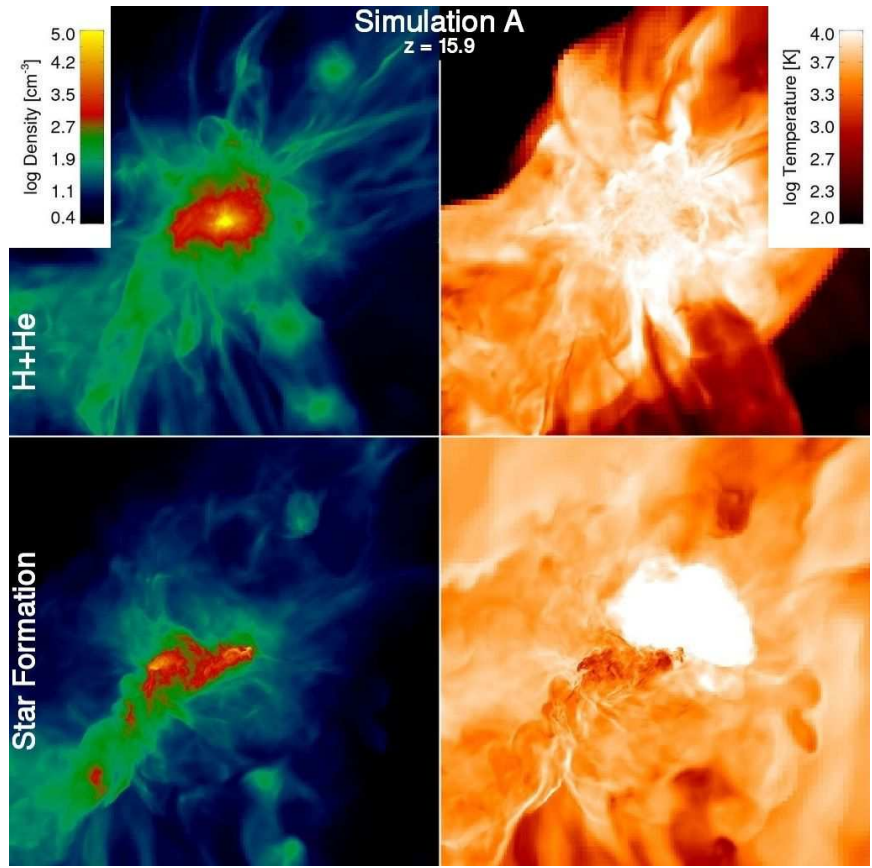}
    \caption[Density and Temperature Projections of Simulation
      A]{\label{fig:proj_a} Density-squared weighted projections of
      gas density (\textit{left}) and temperature (\textit{right}) of
      the most massive halo in simulation A.  The field of view is 1.2
      proper kpc.  The \textit{top} row shows the model without star
      formation and only atomic hydrogen and helium cooling.  The
      \textit{bottom} row shows the same halo affected by primordial
      star formation.}
  \end{center}
\end{figure}

%
%
\begin{figure}
  \begin{center}
    \includegraphics[width=0.75\textwidth]{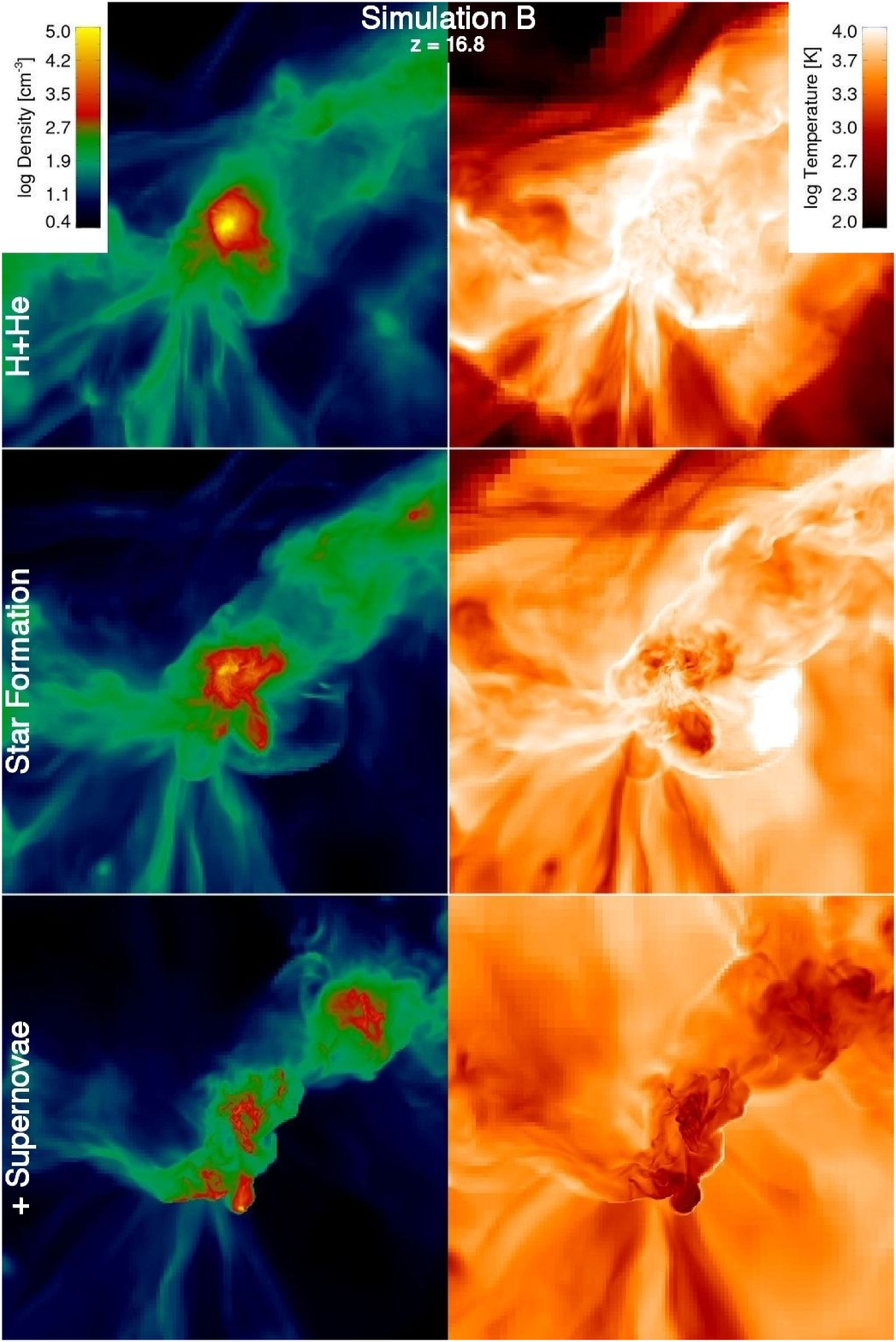}
    \caption[Density and Temperature Projections of Simulation
      B]{\label{fig:proj_b} Same as Figure \ref{fig:proj_a} for
      simulation B.  Here the \textit{bottom} row shows the halo with
      primordial stellar feedback and supernovae.}
  \end{center}
\end{figure}

\subsection{Global Nature of Objects}

In addition to providing the first ionizing photons and metals to the
universe, Pop III stars change the global gas dynamics of \tvir~$\sim$
10$^4$ K star forming halos.  Figures \ref{fig:proj_a} and
\ref{fig:proj_b} compare the structure of the most massive halo in all
simulations, depicting density-squared weighted projections of gas
density and temperature.  All of the halos have a virial mass of $3.5
\times 10^7 \Ms$.  The models with neither star formation nor
\hh~chemistry show a condensing \tvir~= 10$^4$ K halo with its
associated virial heating.  In comparison, feedback from primordial
star formation expels the majority of the gas in low-mass star forming
progenitors.  This induces the formation of an inhomogeneous medium,
where the radiation anisotropically propagates, creating champagne
flows in the directions with lower column densities.  The temperature
projections illustrate both the ultraviolet heated ($\sim10^4$ K) and
optically thick, cool ($\sim 10^3$ K) regions in the host halo and
IGM.  SN explosions alter the gas structure by further stirring and
ejecting material after main sequence.  Next we quantify these visual
features with the baryon fraction of the halo and an inspection of
phase diagrams of density and temperature.

As the gas in the progenitors is mostly evacuated, the baryon
fraction of high-redshift star forming halos are greatly reduced.  In
halos with masses $\lsim 10^6$ \Ms, the baryon fraction lowers to $5
\times 10^{-3}$ ten million years after the star's death without a SN
\citep[cf.][]{Yoshida06a}.  When we include SNe, the baryon fraction
decreases further to $1 \times 10^{-5}$ six million years after the
explosion \citep[cf.][]{Kitayama05}.  This is in stark contrast with
the cosmic fraction \Ob/\Om~= 0.17.

Within these shallow potential wells, outflows from stellar feedback
impede subsequent star formation until sufficient gas is
reincorporated, occurring through mergers and smooth IGM accretion.
After the halo mass surpasses $\sim10^7$ \Ms, total evacuation does
not occur but significant outflows are still generated.  Near the same
mass scale, multiple sites of star formation occur in the same halo.
These stars rarely shine simultaneously since we neglect
\hh~self-shielding, which only affects the timing of star formation
and not the global star formation rate.  Here after the first star
dies, the second star, whose dense, cool core survived the UV heating,
condenses and forms a star a few million years afterwards.  This
scenario of adjacent star formation is similar to the one presented in
\citet{Abel07}, but the multiple sites of star formation are caused by
\hh~and \lya~cooling in central protogalactic shocks \citep[cf.][]
{Shapiro87}, not from residual cores from a recent major merger.

When the most massive halo reaches \tvir~$\sim 10^4$ K, the baryon
fraction within the virial radius has only partially recovered to
0.081, 0.11, and 0.049 in the SimA-RT, SimB-RT, and SimB-SN
calculations.  Without any stellar feedback, these fractions are 0.14
in the HHe runs.

These outflows also create inhomogeneities in and around halos and
increase the baryonic spin parameter
\begin{equation}
\label{eqn:spin}
\lambda_{\rm{g}} = \frac{L_{\rm{g}} \vert E_{\rm{g}}
  \vert^{1/2}}{GM_{\rm{g}}^{5/2}},
\end{equation}
where $L_{\rm{g}}$, $E_{\rm{g}}$, and $M_{\rm{g}}$ are the total
baryonic angular momentum, kinetic energy, and mass of the system.
The DM spin parameter $\lambda$ uses the total DM angular momentum,
kinetic energy, and mass of the system.  At the time of collapse in
the HHe runs, $\lambda$ = 0.022 (0.013) and is marginally lower than
the average $\langle \lambda \rangle \simeq 0.04$ found in numerical
simulations \citep{Barnes87, Eisenstein95}.  It is not affected by
stellar feedback as DM dominates the potential well.  Without star
formation, the baryonic spin parameter $\lambda_{\rm{g}}$ = 0.010
(0.010) and is slightly lower than $\lambda$.  However with stellar
and SNe feedback, $\lambda_{\rm{g}}$ increases up to a factor of 10.
The effect is smaller without SNe but still significant, raising
$\lambda_{\rm{g}}$ to 0.053 (0.022).

The increase of $\lambda_{\rm{g}}$ may be caused by cosmological tidal
forces on the expelled gas, which gains angular momentum when it is at
large radii.  At this point, tidal forces are at its greatest
influence.  Then a fraction of the expelled gas, now with a higher
specific angular momentum, falls back to the dwarf galaxy and thus
increasing its spin parameter.

%
%
\begin{figure}[t]
\begin{center}
\includegraphics[width=\textwidth]{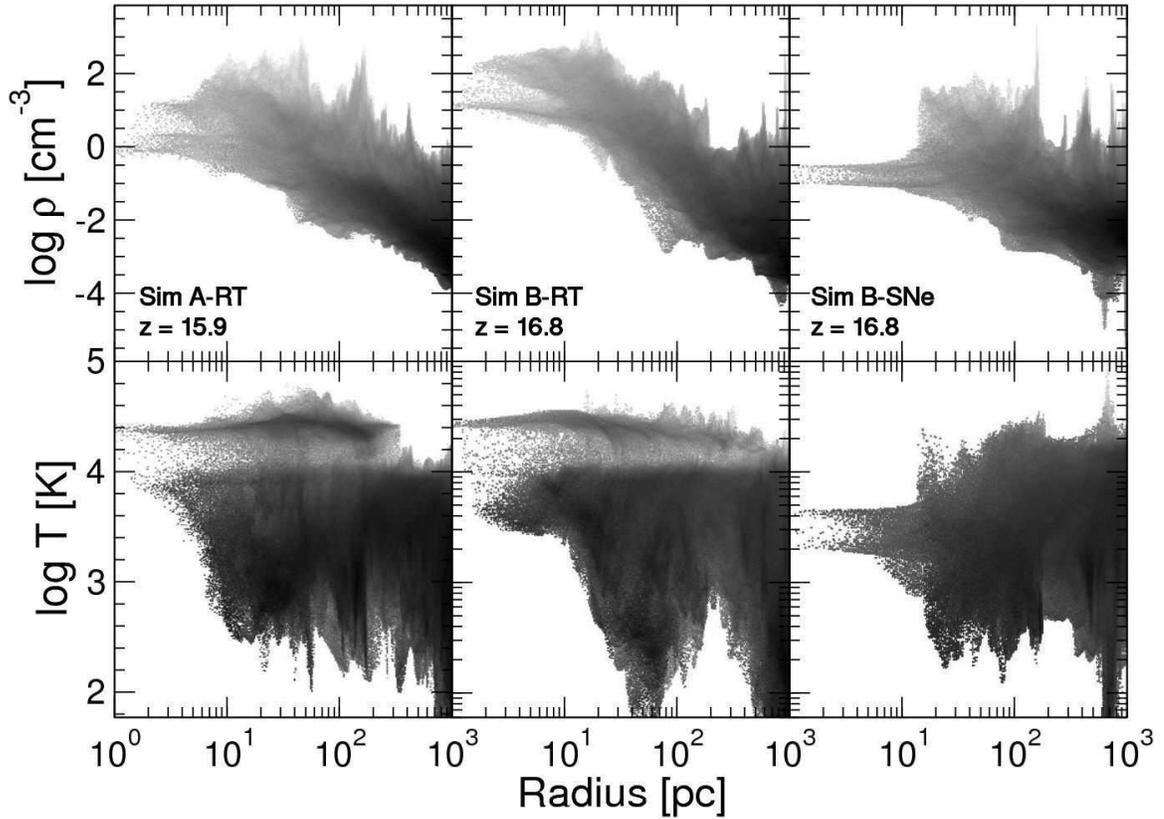}
\caption[Dwarf Galaxy Radial Profiles]{\label{fig:radial}
  Mass-weighted radial profiles of density (\textit{top}) and
  temperature (\textit{bottom}), centered on the densest DM particle.
  The columns show data from SimA-RT (\textit{left}), SimB-RT
  (\textit{middle}), and SimB-SN (\textit{right}).  Note how the cool
  and warm gas phases coexist at similar radii throughout the halo.}
\end{center}
\end{figure}

\subsection{ISM Phases}

\begin{figure}
\begin{center}
\includegraphics[width=0.8\textwidth]{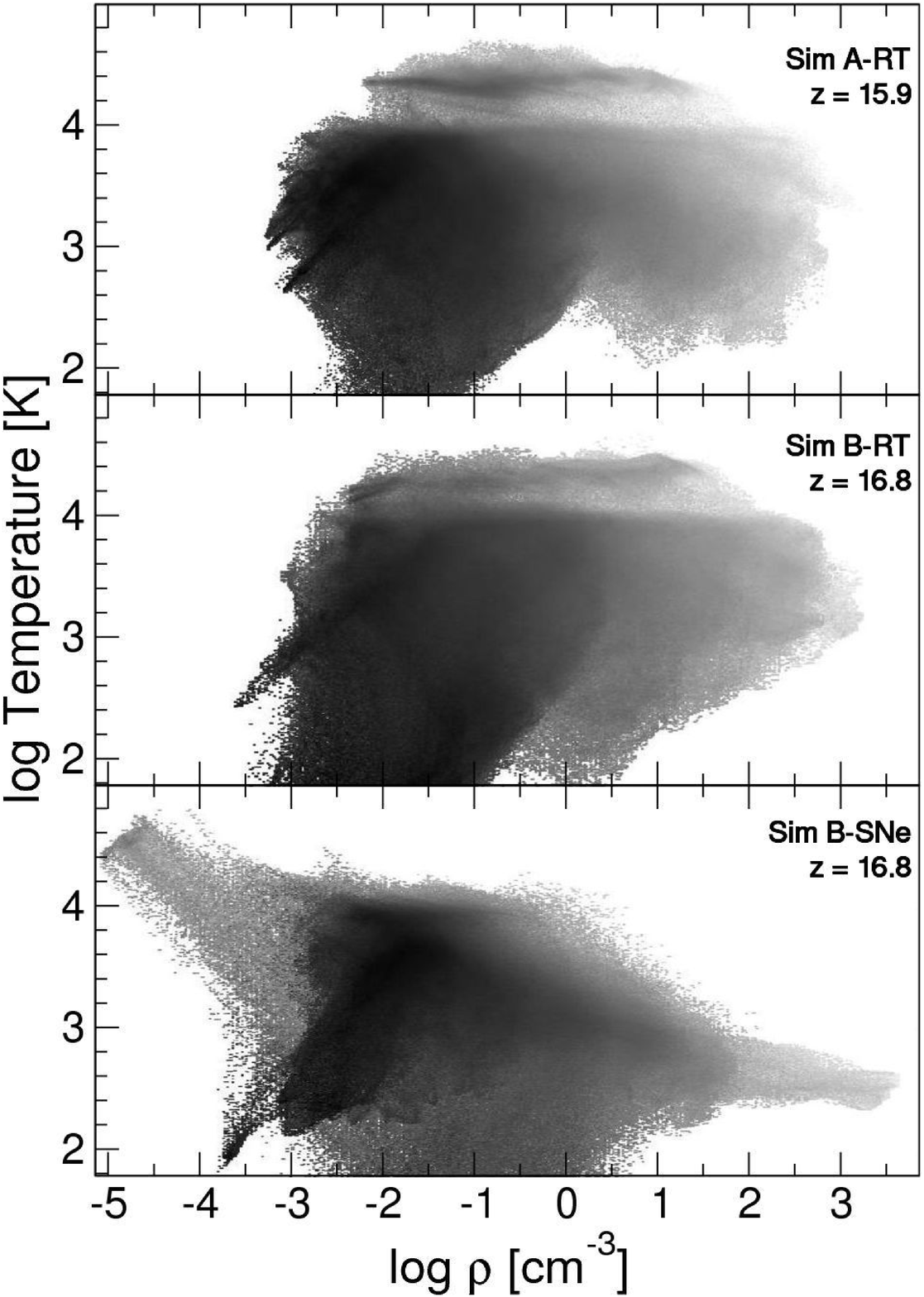}
\caption[Density--Temperature Phase Diagrams]{\label{fig:phase}
  Mass-weighted $\rho$--T phase diagrams of a sphere with radius 1
  kpc, centered on the most massive halo in SimA-RT (\textit{top}),
  SimB-RT (\textit{middle}), and SimB-SN (\textit{bottom}).  At $T >
  10^4$ K, one can see the \ion{H}{2} regions created by current star
  formation.  The warm, low density ($\rho < 10^{-3} \cubecm$) gas in
  SimB-SN are contained in SNe shells.}
\end{center}
\end{figure}

The combination of molecular cooling, stellar feedback, and SN
explosions create a multi-phase ISM in star forming halos.  These
phases are interspersed throughout the halo.  They are marginally seen
in the temperature projections in Figures \ref{fig:proj_a} and
\ref{fig:proj_b}.  However they are better demonstrated by the
mass-weighted radial profiles in Figure \ref{fig:radial} and
density-temperature phase diagrams in Figure \ref{fig:phase}.  The
radial profiles are centered on the densest DM particle.  For a given
radius within the halo, the gas density can span up to 3 orders of
magnitude, and the temperature ranges from $\sim$100 K in the cool
phase to 30,000 K in the warm, ionized phase.  Below we describe the
different ISM phases at redshift 15.9 and 16.8 for simulation A and B,
respectively.

%
%

%
%
\begin{figure}[t]
\begin{center}
\includegraphics[width=0.8\textwidth]{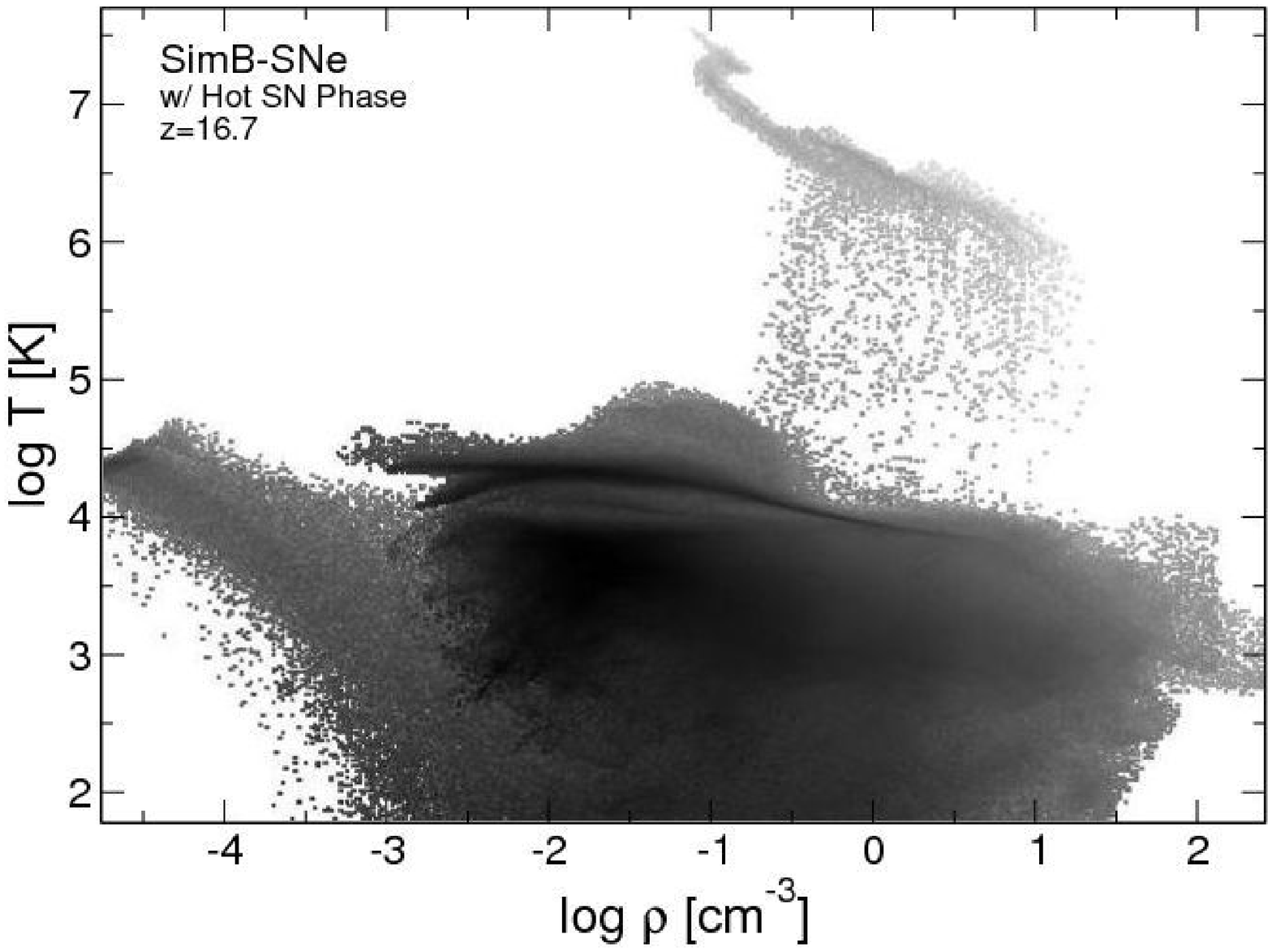}
\caption[Phase Diagram after a Supernova]{\label{fig:hotphase}
  Mass-weighted $\rho$--T phase diagram of a sphere with radius 1 kpc,
  centered on the most massive halo in SimB-SN, 30 kyr after a
  pair-instability SN.  In addition to the ISM phases seen in Figure
  \ref{fig:phase}, the SN ejecta is still hot and X-ray emitting at
  temperatures of 10$^{6-7}$ K. }
\end{center}
\end{figure}

\textit{Cool phase}--- The relatively dense ($\rho > 100 \cubecm$) and
cool ($T > 1000$ K) gas has started to condense by \hh~cooling.
Current star formation dissociates \hh~in nearby condensations through
LW radiation in our simulations as we neglect self-shielding.  In many
cases, especially when \mvir $\gsim 10^7$ \Ms, nearby clumps remain
cool and optically thick.  After the star dies, \hh~formation can
proceed again to form a star in these clumps.  There are two other
sources of cool gas.  First, the filaments are largely shielded from
being photo-heated and provide the galaxy with cold accretion flows.
Second, after SN explosions, the material within the expanding shell
cools through adiabatic expansion and Compton cooling to temperatures
as low as 100 K, which is seen in the $\rho$-T phase diagram at very
low densities.

\textit{Warm, neutral phase}--- Gas that cools by atomic hydrogen line
transitions, but not molecular, has $T \sim 8000$ K and densities
ranging from $10^{-3}$ to 10$^2 \cubecm$.  Gas in relic \ion{H}{2}
regions and virially shock-heated gas compose this phase.

\textit{Warm, ionized phase}--- In the RT simulations at the final
redshift, a Pop III star is creating an \ion{H}{2} region with
temperatures up to 30,000~K.  These regions are approximately 300 and
800 pc in radius with a non-spherical geometry because of the
inhomogeneity of the ISM.

\textit{Hot, X-ray phase}--- The $3 \times 10^{52}$ ergs of energy
deposited by one pair-instability SN in the SimB-SN simulation heats
the gas to over $10^8$ K immediately after the explosion.  Figure
\ref{fig:hotphase} shows the density and temperature of the ISM 30 kyr
after a SN, where the adiabat of the hot phase is clearly visible at
$T > 10^5$ K.  A blastwave initially travelling at 4000\kms~sweeps
through the ambient medium during the free expansion phase.  The gas
behind the shock cools adiabatically and through Compton cooling as
the SN shell expands.

\subsection{Metallicity}

Metallicities of second and later generations of stars depend on the
locality of previous SNe.  In the SimB-SN calculation, outflows
carry most of the SN ejecta to radii up to $\sim$1 proper kpc after 30
Myr.  Interestingly they approximately fill the relic \ion{H}{2}
region and expand little beyond that.  The low density IGM marginally
resists the outflows, and it is preferentially enriched instead of the
surrounding filaments and halos.

It should be noted that this calculation is an upper limit of
metallicities since all stars end with a SN.  Nevertheless the mixing
and transport of the first metals is a fundamental element of the
transition to Pop II stars and is beneficial to study in detail.  All
quoted metallicities are in units of solar metallicity.  The
metallicities also scale linearly with metal yield of each SN because we
treat the metal field as a tracer field that is advected with the
fluid flow.  We quote the metallicities according to this scaling.

%
%
\begin{figure}[t]
\begin{center}
\includegraphics[width=\textwidth]{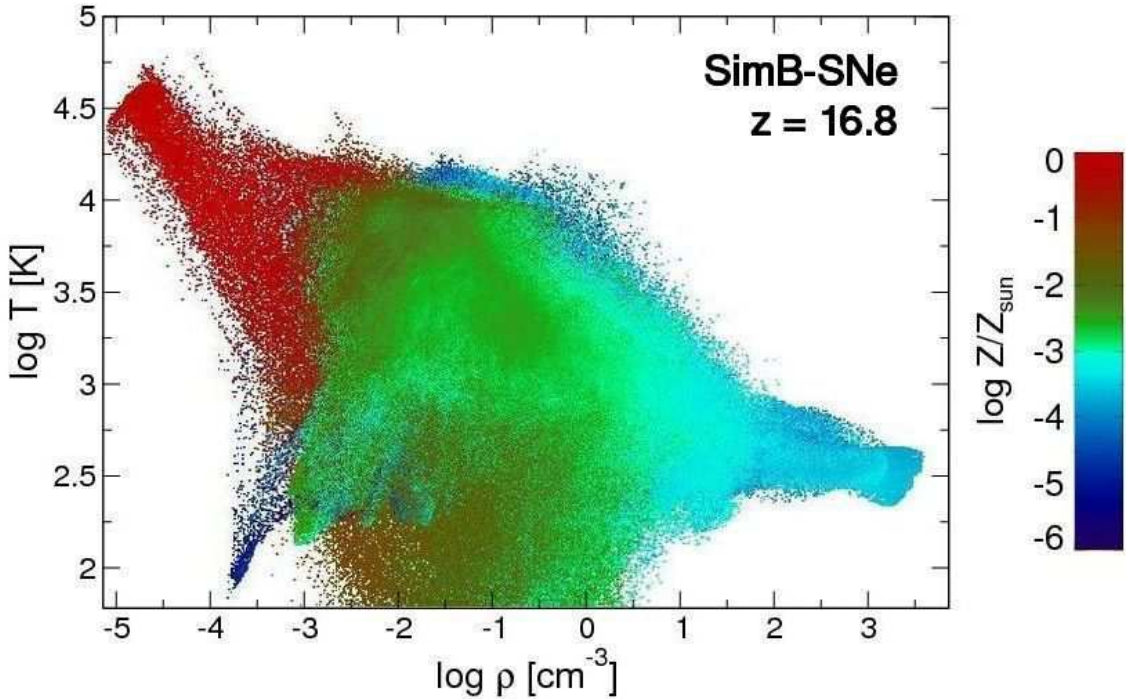}
\caption[Phase Diagram with Metallicity]{\label{fig:Zphase} The same
  phase diagram of SimB-SNe in Figure \ref{fig:phase} but colored by
  mean metallicity.  The SN remnants that are warm and diffuse ($\rho
  < 10^{-3} \cubecm$) have solar metallicities or greater.  The
  majority of the ISM has metallicities $\sim$10$^{-2.5}$ solar.  The
  densest, collapsing material has a metallicity $\sim$10$^{-3.5}$
  solar.}
\end{center}
\end{figure}

%
%
\begin{figure}[t]
\begin{center}
\includegraphics[width=\textwidth]{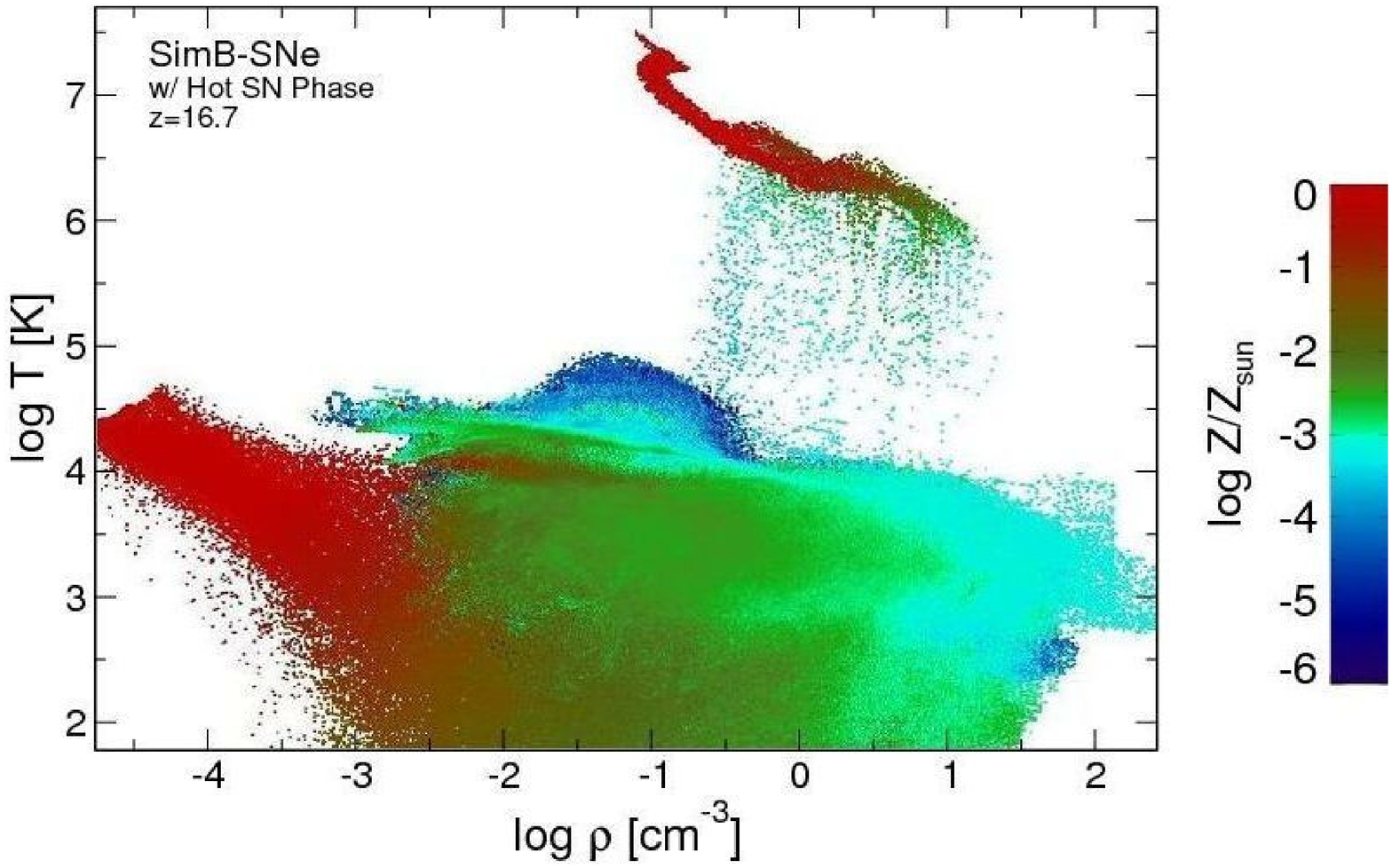}
\caption[Phase Diagram with Metallicity after a
  Supernova]{\label{fig:hotZphase} The same phase diagram of SimB-SNe
  in Figure \ref{fig:hotphase} but colored by mean metallicity.}
\end{center}
\end{figure}

When the most massive halo reaches a virial temperature of $10^4$ K at
z = 16.8, metals are thoroughly mixed in the halo, and its mean
metallicity is \zstar{-4.8}, where $M_{\rm{yield}}$ is the metal
ejecta from one SN in units of solar masses.  The metallicity of this
halo fluctuates around this value because stars continue forming but
ejecting most of their metals into the IGM.  Also the filaments are
still mostly pristine and provide a source of nearly metal-free cold
gas.  The volume averaged metallicity of the enriched IGM ($\delta <
10$) is \zstar{-3.7}, compared to the filaments and halos ($\delta >
10$) that are less enriched with \zstar{-4.5}.  The metal volume
filling fraction is 4.3\% of the volume where we allow star formation
to occur.  This percentage should be higher than the cosmic mean
because this comoving volume of (205 kpc)$^3$ is a biased with an
overdensity $\delta \equiv \rho/\bar{\rho}$ = 1.8.  Thus star
formation rates are greater than the mean since there are more
high-$\sigma$ peaks, and the metal filling fraction should scale with
this bias.

Figure \ref{fig:Zphase} shows the same $\rho$-T phase diagram as in
Figure \ref{fig:phase} but colored by the mean metallicity of the gas.
These data are taken immediately before the formation of a star with a
metallicity of \zstar{-5.5}.  There are three distinct metallicity
states in the halo.  The majority of the gas in the halo has a density
between 10$^{-3}$ and 1\cubecm.  This gas has a mean metallicity of
\zstar{-4.4}.  At higher densities, the metallicity is slightly lower
at \zstar{-5.4}.  The same preferential enrichment of diffuse regions
may have caused the lower metallicities in this dense cloud.  The
third phase is the warm, low density ($\rho < 10^{-4}\cubecm$) gas
that exists in recent SN remnants and has solar metallicities and
greater.  Figure \ref{fig:hotZphase} depicts the state of the most
massive halo 30 kyr after a SN, where the hot phase produced by a SN
is super-solar.  The high-density tail of the ejecta is the SN shell
and is mixing with the lower metallicity ambient medium.  As the
ejecta expands and cools, it will contribute to the warm, low density,
solar material in the lower-left of the phase diagram.

%
%
\begin{figure}[t]
\begin{center}
\includegraphics[width=0.6\textwidth]{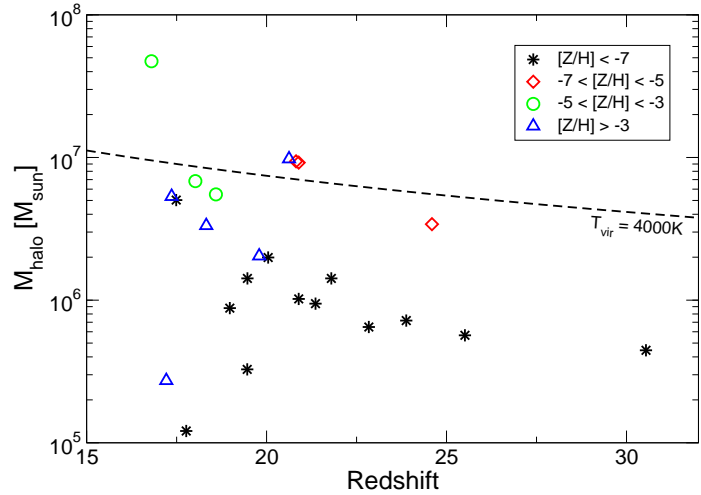}
\caption[Star Formation History with Supernovae]{\label{fig:SF} Star
  formation history of SimB-SN.  The $y$-axis shows the total mass of
  the host halo.  The symbols indicate the different metallicities of
  the stars and are labeled in the legend.  The dashed line marks the
  mass of a halo with a virial temperature of 4000 K, approximately at
  which multiple sites of star formation occur.}
\end{center}
\end{figure}

\subsection{Metal-enriched Star Formation History}

The star formation history of SimB-SN is depicted in Figure
\ref{fig:SF} by plotting the total mass of the host halo versus the
formation redshift.  The different symbols represent the metallicity
of the star.  Before redshift 20, zero metallicity stars form in
low-mass halos, whose masses increase from $5 \times 10^5 \Ms$ to $2
\times 10^6 \Ms$ due to negative feedback from the LW flux and UV
heating from earlier stars.  This agrees with the results of
\citet{Machacek01}, who demonstrated the critical halo mass to cool
increases with the LW radiation background intensity.  There are two
stars that form in halos with masses of $3-5 \times 10^5 \Ms$ because
of the positive feedback from a SN blastwave impacting a satellite
halo and creating excess electrons that induce \hh~cooling
\citep{Ferrara98}.

The first instance of a metal enriched star occurs at $z = 24.6$ with
a metallicity of \zstar{-8.4} in the most massive halo with a mass of
$3.4 \times 10^6 \Ms$.  The gas that was reincorporated into the host
halo was enriched by the star that formed at $z = 31$.  The resulting
SN again expels the gas from the halo.  The gas is reincorporated
again so stars can form in the same halo around redshift 21.  Here
three instances of star form with metallicites of [--8.2, --8.1,
--4.3] $(M_{\rm{yield}} / \Ms)$.  The aggregate energy from these
three SNe expel most of the gas from the potential well yet again, and
the most massive halo does not form any stars until $z = 16.8$.
Similar instances of reincorporation of enriched material happens
for smaller halos at redshifts less than 20.  There is one metal
enriched star with a metallicity of \zstar{-4.8} that forms in a
low-mass halo with a mass of $2.7 \times 10^5 \Ms$ at $z = 17.2$.
This halo was a satellite of a more massive star forming halo that was
enriched by a SN blastwave originating from the parent halo.

\section{Discussion}
\label{sec:reionDiscuss}

We find that the combination of Pop III stellar feedback and continued
\hh~cooling in \tvir~$<$ 10$^4$ K halos alters the landscape of
high-redshift galaxy formation.  The most drastic changes are as
follows:

\medskip

1. \textit{Dynamic assembly of dwarf galaxies}--- A striking
difference when we include Pop III radiative feedback are the outflows
and gas inhomogeneities in the halos and surrounding IGM.  The
outflows enrich the IGM and reduce the baryon fraction of the $10^4$ K
halo as low as 0.05, much lower than the cosmic fraction \Ob/\Om~=
0.17 \citep[cf.][]{Yoshida06a, Abel07}.  This substantially differs
from the current theories of galaxy formation where relaxed isothermal
gas halos hierarchically assemble a dwarf galaxy.  Furthermore, Pop
III feedback increases the total baryonic angular momentum of the
system by a factor of 2--5 without SNe and up to 10 with SNe.

\medskip

2. \textit{Pop III sphere of influence}--- Pop III feedback is mainly
a local phenomenon except its contribution to the UVB.  How far its
\ion{H}{2} region, outflows, and metal ejecta (if any) extend will
predominately determine the characteristics of the next generation of
stars.  Highly biased (clustered) regions are significantly affected
by Pop III feedback, whereas isolated halos will feel little feedback.
The first galaxies will form in these biased regions and thus should
be significantly influenced by their progenitors.

\medskip

3. \textit{Dependence on star forming progenitors}--- Although our
study of calculations with SNe only provided an upper limit of metal
enrichment, it is clear that the metallicity, therefore metal-line and
dust cooling and low-mass star formation, depends on the nature of the
progenitors of the dwarf galaxy.  If the galaxy was assembled by
smaller halos that hosted a Pop III star that did not produce a SN,
the galaxy will continue to have a top-heavy initial mass function
(IMF).

\medskip

4. \textit{Complex protogalactic ISM}--- The interplay between stellar
and SNe feedback, cold inflows, and molecular cooling produce a truly
multi-phase ISM that is reminiscent of local galaxies.  The cool,
warm, and hot phases are interspersed throughout the dwarf galaxy,
whose temperatures and densities can span up to three orders of
magnitude at a given radius.

\medskip

5. \textit{Metallicity floor}--- When the halo is massive enough to
host multiple sites of star formation, the metal ejecta does not
significantly increase the mean metallicity of the host halo.  There
seems to be a balance between galactic outflows produced from SNe,
inflowing metal-enriched and pristine gas, and SNe ejecta that is not
blown out of the system.  In our high yield models, the metallicity
interestingly fluctuates around $10^{-3} Z_\odot$ in the most massive
halo when this balance occurs at and above mass scales
$\sim$10$^7$\Ms.

\medskip

Clearly the first and smallest galaxies are complex entities, contrary
to their low mass and generally assumed simplicity.  Our calculations
reflect the important role of Pop III stellar feedback in early galaxy
formation.  

These high-redshift galaxies have a $\sim$5--15\% chance of being
undisturbed by mergers until the present day, being ``fossils'' of
reionization \citep{Gnedin06}.  Dwarf spheroidals (dSph) galaxies are
some of the darkest galaxies in the universe, having high
mass-to-light ratios up to 100 \citep{Mateo98}.  Gas loss in dSph's
close to the Milky Way or M31 can be explained by gas tidal stripping
during orbital encounters \citep{Mayer07}.  However there are some
galaxies (e.g. Tucana, Cetus) removed from both the Milky Way and
Andromeda galaxies and cannot be explained by tidal stripping.  In
addition to ultraviolet heating from reionization \citep{Bullock00,
  Susa04} and intrinsic star formation \citep{MacLow99}, perhaps
stellar feedback from Pop III stars influenced the gas-poor nature of
dSph's.  Even at the onset of widespread star formation in the objects
studied here, the baryon fraction can be three times lower than the
cosmic mean, and the dwarf galaxy may never fully recover from the
early mass loss.  This initial deficit may play an important role in
future star formation within these low-mass galaxies and could
partially explain the lack of gas in isolated dSph's.

With the radiative and chemical feedback from the progenitors of the
early dwarf galaxies, we have an adequate set of cosmological
``initial conditions'' to study the transition from Pop III to
metal-enriched (Pop II) stars.  In this setup, the current metal
tracer field would actually contribute to the radiative cooling.  To
accomplish this, we need to include a metal-line and dust cooling
model to investigate the dynamical effect of this additional cooling.
However, metal-line cooling might not be important at these low
metallicities.  \citet{Jappsen07} showed that metal-line cooling at
metallicities below 10$^{-2} Z_\odot$ does not significantly affect
the dynamics of a collapsing halo.  Conversely, dust cooling can
induce fragmentation of solar mass fragments at metallicities as low
as $\sim$10$^{-6}$ \citep{Schneider06}.

Perhaps when the protogalactic gas cloud starts to host multiple sites
of star formation, the associated SNe produce sufficient dust in order
for a transition to Pop II.  In lower mass halos, the SN ejecta is
blown out of the halo, and future star formation cannot occur until
additional gas is reincorporated into the halo.  However in these
halos with masses $\gsim$10$^7$\Ms, the SN does not totally disrupt
the halo.  A fraction of the SN ejecta is contained within the halo
and could contribute to subsequent sites of star formation.  Now this
SN ejecta and associated dust could instigate the birth of the first
Pop II stars.


As discussed above, the metallicity of the most massive halo fluctuates
around $10^{-3} Z_\odot$.  This is intriguingly the same value as a
sharp cutoff in stellar metallicities in four local dSph's: Sculptor,
Sextans, Fornax, and Carina \citep{Tolstoy04, Helmi06}.  This is in
contrast with the galactic halo stars, whose metal-poor tail extends
to $Z/Z_\odot = 10^{-4}$ \citep{Beers05}.  We must take care when
comparing our results to observations since we made the simplification
that every Pop III star produces a pair-instability SN.  As discussed
in \citeauthor{Helmi06}, the galactic halo may be composed of remnants
of galaxies that formed from high-$\sigma$ density fluctuations, and
dwarf galaxies originate from low-$\sigma$ peaks.  In this scenario,
the objects (or its remnants) simulated here would most likely reside
in galactic halos at the present day.  If we attempt to match this
metallicity floor of $10^{-4} Z_\odot$ in the galactic halo, this
requires $\sim$8\Ms~of metals produced for every Pop III star or
roughly one in ten Pop III stars ending in a pair-instability SN.

\section{Summary}
\label{sec:summary}

Radiative feedback from Pop III stars play an important role in
shaping the first galaxies.  We studied the effects of this feedback
on the global nature of high-redshift dwarf galaxies, using a set of
five cosmology AMR simulations that accurately model radiative transfer
with adaptive ray tracing.  Additionally, we focused on the metal
enrichment of the star forming halos and their associated star
formation histories.  Our key findings in this paper are listed below.

\medskip

1. Dynamical feedback from Pop III stars expel nearly all of the
baryons from low-mass host halos.  The baryon fractions in star
forming halos never fully recover even when it reaches a virial
temperature of 10$^4$ K.  The baryon fraction is reduced as low as
$\sim$0.05 with SNe feedback, three times lower than the cases without
stellar feedback.

2. The expelled gas gains angular momentum as it exists at large
radii.  When it is reincorporated into the halo, it increases the spin
parameter by a factor of 2--5 without SNe and up to 10 with SNe.

3. The accurate treatment of radiative transfer produces a complex,
multi-phase ISM that has densities and temperatures that span up to 4
orders of magnitude at a given radius.

4. Pair-instability SN preferentially enrich the IGM to a metallicity
an order of magnitude higher than the surrounding overdensities.

5. Once a SN explosion cannot totally disrupt its host halo, the mean
metallicity fluctuates around \zstar{-4.8} as there may be a balance
between SN outflows, cold inflows, and contained SNe ejecta.

\medskip

We conclude that Pop III stars play an integral part in the early
universe as they determine the characteristics of the first galaxies,
which then reionize the universe and may survive until the present
day.  Their feedback is essential to consider in the early universe
because of their large masses and luminosities and being hosted in
shallow potential wells.  Although our high-resolution simulations
included an accurate model of radiative transfer, there are still
uncertainties in the true nature of the first galaxies, arising from
the unknown primordial IMF.  With future observations of early dwarf
galaxies at $z > 6$, it will be possible to infer some details about
the first stars by using theoretical predictions from cosmological
simulations that accurately model their stellar feedback.

\chapter[The Beginning of Reionization]{How Massive Metal-Free Stars
  Start Cosmological Reionization}
\label{chap:reion}

We considered the impact of Pop III star formation on early dwarf
galaxies in the last section.  Now we focus on the global impact of
these stars on cosmological reionization and the nature of star
formation during this epoch.  We show that Pop III stars reionize the
universe to $\sim$10\%, where reionization is completed by
low-luminosity galaxies.  Nevertheless, it is important to gauge the
transition from isolated primordial star forming regions to galaxies.
The heating and reionization of the IGM by Pop III stars suppresses
galaxy formation that are forming from pre-heated gas.  Star formation
in these galaxies are regulated by Jeans smoothing instead of the
ability to radiatively cool through atomic line transitions.

As we noted in Chapter \ref{chap:rates}, the SNe from Pop III stars
will only be detectable by JWST but not the main sequence radiation.
It is timely to establish other constraints on Pop III star formation.
One of these constraints is the thermal history of the universe.  In
these simulations, we can directly track this and perhaps connect
them with observed IGM temperatures at $z \sim 4$.  Furthermore, some
local dwarf galaxies may hold clues about the first stars in their
structures and properties, e.g. mass-to-light ratios, metallicity
gradients, stellar ages, and abundance patterns.

This chapter specifically concentrates on the properties of star
formation in these low-mass halos in the presence of stellar feedback
and the characteristics of the beginning of cosmological reionization
caused by these massive, metal-free stars.  This chapter is in
preparation for publication in \textit{The Astrophysical Journal}.  It
is co-authored by Tom Abel, who developed the original version of the
adaptive ray tracing code used in these simulations.

\section{Motivation}

It is clear that quasars are not responsible to keep the universe
ionized at redshift 6. The very brightest galaxies at those redshifts
alone also provide few photons.  The dominant sources of reionization
so far are observationally unknown despite remarkable advances in
finding sources at high redshift \citep[e.g.][]{Shapiro86, Bouwens04,
  Fan06, Thompson06, Eyles06} and hints for a large number of
unresolved sources at very high redshifts \citep{Spergel06,
  Kashlinsky07}. At the same time, ab initio numerical simulations of
structure formation in the concordance model of structure formation
have found that the first luminous objects in the universe are formed
inside of cold dark matter (CDM) dominated halos of total masses $2
\times 10^5 - 10^6 \Ms$ \citep{Haiman96, Tegmark97, Abel98}.  Fully
cosmological ab initio calculations of \citet{Abel00, Abel02a} and more
recently \citet{Yoshida06b} clearly show that these objects will form
isolated very massive stars. Such stars will be copious emitters of
ultraviolet (UV) radiation and are as such prime suspects to get the
process of cosmological reionization started.  In fact, one
dimensional calculations of \citet{Whalen04} and \citet{Kitayama04}
have already argued that the earliest \ion{H}{2} regions will
evaporate the gas from the host halos and that in fact most of the UV
radiation of such stars would escape into the intergalactic
medium. Recently, \citet{Yoshida06a} and \citet{Abel07} demonstrated
with full three-dimensional radiation hydrodynamical simulations that
indeed the first \ion{H}{2} regions break out of their host halos
quickly and fully disrupt the gaseous component of the cosmological
parent halo.  The initially circumstellar material of the first stars
finds itself radially moving away from the star at $\sim30\kms$ at a
distance of $\sim100$ pc at the end of the stars life. At this time,
the photo-ionized regions have high electron fractions and little
destructive Lyman-Werner band radiation fields, creating ideal
conditions for molecular hydrogen formation which may in fact
stimulate further star formation above levels that would have occurred
without the pre-ionization. Such conclusion have been obtained in
calculations with approximations to multi dimensional radiative
transfer or one dimensional numerical models \citep{Ricotti02a,
  Nagakura05, OShea05, Yoshida06a, Ahn07, Johnson07}. These early
stars may also explode in supernovae and rapidly enrich the
surrounding material with heavy elements, deposit kinetic energy and
entropy to the gas out of which subsequent structure is to form. This
illustrates some of the complex interplay of star formation,
primordial gas chemistry, radiative and supernova feedback and readily
explains why any reliable results will only be obtained using full ab
initio three dimensional hydrodynamical simulations. In this paper, we
present the most detailed such calculations yet carried out to date
and discuss issues important to the understanding of the process of
cosmological reionization.

It is timely to develop direct numerical models of early structure
formation and cosmological reionization as considerable efforts are
underway to
\begin{enumerate}
\item Observationally find the earliest galaxies with the James Webb
  Space Telescope \citep[JWST;][]{Gardner06} and the Atacama Large
  Millimeter Array \citep[ALMA;][]{Wilson05},
\item Further constrain the amount and spatial non-uniformity of the
  polarization of the cosmic microwave background radiation
  \citep{Page06},
\item Measure the surface of reionization with LOFAR
  \citep{Rottgering06}, MWA \citep{Bowman07}, GMRT \citep{Swarup91}
  and the Square Kilometer Array \citep[SKA;][]{Schilizzi04}, and
\item Find high redshift gamma ray bursts with SWIFT \citep{Gehrels04}
  and their infrared follow up observations.
\end{enumerate}

We begin by describing the cosmological simulations that include
primordial star formation and accurate radiative transfer.  In
\S\ref{sec:SF}, we report the details of the star formation
environments and host halos in our calculations.  Then in
\S\ref{sec:reion}, we describe the resulting cosmological
reionization, and investigate the environments in which these
primordial stars form and the evolution of the clumping factor.  We
compare our results to previous calculations and further describe the
nature of the primordial star formation and feedback in
\S\ref{sec:discussion}.  Finally we summarize our results in the last
section.


\section{Radiation Hydrodynamical Simulations}
\label{sec:simulations}

We use radiation hydrodynamical simulations with a modified version of
the cosmological AMR code \enzo~to study the radiative effects from
the first stars \citep{Bryan97, Bryan99}.  We have integrated adaptive
ray tracing \citep{Abel02b} into the chemistry, energy, and
hydrodynamics solvers in \enzo~that accurately follow the evolution of
the \ion{H}{2} regions from stellar sources and their relevance during
structure formation and cosmic reionization.

Seven different simulations are discussed here.  Table \ref{tab:sims}
gives an overview of the parameters and the physics included in these
calculations.  We perform two cosmological realizations, Sim A and B,
with three sets of assumptions about the primordial gas chemistry.
The simplest calculations here assume only adiabatic gas physics and
provide the benchmark against which the more involved calculations are
compared.  We compare this to one model with atomic hydrogen and
helium cooling only and one that includes \hh~cooling.  Massive,
metal-free star formation is included only in the \hh~cooling models.

%
%
\begin{landscape}
\begin{center}
\begin{longtable}{lccccccc}

\caption{Simulation Parameters} \label{tab:sims6} \\

\hline\hline \\[-3ex]
  Name & $l$ & Cooling model & SF & SNe & N$_{\rm{part}}$ &
  N$_{\rm{grid}}$ & N$_{\rm{cell}}$ \\
   & [Mpc] &  &  & &  &  & \\ 
\hline
\endhead

SimA-Adb & 1.0 & Adiabatic & No & No & 2.22 $\times$ 10$^7$ & 30230 & 9.31
$\times$ 10$^7$ (453$^3$) \\
SimA-HHe & 1.0 & H, He & No & No & 2.22 $\times$ 10$^7$ & 40601 & 1.20
$\times$ 10$^8$ (494$^3$) \\
SimA-RT & 1.0 & H, He, \hh & Yes & No & 2.22 $\times$ 10$^7$ & 44664 & 1.19
$\times$ 10$^8$ (493$^3$) \\
SimB-Adb & 1.5 & Adiabatic & No & No & 1.26 $\times$ 10$^7$ & 23227 & 6.47
$\times$ 10$^7$ (402$^3$) \\
SimB-HHe & 1.5 & H, He & No & No & 1.26 $\times$ 10$^7$ & 21409 & 6.51
$\times$ 10$^7$ (402$^3$) \\
SimB-RT & 1.5 & H, He, \hh & Yes & No & 1.26 $\times$ 10$^7$ & 24013 & 6.54
$\times$ 10$^7$ (403$^3$) \\
SimB-SN & 1.5 & H, He, \hh & Yes & Yes & 1.26 $\times$ 10$^7$ & 24996 & 6.39
$\times$ 10$^7$ (400$^3$) \\

\hline
\end{longtable}

\tablecomments{Col. (1): Simulation name. Col. (2): Box
  size. Col. (3): Cooling model.  Col. (4): Star formation. Col. (5):
  Supernova feedback. Col. (6): Number of dark matter
  particles. Col. (7): Number of AMR grids. Col. (8): Number of unique
  grid cells.}
\end{center}
\end{landscape}

These calculations are initialized at redshift $z$ = 130
(120)%
\renewcommand{\thefootnote}{\fnsymbol{footnote}}%
\footnote{To simplify the discussion, simulation A will always be
  quoted first with the value from simulation B in parentheses.}%
\renewcommand{\thefootnote}{\arabic{footnote}}
when the intergalactic medium has a temperature of 325
(280) K in box sizes 1 comoving Mpc (1.5 Mpc) for Sim A (B).  We use
the cosmological parameters of $(\Omega_B\,h^2, \ \Omega_M, \ h, \
\sigma_8, \ n)=(0.024,\ 0.27,\ 0.72,\ 0.9,\ 1)$ from first year WMAP
results, where the constants have the usual meaning \citep{Spergel03}.
The changes in the third year WMAP results \citep{Spergel06} does not
affect the evolution of individual halos studied here but only delays
structure formation by $\sim$40\% \citep{Alvarez06b}.  The adiabatic
simulations as well as the atomic hydrogen and helium cooling only
calculations are described in \citet{Wise07a}. The new models
presented here have the exact same setup and random phases in the
initial density perturbation and only differ in that they include star
formation as well as follow the full radiation hydrodynamical
evolution of the \ion{H}{2} regions and supernova feedback in Sim B.
We use the designations RT and SN to distinguish cases in which only
star formation and radiation transport were included (RT) and the one
model which also includes supernovae (SN) in Sim B.  We use the same
refinement criteria as in our previous work, where we refine if the DM
(gas) density becomes three times greater than the mean DM (gas)
density times a factor of $2^l$, where $l$ is the AMR refinement
level.  We also refine to resolve the local Jeans length by at least
16 cells.  Cells are refined to a maximum AMR level of 12 that
translates to a spatial resolution of 1.9 (2.9) comoving parsecs.
Refinement is restricted to the innermost initial nested grid that has
a side length of 250 (300) comoving kpc.

The star formation recipe and radiation transport are detailed in
\citet{Wise07c}.  Here we overview the basics about our method.  Star
formation is modelled using the \citet{Cen92b} algorithm with the
additional requirement that an \hh~fraction of $5 \times 10^{-4}$ must
exist before a star forms.  We allow star formation to occur in the
Lagrangian volume of the surrounding region out to three virial radii
from the most massive halo at $z = 10$ in the dark matter only runs as
discussed in \citet{Wise07a}.  This volume that has a side length of
195 (225) comoving kpc at $z = 30$ and 145 (160) comoving kpc at the
end of the calculation.  The calculations with SNe use a stellar mass
$M_\star$ of 170\Ms, whereas the ones without SNe use a mass of
100\Ms.  The ionizing luminosities are taken from no mass loss models
of \citet{Schaerer02}, and we employ the SN energies from
\citet{Heger02}.  Star particles after main sequence are tracked but
are inert.  There is evidence of lower mass primordial stars forming
within relic \ion{H}{2} regions \citep{OShea05, Yoshida06a}, but we
neglect this to avoid additional uncertain parameters.  This is a
desired future improvement, however.

We use adaptive ray tracing \citep{Abel02b} to calculate the
photo-ionization and heating rates caused by stellar radiation.  We
consider photo-ionization from photons with an energy of 28.4 (29.2)
eV that is the mean energy of ionizing radiation from a metal-free
star with 100 (170) \Ms.  We account for \hh~photo-dissociation with a
1/$r^2$ Lyman-Werner radiation field without self-shielding.  We use a
non-equilibrium, nine-species (H, H$^{\rm +}$, He, He$^{\rm +}$,
He$^{\rm ++}$, e$^{\rm -}$, H$_2$, H$_2^{\rm +}$, H$^{\rm -}$)
chemistry solver in \enzo~\citep{Abel97, Anninos97} that takes into
account the additional photo-ionization from the radiation transport.

We present results from SimA to a redshift of 20, which we stop
because of computing constraints.  In SimB, we end the simulations
when the most massive halo begins to rapidly collapse (i.e. \tcool~$<$
\tdyn) in the hydrogen and helium cooling only runs at redshift 16.8.
The virial temperature \tvir~of the halo is $\sim$10$^4$ K at this
redshifts.

%
%
\begin{figure}[t]
\begin{center}
\includegraphics[width=0.6\textwidth]{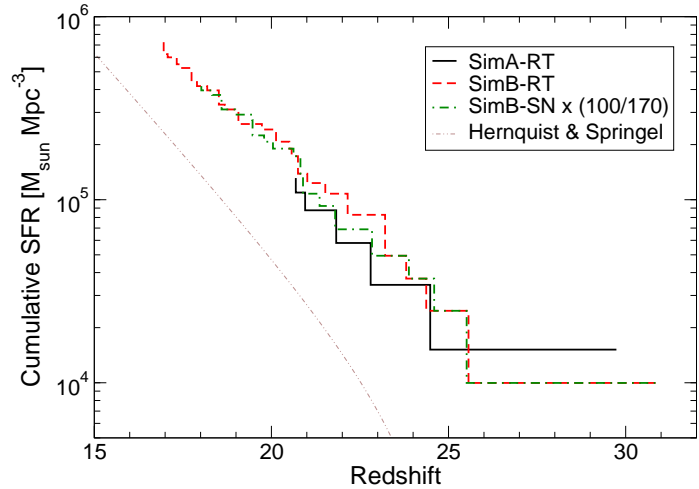}
\caption[Cumulative star formation rate]{\label{fig:cumulSF}
  Cumulative star formation rate in units of comoving \Ms~Mpc$^{-3}$
  of SimA-RT (\textit{black solid}), SimB-RT (\textit{red dashed}),
  and SimB-SN (\textit{green dot-dashed}).  The star formation rate of
  SimB-SN has been scaled by 100/170, which is the ratio of Pop III
  stellar masses used in SimB-RT and SimB-SN, in order to make a
  direct comparison between the two simulations.  The \textit{brown
    dot-dot-dashed} line represents the cumulative star formation rate
  in atomic hydrogen cooling halos from \citet{Hernquist03}.}
\end{center}
\end{figure}

\section{Star Formation}
\label{sec:SF}

Here we describe the aspects of massive metal-free star formation in
our simulations.  The first star forms at redshift 29.7 (30.8) in halo
typical of Pop III star formation without any feedback that has a mass
of $\sim 5 \times 10^5 \Ms$ \citep[cf.][]{Abel00, Abel02a, Machacek01,
  Yoshida03, Yoshida06a}.  Afterwards there are a total of 6, 27, and
20 instances of star formation in SimA-RT, SimB-RT, and SimB-SN,
respectively.

\subsection{Star Formation Rate}

We show the cumulative star formation rate (SFR) in units of comoving
\Ms~Mpc$^{-3}$ in Figure \ref{fig:cumulSF}.  This quantity is simply
calculated by taking the total mass of stars formed at a given
redshift divided by the comoving volume where stars are allowed to
form (see \S\ref{sec:simulations}).  In this figure, we decrease the
SFR of SimB-SN by a factor of 1.7 in order to directly compare the
rates from the other two simulations.  This minimizes some of the
uncertainties entered into our calculations when we chose the free
parameter of Pop III stellar mass.  The cumulative rates are very
similar in both realizations.  The refined volume of Sim A (Sim B) has
an average overdensity $\delta \equiv \rho / \bar{\rho}$ = 1.4 (1.8).
The more biased regions in Sim B allows for a higher density of
star-forming halos that leads to the increased cumulative SFR.

We also overplot the cumulative SFR in atomic hydrogen cooling halos
from \citet{Hernquist03} in this figure.  It is up to an order of
magnitude lower than the rates seen in our calculations up to redshift
20.  They only focused on larger mass halos in their simulations.  The
disparity between the rates is caused by our simulations only sampling
a highly biased region, where we focus on a region containing a
3-$\sigma$ density fluctuation, and from the contribution from Pop III
stars.  The rates of \citet{Hernquist03} are calculated from an
extensive suite of smoothed particle hydrodynamics simulations that
encompasses both large and small simulation volumes and give a more
representative global SFR due to their larger sampled volumes.
However, our adaptive spatial resolution allows us to study both the
small- and large-scale radiative feedback from Pop III stars, which is
the main focus of the paper, in addition to the quantitative measures
such as a SFR.

%
%
\begin{figure}[t]
\begin{center}
\includegraphics[width=0.6\textwidth]{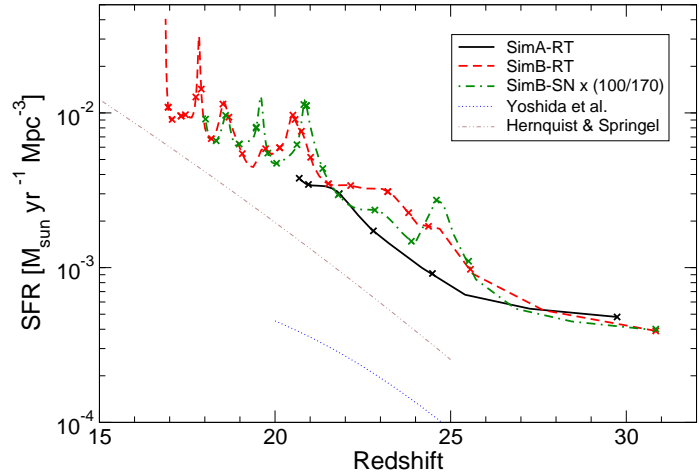}
\caption[Star formation rate]{\label{fig:SF6} Comoving star formation
  rate in units of \sfr.  The lines representing the simulation data
  have the same meaning as in Figure \ref{fig:cumulSF}.  The crosses
  show at which redshifts stars form.  The rates in SimB-SN are scaled
  for the same reason as in Figure \ref{fig:cumulSF}.  For comparison,
  we overplot the star formation rates from \citet{Hernquist03} in
  atomic hydrogen line cooling halos and \citet{Yoshida03} for
  100\Ms~Pop III stars.}
\end{center}
\end{figure}

To calculate a true SFR \citep[i.e.][in units of comoving \sfr]
{Madau96} from the cumulative SFR, we first fit the cumulative SFR
with a cubic spline with 10 times the temporal resolution.  Then we
smooth the data back to its original time resolution and evaluate its
time derivative to obtain the SFR that we show in Figure
\ref{fig:SF6}.  We also mark the redshifts of star formation with
crosses.  We again compare our rates to ones calculated in
\citet{Hernquist03} for metal-enriched stars and \citet{Yoshida03} for
Pop III stars with a mass of 100 \Ms.  Our rates are higher for
reasons discussed previously.  We do not advocate these SFRs as cosmic
averages but give them as a useful diagnostic of the performed
simulations.

We see an increasing function from $4 \times 10^{-4}$ \sfr~at redshift
30 to $\sim$$6 \times 10^{-3}$ \sfr~at redshift 20.  Here only one
star per halo forms in objects with masses $\lsim 5 \times 10^6 \Ms$.
Above this mass scale, star formation is no longer isolated in nature
and can be seen by the bursting nature of the star formation after
redshift 20, where the SFR fluctuates around $10^{-2}$
\sfr~\citep[cf.][]{Ricotti02b}.  Since we neglect \hh~self-shielding,
the strong Lyman-Werner (LW) radiation dissociates almost all \hh~in
the host halo and surrounding regions.  Thus we rarely see
simultaneous instances of star formation.  However, the regions that
were beginning to collapse when a nearby star ignites form a star 3 --
10 million years after the nearby star dies.  This only results in a
minor change in the timing of star formation.  Furthermore this delay
is minimal compared to the Hubble time and does not affect SFRs.

%
%
\begin{figure}[t]
\begin{center}
\includegraphics[width=0.6\textwidth]{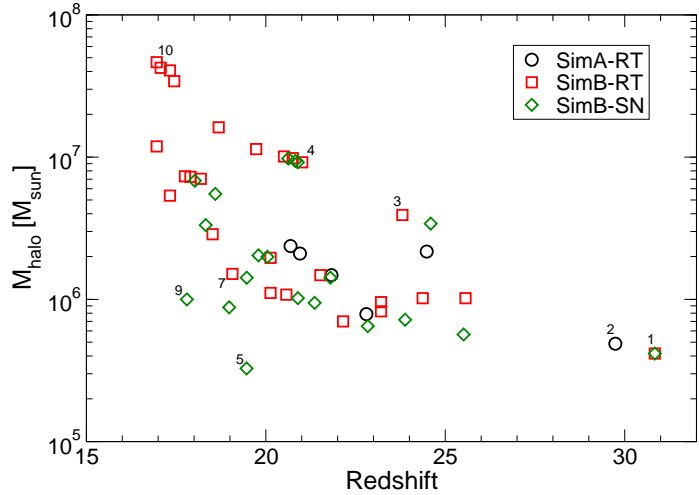}
\caption[Star formation history versus halo mass]
{\label{fig:SFhistory} Star formation history versus host halo DM mass
  for SimA-RT (\textit{black circles}), SimB-RT (\textit{red
    squares}), and SimB-SN (\textit{green triangles}).  One symbol
  represents one star.  The numbers correspond to the halo numbers
  listed in Table \ref{tab:halos6}.}
\end{center}
\end{figure}

\subsection{Star Forming Halo Masses}
\label{sec:haloMass}


We show the star formation history versus the host halo DM masses as a
function of redshift in Figure \ref{fig:SFhistory}.  The DM halo
masses are calculated with the HOP algorithm \citep{Eisenstein98}.
First we focus on star formation in the largest halo.  Around redshift
30, the first star forms in all three simulations within a halo with a
mass $\sim$$5 \times 10^5 \Ms$.  The stellar radiation drives a
$\sim$30\kms~shock wave that removes almost all of the gas from the
shallow potential well.  It takes approximately 40 million years for
gas to reincorporate into the potential well from smooth IGM accretion
and mergers.  At $z \sim 24$, the second star forms in the most
massive progenitor that now has a mass of $\sim$$3 \times 10^6 \Ms$.
Yet again, the stellar feedback expels most of the gas from its host.
For Sim A (Sim B), another 50 (30) million years passes before the
next star forms in this halo.  Now it has a mass of 10$^7$ \Ms,
which is a sufficient amount of potential energy to confine most of
the stellar and SNe outflows.  In SimA-RT and SimB-RT, the halo now
hosts multiple sites of star formation that is seen in the nearly
continuous bursts of star formation in the most massive halo.  However
in SimB-SN at $z = 20.8$, three stars form in succession in the most
massive halo.  Their aggregate stellar and SNe feedback expels the gas
from its halo one more time.  


Most of the stars form in low-mass halos with masses $\sim$10$^6$
\Ms~that are forming its first star between redshifts 18 -- 25 in our
calculations.  A slight increase in host halo masses with respect to
redshift occurs because of the negative feedback from ionization,
ultraviolet heating, and \hh~dissociation from previous stars that
increases the critical halo mass in which gas can cool and condense
\citep[e.g.][]{Machacek01, Yoshida03, Mesigner06}.

One interesting difference in SimB-SN from the other calculations is
that it forms stars in halos with masses with $\lsim$$5 \times 10^5
\Ms$ when stars are mainly forming in more massive halos.  In these
cases, the SN shell expands into the ambient medium and impacts the
surrounding satellite halos.  The blast wave induces star formation in
low-mass halos that otherwise would not have formed a star
\citep{Ferrara98}.

%
%
\begin{figure}
\begin{center}
\plottwo{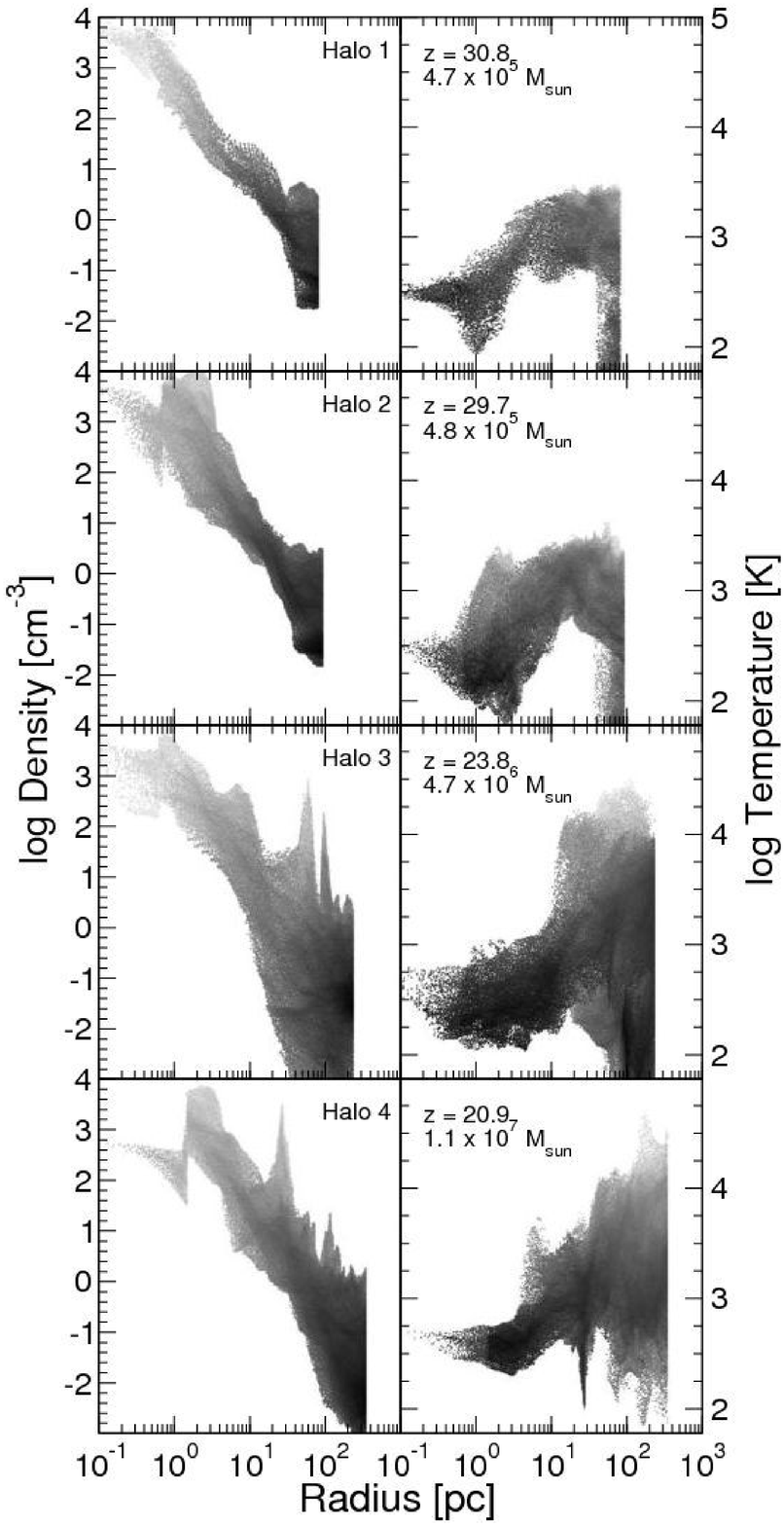}{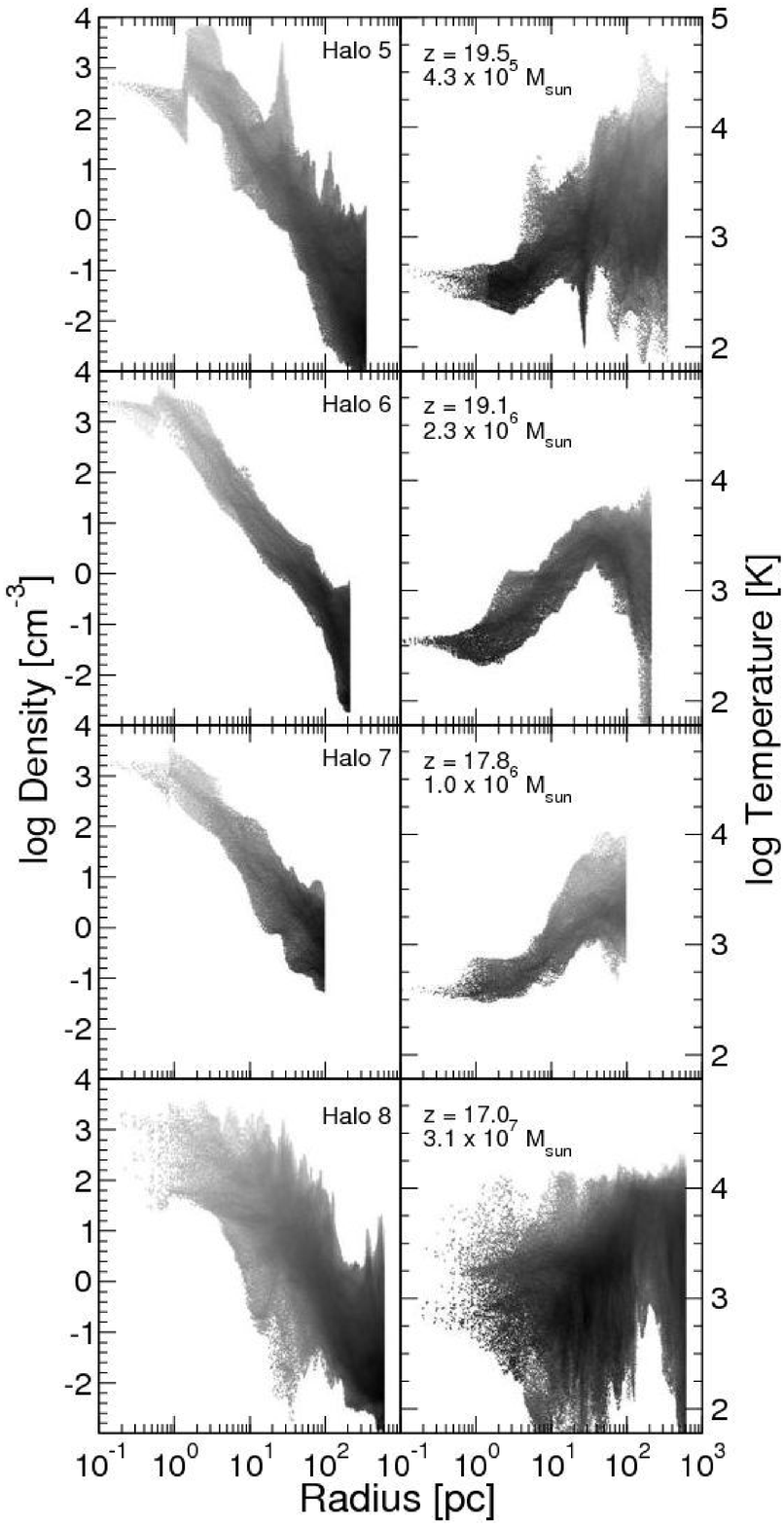}
\caption[Radial profiles of star forming regions]{\label{fig:radial6}
  \footnotesize{Radial profiles of number density (\textit{left column}) and
  temperature (\textit{right column}) for selected star forming halos
  inside the virial radius.  These data represent the state of the
  region immediately before star formation.  Notice the added
  complexity (range) in the density and temperature with increasing
  host halo mass, especially if the region has been affected by
  stellar radiation, as in Halos 3, 4, 5, 10, 11, and 12.  The
  discontinuities in density in the inner parsec arise from our star
  formation recipe when we remove half the mass in a sphere containing
  twice the stellar mass.}}
\end{center}
\end{figure}

%
%
\begin{center}
\begin{longtable}{ccccccc}
\caption{Selected Star Forming Halo Parameters} \label{tab:halos6} \\

\hline\hline \\[-3ex]
  \# & Sim & $z$ & M$_{{\rm vir}}$ & $f_b$ & $\rho_c$ & $T_c$ \\
   &  &  & [\Ms] & & [cm$^{-3}$] & [K] \\
\hline
\endhead

 1 & SimB-RT & 30.8 & $4.7 \times 10^5$ & 0.089  & 3000 &  330 \\
 2 & SimA-RT & 29.7 & $4.8 \times 10^5$ & 0.094 & 3000 &  310 \\
 3 & SimB-RT & 23.8 & $4.7 \times 10^6$ & 0.036 & 1400 &  310 \\
 4 & SimB-SN & 20.9 & $1.1 \times 10^7$ & 0.051 &  350 &  460 \\
 5 & SimB-SN & 19.5 & $4.3 \times 10^5$ & 0.017 &  470 &  170 \\
 6 & SimB-RT & 19.1 & $2.3 \times 10^6$ & 0.13  & 1500 &  360 \\
 7 & SimB-SN & 17.8 & $1.0 \times 10^6$ & 0.18  & 1100 &  390 \\
 8 & SimB-RT & 17.0 & $3.1 \times 10^7$ & 0.10  &  440 & 1000 \\
\hline
\end{longtable}

\tablecomments{Col. (1): Halo number. Col. (2): Simulation
  source. Col. (3): Redshift. Col. (4): Virial mass.  Col. (5): Baryon
  mass fraction. Col. (6): Central number density. Col. (7): Central
  temperature.}

\end{center}

\subsection{Star Formation Environments}


We further study the nature of high-redshift star formation by
selecting eight star forming regions and studying the surrounding
interstellar medium (ISM) prior to star formation.  The ISM in the
10$^4$ K halos are described in more detail in \citet{Wise07c}.  The
sample of regions are chosen in order to compare different star
formation environments.  These regions can be categorized into (1)
first star inside an undisturbed halo, (2) first star that is delayed
by LW radiation, (3) induced star formation by positive feedback, (4)
star formation after gas reincorporation, and (5) star formation in a
halo with a virial temperature over 10$^4$ K.  The represented halos
and their parameters are listed in Table \ref{tab:halos6} and
annotated in Figure \ref{fig:SFhistory}.

We plot the mass-weighted radial profiles of number density
(\textit{left columns}) and temperature (\textit{right columns})
within the virial radius for these eight halos in Figure
\ref{fig:radial6} and describe them below.

\medskip

1. \textit{First star} (Halo 1, 2)--- These stars are the first to
form in the simulation volume.  The structure of the host halos within
our resolution limit exhibit similar characteristics, e.g., a
self-similar collapse and central temperatures of 300 K, as in
previous studies \citep{Abel00, Abel02a, Bromm02a, Yoshida06b}.  The
halo masses are $4.8 (4.7) \times 10^5$ \Ms.  Heating from
virialization raises gas temperatures to 3000 K, and in the central
parsec, \hh~cooling becomes effective and cools the gas down to 100 K
that drives the free-fall collapse.  The central gas densities and
temperatures are approximately 3000\cubecm~and 320 K, respectively.

\medskip

2. \textit{Delayed first star} (Halo 6, 7)--- The host halos have
similar radial profiles as the halos that hosted the first stars but
with masses of $1-3 \times 10^6 \Ms$.  Here the \hh~cooling has been
stifled by the LW radiation from nearby star formation.  Only when the
halo mass passes a critical mass, the core can cool and condense by
\hh~formation \citep{Machacek01, Yoshida03}.  The central densities
are slightly lower than the first stars with 1500 and 1100\cubecm~in
SimB-RT and SimB-SN, respectively.  The central temperatures are
marginally higher at 360 and 390 K.

\medskip

3. \textit{Induced star formation} (Halo 5)--- At $z = 19.61$, a
massive star explodes in a SN, whose shell initially propagates
outward at 4000\kms.  After two million years, the shell passes a
satellite halo with $4 \times 10^5 \Ms$ that is 480 pc from the origin
of the SN.  The combination of the shock passage and excess free
electrons in the relic \ion{H}{2} catalyze \hh~formation in this
low-mass halo \citep[e.g.][]{Ferrara98, OShea05, Mesigner06}.  The SN
blast wave heats the gas over 10$^4$ K to radii as low as five
parsecs.  In the density profile, both low and high density gas exists
at similar radii.  Here the shock passage creates a tail of gas
streaming from the central core, whose asymmetries can be seen in the
density profile.  However, the core survives and benefits from the
excess electrons created during this event.  The central temperature
is interestingly a factor of two lower than the previous cases at 170
K.  The \hh~criterion for star formation is reached faster because of
the excess electrons, which creates a star particle at a lower density
(470\cubecm).

\medskip

4. \textit{Star formation after reincorporation} (Halo 3, 4)--- After
a sufficient amount of gas that was expelled by dynamical feedback of
the first star is reincorporated into the halo, star formation is
initiated again.  Here virial temperatures of the halos are under
10$^4$~K.  Halo 3 is the second instance of star formation in this
halo, whereas the other halos are forming their third star.  These
halos have a larger spread in gas densities and temperatures than the
halos forming their first star.  Gas is heated by virialization and
prior stars to over 10$^4$ K outside 10 pc.  The central quantities in
halo 3 are similar to the regions described in the delayed star
formation section.  Halos 4 shows a more diffuse and warmer core with
densities of 350 \cubecm~and temperatures of 460 K.

\medskip

5. \textit{Star formation in 10$^{\,4}$K halos} (Halo 8)--- In this
halo, \hh~formation is aided by atomic hydrogen cooling.  The ISM
becomes increasingly complex as more stars form in the halo.  The
temperatures range from 100 K to 20,000 K throughout the halo.  It has
hosted eight massive stars since it started to continually form stars
at redshift 21.  The densities are similar to the other regions that
have been affected by other stars.

%
%
\begin{figure}
  \begin{center}
    \includegraphics[width=0.65\textwidth]{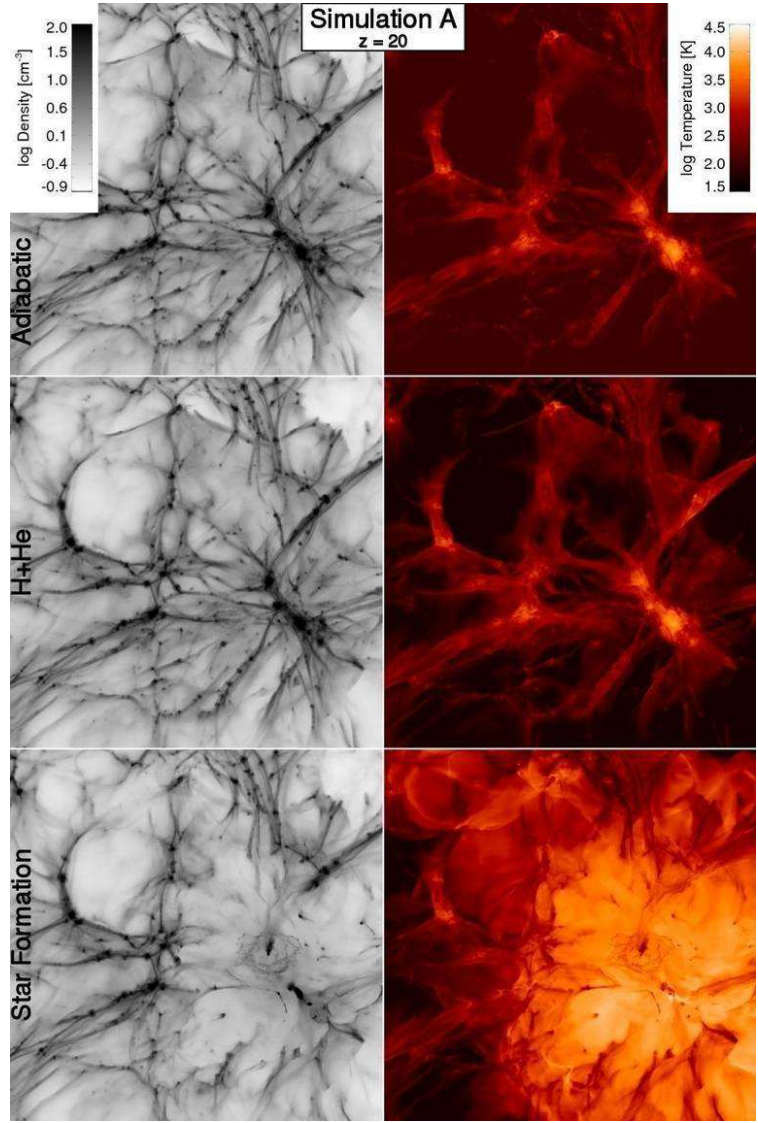}
    \caption[Density and temperature profiles of simulation A]
    {\label{fig:projA} Density-squared weighted projections of gas
      density (\textit{left}) and temperature (\textit{right}) of Sim
      A.  The field of view is 8.5 proper kpc (1/216 of the simulation
      volume) and the color scale is the same for all simulations.}
\end{center}
\end{figure}

%
%
\begin{figure}
  \begin{center}
    \includegraphics[width=0.65\textwidth]{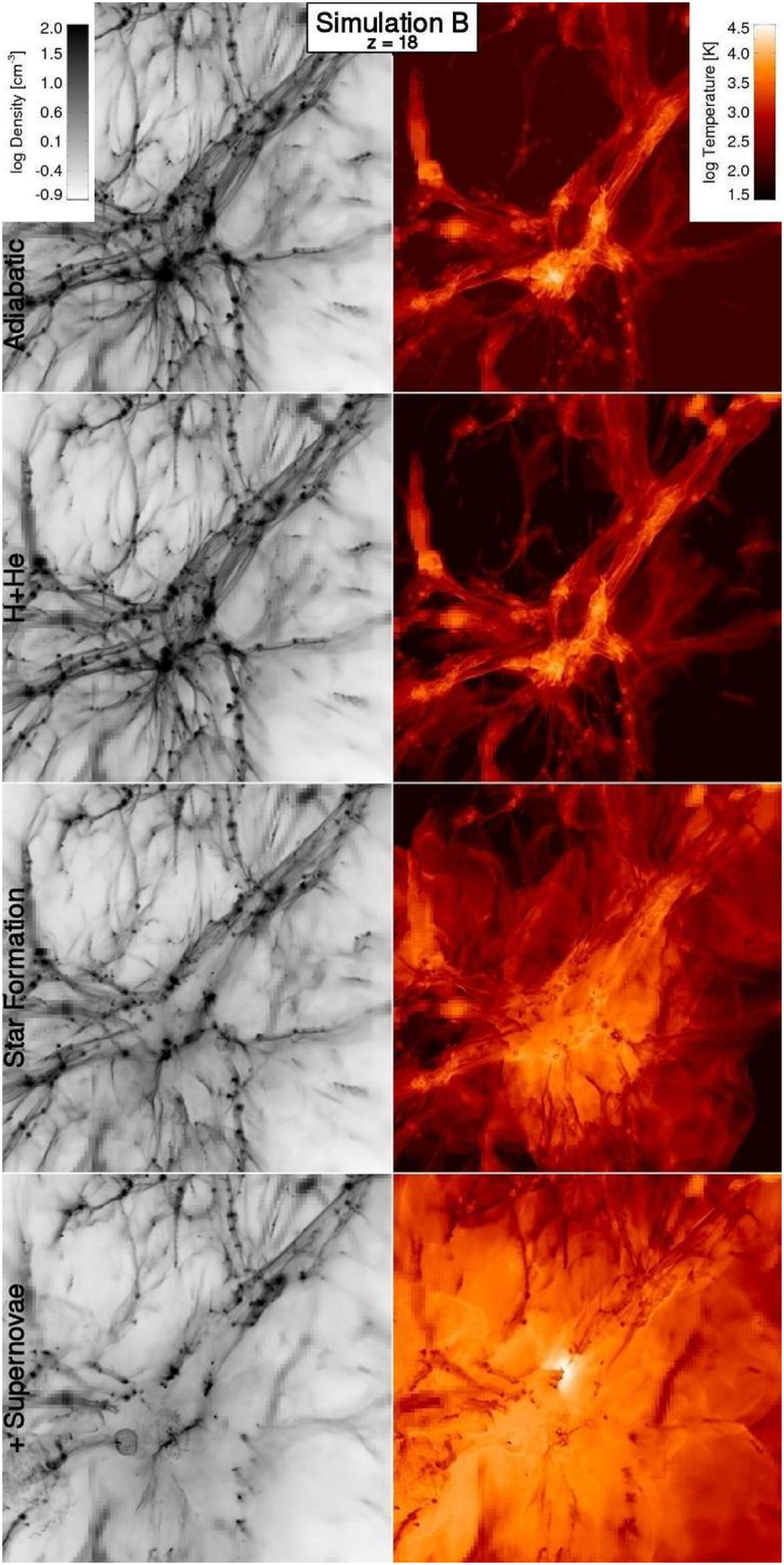}
    \caption[Density and temperature profiles of simulation B]
    {\label{fig:projB} Same as Figure \ref{fig:projA} but for Sim B.}
\end{center}
\end{figure}

\section{Starting Cosmological Reionization}
\label{sec:reion}

In this section, we first describe the ionizing radiation from massive
stars that starts cosmological reionization.  Then we discuss the
effects of recombinations in the inhomogeneous IGM and kinetic energy
feedback from Pop III stars.  Lastly the evolution of the average IGM
thermal energy is examined.

To illustratively demonstrate radiative feedback from massive stars on
the host halos and IGM, we show projections of gas density and
temperature that are density-squared weighted in Figures
\ref{fig:projA} and \ref{fig:projB} for all of the simulations at
redshift 20 and 18, respectively.  These projections have the same
field of view of 8.5 proper kpc and the same color maps.  The
large-scale density structure is largely unchanged by the stellar
feedback, and the adjacent filaments remain cool since they are
optically thick to the incident radiation.  \hh~cooling produces more
centrally concentrated objects, and the dynamical stellar feedback
destroys the baryonic structure in $\sim$10$^6$\Ms~halos.  Kinematic
feedback from SNe has an even larger effect on the surrounding gas.
In SimB-SN, this effect is seen in the reduced small-scale structure
and low-mass halos with no gas counterparts.  However, the most
apparent difference in the radiative simulations is the IGM heating by
Pop III stars, especially in SimB-SN.

%
%
\begin{figure}[t]
\begin{center}
\vspace{0.4cm}
\includegraphics[width=0.6\textwidth]{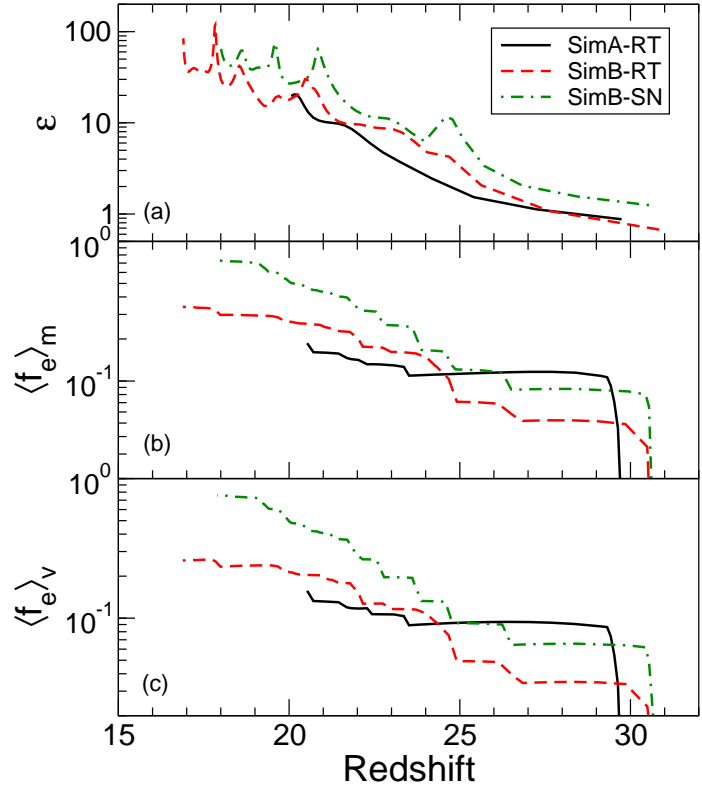}
\caption[Volume-averaged emissivity and ionization fractions]
{\label{fig:emis} (a) Averaged emissivity in units of ionizing photons
  per baryon per Hubble time that is calculated from the star
  formation rate in Figure \ref{fig:SF6}.  (b) Mass-averaged ionization
  fraction of the inner 250 (300) comoving kpc for SimA (SimB).  (c)
  Volume-averaged ionization fraction for the same runs.}
\end{center}
\end{figure}

\subsection{UV Emissivity}


A key quantity in reionization models is volume-averaged emissivity of
ionizing radiation.  We utilize the comoving SFR $\dot{\rho}_\star$ to
calculate the proper volume-averaged UV emissivity
\begin{equation}
  \label{eqn:emis}
  \epsilon = \frac{\dot{\rho}_\star Q_{\rm{HI}} t_{\rm{H}}}{\bar{\rho}_b}
\end{equation}
in units of ionizing photons per baryon per Hubble time.  Here
$Q_{\rm{HI}}$ is the number of ionizing photons emitted in the
lifetime of a star per solar mass, $\bar{\rho}_b \simeq 2 \times
10^{-7}$ \cubecm~is the comoving mean number density, and
\begin{equation}
  \label{eqn:hubbletime}
  t_{\rm{H}} \approx \frac{2}{3 H_0 \sqrt{\Omega_m}} (1+z)^{-3/2}
\end{equation}
is the Hubble time in a Einstein de-Sitter universe, which is valid
for $\Lambda$CDM cosmology at $z \gg 1$.  For Pop III stellar masses
greater than 100 \Ms, $Q_{\rm{HI}} \approx 10^{62}$ photons per solar
mass, corresponding to 84000 ionizing photons per stellar proton
\citep{Schaerer02}.  We plot the emissivity $\epsilon$ in Figure
\ref{fig:emis}a.  It follows the same behavior as the SFR, but now can
be directly used in semi-analytic reionization models.  The emissivity
increases from unity at redshift 30 to $\sim$100 at the end of our
simulations.  Our results agree with the emissivity calculated in
semi-analytic models that include Pop III stars
\citep[e.g.][]{Onken04}.

\subsection{Effective Number of Ionizations per UV Photon}

Although there are approximately 50 photons emitted per baryon at the
end of the RT simulations, the simulation volumes remain are
approximately 25\% ionized.  We show the mass-averaged and
volume-averaged ionization fraction within the innermost initial grid
in Figures \ref{fig:emis}b and \ref{fig:emis}c.  At $z > 20$, the
ionization fraction is $\sim$10\% with occasional sharp increases that
happen when a star ignites.  In the star formation only runs, the
volume only becomes slightly more ionized even though the emissivity
is increasing at a faster rate.  This happens because of the strong
recombinations in the inhomogeneous IGM and the partial containment of
\ion{H}{2} regions in host halos with masses greater than 10$^7$ \Ms.
\citet{Kitayama04} provided a useful approximation of the critical
halo mass
\begin{equation}
  \label{eqn:ionCritMass}
  M_{\rm{crit}}^{\rm{ion}} \sim 2.5 \times 10^6
  \left(\frac{M_\star}{200 \Ms}\right)^{3/4}
  \left(\frac{1+z}{20}\right)^{-3/2} \Ms,
\end{equation}
in which an ionization front (I-front) cannot escape.  This
approximation is valid for stellar masses between 80 and 500 \Ms,
redshifts between 10 and 30, and singular isothermal spheres.  Our
simulations exhibit this same trait in which I-fronts only partially
breakout from the host halo above this mass scale.

In the SN runs, the mass- and volume-averaged ionization fractions
dramatically increase to 0.2 and 0.75 between $z$ = 21 and 18.  At
this time, three stars form in succession in the most massive halo.
After the first star goes SN, a diffuse and hot medium is left behind,
but the blastwave has not completely disrupted two other nearby
condensing clumps.  The radiation from the second star now does not
have to ionize its host halo and has an escape fraction of near unity.
The same happens for the third star in this halo.  The combination of
the radiation and SN explosions from this stars create a sharp IGM
ionization transition that is three times more ionized than before
this event.

%
%
\begin{figure}[t]
\begin{center}
\includegraphics[width=0.6\textwidth]{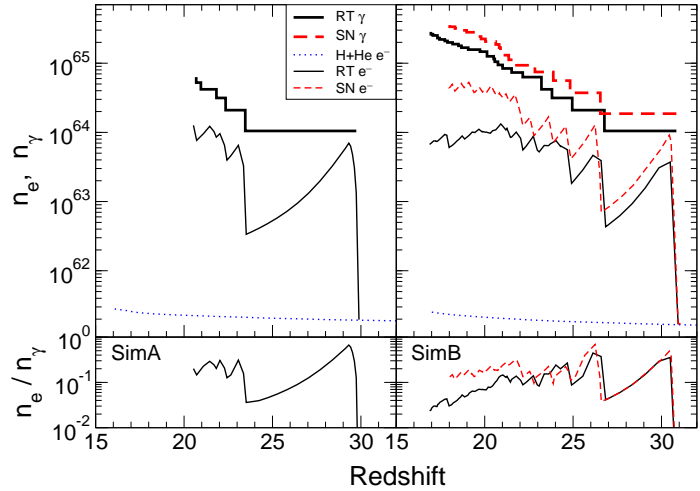}
\caption[Effective ionizations per UV photon]{\label{fig:elec}
  \textit{Top panels}: Total number of ionizing photons emitted
  (\textit{thick lines}) and total number of electrons (\textit{thin
    lines}) for simulations with cooling only (\textit{dotted blue}),
  star formation only (\textit{solid black}), and supernovae
  (\textit{dashed red}) in the inner 250 and 300 comoving kpc for SimA
  (\textit{left}) and SimB (\textit{right}).  The \ion{H}{2} regions
  are completely contained in these volumes.  \textit{Bottom panels}:
  The ratio of total number of electrons to the total number of
  ionized photons emitted.}
\end{center}
\end{figure}

To examine the strength of recombinations, we compare the total number
of electrons to the total number of ionizing photons emitted in Figure
\ref{fig:elec}.  The ratio of these two quantities is the number of UV
photons needed for one effective ionization.  This ratio is
approximately two after the first star dies.  The total number of
electrons remains basically constant with the increasing SFR that
drives this ratio to $\sim$50 at redshifts less than 20.  The effects
of SNe as previously discussed increases the number of electrons by a
factor of 3, but effective ionizations per UV photons is still quite
low $\sim$1/10.

%
%
\begin{figure}[t]
\begin{center}
\includegraphics[width=0.6\textwidth]{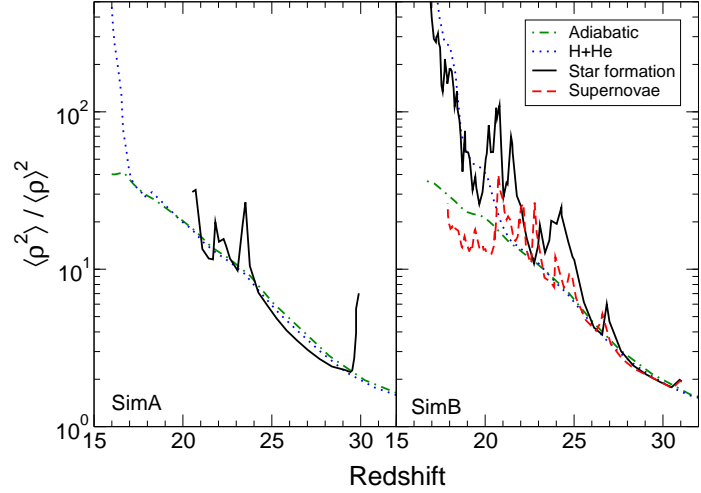}
\caption[Clumping factor evolution]{\label{fig:clumping} Clumping
  factor $C = \langle\rho^2\rangle / \langle\rho\rangle^2$ for SimA
  (\textit{left}) and SimB (\textit{right}), comparing the cases of
  the adiabatic equation of state (\textit{dot-dashed}), atomic
  hydrogen and helium cooling (\textit{dotted}), star formation only
  (\textit{solid}), and supernovae (\textit{dashed}).}
\end{center}
\end{figure}

\subsection{Clumping Factor Evolution}

Volume averaged recombination rates in an inhomogeneous IGM scale with
the clumping factor $C = \langle\rho^2\rangle / \langle\rho\rangle^2$,
where the angled brackets denote volume averaged quantities.  The
recombination rate for hydrogen, e.g., is simply
\begin{equation}
  \label{eqn:clumping}
  \left(\frac{dn_{\rm{HII}}}{dt}\right)_{\rm{rec}} = C k_{\rm{rec}}
  f_e \bar{\rho}_b (1+z)^3 ,
\end{equation}
where $k_{\rm{rec}}$ is the case B recombination rate for hydrogen at
$T \approx 10^4$ K.  Both the increased recombinations in overdense
regions and photon escape fractions lower than unity result in the
high number of UV photons needed for one effective ionization that we
see in our simulations.

Figure \ref{fig:clumping} compares the clumping factor in the
adiabatic, cooling only, star formation, and supernovae calculations.
Since we resolve the local Jeans length by at least 4 cells in all
simulations, the clumping factor is not underestimated, given our
assumptions about gas cooling in each model.  The RT and SN
calculations capture the full evolution of the clumping factor since
gas can fully condense by \hh~cooling in the pristine gas, accurately
following the small-scale structure at low metallicities and high
redshifts.

The clumping factor in the adiabatic case smoothly increases to
$\sim$40 at $z = 17$ from unity at $z > 30$ because of the increase in
number density of halos with masses above the cosmological Jeans mass.
The cooling run only deviates from the adiabatic case when the most
massive halo can start cooling by \lya~cooling, and the center begins
a free-fall collapse, which causes the rapid increase in $C$.  The
clumping factor in the star formation only simulations become larger
than the other simulations as several halos start to condense by
\hh~cooling.  The clumping decreases as these central concentrations
are disbanded by stellar radiation.  The combination of collapsing
halos and stellar radiation generates fluctuations in the clumping
factor around twice the value in the adiabatic case.  SN explosions
disperse gas more effectively than radiative feedback alone in larger
halos and can have a bigger impact on the clumping factor.  At
redshift 20, the three stars and their SNe energy in the most massive
halo destroy the surrounding baryonic structures and reduce the
clumping back to the values seen in non-radiative cases.

\begin{figure}[t]
\begin{center}
  \includegraphics[width=0.6\textwidth]{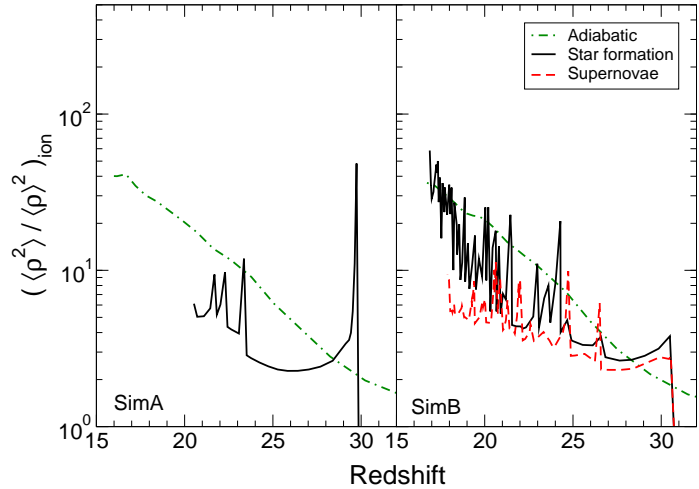}
\caption[Ionized clumping factor evolution]
{\label{fig:clump_ion} Clumping factor in ionized regions with $f_e >
  10^{-3}$ for SimA (\textit{left}) and SimB (\textit{right}).  The
  line styles are the same as in Figure \ref{fig:clumping}.}
\end{center}
\end{figure}

We show the clumping factor $C_{\rm{ion}}$ in ionized regions above
$f_e > 10^{-3}$ in Figure \ref{fig:clump_ion}.  It fluctuates around
and slightly below the values found in adiabatic simulations.  When an
\ion{H}{2} is still confined within its parent halo, the ionized
material is still at high densities that causes the spikes in
$C_{\rm{ion}}$.  As the shock wave caused by the stellar radiation
propagates into the IGM, baryon expulsion and photo-evaporation of
small gas clumps in the \ion{H}{2} regions cause $C_{\rm{ion}}$ to
decrease.  This repeats as star formation ensues and causes the
fluctuations in $C_{\rm{ion}}$.



%
%

\subsection{Kinetic Energy Feedback}
\label{sec:kinetic}

SN explosion energy and kinetic energy generated in D-type I-front
play a key role in star formation in low-mass halos, which are easily
affected due to their shallow potential well \citep[e.g.][]
{Haehnelt95, Bromm03, Whalen04, Kitayama04, Kitayama05}.  The kinetic
energy created by SNe is sufficient to expel the gas from these
low-mass halos.  For example, the binding energy of a $10^6 \Ms$ halo
is only $2.8 \times 10^{50}$ erg at $z = 20$, which is two orders of
magnitude smaller than a typical energy output of a pair instability
SN \citep{Heger02}.  For a \tvir~$>$ 10$^4$ K halo at the same
redshift, it is $9.4 \times 10^{52}$ erg.  With our chosen stellar
mass of 170 \Ms, it takes 3 -- 4 SNe to overcome this potential
energy.

The shock wave created by the D-type I-front travels at a velocity
$v_s$ = 25 -- 35\kms~for density gradients (i.e. $\rho(r) \propto
r^{-w}$) with slopes between 1.5 and 2.25 \citep{Shu02, Whalen04,
  Kitayama04}.  This velocity is the escape velocity for halos with
masses greater than $3 \times 10^8 \Ms$ at $z = 15$, which is an order
of magnitude greater than the most massive halos studied here.
However less massive halos can contain these I-fronts because pressure
forces slow the I-front after the star dies.  

Using the position of the shock wave when the star dies
(eq. \ref{eqn:shockPos}) and energy arguments, we can estimate the
critical halo mass where the material in the D-type I-front can escape
from the halo by comparing the binding energy $E_b$ of the halo and
kinetic energy in the shell.  For most massive stars, the shock wave
never reaches the final Str{\"o}mgren radius,
\begin{equation}
  \label{eqn:stroemgren}
  R_{\rm{str}} = 150 \left(\frac{\dot{N}_{\rm{HI}}}{10^{50} \mathrm{ph \;
        s}^{-1}}\right)^{1/3}
  \left(\frac{n_f}{1\cubecm}\right)^{-2/3} \; \mathrm{pc} ,
\end{equation}
before the star dies.  Here $\dot{N}_{\rm{HI}}$ is the ionizing photon
rate of the star, and $n_f$ is the number density contained in this
radius.  After the lifetime of the star, the shock reaches a
radius
\begin{equation}
  \label{eqn:shockPos}
  R_s = 83 \left(\frac{v_s}{30\kms}\right)
  \left(\frac{t_\star}{2.7 \mathrm{Myr}}\right) \; \mathrm{pc} ,
\end{equation}
where $t_\star$ is the stellar lifetime \citep[see
also][]{Kitayama04}.  We can neglect isolated, lower mass (M $\lsim$
30\Ms) Pop III stars whose shock wave reaches $R_{\rm{str}}$ within
its lifetime.  In this case, the I-front stops at $R_{\rm{str}}$, and
the shock wave becomes a pressure wave that has no associated density
contrast in the neutral medium \citep{Shu92}.  Thus we can safely
ignore these stars in this estimate.


Assume that the source is embedded in a single isothermal sphere.  The
mass contained in the shell is
\begin{equation}
  \label{eqn:massswept}
  M_{\rm{sw}} = \frac{(\Omega_b/\Omega_M) \> M_{\rm{vir}} \>
    R_s}{r_{\rm{vir}}} - V_s \rho_i
\end{equation}
that is the mass enclosed in the radius $R_s$ in an isothermal sphere,
corrected for the warm, ionized medium behind the I-front.  Here $V_s$
is the volume contained in a sphere of radius $R_s$, and $\rho_i$ is
the gas density of the ionized medium, whose typical number density is
1\cubecm~for stellar feedback from a massive primordial star
\citep{Whalen04, Kitayama04, Yoshida06a, Abel07}.  For massive stars
($M_\star \gsim 30 \Ms$), the mass of the central homogeneous medium
is small (i.e. 10\%) compared to the shell.  We compensate for this
interior mass by introducing the fraction $\eta$, so the shell mass is
simply
\begin{equation}
  \label{eqn:swept2}
  M_{\rm{sw}} = \eta \; \frac{(\Omega_b/\Omega_M) \> M_{\rm{vir}} \>
    r_s}{r_{\rm{vir}}}.
\end{equation}
For these outflows to escape from the halo, the kinetic energy
contained in the shell must be larger than the binding energy, which
is
\begin{equation}
  \label{eqn:outflowE}
  \frac{1}{2} M_{\rm{sw}} v_s^2 \; > \; \frac{G M_{\rm{vir}}^2}{2 r_{\rm{vir}}}.
\end{equation}
Using equations (\ref{eqn:shockPos}) and (\ref{eqn:swept2}) in this
condition, we obtain the maximum mass
\begin{eqnarray}
  \label{eqn:criticalMass}
  M_{\rm{max}} &\; \sim \;& \frac{r_s \> v_s^2 \> \Omega_b}{G \>
    \Omega_M} \nonumber\\
  M_{\rm{max}} & \sim & 3.20 \times 10^6 \left(\frac{r_s}{100 \mathrm{pc}}\right)
  \left(\frac{v_s}{30 \kms}\right)^2 \nonumber\\
  &   & \times \left(\frac{\eta}{0.9}\right)
  \left(\frac{\Omega_b/\Omega_M}{0.17}\right) \Ms
\end{eqnarray}
of a halo where the material in the shock wave becomes unbound,
expelling the majority of the gas from the halo.

%
%
\begin{figure}[t]
\begin{center}
\includegraphics[width=0.6\textwidth]{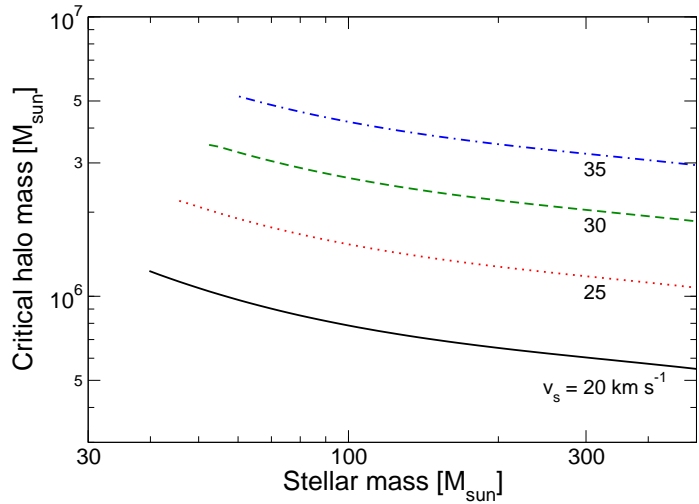}
\caption[Maximum halo mass for baryon expulsion]{\label{fig:critMass}
  Maximum halo mass in which a D-type ionization front can create
  outflows as a function of primordial stellar mass for shock
  velocities $v_s$ of 20, 25, 30, and 35\kms.  Here the fraction
  $\eta$ of mass contained in the shell is 0.9.}
\end{center}
\end{figure}

In Figure \ref{fig:critMass}, we use the stellar lifetimes and
ionizing luminosities from \citet{Schaerer02} to calculate the
critical halo mass for outflows for stellar masses 5 -- 500\Ms~and for
shock velocities of 20, 25, 30, and 35\kms~with $\eta = 0.9$.  For
stellar masses smaller than 30\Ms, the D-type I-front reaches the
final Str{\"o}mgren sphere and cannot expel any material from the
host.  Hence they are not plotted in this figure.  For the more
massive stars, the star dies before the D-type I-front can reach the
Str{\"o}mgren radius, thus being limited by $t_\star$.  This maximum
halo mass is in good agreement with our simulations as we see halos
with masses greater than $5 \times 10^6\Ms$ retaining most of their
gas in the star formation only cases.  However in larger halos,
stellar sources still generate champagne flows, but this material is
still bound to the halo and returns in tens of million years.

\subsection{Thermal Energy}

Thermal feedback is yet another mechanism by which Pop III stars leave
their imprint on the universe.  The initial heating of the IGM will
continue and intensify from higher SFRs at lower redshifts
\citep[e.g.][]{Hernquist03, Onken04}.  It is possible to constrain the
reionization history by comparing temperatures in the \lya~forest to
different reionization scenarios \citep{Hui03}.  Temperatures in the
\lya~forest are approximately 20,000~K at $z = 3 - 5$ \citep{Schaye00,
  Zaldarriaga01}.  Although our focus was not on redshifts below 15
due to the uncertainty of the transition to the first low-mass
metal-enriched (Pop II) stars, we can utilize the thermal data in our
radiation hydrodynamical simulations to infer the thermal history of
the IGM at lower redshifts.

The excess energy from hydrogen ionizing photons over 13.6 eV
photo-heat the gas in the \ion{H}{2} region.  The mean temperature
within \ion{H}{2} regions in our calculations is $\sim$30,000 K.  When
the short lifetime of a Pop III star is over, the \ion{H}{2} region
cools mainly through Compton cooling off the cosmic microwave
background.  The same framework applies to SNe remnants as well.  The
timescale for Compton cooling is
\begin{equation}
  \label{eqn:compton}
  t_C = 1.4 \times 10^7 \left(\frac{1+z}{20}\right)^{-4} f_e^{-1} \; \mathrm{yr}.
\end{equation}
This process continues until the gas recombines, and Compton cooling
is no longer efficient because of its dependence on electron fraction.
Radiation preferentially propagates into the voids and leaves the
adjacent filaments and its embedded halos virtually untouched.  Hence
we can restrict the importance of Compton cooling to the diffuse IGM
since Compton cooling cools the gas to low temperatures without being
impeded by recombinations that are proportional to $n_e^2$.  This
causes the relic \ion{H}{2} region to cool to temperatures down to
300~K.  The temperature evolution in our radiative calculations agrees
with the analytic models of relic \ion{H}{2} \citep{Oh03}.

%
%
\begin{figure}[t]
\begin{center}
\includegraphics[width=0.6\textwidth]{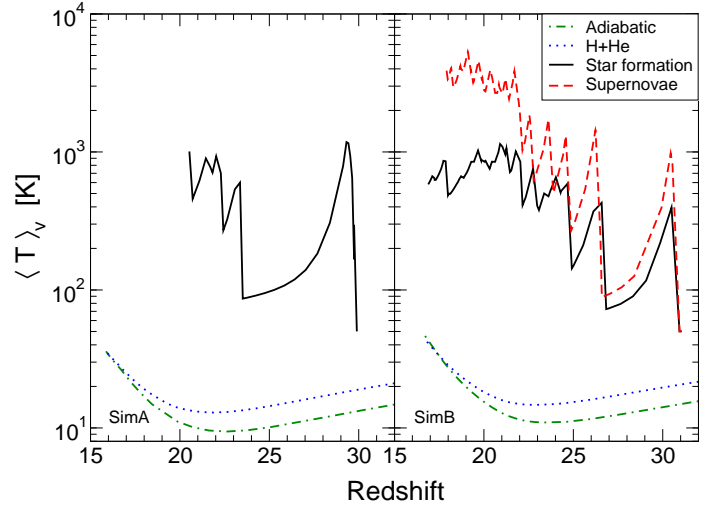}
\caption[Volume-averaged temperature]{\label{fig:volTemp} Evolution of
  the volume-averaged temperature in the inner 250 and 300 comoving
  kpc for Sim A (\textit{left}) and Sim B (\textit{right}),
  respectively.  The simulations for the adiabatic (\textit{green
    dot-dashed}), cooling only (\textit{blue dotted}), star formation
  only (\textit{black solid}), and supernovae (\textit{red dashed})
  simulations are plotted.}
\end{center}
\end{figure}

\begin{figure}[t]
  \begin{center}
    \includegraphics[width=0.6\textwidth]{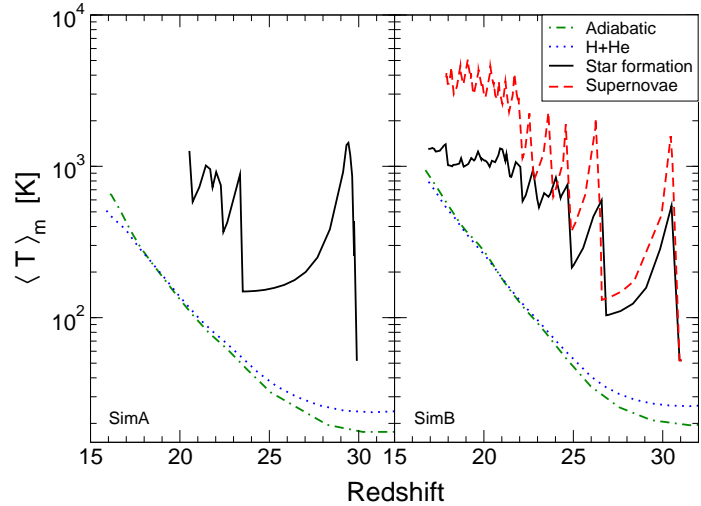}
  \caption[Mass-averaged temperature]{\label{fig:massTemp} Same as
    Figure \ref{fig:volTemp} but for the mass-averaged temperature.}
\end{center}
\end{figure}

We plot the volume-averaged temperature \tv~and
mass-averaged temperature \tm~in the volume where we
allow star formation, i.e. the inner 250 (300) comoving kpc, in
Figures \ref{fig:volTemp} and \ref{fig:massTemp}.  The first star in
the calculations raises the volume averaged temperatures to
$\sim$100~K.  The mass-averaged temperatures are slightly higher at
$\sim$125~K since the star has heated a larger fraction of mass, its
host halo, when compared to the volume of the \ion{H}{2} region.  The
supernovae calculations are even higher due to the hot SN bubble that
has an initial temperature of 10$^8$~K.  The high initial temperature
causes the mass-averaged temperature in the SN simulations to spike
when SNe occur to several times higher than the RT simulations.
Afterwards the remnant cools from Compton and adiabatic processes as it
expands to temperatures similar to the RT simulations.

Because photo-heating is confined to the \ion{H}{2} regions, the
trends seen in average temperatures follows the same behavior as the
ionization fraction with the exception of the spikes associated with
SNe.  In the RT simulations, the volume-averaged temperature rises
gradually from 60~K to $\sim$200~K from redshifts 25 to 15.  The
mass-averaged temperature increases more than \tv~because of the
photo-heating of the host halo and virial heating of the halos, which
is the cause of the increase in the simulations without star
formation.  Without SNe, \tm~is only up to two times the
temperatures in the no star formation runs.  Both \tv~and \tm~in the
SN calculations exhibit a sharp transition to higher temperatures
around 4000~K at $z = 21$, corresponding to the same transition to a
more ionized state of $f_e = 0.75$.  

\begin{figure}[t]
  \begin{center}
    \includegraphics[width=0.6\textwidth]{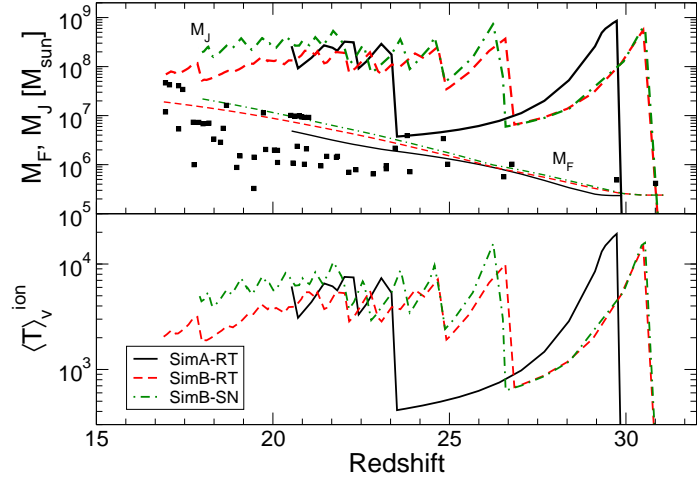}
  \caption[Jeans filtering mass and averaged temperature in ionized
  regions]{\label{fig:filtering} (a) The Jeans mass $M_J$ and
    filtering mass $M_F$ that can form bound objects.  The squares
    denote the total mass of star forming halos in all three
    simulations.  (b) Volume-averaged temperature in the ionized
    regions ($f_e > 10^{-3}$) that is used in computing the filtering
    mass.}
\end{center}
\end{figure}

The photo-heating is better represented by the average temperature
$\langle T \rangle_{\rm{v}}^{\rm{ion}}$ in ionized regions, which is
plotted in Figure \ref{fig:filtering}b.  The first few stars heat the
gas up to 20,000~K that then cool by adiabatic expansion and Compton
cooling.  When star formation occurs in several halos and the ionized
filling fraction increases, the average temperature fluctuates around
4000~K because there are both active and relic \ion{H}{2} in the
simulation, causing $\langle T \rangle_{\rm{v}}^{\rm{ion}}$ to be
lower than 20,000~K that happens during the formation of the first few
\ion{H}{2} regions.  The increased temperatures cause the
photo-evaporated and Jeans smoothing of the gas in the relic
\ion{H}{2} regions.  We discuss these effects in the next section.





%
%

\section{Discussion}
\label{sec:discussion}

We have studied the details of massive metal-free star formation and
its role in the start of cosmological reionization.  We have treated
star formation and radiation in a self-consistent manner, allowing for
an accurate investigation of the evolution of cosmic structure under
the influence of early Pop III stars.  Stellar radiation from these
stars provides thermal, dynamical, and ionizing feedback to the host
halos and IGM.  Although Pop III stars do not provide the majority of
ionizing photons needed for cosmological reionization, they play a key
role in the early universe because early galaxies that form in these
relic \ion{H}{2} regions are significantly affected by Pop III
feedback.  Hence it is important to consider primordial stellar
feedback while studying early galaxy formation.  In this section, we
compare our results to previous numerical simulations and
semi-analytic models of reionization and then discuss any potential
caveats of our methods and possible future directions of this line of
research.

%
%

\subsection{Comparison to Previous Models}

\subsubsection{Filtering Mass}
\label{sec:filtering}

One source of negative feedback is the suppression of gas accretion
into potential wells when the IGM is preheated.  The lower limit of
the mass of a star forming halo is the Jeans filtering mass
\begin{equation}
  \label{eqn:filtering}
  M_F^{2/3}(a) = \frac{3}{a} \int_0^a \> da^\prime
  M_J^{2/3}(a^\prime) \left[ 1 - \left(\frac{a^\prime}{a}\right)
  \right],
\end{equation}
where $a$ and $M_J$ are the scale factor and time dependent Jeans mass
in the \ion{H}{2} region \citep{Gnedin98, Gnedin00b}.  Additionally,
the virial shocks are weakened if the accreting gas is preheated and
will reduce the collisional ionization in halos with $\tvir \gsim
10^4$ K.  To illustrate the effect of Jeans smoothing, we take the
large \ion{H}{2} region of SimB-SN because it has the largest ionized
filling fraction, which is constantly being heated after $z = 21$.
Temperatures in this region fluctuates between 1,000~K and 30,000~K,
depending on the proximity of the currently living stars.  In Figure
\ref{fig:filtering}, we show the volume-averaged temperature and the
resulting filtering mass of regions with an ionization fraction
greater than $10^{-3}$ along with the total mass of star forming
halos.

\citet{Gnedin00b} found the minimum mass of a star forming halo is
better described by $M_F$ instead of $M_J$.  Our simulations are in
excellent agreement for halos that are experiencing star formation
after reincorporation of their previously expelled gas.  The filtering
mass is the appropriate choice for a minimum mass in this case as the
halo forms from preheated gas.  However for halos that have already
assembled before they become embedded in a relic \ion{H}{2} region,
the appropriate minimum mass $M_{\rm{min}}$ is one that is regulated
by the LW background \citep{Machacek01, Wise05}.  This is evident in
the multitude of star forming halos below $M_F$.  With the exception
of star formation induced by SN blast waves or I-fronts, this verifies
the justification of using $M_{\rm{min}}$ and $M_F$ for Pop III and
galaxy formation, respectively, as a criterion for star forming halos
in semi-analytic models.


\subsubsection{Star Formation Efficiency}
\label{sec:SFeff}

Semi-analytic models rely on a star formation efficiency $f_\star$,
which is the fraction of collapsed gas that forms stars, to calculate
quantities such as emissivities, chemical enrichment, and IGM
temperatures.  Low-mass halos that form a central star have $f_\star
\sim 10^{-3}$ whose value originates from a single 100~\Ms~star
forming in a dark matter halo of mass 10$^6$~\Ms~\citep{Abel02a,
  Bromm02a, Yoshida06b}.  Pop II star forming halos are usually
calibrated with star formation efficiencies from local dwarf and
high-redshift starburst galaxies and are usually on the order of a few
percent \citep[e.g.][]{Taylor99, Gnedin00a}.

This leads to the question: how efficient is star formation in these
high-redshift halos while explicitly considering feedback?  This is
especially important when halos start to form multiple massive stars
and when metallicities are not sufficient to induce Pop II star
formation.  The critical metallicity for a transition to Pop II is
still unclear.  Recently, \citet{Jappsen07} showed that metal line
cooling is dynamically unimportant until metallicities of $10^{-2} \>
Z_\odot$.  On the other hand, dust that is produced in SNe can
generate efficient cooling down in gas with $10^{-6} \> Z_\odot$
\citep{Schneider06}.  If the progenitors of the more massive halos did
not result in a pair-instability SN, massive star formation can
continue until it becomes sufficiently enriched.  Hence our
simulations can probe the efficiency of this scenario of massive
metal-free star formation.

We calculate $f_\star$ with the ratio of the sum of the stellar masses
to the total gas mass of unique star-forming halos.  For example at
the final redshift of 15.9 in SimA-RT, the most massive halo and its
progenitors had hosted 14 stars and the gas mass of this halo is $2.6
\times 10^6 \Ms$, which results in $f_\star = 5.4 \times 10^{-4}$ for
this particular halo.  Expanding this quantity to all star forming
halos, $f_\star/10^{-4} = 5.0, 4.1, 6.3$ for SimA-RT, SimB-RT, and
SimB-SN, respectively.  Our efficiencies are smaller than the isolated
Pop III case because halos cannot form any stars after the first star
expels the gas, and 30 -- 60 million years must pass until stars can
form again when the gas is reincorporated into the halo.  $f_\star$ is
larger in the SN case because more stars form in lower mass halos.
There are two reasons for this occurring.  First, the positive
feedback happens when SN blast waves induces star formation in
low-mass satellite halos.  Second, the local LW radiation field
originating from the most massive halos has been reduced since star
formation is suppressed from the SNe kinetic energy feedback.

A better comparison can be made by using the same techniques used in
semi-analytic models.  Most models calculate the SFR by taking the
product of $f_\star$ and the collapsed mass fraction $\psi$ in halos
above some minimum mass.  We can calculate $f_\star$ by taking the
same approach but the inverse: determine $f_\star$ from the SFRs in
our calculations and by assuming a critical star-forming halo mass.
We take the cumulative SFR $\rho_\star^{\rm{tot}}$
(Fig. \ref{fig:cumulSF}) to calculate
\begin{equation}
  \label{eqn:fstar}
  f_\star(z) = \frac{\rho_\star^{\rm{tot}}(z)}{\psi(z, M_{\rm{min}}) \> \rho_b},
\end{equation}
where $\rho_b$ is the comoving mean gas density.  As a simple example
of a fixed minimum mass of a star forming halo, the collapsed mass
fraction $\psi = 0.07$ in 10$^6$ \Ms~halos at redshift 15, using WMAP
first year parameters and Press-Schetcher formalism \citep{Press74,
  Sheth02}.  This results in $f_\star = 1.5 \times 10^{-3}$ and $2.6
\times 10^{-3}$ for the RT and SN calculations, respectively.  The
difference between the two fractions merely comes from the different
choices of primordial stellar masses.  The efficiencies are lower than
the explicitly calculated values because of our fiducial choice of the
minimum mass of a star forming halo, which should be $\sim$$3 \times
10^5 \Ms$ when the UV background is insignificant and increasing with
redshift by an order of magnitude \citep{Machacek01, Wise05}.  It
should be noted that this approach is sensitive to the choice of
minimum mass, arising from the exponential nature of the collapsed
mass fraction of high-$\sigma$ peaks in Press-Schetcher formalism.


By regarding the feedback created by Pop III stars and associated
complexities during the assembly of these halos, the $f_\star$ values
of $\sim$$5 \times 10^{-4}$ that are explicitly determined from our
radiation hydrodynamical simulations provide a more accurate estimate
on the star formation efficiencies.

\subsubsection{Intermittent \& Anisotropic Sources}

Our treatment of star formation and feedback produces intermittent
star formation, especially in low-mass halos.  If one does not account
for this, star formation rates might be overestimated in this phase of
star formation.  Kinetic energy feedback is the main cause of this
behavior.  As discussed in sections \ref{sec:haloMass} and
\ref{sec:kinetic}, shock waves created by D-type I-fronts and SN
explosions expel most of the gas in halos with masses $\lsim 10^7$
\Ms.  A period of quiescence follows these instances of star
formation.  Then stars are able to form after enough material has
accreted back into the halo.  Only when the halo becomes massive
enough to retain most of the outflows and cool efficiently through
\lya~and \hh~radiative processes does star formation becomes more
regular with successive stars forming.

%
%
\begin{figure}[t]
  \begin{center}
    \includegraphics[width=\textwidth]{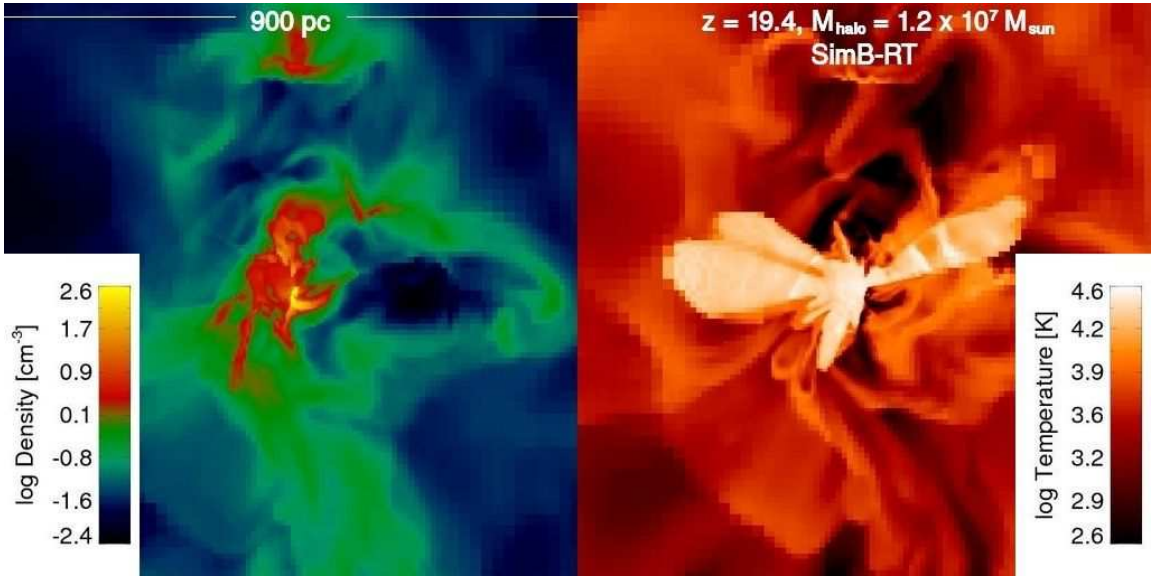}
  \caption[Anisotropic \ion{H}{2} region]{\label{fig:aniso} Density
    (\textit{left}) and temperature (\textit{right}) slices of an
    anisotropic \ion{H}{2} region in the most massive halo of SimB-RT.
    The star has lived for 2.5 Myr out of its 2.7 Myr lifetime.  The
    field of view is 900 proper parsecs.}
\end{center}
\end{figure}

The central gas structures in the host halo are usually anisotropic as
it is acquiring material through smooth IGM accretion and mergers.  At
scales smaller than 10 pc, the most optically thick regions produce
shadows where the gas radially behind the dense clump is not
photo-ionized or photo-heated by the source.  This produces cometary
and so-called elephant trunk structures that are also seen in local
star forming regions and have been discussed in detail since
\citet{Pottasch58}.  At a larger distance, the surrounding cosmic
structure is composed of intersecting or adjacent filaments and
satellite halos that break spherical symmetry.  The filaments and
nearby halos are optically thick and remain cool and thus the density
structures are largely unchanged.  The entropy of dense regions is
not increased by stellar radiation and will feel little negative
feedback from an entropy floor that only exists in the ionized IGM
\citep[cf.][]{Oh03}.  Ray-tracing allows for accurate tracking of
I-fronts in this inhomogeneous medium.  Radiation propagates through
the least optically thick path and generates champagne flows that have
been studied extensively in the context of present day star formation
\citep[e.g.][]{Franco90, Churchwell02, Shu02, Arthur06}.  In the
context of massive primordial stars, these champagne flows spread into
the voids and are impeded by the inflowing filaments.  The resulting
\ion{H}{2} regions have ``butterfly'' morphologies \citep{Abel99,
  Abel07, Alvarez06a, Mellema06, Yoshida06a}.  We also point out that
sources embedded in relic \ion{H}{2} largely maintain or increase the
ionization fraction, depending on the overdensities in the nearby IGM.
Here the already low optical depth of the recently ionized medium
(within a recombination time) allows the radiation to travel to
greater distances than a halo embedded in a completely neutral IGM.
The \ion{H}{2} regions become increasingly anisotropic in higher mass
halos.  We show an example of an \ion{H}{2} region near the end of the
star's lifetime in a dark matter halo with mass $1.2 \times 10^7 \Ms$
in Figure \ref{fig:aniso}.

\subsection{Potential Caveats and Future Directions}

Although we have simulated the first generations of stars with
radiation hydrodynamic simulations, our methods have neglected some
potentially important processes and made an assumption about the Pop
III stellar masses.

One clear shortcoming of our simulations is the small volume and
limited statistics of the objects studied here.  However, it was our
intention to focus on the effects of Pop III star formation on
cosmological reionization and on the formation of an early dwarf
galaxy instead of global statistics.  We have verified even in a
3-$\sigma$ peak that Pop III stars cannot fully reionize the universe,
which verified previous conclusions that low-luminosity galaxies
provide the majority of ionizing photons.  Furthermore, it is
beneficial to study Pop III stellar feedback because it regulates the
nature of star formation in these galaxies that form from pre-heated
material.  Further radiation hydrodynamics simulations of primordial
star and galaxy formation with larger volumes while still resolving
the first star forming halos of mass $\sim$$3 \times 10^5 \Ms$ will
improve the statistics of early star formation, especially in more
typical overdensities, i.e. 1-$\sigma$ peaks that may be survive to
become dwarf spheroidal galaxies at $z = 0$.

In this work, we treated the LW radiation field as optically thin, but
in reality, \hh~produces a non-zero optical depth above column
densities of $10^{14} \> \mathrm{cm}^{-2}$ \citep{Draine96}.
Conversely, Doppler shifts of the LW lines arising from large velocity
anisotropies and gradient may render \hh~self-shielding unimportant up
to column densities of $10^{20} - 10^{21} \> \mathrm{cm}^{-2}$
\citep{Glover01}.  If self-shielding is important, it will lead to
increased star formation in low-mass halos even when a nearby source
is shining.  Moreover, \hh~production can also be catalyzed ahead of
I-fronts \citep{Ricotti01, Ahn07}.  In these halos, LW radiation will
be absorbed before it can dissociate the central \hh~core.  On the
same topic, we neglect any type of soft UV or LW background that is
created by sources that are cosmologically nearby ($\Delta z / z \sim
0.1$).  A soft UV background either creates positive or negative
feedback, depending on its strength \citep{Mesigner06}, and a LW
background increases the minimum halo mass of a star-forming halo
\citep{Machacek01, Yoshida03, Wise07b}.  However in our calculations,
the lack of self-shielding, which suppresses star formation in
low-mass halos, and the neglect of a LW background, which allows star
formation in these halos, virtually counterbalance each other.  Hence
we do not expect any significant deviations in the SFRs and
reionization history if we treat these processes explicitly.

To address the incident radiation and the resulting UV background from
more rare density fluctuations outside of our simulation volume, it
will be useful to bridge the gap between the start of reionization on
Mpc scales to larger scale (10 -- 100 Mpc) simulations of
reionization, such as the work of \citet{Sokasian03}, \citet{Iliev06},
\citet{Zahn07}, and \citet{Kohler07}.  Radiation characteristics from
a volume that has similar overdensities as our Mpc-scale simulations
can be sampled from such larger volumes to create a radiation
background that inflicts the structures in our Mpc scale simulations.
Inversely, perhaps the small-scale evolution of the clumping factor,
filtering mass, and average temperature and ionization states can be
used to create an accurate subgrid model in large volume reionization
simulations.

Another potential caveat is the continued use of primordial gas
chemistry in metal enriched regions in the SN runs.  As discussed in
\S\ref{sec:SFeff}, metal line cooling may be dynamically unimportant
until metallicities of $10^{-2} Z_\odot$, but dust cooling could be
important as low as $10^{-6} Z_\odot$.  Our simulations with SNe give
excellent initial conditions to self-consistently treat the transition
to low-mass star formation.  In future work, we plan to introduce
metal-line and dust cooling models \citep[e.g. from][] {Glover07} to
study this transition.

The one main assumption about Pop III stars in our calculations is the
fixed, user-defined stellar mass.  The initial mass function (IMF) of
these stars is largely unknown, therefore we did not want to introduce
an uncertainty by choosing a fiducial IMF.  It is possible to
calculate a rough estimate of the stellar mass by comparing the
accretion rates and Kelvin-Helmholtz time of the contracting molecular
cloud \citep{Abel02a, OShea05}.  Protostellar models of primordial
stars have also shown that the zero-age main sequence (ZAMS) is
reached at 100 \Ms~for typical accretion histories after the star
halts its adiabatic contraction \citep{Omukai03, Yoshida06b}.  Based
on accretion histories of star forming halos, one can estimate the
ZAMS stellar mass for each halo and create a more self-consistent and
ab initio treatment of Pop III star formation and feedback.

\section{Summary}
\label{sec:summary6}

We conducted three radiation hydrodynamical, adaptive mesh refinement
simulations that supplement our previous cosmological simulations that
focused on the hydrodynamics and cooling during early galaxy
formation.  These new simulations concentrated on the formation and
feedback of massive, metal-free stars.  We used adaptive ray tracing
to accurate track the resulting \ion{H}{2} regions and followed the
evolution of the photo-ionized and photo-heated IGM.  We also explored
the details of early star formation in these simulations.  Theories of
early galaxy formation and reionization and large scale reionization
simulations can benefit from the useful quantities and characteristics
of the high redshift universe, such as SFR and IGM temperatures and
ionization states, calculated in our simulations.  The key results
from this work are listed below.

\medskip

1. SFRs increase from $4 \times 10^{-4}$ at redshift 30 to $6 \times
10^{-3}$ \sfr~at redshift 20.  Afterwards the SFR begins to have a
bursting nature in halos more massive than $10^7 \Ms$ and fluctuates
around $10^{-2}$ \sfr.  These rates are larger than the ones
calculated in \citet{Hernquist03} because our simulation volume
samples a highly biased region that contains a 3-$\sigma$ density
fluctuation.  The associated emissivity from these stars increase from
1 to $\sim$100 ionizing photons per baryon per Hubble time between
redshifts 15 and 30.

2. In order to provide a comparison to semi-analytic models, we
calculate the star formation efficiency to be $\sim$$5 \times 10^{-4}$
averaged over all redshifts and the simulation volume.  For Pop III
star formation, this is a factor of two lower than stars that are not
affected by feedback \citep{Abel02a, Bromm02a, Yoshida06b, OShea07}.

3. Shock waves created by D-type I-fronts expel most of the gas in the
host halos below $\sim$$5 \times 10^6 \Ms$.  Above this mass,
significant outflows that are still bound to the halo are generated.
This feedback creates a dynamical picture of early structure
formation, where star formation is suppressed in halos because of this
baryon depletion, which is more effective than UV heating or the
radiative dissociation of \hh.

4. We see two instances of induced star formation in low-mass
($\sim$$5 \times 10^5 \Ms$) satellite halos that recently had a SN
blastwave overtake it.  Central temperatures in these halos are half
the central temperatures of star forming halos with similar masses.

5. Although the volume-averaged emissivities are high, only 1 in 50 UV
photons result in a sustainable ionization in the RT simulations.
Interestingly, the SN calculations contain a sharp transition where
the volume filling fraction increases by a factor of three in a mere
25 million years.  Multiple star formation in a single halo and their
SNe increase the photon escape fraction to nearly unity after the
first star dies, which leads to the dramatic increase in ionized IGM.
In this simulation, the effective number of ionizations per UV photon
is 1 in 10, which is still small but significantly higher than its
star formation only counterpart.

6. Our simulations that include star formation and \hh~formation
capture the entire evolution of the clumping factor that is used in
semi-analytic models to calculate the effective enhancement of
recombinations in the IGM.  We showed that clumping factors in the
ionized medium fluctuate around the values found in adiabatic
simulations.  They evolve from unity at high redshifts and steadily
increase to $\sim$100 at redshift 15.  Dynamical stellar feedback
causes the fluctuations of the clumping factor.

7. We calculated the Jeans filtering mass with the volume-averaged
temperature only in fully and partially ionized regions, which yields
a better estimate than the temperature averaged over both ionized and
neutral regions.  The filtering mass depends on the thermal history of
the IGM, which mainly cools through Compton cooling.  It increases by
two orders of magnitude to $\sim$$3 \times 10^7 \Ms$ at $z \sim 15$.
It describes the minimum mass a halo requires to collapse after
hosting a Pop III star.  For halos forming their first star, the
minimum mass found in \citet{Machacek01} is appropriate, which is
regulated by a LW background.

\medskip

Pop III stellar feedback plays a key role in early star formation and
the beginning of cosmological reionization.  The shallow potential
wells of their host halos only amplify their radiative feedback.  Our
understanding of the formation of the oldest galaxies and the
characteristics of isolated dwarf galaxies may benefit from including
the earliest stars and their feedback in galaxy formation models.
Although these massive stars only partially reionized the universe,
their feedback on the IGM and galaxies is crucial to include since it
affects the characteristics of low-mass galaxies that are primarily
responsible for cosmological reionization.  Harnessing observational
clues about reionization, observations of local dwarf spheroidal
galaxies, and numerical simulations that accurately handle star
formation and feedback may provide great insight on the formation of
the first galaxies, their properties, and how they completed
cosmological reionization.

\chapter{Conclusions and Future Work}
\label{chap:conclusion}

The first stars in the universe form in isolation in their parent dark
matter halos with masses $\sim$$10^6 \Ms$, starting as early as 100
million years after the Big Bang \citep{Tegmark97, Abel02a,
  Yoshida06b}.  Because big bang nucleosynthesis does not create
anything heavier than lithium, the first stars are mainly composed of
hydrogen and helium.  The most efficient coolant in this pristine gas
is \hh, forming in the gas-phase.  Because of the differences in
radiative cooling of metal-free gas, the first stars are approximately
100 times more massive than the Sun \citep{Abel02a} and have a large
impact on their surroundings and future structure formation.

The purpose of this thesis has been to assess this impact on early
galaxies.  We have methodically approached this problem by introducing
each piece of the puzzle progressively.  In general, this allows one
to judge the relevance of each physical process; in our case, we have
learnt about the importance of turbulence, radiative feedback,
\hh~formation, and Pop III stars in the early universe and how they
affect the first galaxies.  Low-luminosity galaxies have been known to
be responsible to reionize the universe \citep[e.g.][]{Shapiro86}.
Thus their properties and formation have been of great interest for
the past few decades.  Only in the last decade has it been possible to
study these galaxies in numerical simulations with adequate
resolution.  The cosmological simulations of \citet{Abel00, Abel02a}
first resolved the formation of the first protostar with AMR
calculations that covered 10$^{12}$ orders of magnitude in length
scale.  Star formation and feedback in galaxies before reionization
($z > 6$) have been explored by \citet{Gnedin97}, \citet{Ricotti02b},
and many others.  Furthermore, some have argued that \hh~cooling is
unimportant \citep{Oh02, Ciardi05, Haiman06} in early structure
formation as it can be dissociated from large distances and have thus
ignored the effects of primordial stellar feedback.  This thesis
attempts to bridge the gap in between these calculations of the first
stars and early galaxies and how the isolated mode of primordial star
formation transitions to dwarf galaxy formation.  Additionally, we
have explored the validity of neglecting \hh~formation and the ensuing
scenario of galaxy formation without \hh~chemistry, which has been
frequently used in galaxy formation models in the early universe.

\section{Supernovae from Primordial Stars}

We developed a semi-analytical method that calculates the abundances
of primordial stars and the observed SNe rates.  This method uses
results from AMR simulations that determines the minimum halo mass of
an object that can host a cool, dense core, which then forms a Pop III
star \citep{Machacek01}.  We find that star formation in early
galaxies contribute the most to cosmological reionization, which is in
agreement with other studies \citep{Cen03, Somerville03, Ciardi03}.
The main conclusions of this study are:
\begin{itemize}
\item Radiation from star formation in protogalaxies suppresses
  primordial star formation.  We expect $\sim$0.34 primordial SNe per
  square degree per year.  These rates can vary from 0.1 to $>$1.5
  deg$^{-2}$ yr$^{-1}$ depending on the choice of primordial stellar
  mass and protogalaxy parameters while still constrained by WMAP
  results.  The peak of SNe rates occurs earlier with increasing
  primordial stellar masses.  These results are upper limits since the
  rate of visible primordial SNe depends on the IMF because only a
  fraction will lie in the pair instability SNe mass range.  The other
  massive primordial stars might result in jet-driven SNe or long
  duration GRBs.
\item Stellar metal abundances and star formation rates in local dwarf
  galaxies aid in estimating the protogalaxy properties.  Our best
  models find that 4\% of the gas form stars and 5\% of their
  radiation escapes into and reionizes the IGM.
\item Primordial stars enrich the IGM to a maximum volume-averaged
  [Fe/H] $\simeq$ --4.1 if the IMF is skewed toward the
  pair-instability mass range.  A proper IMF will lessen this
  metallicity since only a fraction of stars will exist in this mass
  range.
\item The entire error bar in the first year WMAP measurement of
  the optical depth to electron scattering can be explained by a
  higher/lower star formation efficiency and photon escape fraction.
  Massive primordial stars provides $\sim$10\% of the necessary
  ionizing photons to achieve cosmological reionization.  No exotic
  processes or objects are necessary.
\item Only the upper mass range of pair-instability SNe will be
  observable by JWST since the low mass counterparts do not produce
  enough $^{56}$Ni to be very luminous.
\end{itemize}

\section{Galaxy Formation without Molecular Hydrogen}

Our first cosmological simulations in this thesis involved the
scenario of galaxy formation without \hh~cooling.  The first objects
to form have a virial temperature of 10$^4$ K, which translates to
$10^8 \Ms$ at redshift 10 \citep{Rees77}.  Previous works concluded
that the angular momentum acquired from tidal torques induces disc
formation in the initial collapse of the gas cloud \citep{Fall80,
  Mo98, Begelman06}.  We investigate this scenario with AMR
simulations that first focus on the virialization of such objects and
then the central gaseous collapse.  Our main results are as follows:
\begin{itemize}
\item Baryons cannot virialize through heating alone but must gain
  kinetic energy that manifests in a faster bulk inflow and, in the
  case of efficient gas cooling, supersonic turbulent motions up to
  Mach numbers of three.  We expect turbulence in larger galaxies, up
  to $10^{12} \Ms$, to be even more supersonic.
\item Baryonic velocity distributions are Maxwellian, which shows
  violent relaxation occurs for gas as well as dark matter.  Turbulent
  velocities exceed typical rotational speeds, and these halos are
  only poorly modeled as solid body rotators.
\item Turbulence generated during virialization mixes angular momentum
  so that it redistributes to a radially increasing function.  This
  allows the gas with the lowest specific angular momentum to infall
  and form a $\sim$$10^5 \Ms$ central dense object within the central
  five parsecs.  It forms before any multiple fragmentation occurs in
  the global disc.
\item The final characteristics of the central object depend on the
  merger history of the parent halo because these mergers influence
  the amount of turbulence and morphologies in the halo.
\item Our simulations of the central collapse highlight the relevance
  of secular bar-like instabilities in galaxy formation, as well as
  star formation.  Similar bar structures are witnessed in primordial
  star formation simulations.  These structures transport angular
  momentum outwards, so the collapse can continue.  Smaller scale bars
  form within the larger bars, which shows the ``bars within bars''
  scenario is also applicable to a gaseous collapse.  We follow the
  collapse to length scales smaller than a solar radius and show that
  it follows a self-similar collapse with $\rho \propto r^{-12/5}$.
\end{itemize}

However, we do not advocate that these objects form in this way
because we have neglected \hh~cooling and primordial stellar
feedback.  We focused next on the key role of \hh~cooling in early
galaxy formation.

\section{Importance of Molecular Hydrogen}

We conducted a suite of fourteen simulations that focused on the
importance of \hh~cooling with various degrees of negative feedback.
Our work is an extension of \citet{Machacek01}, who showed that a LW
radiation background only increases the minimum mass a halo must
obtain before forming stars but never fully suppressing primordial
star formation.  We made the following conclusions in these
calculations.
\begin{itemize}
\item Our simulations with an UVB of \flw~= (0, $10^{-22}$,
  $10^{-21}$) agree with previous results of \citet{Machacek01}. Above
  \flw~= $10^{-21}$, it had been argued that a LW background would
  inhibit any \hh~formation until the halo could cool through atomic
  transitions.  We showed that central shocks provide sufficient free
  electrons from collisional ionization \citep{Shapiro87} to drive
  \hh~formation faster than dissociation rates even in a \flw~=
  $10^{-20}$ background.
\item We investigated at what halo mass collisional ionization becomes
  conducive for \hh~formation.  This occurs at a virial temperature
  $\sim$4000 K.  Here we remove all residual free electrons from
  recombination, in which \hh~formation can only occur once
  collisional ionizations induce \hh~cooling.
\item Recent major mergers above this mass scale create complex
  cooling structures, unlike the non-fragmented central cores in
  smaller halos.
\item Even our most extreme assumptions of a large radiation
  background (\flw~= $10^{-20}$) and no residual free electrons cannot
  defeat the importance of \hh~cooling in the early universe.
\end{itemize}

Molecular hydrogen cooling triggers collapses in halos with virial
temperatures well below $10^4$ K.  This increases the mass fraction
contained in these halos by three times at redshift 20 and the
abundance of high-redshift star forming halos by an order of
magnitude!  This chapter has strengthened the results of
\citet{Machacek01} that \hh~cooling plays a key role in high-redshift
structure formation.

\section{Primordial Stellar Feedback and the First Galaxies}

The last iteration of the simulations presented in this thesis include
accurate radiative transfer, using adaptive ray tracing
\citep{Abel02b}.  We include self-consistent star formation and
feedback in our simulations and use the same initial conditions as
before but only including these additional processes.  We also track
the SNe ejecta in one simulation and the propagation of the first
metals in the universe.  We find that primordial stars have a large
impact on future galaxy formation and conclude with the following key
points:
\begin{itemize}
\item Dynamical feedback from Pop III stars expel nearly all of the
  baryons from low-mass host halos.  The baryon fractions in star
  forming halos never fully recover even when it reaches a virial
  temperature of $10^4$ K.  The baryon fraction is reduced as low as
  $\sim$0.05 with SNe feedback, three times lower than the cases
  without stellar feedback.
\item The expelled gas gains angular momentum by tidal torques as it
  exists at large radii.  When it is reincorporated into the halo, it
  increases the spin parameter by a factor of 2--5 without SNe and up
  to 10 with SNe.
\item Radiative and SNe feedback produces a complex, multi-phase
  interstellar medium that resembles the conditions in larger
  galaxies.  It is clear that the first and earliest galaxies are
  complex entities, contrary to their generally assumed simplicity due
  to their low masses.
\item Pair-instability SNe preferentially enrich the IGM to a
  metallicity an order of magnitude higher than the surrounding halos
  and filaments.
\item Once a SN explosion cannot totally disrupt its host halo, the
  mean metallicity fluctuates around a mean value as there may be a
  balance between SN outflows, cold inflows, and contained SN ejecta.
\end{itemize}

\section{Starting Cosmological Reionization}

We then addressed the impact of Pop III stellar radiation on
cosmological reionization.  Previous studies have shown that
low-luminosity galaxies are mainly responsible for this, but the first
stars must have at least started reionization.  Our main conclusions
about reionization by Pop III stars are listed below.
\begin{itemize}
\item The comoving star formation rates (SFR) increase from $4 \times
  10^{-4}$ at redshift 30 to $6 \times 10^{-3}$ \sfr~at redshift 20.
  When halos become more massive than $10^7 \Ms$, stars form in a more
  bursting nature, where the SFR fluctuates around $10^{-2}$ \sfr.
  These rates are larger than \citet{Hernquist03} because our
  simulation volume samples a highly biased region that contains a
  3$\sigma$ density fluctuation.
\item In order to provide a comparison to semi-analytic models, we
  calculate the star formation efficiency to be $\sim$$5 \times
  10^{-4}$ averaged over all redshifts and the simulation volume.  For
  Pop III star formation, this is a factor of two lower than stars
  that are not affected by feedback.
\item We see two instances of induced star formation in low-mass
  ($\sim$$5 \times 10^5 \Ms$) satellite halos that recently had a SN
  blastwave overtake it.  Central temperatures in these halos are half
  the central temperatures of star forming halos with similar masses.
\item Only 1 in 50 UV photons result in a sustainable ionization in
  our simulations without SNe.  This ratio interestingly increases to
  1 in 10 when we consider SNe.  Here multiple star formation and
  their SNe increase the photon escape fraction to nearly unity after
  the first star dies, which allows later stars to more effectively
  ionize the IGM.
\item We showed that the Jeans filtering mass \citep{Gnedin98} is in
  excellent agreement with the star forming halos that are forming
  from reincorporated gas that was previously expelled by stellar and
  SNe feedback.  This increases by two orders of magnitude to $\sim$$3
  \times 10^7 \Ms$ at redshift 15.  For halos forming their first
  star, the minimum mass found in \citet{Machacek01} is appropriate.
\end{itemize}

\section{Future Work}

We have included an accurate treatment of the formation and feedback
of the first stars in the universe; however, there are still several
important processes that we have neglected and will include in future
studies.  In this thesis, we have treated the \hh~dissociating
radiation as optically thin, but in reality, \hh~produces a non-zero
optical depth above column densities above $10^{14}$ cm$^{-2}$
\citep{Draine96}.  Currently it is computationally prohibitive to
ray-trace optically thin radiation, even with adaptive techniques, in
cosmological simulations.  This will undoubtedly improve as
computational power relentlessly increases.  However, this does not
preclude us from devising an ingenious scheme to treat small- and
large-scale radiation scales accurately and efficiently.  This will
allow the study of many radiation sources in a cosmological or
galactic setting.  Perhaps this new method will be abstractly
comparable to the adaptive P$^3$M algorithm that efficiently computes
the gravitational field at all length scales.

Besides considering larger volumes with the same resolution, it will
be useful to connect our simulations on Mpc-scales to larger volume
simulations (10 -- 100 Mpc) that focus on cosmological reionization.
The properties of the radiation field in large-scale calculations can
be sampled to a radiation background in our simulations.  One must
ensure that the sampled region in the large-scale calculations has
similar overdensities as the Mpc-scale ones.  We can also regard the
inverse process of composing an accurate sub-grid model with the
clumping factor, filtering mass, and average temperatures from our
simulations to reionization simulations.

Another necessary improvement to our simulations is the inclusion of
metal-line and dust cooling.  We already have the appropriate ``initial
conditions'' for this as we track the heavy elements from SNe of
massive, metal-free stars.  Perhaps it will be possible to follow the
transition from a top-heavy IMF of Pop III to a more regular Salpeter
IMF, associated with metal-enriched Pop II stars.  This requires the
introduction of metal-enriched cooling models, such as the ones
described in \citet{Glover07} and \citet{Smith07}.

We note one last possible improvement in our simulations of removing
the assumption of a fixed Pop III stellar mass.  We can estimate the
stellar mass from the accretion rates and the Kelvin-Helmholtz time
\citep{Abel02a, OShea05}.  Alternatively, we could follow the collapse
to stellar scales and include protostellar models to calculate where
its radiation effectively suppresses further accretion, thus giving an
accurate estimate of the stellar mass.

\vspace{2em}

Even after centuries of development, there are still many unanswered
questions in galaxy formation.  In this thesis, we have demonstrated
the importance of \hh~cooling and primordial star formation and
feedback in early galaxy formation.  Models of the first galaxies are
necessary to interpret future observations of the earliest galaxies
that should be available within the next decade.  With the
observations in hand, both theorists and observers will benefit from
the wealth of knowledge contained in these data and their comparisons
to accurate models of early galaxy formation.


%

\bibliography{suthesis}
{}

\end{document}